\documentclass[12pt]{article}
\usepackage[utf8]{inputenc}
\usepackage[english]{babel}
\usepackage[centertags]{amsmath}
\usepackage{amssymb}
\usepackage{bm}
\usepackage{amsbsy}
\usepackage{graphicx}
\usepackage[labelformat = empty]{subcaption}
\usepackage{appendix}
\usepackage{mathrsfs}
\usepackage{epsfig}
\usepackage{color}
\usepackage{euscript}
\usepackage{ulem}
\usepackage{diagbox} 

\usepackage{multirow}
\usepackage{cite}

\newcommand{\GG}{GeV$^2$\,}

\definecolor{dgreen}{rgb}{0,0.6,0}

\definecolor{darkblue}{rgb}{0., 0, 1}

\definecolor{purple}{rgb}{0.65,0.,0.78}

\definecolor{orange}{rgb}{0.89,0.46,0.15}

\definecolor{darkyellow}{rgb}{0.7, 0.6, 0.0}

\usepackage{jheppubm}

\newcommand{\be}{\begin{equation}}
\newcommand{\ee}{\end{equation}}
\newcommand{\bea}{\begin{eqnarray}}
\newcommand{\eea}{\end{eqnarray}}

\newcommand{\fb}{\mathfrak{b}}

\newcommand{\fg}{\mathfrak{g}}

\newcommand{\cA}{{\cal A}}
\newcommand{\cL}{{\cal L}}

\numberwithin{equation}{section}

\title{Jet Quenching in Holographic QCD as an Indicator of  Phase Transitions in Anisotropic Regimes}

\author{Irina Ya. Aref'eva$^a$, Ali Hajilou$^a$, Alexander Nikolaev$^{a}$, Pavel Slepov$^a$}

\affiliation{$^a$Steklov Mathematical Institute, Russian Academy of  Sciences, \\ Gubkina str. 8, 119991, Moscow, Russia}

\emailAdd{arefeva@mi-ras.ru}
\emailAdd{ma.hajilou@gmail.com}
\emailAdd{alexn99@gmail.com}
\emailAdd{slepov@mi-ras.ru}

 \abstract{In this paper, we employ the gauge/gravity duality to study jet quenching (JQ) phenomena in  the quark-gluon plasma. For this purpose, we implement  holographic QCD models constructed from an Einstein-Maxwell-dilaton gravity at finite temperature and finite chemical potential for light and heavy quarks. The models capture both the confinement and deconfinement phases of QCD and the first-order phase transitions. We calculate  the JQ  parameter
in different  models and  compare them with the experimental data obtained in heavy-ions studies.
In particular, we investigate how JQ, as a function of temperature $T$, chemical potential $\mu$, and magnetic field $c_B$, serves as a probe for identifying first-order phase transitions within the $(T,\mu,c_B)$ parameter space of holographic QCD. Particular attention is paid to the dependence of JQ on the parameter $\nu$, which  characterizes longitudinal versus transverse anisotropy relative to the heavy-ion collision axis.
By analyzing the dependence of the JQ parameters on these thermodynamic variables, we map critical regions associated with phase boundaries. We compare our findings to earlier studies of the running coupling constant's behavior within the gauge-gravity duality framework. This approach provides new insights into the interplay between non-perturbative dynamics and phase structure in strongly coupled systems.}


\begin{document}

\maketitle

\newpage
\section{Introduction}

It is well known that collisions of high-energy particles produce jets of elementary particles. In particular, collisions of ultra-relativistic heavy-ion beams create a hot, dense medium comparable to conditions in the early universe. The resulting jets interact strongly with this medium, leading to a significant energy decrease known as JQ. JQ phenomenon is studied in high-energy heavy-ion collisions, particularly in the context of quark-gluon plasma (QGP) formation. It refers to the energy loss of high-energy partons (quarks and gluons) as they traverse the QGP, a hot and dense medium created in such collisions. The JQ parameter,  denoted as \(\hat{q}\), quantifies the average transverse momentum squared transferred from the parton to the medium per unit path length. It is a key observable for understanding the properties of the QGP.
\\

The concept of the JQ parameter emerged from studies of parton energy loss in dense media by Baier, Dokshitzer, Mueller, Peign\'e, Schiff, and Zakharov \cite{Baier:1996kr,Baier:2000mf} (see also \cite{Wiedemann:2000za,Kovner:2003zj}). Quantitatively, the JQ parameter is characterized by the parameter $\hat{q}$, defined as the average squared transverse momentum transfer per unit path length,
$
\hat{q} = \langle p_T^2 \rangle / L \, $.
This parameter measures how energetic partons lose energy via medium-induced gluon radiation. Crucially, $\hat{q}$ links microscopic parton-medium interactions to observable jet suppression. For a quark with energy $E$, the average energy loss scales as
$
\langle \Delta E \rangle \propto \alpha_s  \hat{q}  L^2 $,
where $L$ is the medium length. Experimental results from RHIC and LHC show significant jet suppression through the nuclear modification factor $R_{AA} < 1$ and dijet asymmetry, yielding key values:
$\hat{q} \approx 1.2 \pm 0.3 \,\text{GeV}^2/\text{fm}$ for $\sqrt{s_{NN}} = 200$ GeV Au+Au collisions (RHIC);
$\hat{q} \approx 1.5 \pm 0.4  \text{GeV}^2/\text{fm}$ for $\sqrt{s_{NN}} = 2.76$ TeV Pb+Pb collisions (LHC) \cite{Eskola:2004cr,Dainese:2004te,AbdulKhalek:2022hcn}.
\\

A non-perturbative calculation scheme for $\hat{q}$ using gauge-gravity duality was proposed by Liu, Rajagopal, and Wiedemann \cite{Liu:2006ug}. They related $\hat{q}$ to a lightlike Wilson loop in the adjoint representation (denoted $W^A$) of $\mathcal{N}=4$ SYM theory:
\begin{equation}\label{LLWL}
\langle W^A(\mathcal{C}) \rangle \sim e^{-\frac{1}{4\sqrt{2}} \hat{q}  L^{-} L_{\perp}^2} \, ,
\end{equation}
where $\mathcal{C}$ is a rectangular loop with light-cone ($L^{-}$) and transverse ($L_{\perp}$) extensions. Subsequent dual calculations using the Nambu-Goto action in AdS-Schwarzschild spacetime established $\hat{q}$'s explicit dependence on the 't Hooft coupling $\lambda$ and horizon position $z_h$ (for general holographic applications to HIC; see \cite{Casalderrey-Solana:2011dxg,Arefeva:2014kyw,deWolf}). Later work incorporated chemical potential $\mu$, magnetic fields $B$, and other parameters \cite{Argyres:2006yz,Gubser:2008as,Caceres:2012ii,1203.0561,1205.4684,Muller:2012uu,Betz:2012hv,Giataganas:2012zy,Yee:2012yc,Arnold:2012qg,JET:2013cls,
1308.5991, Casalderrey-Solana:2016iee,
Pasechnik:2016wkt,Ageev:2016gtl,Giataganas:2017koz,BitaghsirFadafan:2017tci,Zhu:2019ujc,Rougemont:2020had,Zhu:2020qyw,Banerjee:2021sjm,Grefa:2022sav,Qin:2015srf,Li:2016bbh,Ageev:2017qpa, Arefeva:2020bjk, Zhu:2023aaq,Chen:2022goa,Zhou:2022izh,Heshmatian:2023yzz,Arefeva:2024fpv,Xie:2024xbn,Cao:2024jgt}, revealing $\hat{q}$'s thermodynamic dependence in holographic QCD. Nonperturbative calculations have also been performed on the lattice \cite{Panero:2013pla,Panero:2014sua}.
\\

In this paper, we study $\hat{q}$'s dependence on temperature, chemical potential, external magnetic field, and anisotropy parameter $\nu$ across multiple holographic models---including those sensitive to quark masses. Special attention is paid to $\hat{q}$'s behavior near phase transitions: it varies smoothly across second-order transitions but exhibits discontinuities at first-order transitions. We compare our results with those in \cite{Heshmatian:2023yzz,Zhu:2023aaq,Arefeva:2024fpv}.
\\

The paper is organized as follows. 
In Sect.\,\ref{sec:setup}, we first briefly remind the 5-dimensional  anisotropic holographic models in the presence of a non-zero magnetic field for light and heavy quarks, Sect.\, \ref{sec:anm}, and then in Sect.\ref{sect.jq} we present the analytical expressions for the JQ parameters for these models. In Sect.\ref{sect:LQ-NR} we present numerical results of studies of the JQ parameter for light quarks: in Sect.\ref{NR-LQ-zero-nu1-15-45} results for zero magnetic fields and $\nu=$1, 1.5, 3, 4.5, and in Sect.\ref{sect:NR-LQ-nzero} for nonzero magnetic fields. In Sect.\ref{sect:HQ-NR} we present numerical results of studies of the JQ parameter for heavy quarks: in Sect.\ref{sect:HQ-NR-zero} results for zero magnetic fields and $\nu=$1, 1.5, 4.5, and in Sect.\ref{sect:HQ-NR-nzero} for nonzero magnetic fields.
In Sect.\,\ref{Sect:Conc}, we summarize our numerical results obtained for the LQ (light quark) and HQ (heavy quark) models by discussing  their dependence of the anisotropic parameter $\nu$ and parameter $c_B$ specifying the magnetic field.
The paper is complemented by an Appendix \ref{app0}, which describes how the equations of motion (EOMs) were solved and which boundary conditions were applied. In Appendices \ref{app2}, and \ref{app3} some complimentary plots are presented for LQ and HQ models, respectively.

\section{Setup}\label{sec:setup}

\subsection{Anisotropic holographic models in external magnetic field}
\label{sec:anm}

In this paper we deal with the Lagrangian in the Einstein frame used in previous papers
\cite{Arefeva:2018hyo,Arefeva:2022avn,Arefeva:2023jjh}:
\be
  {\cL} = \sqrt{-\fg} \left[ R 
    - \cfrac{f_0(\phi)}{4} \, F_0^2 
    - \cfrac{f_1(\phi)}{4} \, F_1^2
    - \cfrac{f_3(\phi)}{4} \, F_3^2
    - \cfrac{1}{2} \, \partial_{\mu} \phi \, \partial^{\mu} \phi
    - V(\phi) \right], \label{eq:2.01}
\ee
where  $\fg$ is the determinant of the metric tensor, 
$R$ is Ricci scalar, $\phi=\phi(z)$ is the dilaton field, $f_0(\phi)$, $f_1(\phi)$ and
$f_3(\phi)$ are the coupling functions associated with stresses $F_0$,
$F_1$ and $F_3$  of Maxwell fields, $F_{\rho\sigma}=\partial_{\rho}A_{\sigma}-\partial_{\sigma}A_{\rho}$. The indexes  $\rho$ and $\sigma$ numerate the spacetime coordinates $(t, x_1, x_2, x_3, z)$, with $z$ being the holographic radial coordinate. $V(\phi)$ is the  potential of the dilaton field. In this paper, we considered $F_0$, $F_1$ and $F_3$ as the first, second, and third
Maxwell fields, respectively.

In these cases, the metric is 
\bea
\label{an-met}
  ds^2 &=& \cfrac{L^2\fb(z)}{z^2} \  \left[
    - \, g(z) \, dt^2 + dx_1^2 
    + \left( \cfrac{z}{L} \right)^{2-\frac{2}{\nu}}\left( dx_2^2
    + e^{c_B z^2}  dx_3^2\right)
    + \cfrac{dz^2}{g(z)} \right] \! , \\\label{fb}
  \fb(z) &=& e^{2{\cA}(z)},\quad  \cA_{s}(z)=\cA(z)+\sqrt{\frac{1}{6}} \phi(z),
\eea
and matter fields\footnote{Also, we can add a new Maxwell field
  $F_2$ with magnetic ansatz $F_2 = q_2 \, dx^1 \wedge dx^3$ to our
  model \cite{Arefeva:2024mtl}.} are 
\begin{gather}
  \phi = \phi(z), \quad \, \label{eq:2.02} \\
  \begin{split}
    F_0\, -\, \mbox{electric ansatz}, \,\, A_0 &= A_t(z), \quad 
    A_{i}=0,\,\,i= 1,2,3,4, \\
    F_k\,- \, \mbox{magnetic ansatz}, \quad
    F_1 &= q_1 \, dx^2 \wedge dx^3, \quad 
    F_3 = q_3 \, dx^1 \wedge dx^2\, . 
\end{split}\label{eq:2.03}
\end{gather}
In \eqref{an-met} $L$ is the length parameter, $\fb(z)$ is the warp factor
set by ${\cA}(z)$, $g(z)$ is the blackening function, $\nu$ is the
parameter of primary anisotropy caused by non-symmetry of heavy-ion
collision (HIC), and $c_B$ is the coefficient of secondary anisotropy
related to the magnetic field $F_3$. The Choice of ${\cA}(z)$ determines
the light/heavy quarks description of the model. In previous works we
considered  ${\cA}_{LQ}(z) = - \, \mathrm{a} \, \ln (
\mathrm{b}z^2 + 1)$
for light quarks \cite{Arefeva:2022avn,Li:2017tdz} with  $\mathrm{a}=4.046$, $\mathrm{b}= 0.01613$\,\GG.  In this paper, we also accept the same values for parameters $\mathrm{a}$ and $\mathrm{b}$ \footnote{The authors of \cite{Zhou:2022izh} considered the JQ parameter for the LQ model with slightly different parameters: 
$\mathrm{a}=3.943$ and $\mathrm{b}=0.0158$ \GG. 
The authors of \cite{Li:2025ugv} explored various LQ models defined by ${\cA}=-\mathrm{a}\ln (
\mathrm{b} z^2+1)-\mathrm{a} \ln (\mathrm{d} z^4+1)$ for different parameters and flavor numbers. Specifically, for $N_f=2+1$, they used $\mathrm{a}=0.173$, $\mathrm{b}=0.204$ GeV$^2$, and $\mathrm{d}=0.013$ GeV$^4$.}. ${\cA}_{HQ}(z) = - \, \mathrm{c} z^2/4$ for heavy quarks \cite{Andreev:2006ct,He:2010ye,Arefeva:2018hyo} and ${{\cA}_{HQ}(z)} = -\frac{\mathrm{s}}{3}z^2- (\mathrm{p}-c_B q_3)\, z^4,$ for  heavy quarks with magnetic catalysis \cite{Arefeva:2023jjh}.  
$\mathrm{s}$ and $\mathrm{p}$ are parameters that can be fitted to the experimental data such as $\mathrm{s}= 1.16\mbox\,{GeV}^2$ and $\mathrm{p} = 0.273\mbox\,{GeV}^4$. We take $f_1$ in the case of light quarks as  \cite{Arefeva:2022avn}
$
  f_1 = 
   (1 + \mathrm{b} z^2)^\mathrm{a} \, e^{- \mathrm{c} z^2} z^{-2+\frac{2}{\nu}}
$
where $\mathrm{a}=4.046$, $\mathrm{b}= 0.01613\mbox\,{GeV}^2$  and $\mathrm{c}= 0.227$\,\GG.
To respect the linear Regge trajectories, the gauge kinetic function for the HQ  model is chosen in the form \cite{Yang:2015aia}
$f_0(z)=e^{-\mathrm{s} \, z^2-
\cA(z)}$. In \eqref{eq:2.03} $q_1$ and $q_3$ are constant
``charges''\footnote{Roughly speaking, the value of the physical magnetic field scales as $B_{\text{phys}} \sim q_3 e^{-L^2 c_B}$; see the discussion in  \cite{0908.3875, 1505.07894,
Bohra:2019ebj,
 Arefeva:2023jjh, Arefeva:2024mtl}. }. See \cite{Chen:2024mmd} for a discussion of the modifications 
of the LQ model.

The explicit form of the EOM's with ansatz
(\ref{eq:2.02})--(\ref{eq:2.03}) is given in Appendix
(\ref{eq:2.16}--\ref{eq:2.22}). Investigation of their
self-consistency shows that there is one dependent equation in the
system and that all other equations are independent. Thus, the system
(\ref{eq:2.16}--\ref{eq:2.22}) is self-consistent and the dilaton
field equation (\ref{eq:2.16}) serves as a constraint.

\subsection{Jet Quenching}\label{sect.jq}
\subsubsection{Isotropic case}
The JQ parameter in holographic models
described by the isotropic metric
\bea
 ds_{s}^2= \frac{L^2 e^{2 \cA_{s}(z)}}{z^2}\biggl(-g(z)dt^2 + \frac{dz^2}{g(z)} + dx_{1}^2+dx_{2}^2+dx_{3}^2 \biggr)\, ,
\label{stringmetric}
\eea
is given by \cite{Liu:2006ug}; see also \cite{Ageev:2016gtl,Heshmatian:2023yzz,Zhu:2023aaq,Arefeva:2024fpv}
\bea\label{hatq}
\hat{q}&=&\frac{L^2}{\pi \alpha' a},
\label{eq:qfinal}
\eea
where  $a$ is defined by the integral 
\be\label{a-mu}
a=\int^{z_h}_{0} d z \frac{z^2 e^{-2\cA_s(z)}}{\sqrt{g(z) \left(1-g(z)\right)}}\,.
\ee
Note that considering dimensionality we have $[a] = [L^3]$ and $[\alpha'] = [L^2]$. In subsequent analysis, we study the dependence of $\log a$ on various parameters, ignoring universal prefactors ($2$, $\pi$) in the $\hat{q}$ definition and setting $L = 1$,  The $\alpha'$-dependence  related to the 't Hooft coupling $\lambda$ through $\sqrt{\lambda} = L^2 / \alpha'$  is incorporated into $\cA_s$ via \eqref{fb}.
 \\

 In what follows, we refer to $a$ as the IJQ parameter (denoting the inverse jet quenching), in direct contrast to the standard JQ parameter ($\hat{q}$) for jet quenching.
 $\log a$ and $-\log \hat q$ differ by an additive constant. 

We considered the jet quenching parameter $\hat{q}$ in holographic models that effectively describe heavy and light quarks. The holographic  expression for jet quenching \eqref{eq:qfinal} includes some free parameters such as AdS radius $L$, Newton constant $G_5$, and Regge slope $\alpha'$. In the numerical calculations we set $L=1$, $8\pi G_5=1$, $\pi \alpha'=1$, but we keep them in our analytical calculations. These parameters can be changed to obtain some results for $\hat{q}$ to be in correspondence with experimental results at RHIC and LHC \cite{JET:2013cls, Edelstein:2008cp}. The experimental results for jet quenching parameter have been obtained for small values of chemical potential. The location of first-order phase transition in $(\mu, T)$-plane is independent of the choice of the boundary condition for dilaton field and parameter $\alpha'$.  

We can compare our results for JQ parameter with the results from \cite{Zhu:2023aaq}: $\mu=0$, $\hat{q}=6$ $GeV^2/fm$, $T=0.3$ GeV. For these parameters we obtain for LQ model $\log a=10$ and for HQ model $\log a=5$. Using equation \eqref{eq:qfinal}, we can estimate $\alpha'$ for both LQ and HQ models 
\bea
\alpha'_{LQ}&=&\frac{5.07 \, L^2}{3.14 \cdot 6 \cdot e^{10} }\approx 1.2\cdot 10^{-5} L^2,
\eea
\bea
\alpha'_{HQ}&=&\frac{5.07 \, L^2}{3.14\cdot 6\cdot e^{5} }\approx 1.8\cdot 10^{-3} L^2.
\eea

\subsubsection{Anisotropic case}

We consider metrics with two types of anisotropy \eqref{an-met}:   the
parameter $\nu$ of primary anisotropy and secondary anisotropy
related to the magnetic field $F_3$.
In this case, the JQ parameter depends on the orientation. Here we consider a jet moving along the $x_{1}$  direction with a momentum broadening along $x_{2}$ or $x_{3}$. The corresponding JQ parameters are defined as $\hat{q_2}$ and $\hat{q_3}$. Standard holographic calculations for anisotropic cases give \cite{Giataganas:2012zy,Zhu:2023aaq}
\bea\label{qi}
\hat{q_i}=\frac{L^2}{\pi \alpha'a_i},
\eea
where
\bea\label{a2}
a_2&=& \int ^{z_h}_0 \frac{{  e^{-2 A_s(z)}   \left(\frac{z}{L}\right)^{2/\nu }}}{
   \sqrt{g(z)(1-g(z))} 
   } dz,
\\\label{a3}
a_{3}&=& \int ^{z_h}_0 \frac{{  
e^{-2 A_s(z)-c_B z^2}   \left(\frac{z}{L}\right)^{2/\nu }}}{
   \sqrt{g(z)(1-g(z))} 
   } dz.
   \eea
  For $\nu=1$ and $c_B=0$  \eqref{a2}
  and \eqref{a3} coincide with \eqref{a-mu}. As mentioned above, we call $a_2$ and $a_3$ as the IJQ parameters.

\section{JQ parameter for light quarks: numerical results}
\label{sect:LQ-NR}

Since holographic calculations involve an explicit dependence of physical quantities on the location of the horizon \(z_h\), we depict the phase structure in the \((\mu, z_h)\)-plane.  Fig.\,\ref{Fig:LQ-Iso-zh-mu} shows this diagram for isotropic media composed of light quarks (LQ); see \cite{Arefeva:2024xmg,Arefeva:2024vom}. 
\\

For anisotropic media in the presence of magnetic fields, the phase structure changes.
\\

We will calculate the JQ parameter $a$ along lines of constant chemical potential and present density plots of $a$ in the physical parameter space $(\mu, T)$. These fixed-$\mu$ calculations provide detailed profiles of $a$, while the density plots illustrate the overall structure of $a$ in the $(\mu, T)$-plane and help identify potential phase transitions.

\subsection{Zero magnetic field}\label{NR-LQ-zero-nu1-15-45}

\subsubsection{Zero magnetic field, $\nu=1$}

In this subsection, we study the dependence of  $a$ on $\mu$ and $T$ for the isotropic case, i.e. for  $\nu=1$. 
The explicit formulas \eqref{a2} and \eqref{a3} give the values of the IJQ parameter as a function of temperature $T$ and chemical potential $\mu$. In the case of absence of the magnetic field, these two values  $a_2$ and $a_3$ are identical.  

\begin{figure}[h!]
  \centering
  \includegraphics[scale=0.25]{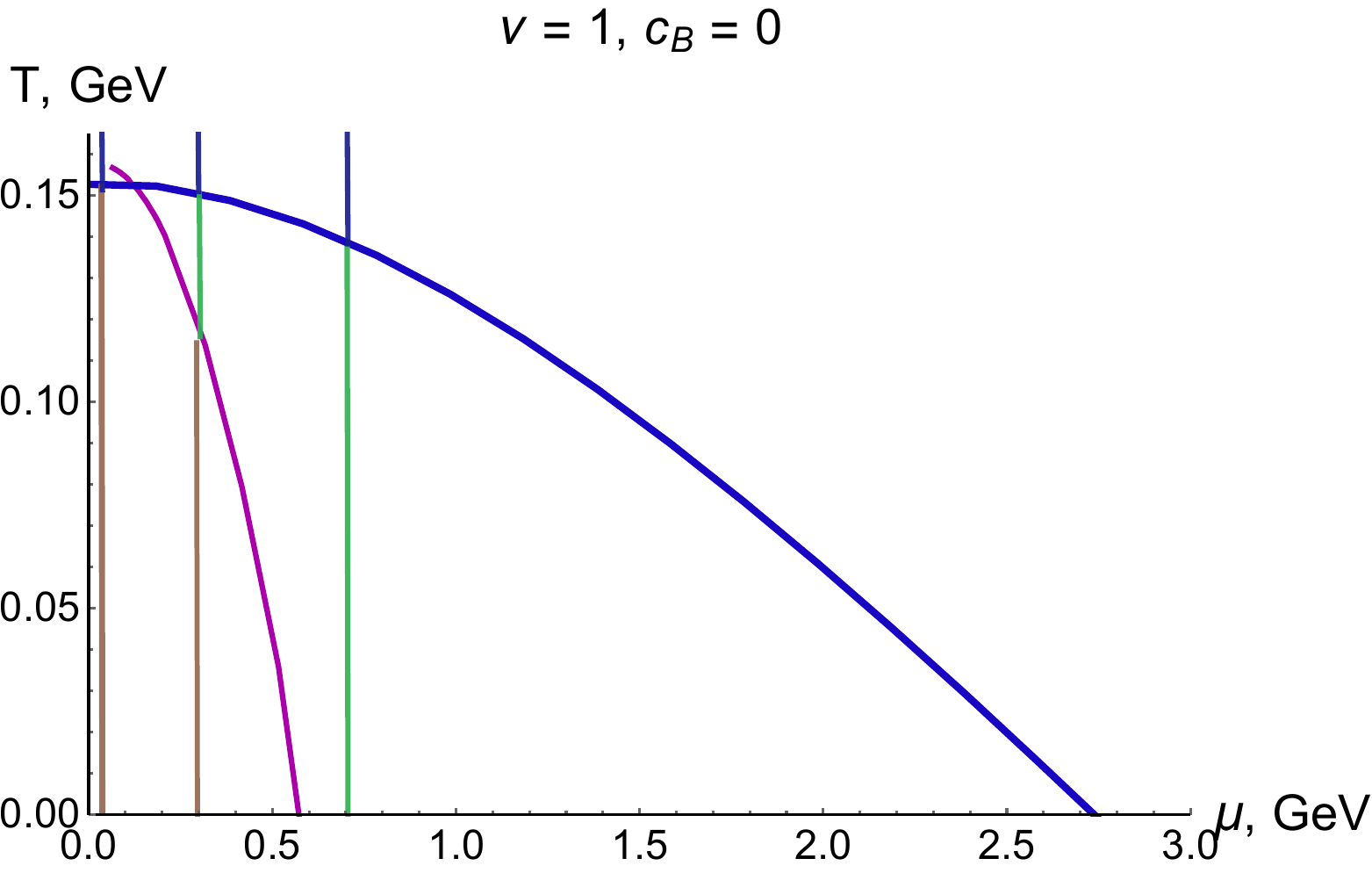}\\
A\\
\includegraphics[scale=0.3]
 {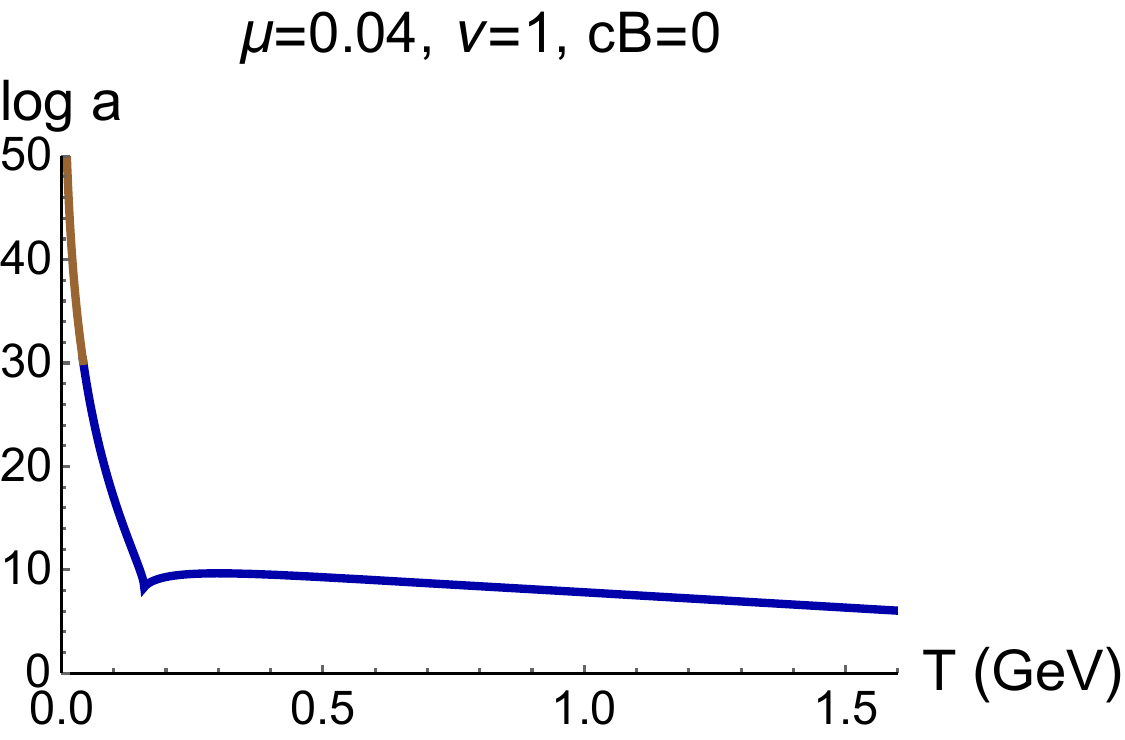}
  \includegraphics[scale=0.25]
   {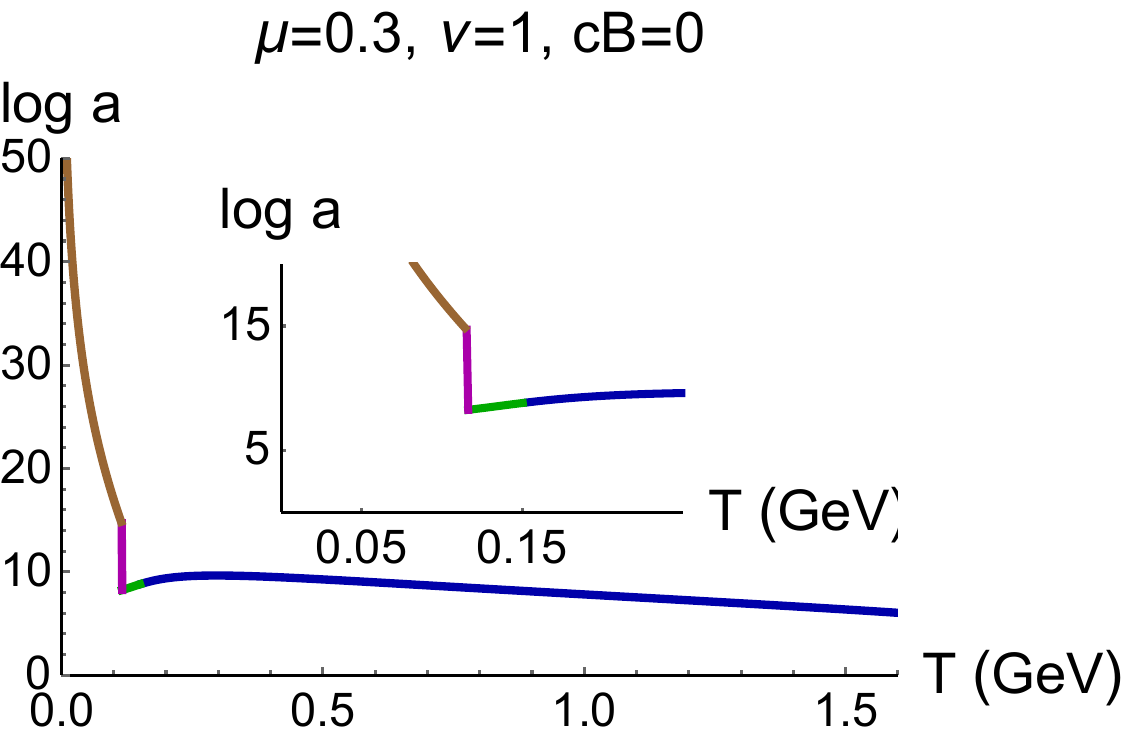} 
\includegraphics[scale=0.25]
   {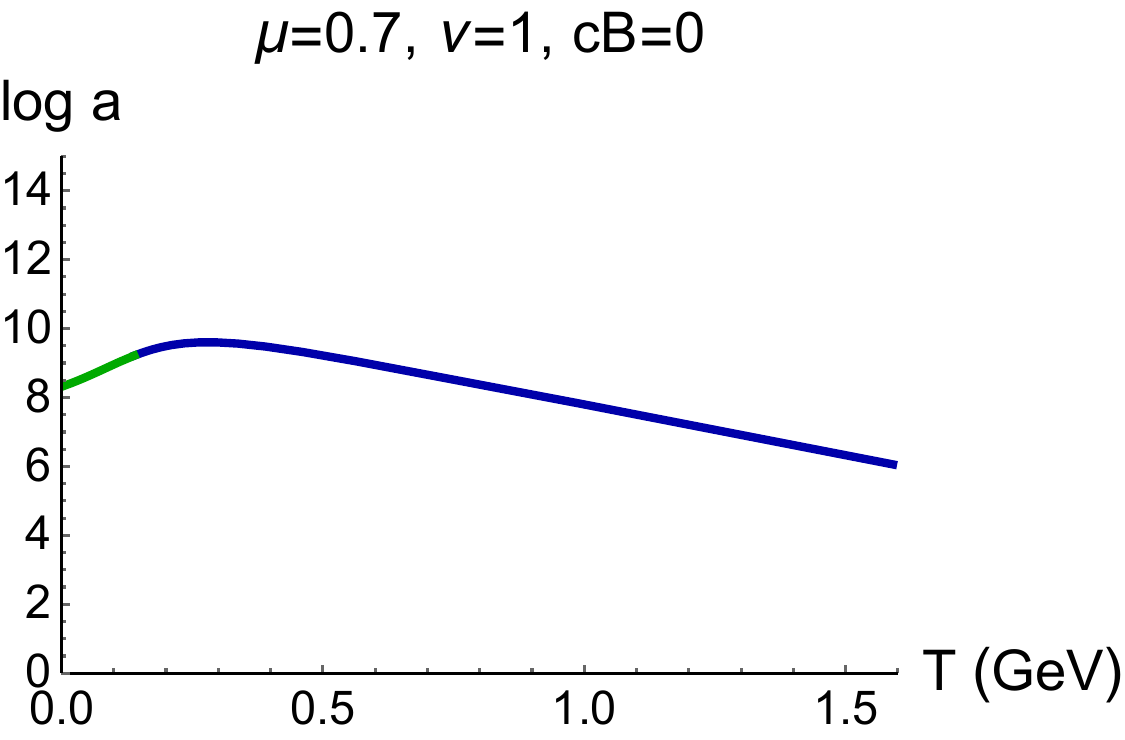}
\\B\hspace{150pt}C\hspace{150pt}D
\caption{ (A) We calculate the JQ parameter for the LQ model, with  $\nu=1$ and $c_B=0$,  along vertical lines (constant $\mu$) in the $(\mu,T)$-plane\protect\footnotemark\
at $\mu = 0.04$, $0.3$, and $0.7$ (GeV). Segments of these lines are colored blue (QGP), brown (hadronic), and green (quarkyonic) according to the phase traversed. The resulting plots of $\log a$ versus temperature are displayed in the bottom panels (B, C, D), using the same color scheme.}
\label{Fig:LQnu1cB0}
\end{figure}

Fig.\,\ref{Fig:LQnu1cB0} depicts the $\log a$ as a function of the temperature $T$, along vertical lines (see Fig.\,\ref{Fig:LQnu1cB0}A) in the $(\mu,T)$-plane at fixed $\mu=0.04$ GeV, $\mu=0.3$ GeV, and $\mu=0.7$ GeV in panels (B, C, D), respectively. The blue and brown lines correspond to the QGP and hadronic phases, respectively.
This synthesis reveals discontinuities in the JQ parameters at first-order phase transitions. Although no abrupt changes appear along the second-order transition line, detailed analysis uncovers a behavioral shift near this boundary: shortly after transitioning to the QGP phase, $\log a$ decreases with increasing temperature, reversing its previous upward trend up to temperature $T\approx 0.25-0.3$ GeV.
Note that, Fig.\,\ref{Fig:LQnu1cB0} can be obtained as a comprehensive picture by combining Figs.\,\ref{Fig:LQQmu004}, \ref{Fig:LQQmu03}, and \ref{Fig:LQQmu07} (see Appendix \ref{app2}), where higher values of $T$ are admissible.

\begin{figure}[h!]
  \centering
\includegraphics[scale=0.25]{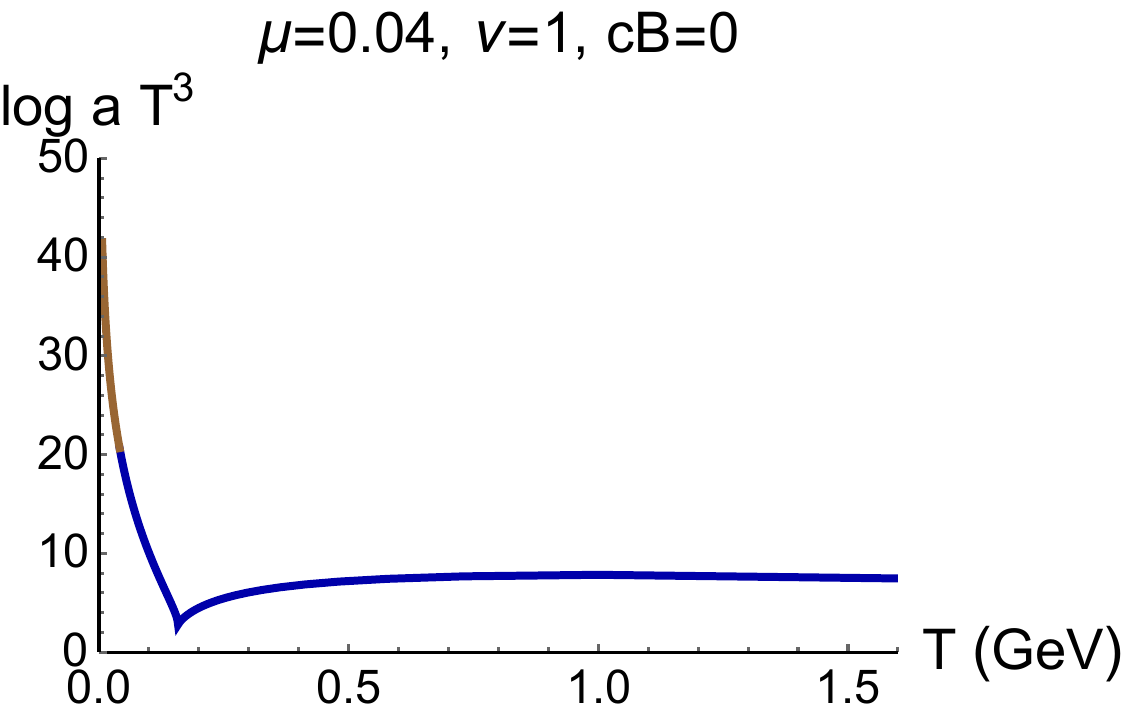}
\includegraphics[scale=0.25]{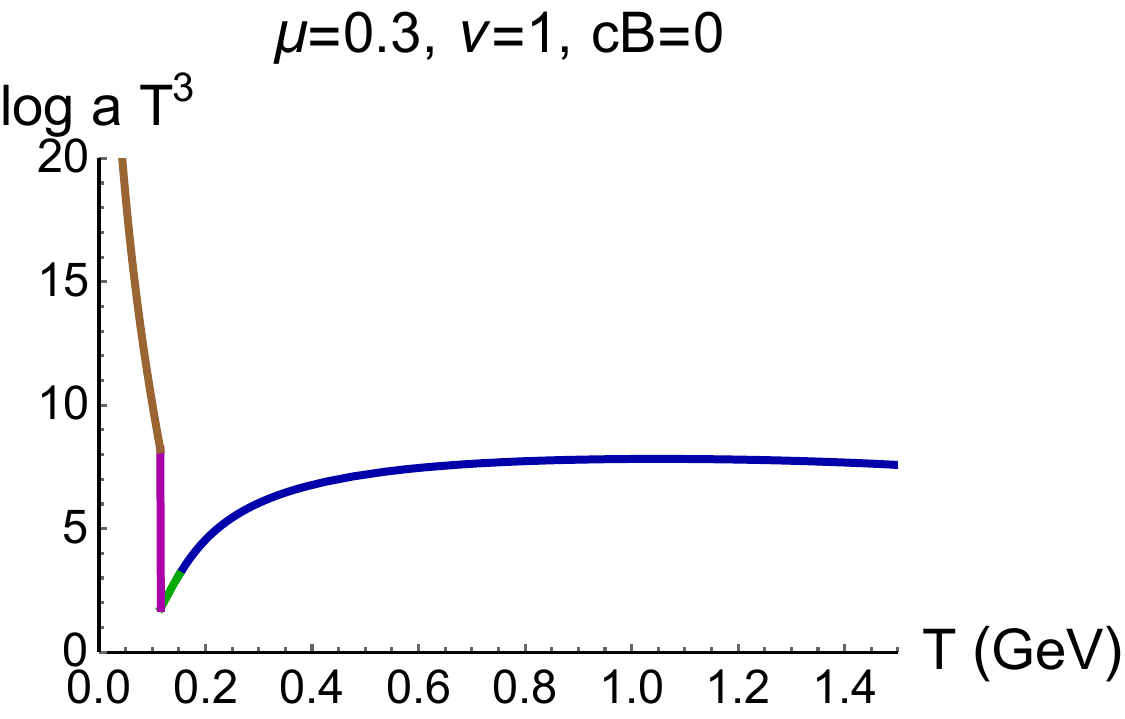}
\includegraphics[scale=0.26]{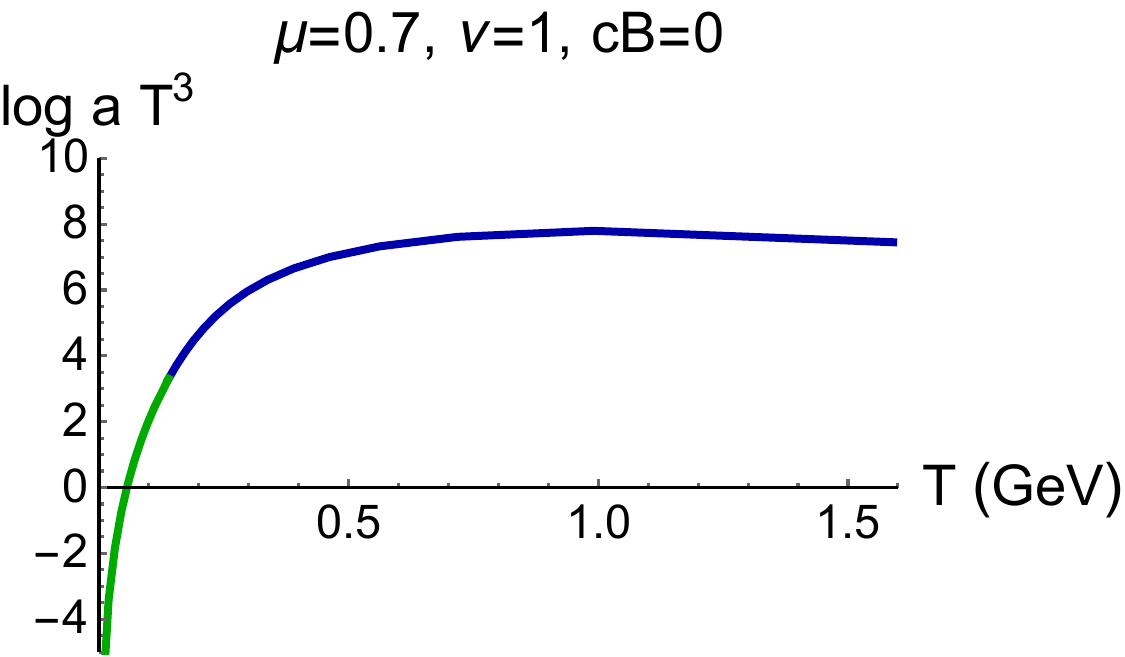}\\
A\hspace{140pt}B\hspace{140pt}
 C  \caption{ The dependence of $\log a T^3$ on $T$ for the isotropic LQ model, with $\nu=1$ and $c_B=0$, at  fixed chemical potentials: (A) $\mu=0.04$ GeV, (B) $\mu=0.3$ GeV, (C) $\mu=0.7$ GeV. }
\label{Fig:LQT3nu1cB0}
\end{figure}
$\,$

The dependence of $\log a T^3$ on $T$ for the isotropic LQ  model for different values of $\mu$ is presented in Fig.\,\ref{Fig:LQT3nu1cB0}. We see that for large $T$, $\log a T^3$ does not depend on $\mu$.

\footnotetext[\thefootnote]{Throughout this paper the dimensionality of quantities is as follows: $[T]= [\mu]=[z]^{-1}= [z_h]^{-1}=  [c_B]^{\frac{1}{2}} =$ GeV.}

\begin{figure}[t!]
  \centering
\includegraphics[scale=0.1]{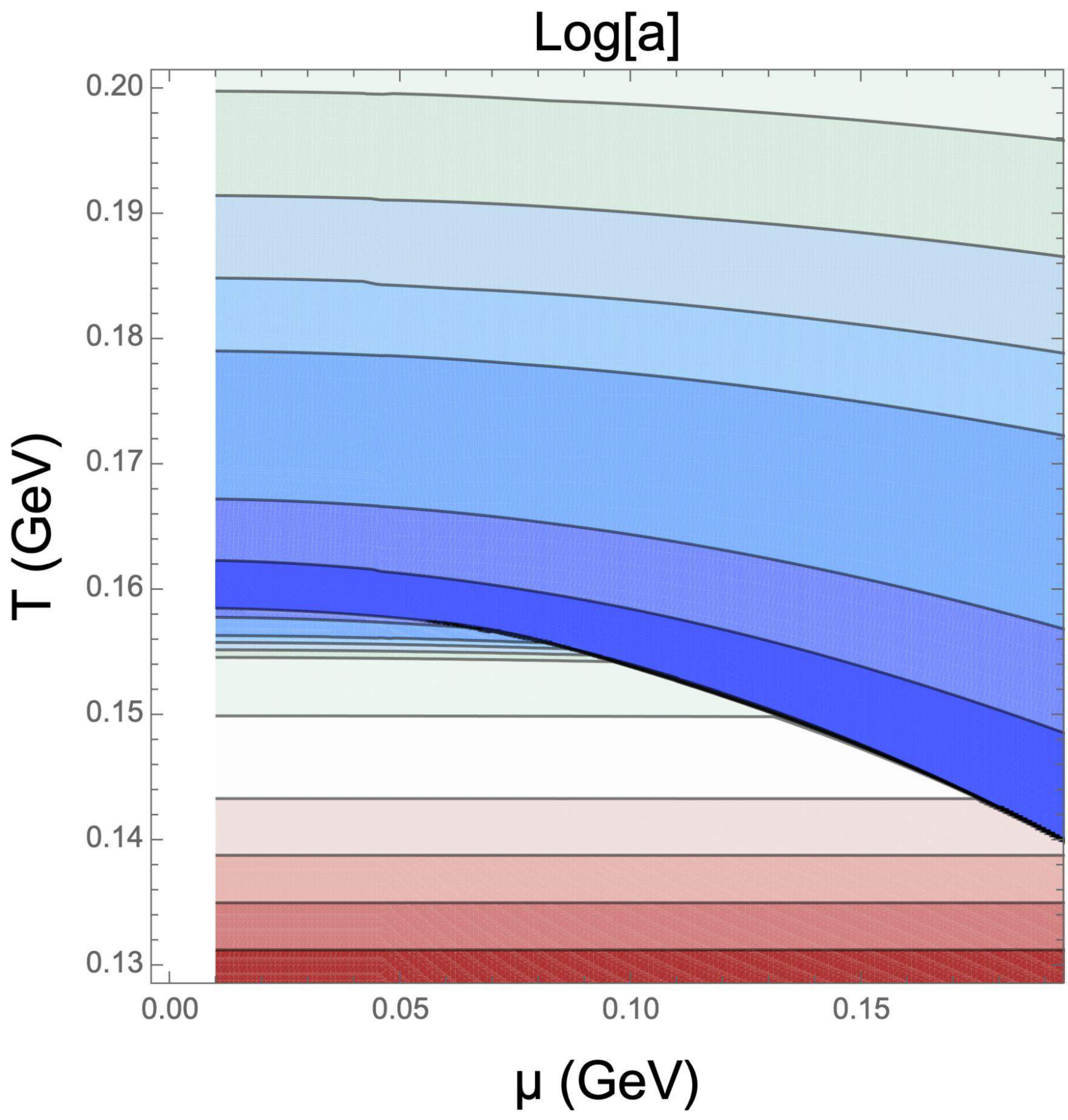}
\includegraphics[scale=0.7] {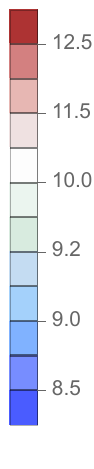}
\includegraphics[scale=0.1]{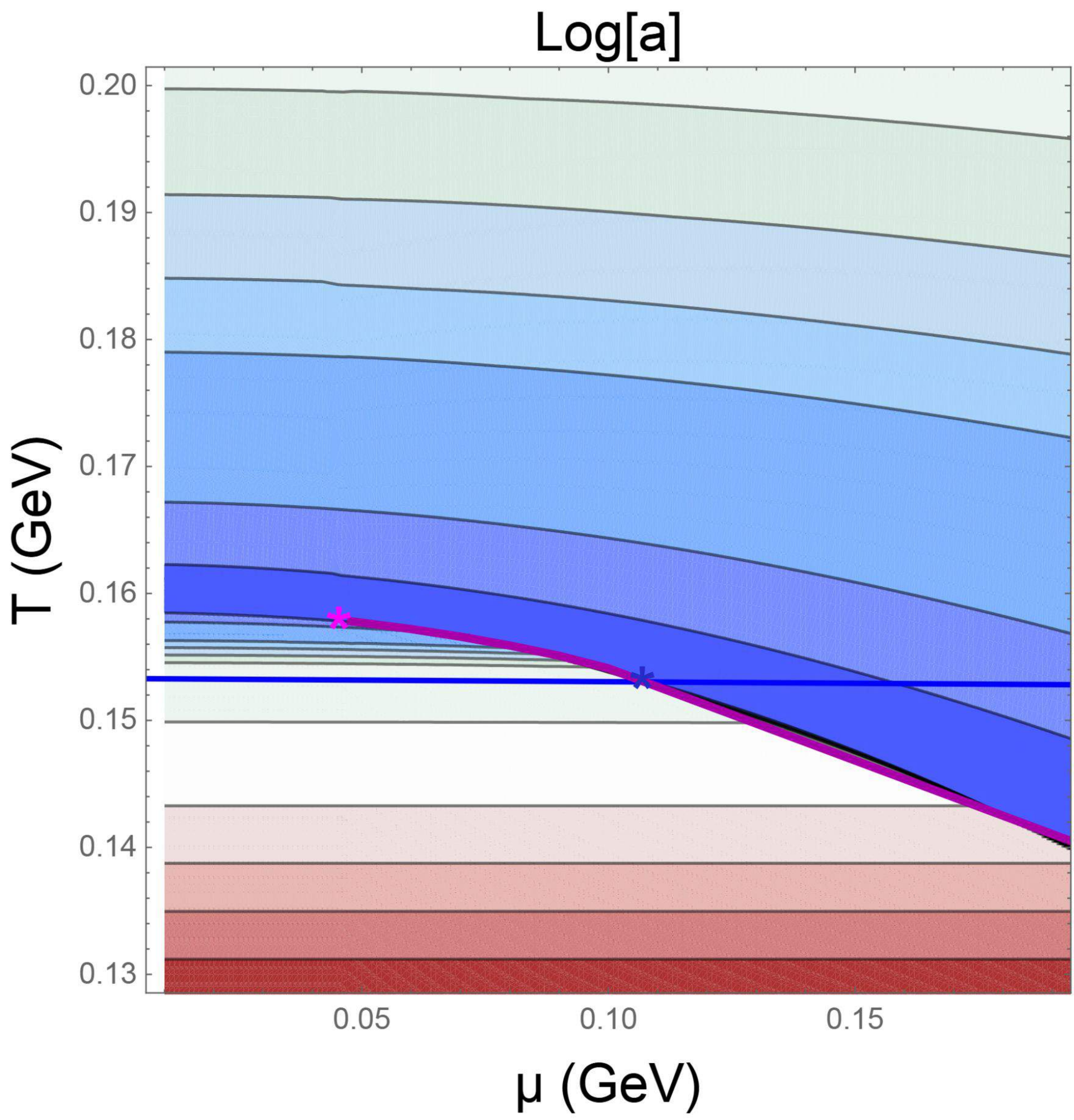}
\\
A\hspace{210pt}B\\$\,$\\
\includegraphics[scale=0.35]{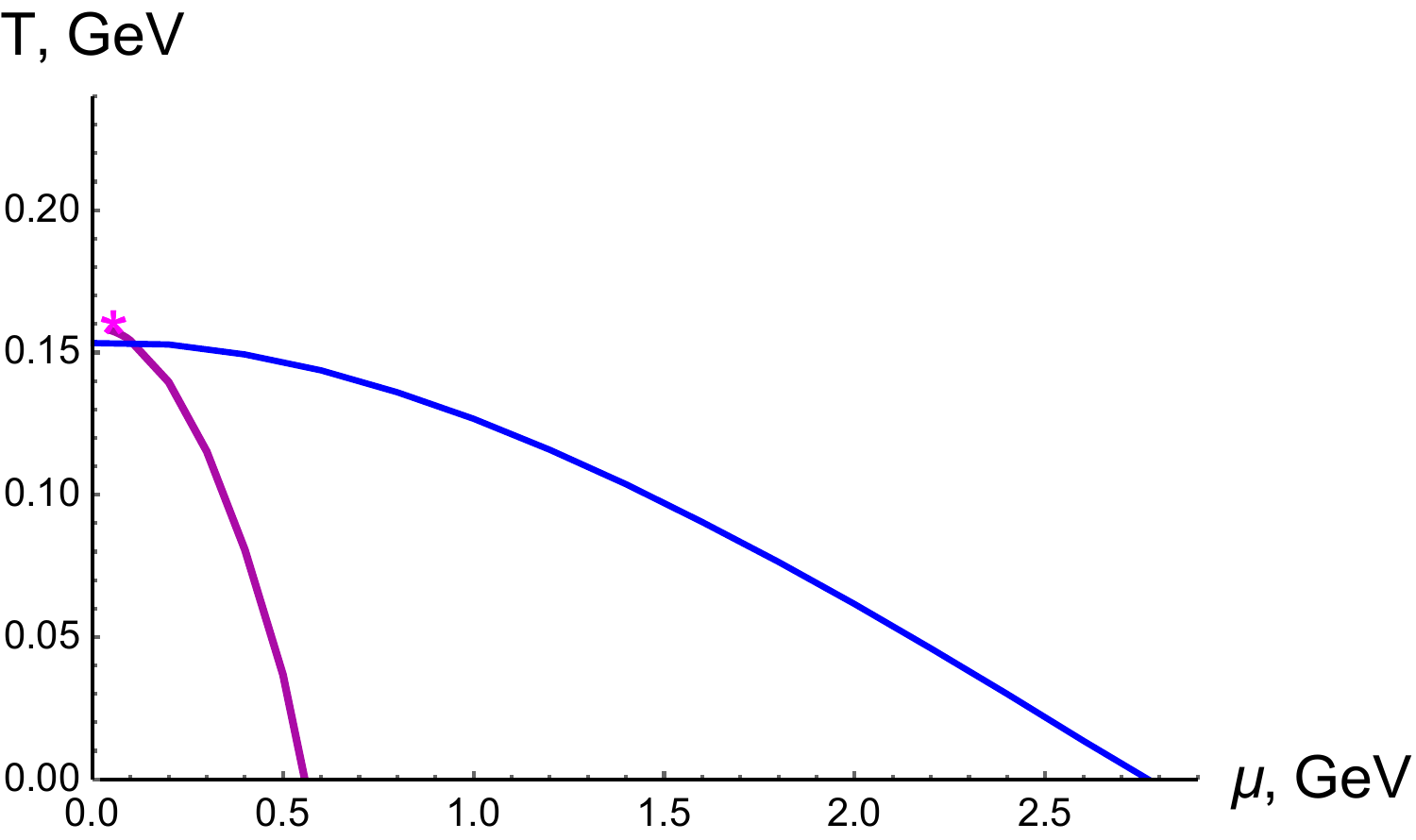}\\
   C
\caption{ (A) Density plots of $\log a$  for the LQ model, with $c_B=0$ and $\nu = 1$. (B) The same with phase transition lines: i) the magenta line shows the first-order transition line that starts at CEP with coordinates ($\mu_{CEP}$,$T_{CEP}$)$\simeq$(0.046 GeV, 0.158 GeV) shown by magenta star; ii) the blue line shows the second-order phase transition line between hadron and QGP phases (for $0<\mu<0.095$ GeV) and between quarkyonic and  QGP phases ($\mu>$0.095 GeV).  (C) The light quark phase diagram, showing the hadronic phase below the magenta line, the quarkyonic phase between the magenta and blue lines, and the QGP phase above the blue line.} 
\label{Fig:DensityLQnu1cB0}
\end{figure}

\newpage
The Density plots of $\log a$ for light quarks at $c_B=0$ and $\nu=1$ are presented in Fig.\,\ref{Fig:DensityLQnu1cB0}A, 
while Fig.\,\ref{Fig:DensityLQnu1cB0}B 
includes phase transition lines. The magenta line in panel (B) marks the first-order transition, starting at the critical endpoint (CEP) located at ($\mu_{\text{CEP}}$, $T_{\text{CEP}}$) $\simeq$ (0.046 GeV, 0.158 GeV) and indicated by a magenta star. The blue line denotes the second-order transition: for $0 < \mu < 0.095$ GeV, it separates the hadronic and QGP phases, while for $\mu > 0.095$ GeV, it divides the quarkyonic and QGP phases. Fig.\,
\ref{Fig:DensityLQnu1cB0}C 
(adapted from \cite{Arefeva:2022avn}) shows the phase diagram, with the hadronic phase below the magenta line, quarkyonic phase between the magenta and blue lines, and QGP phase above the blue line.
\\

Figs.\,\ref{Fig:DensityLQnu1cB0}A and \ref{Fig:DensityLQnu1cB0}B 
reveal discontinuities in the JQ parameter along the first-order phase transition line, confirming the results shown in 
Fig.\,\ref{Fig:LQnu1cB0}.
In the hadronic phase, this parameter depends primarily on temperature $T$ but shows a negligible dependence on the chemical potential $\mu$. In contrast, in both the QGP and quarkyonic phases, it exhibits a significant dependence on both $T$ and $\mu$. A significant dependence of $\log a$ on the chemical potential $\mu$ in the quarkyonic  phase is evident by examining the region between the blue and magenta lines in 
Fig.\,\ref{Fig:DensityLQnu1cB0}. 

Here, in contrast to the region of the hadronic phase, where the lines of constant $\log a$ lie predominantly horizontally, a significant slope of the lines of constant $\log a$ is observed.

Notably, the JQ parameter shows no anomalous behavior along the confinement/deconfinement phase transition line. This is clearly demonstrated by comparing Figs.\,\ref{Fig:DensityLQnu1cB0}A and  \ref{Fig:DensityLQnu1cB0}C.

\subsubsection{Zero magnetic field with spatial anisotropy, $\nu= 1.5,3,4.5$}
\label{sect:LQ-NR-zeromf-nu15-45}

\begin{figure}[h!]
  \centering
\includegraphics[scale=0.25]
{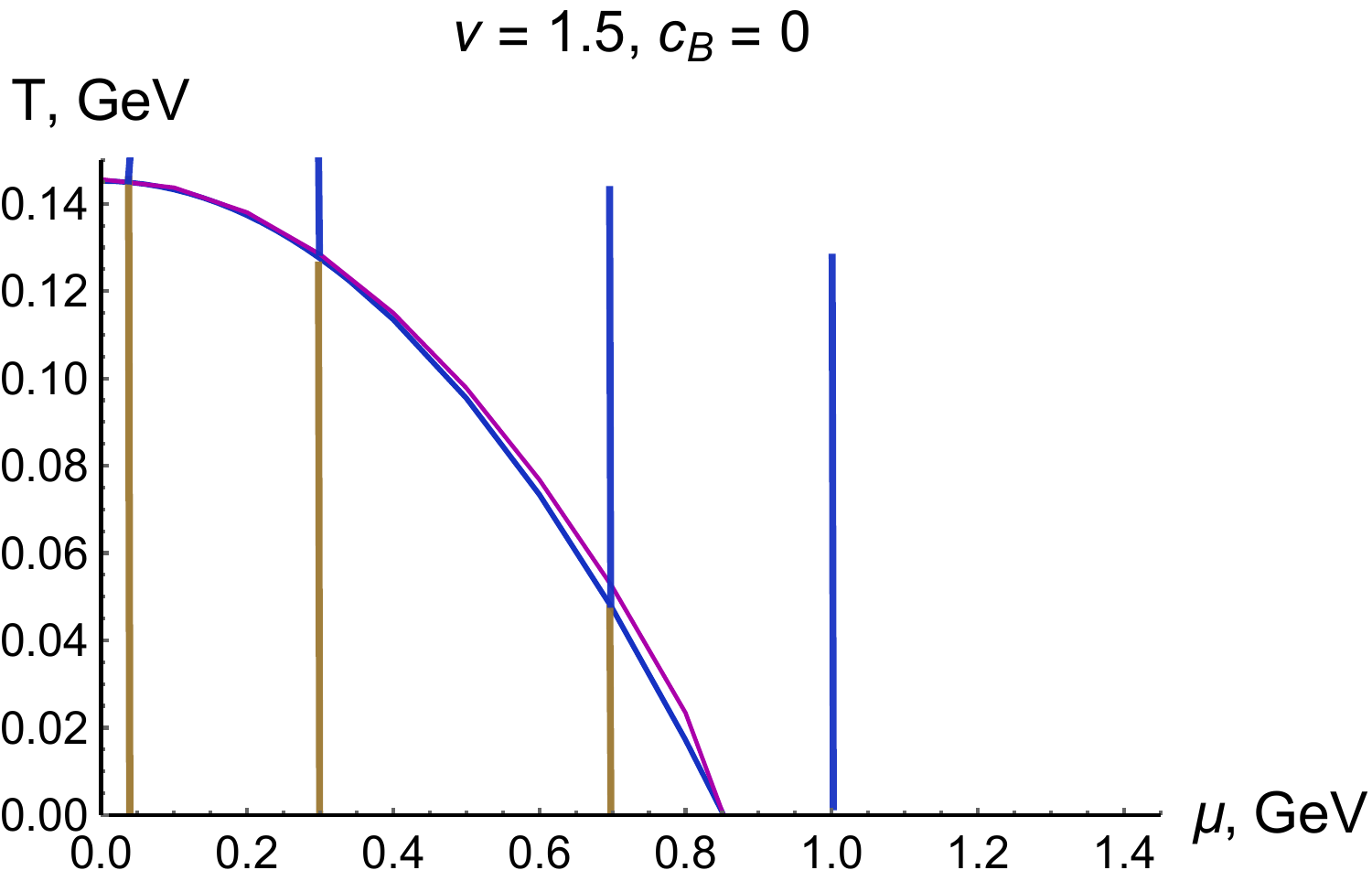}
   \\A\\
\includegraphics[scale=0.15]
   {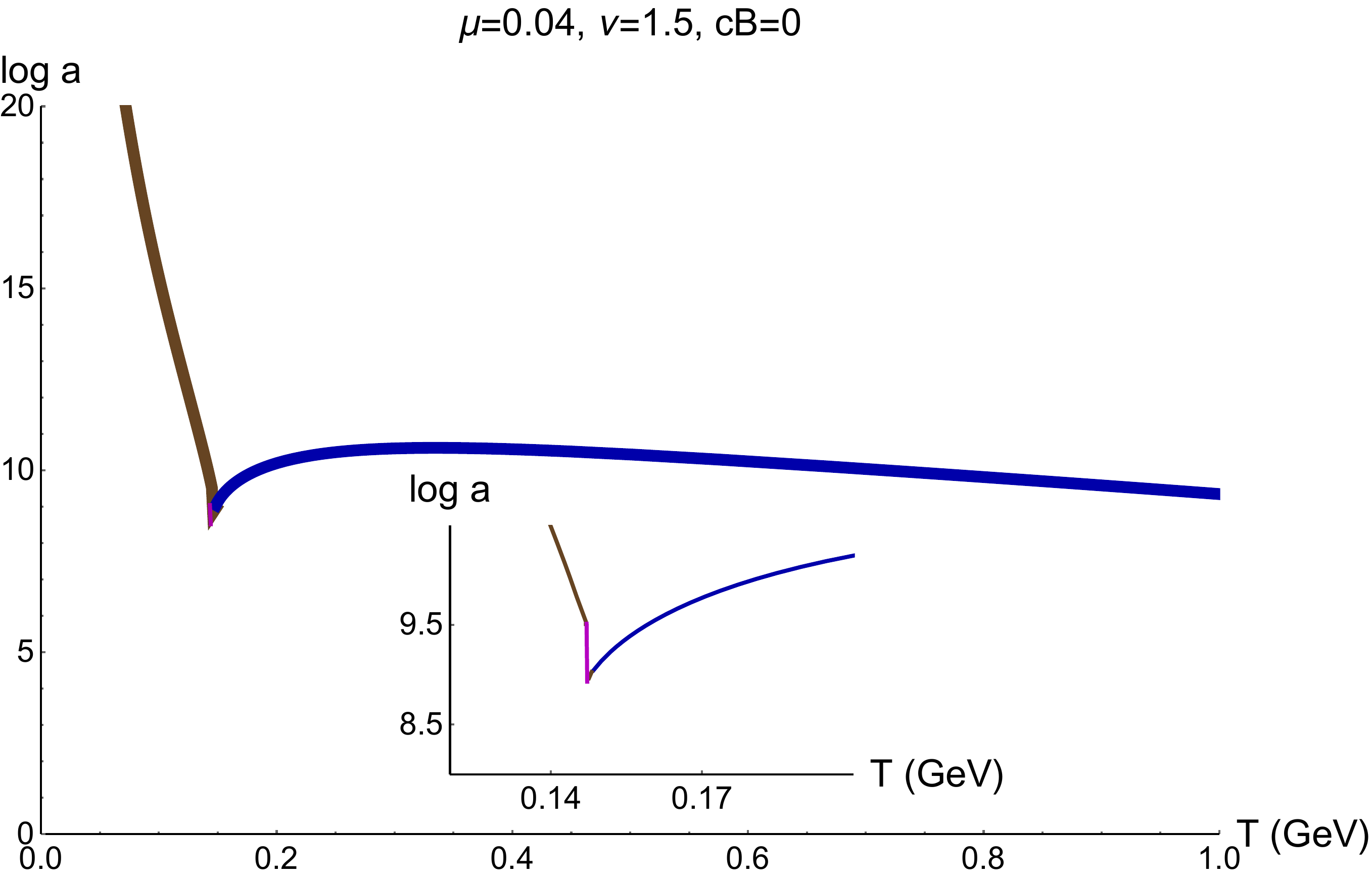}
\includegraphics[scale=0.3]
   {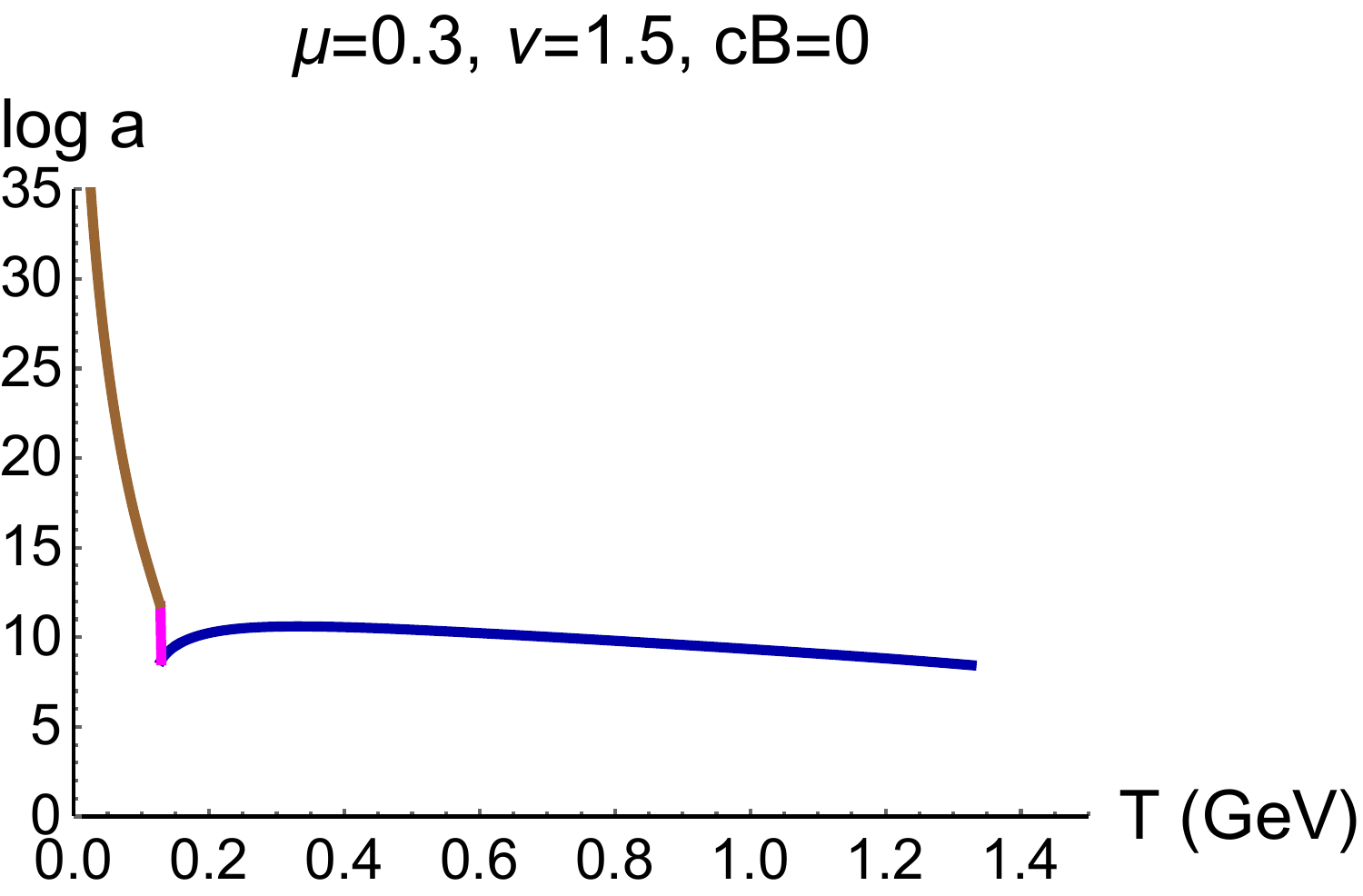}
      \\B\hspace{190pt}C\\
      $\,$\\
\includegraphics[scale=0.23]
   {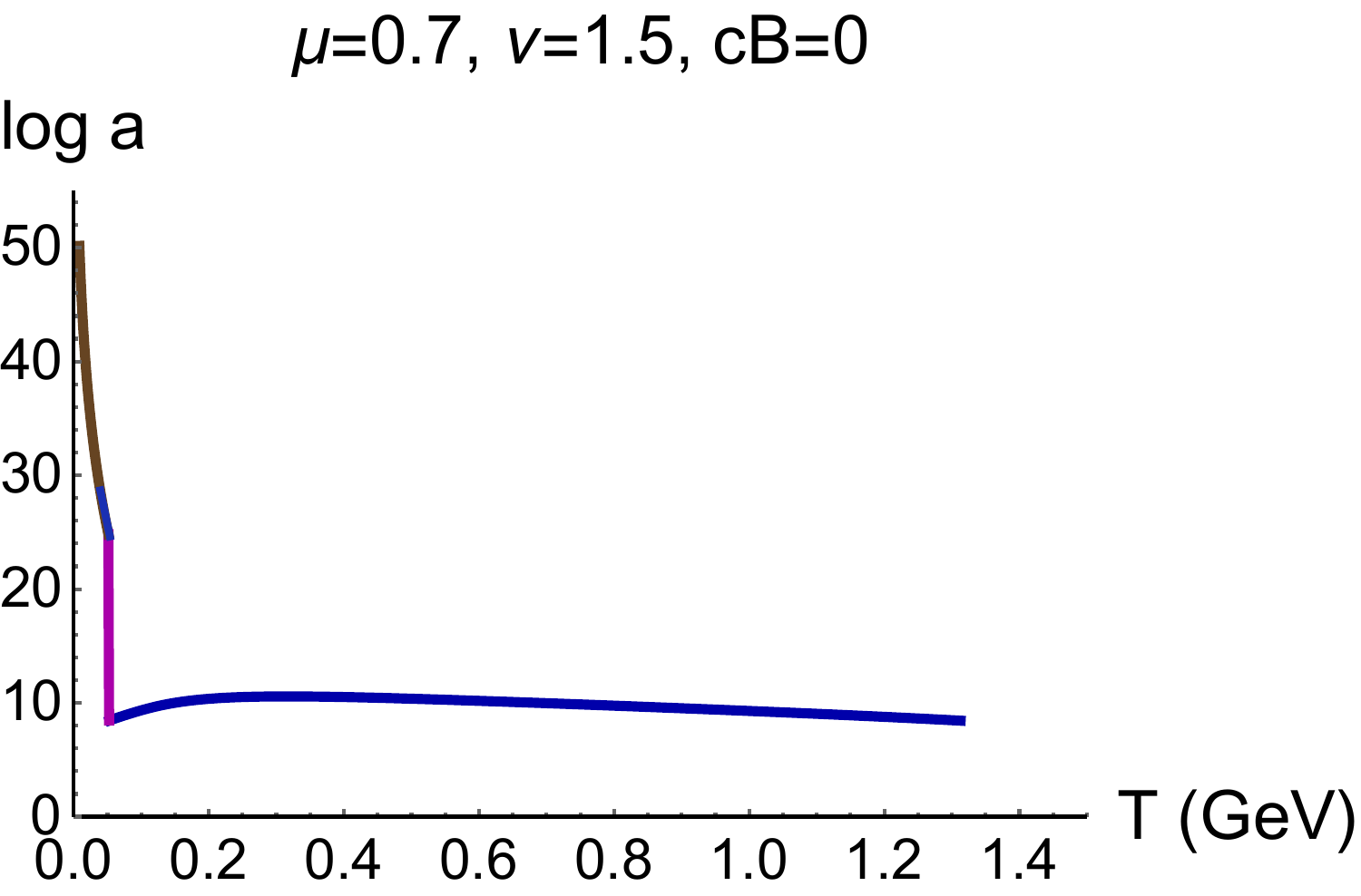}\qquad\qquad
    \includegraphics[scale=0.23]
   {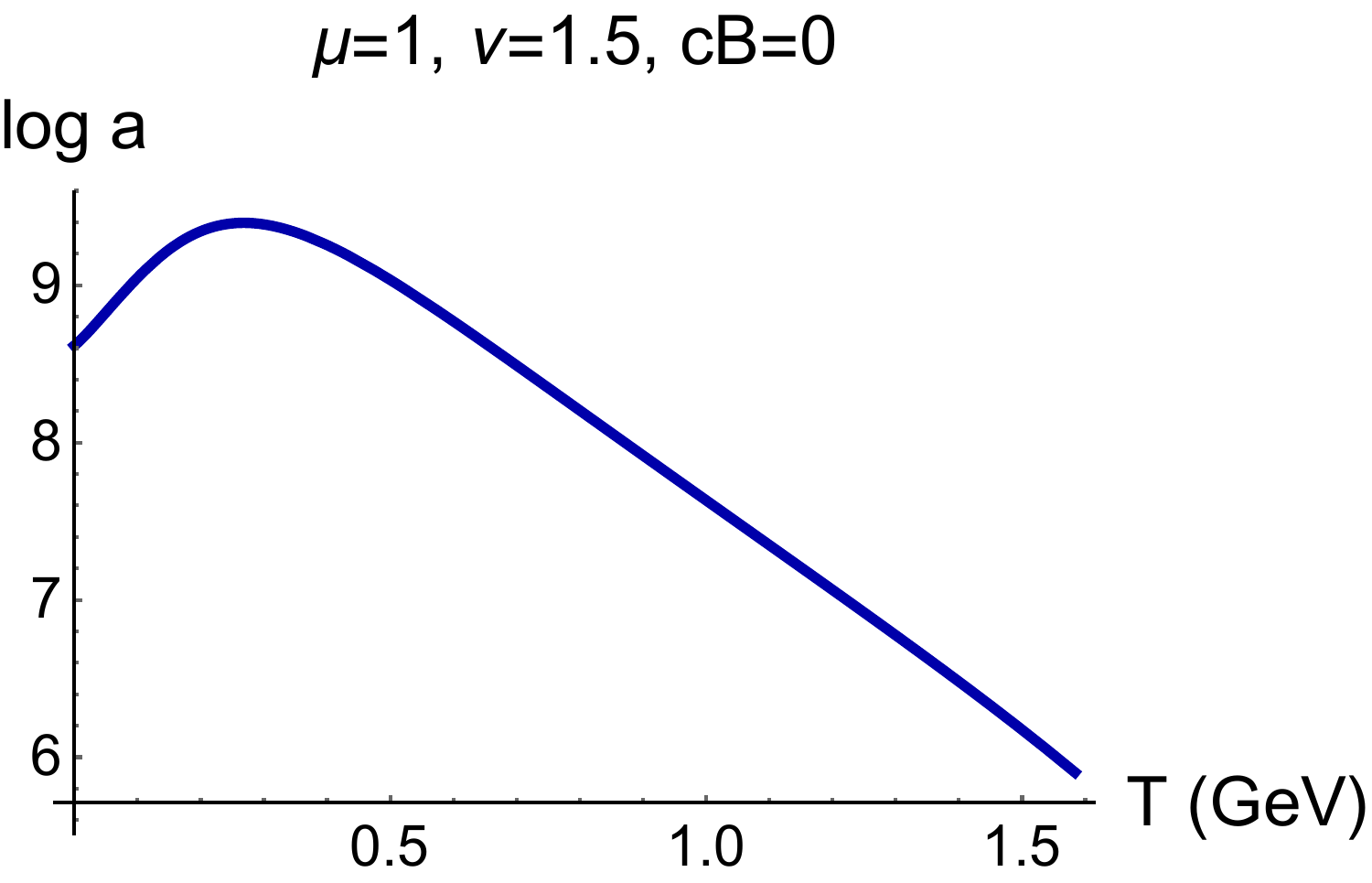}
   \\D\hspace{190pt}E
  \caption{ (A) We calculate the JQ parameter for the LQ model, with  $\nu=1.5$ and $c_B=0$, along vertical lines (constant $\mu$) in the $(\mu,T)$-plane at $\mu = 0.04$, $0.3$, $0.7$ and $\mu=1$ (GeV), shown in this panel. Segments of these lines are colored blue (QGP), brown (hadronic), and green (quarkyonic) according to the phase traversed. The resulting plots of $\log a$ versus temperature are displayed in the bottom panels (B, C, D, E), using the same color scheme.
  }
\label{Fig:LQnu15cB0}
\end{figure}

The phase diagram in $(\mu,T)$-plane for $c_B=0$ and $\nu=1.5$ is presented in Fig.\,\ref{Fig:LQnu15cB0}A.  The magenta and blue lines describe the first-order and confinement/deconfinement phase transitions, respectively. We also draw here the vertical lines with fixed  $\mu=(0.04,0.3,0.7,1)$ (GeV). The Segments of these lines are colored blue (QGP), brown (hadronic), and green (quarkyonic) according to the phase traversed. The Plots of $\log a$ versus temperature are displayed in the bottom panels (B, C, D, E) using the same color scheme. Specifically, $\log a$ versus $T$ is shown for the chemical potential $\mu = 0.04$ GeV in Fig.\,\ref{Fig:LQnu15cB0}B, for $\mu = 0.3$ GeV in Fig.\,\ref{Fig:LQnu15cB0}C, for $\mu = 0.7$ GeV in Fig.\,\ref{Fig:LQnu15cB0}D, and for $\mu = 1$ GeV in Fig.\,\ref{Fig:LQnu15cB0}E.
\\

Fig.\,\ref{Fig:LQnu15cB0}B shows that at $\mu=0.04$ GeV, $\log a$ decreases monotonically in the hadronic phase, indicating enhanced JQ. Following a continuous transition to the QGP phase at $T\sim 0.153$ GeV and a subsequent first-order phase transition, $\log a$ undergoes a jump (marked by the magenta line in the inset of Fig.\,\ref{Fig:LQnu15cB0}B). After this transition, $\log a$ increases until $T \approx 0.3$ GeV, then decreases with further increases in temperature.
Fig.\,\ref{Fig:LQnu15cB0}C shows that $\log a$ at $\mu=0.3$ GeV exhibits similar behavior to Fig.\,\ref{Fig:LQnu15cB0}B, differing only in the magnitude of the discontinuity at the first-order phase transition, which is larger in panel (C).
In Fig.\,\ref{Fig:LQnu15cB0}D ($\mu=0.7$ GeV), $\log a$ decreases in the hadronic phase (indicating strengthening JQ) up to $T\approx 0.025$ GeV, where a phase transition to the QGP occurs. Following this transition, $\log a$ continues to decrease (further strengthening JQ) until $T\sim 0.05$ GeV, where a first-order phase transition occurs.  After the discontinuity at this transition, $\log a$ increases (weakening JQ) until $T \approx 0.25$ GeV, then slowly decreases (strengthening the JQ parameter again). Thus, at higher temperatures in the QGP phase, the JQ parameter strengthens with increasing $T$.
In Fig.\,\ref{Fig:LQnu15cB0}F ($\mu=1$ GeV), the JQ parameters weakens until $T \sim 0.4$ GeV, then strengthens with further increases in temperature.
\\

\begin{figure}[h!]
  \centering
\includegraphics[scale=0.25]{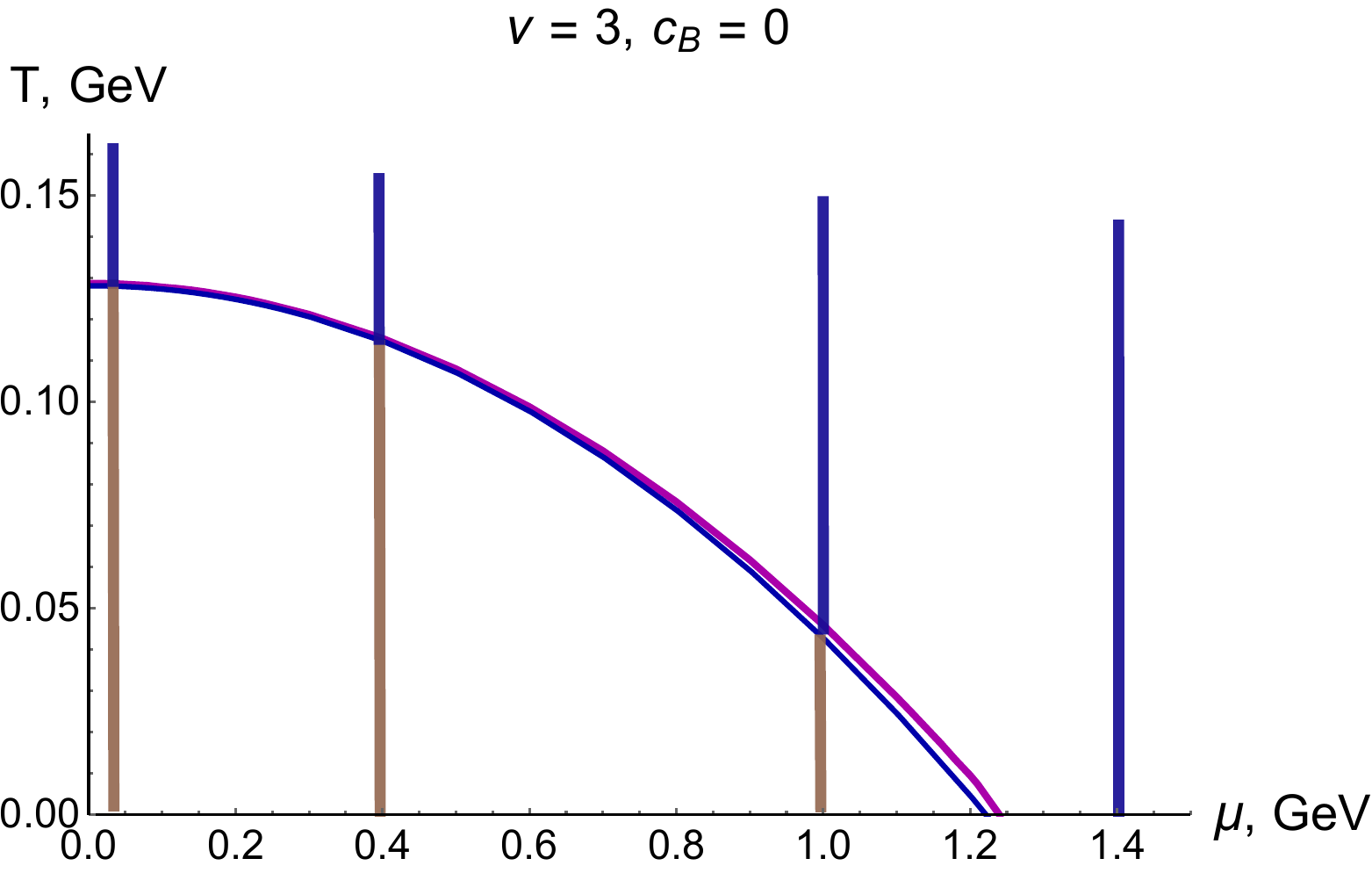}\\A\\
\includegraphics[scale=0.3]
  {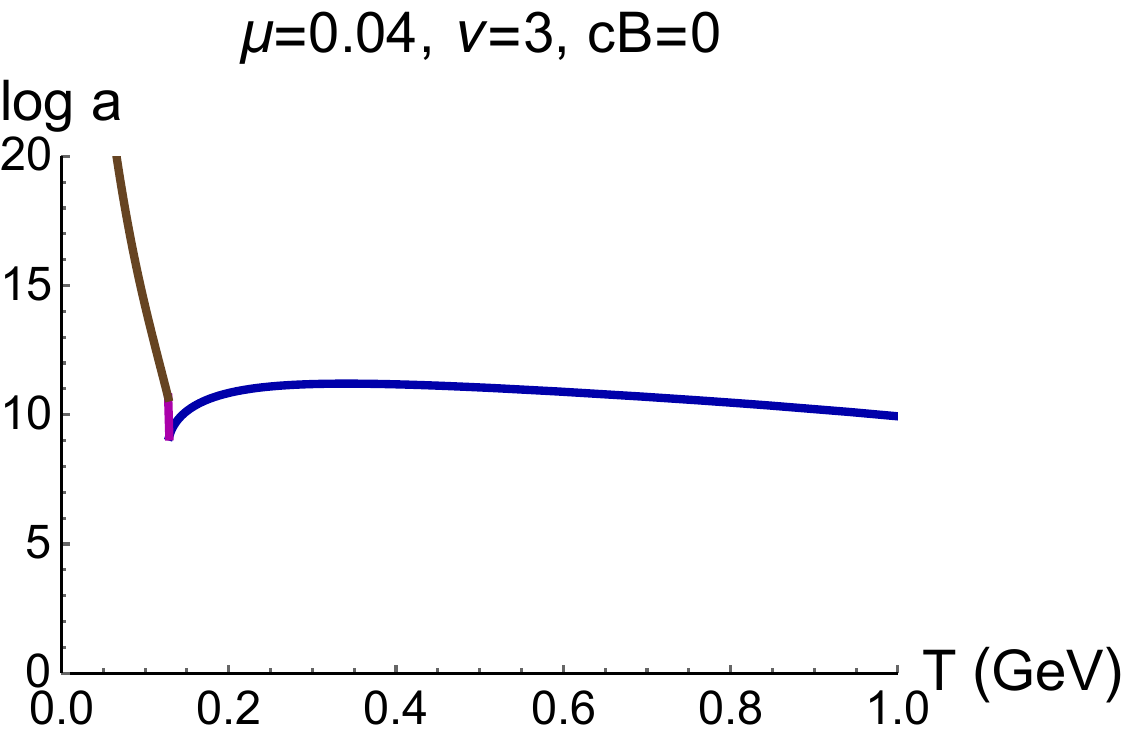}\qquad
  \includegraphics[scale=0.3]
  {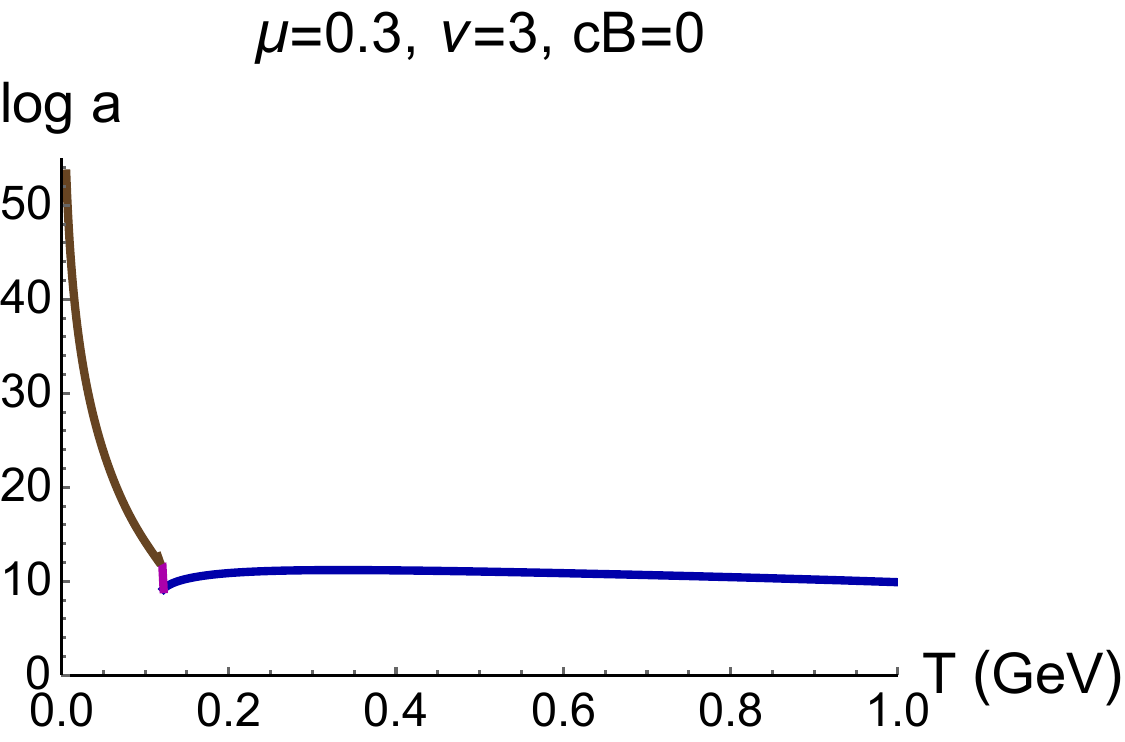}
  \\B\hspace{180pt}C\\$\,$\\
  \includegraphics[scale=0.3]
  {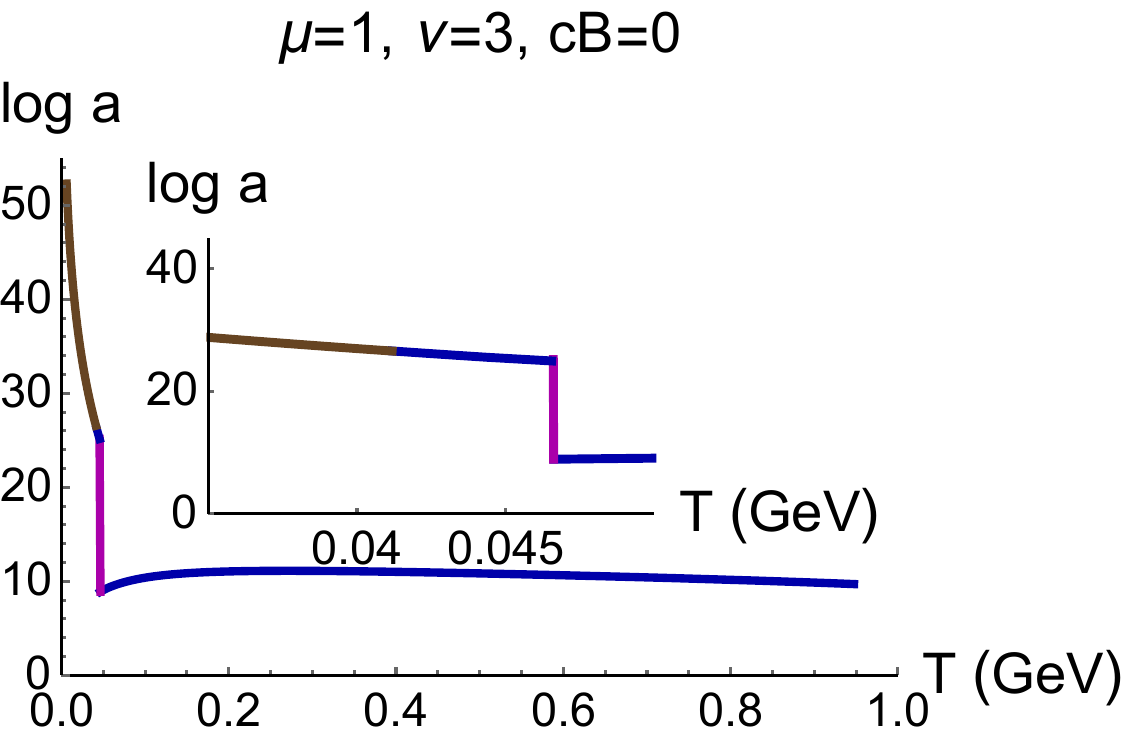}\qquad
  \includegraphics[scale=0.3]
  {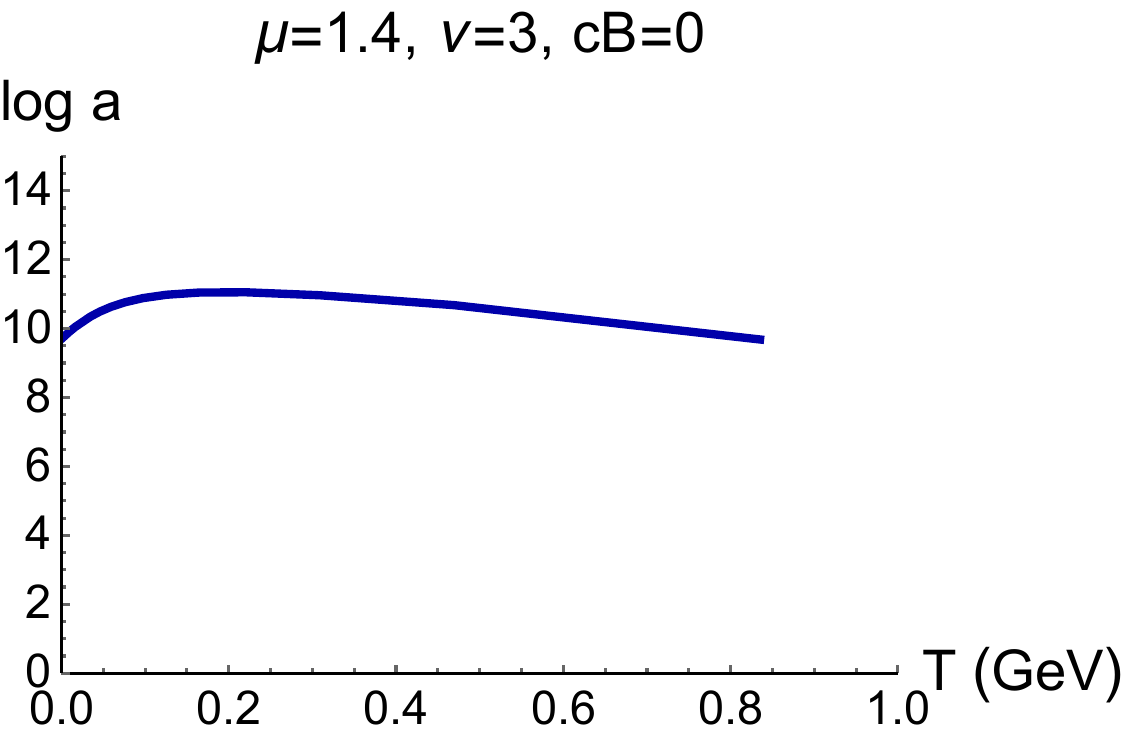}
  \\D\hspace{180pt}E
   \caption{(A) We calculate the JQ parameter for the LQ model, with   $\nu=3$ and $c_B=0$, along vertical lines at fixed $\mu = 0.04$, $0.3$, $1$, and $\mu =1.4$ (GeV) presented at the panel. Segments of these lines are colored blue, brown, and green corresponding to the QGP, hadronic, and quarkyonic phases they traverse, respectively. The resulting  plots for log of the JQ parameter verse  temperature are presented on two bottom panels  (B, C, D, E) are colored using the same scheme.
 }
\label{Fig:LQnu3cB0}
\end{figure}

\begin{figure}[h!]
  \centering
  \includegraphics[scale=0.3]{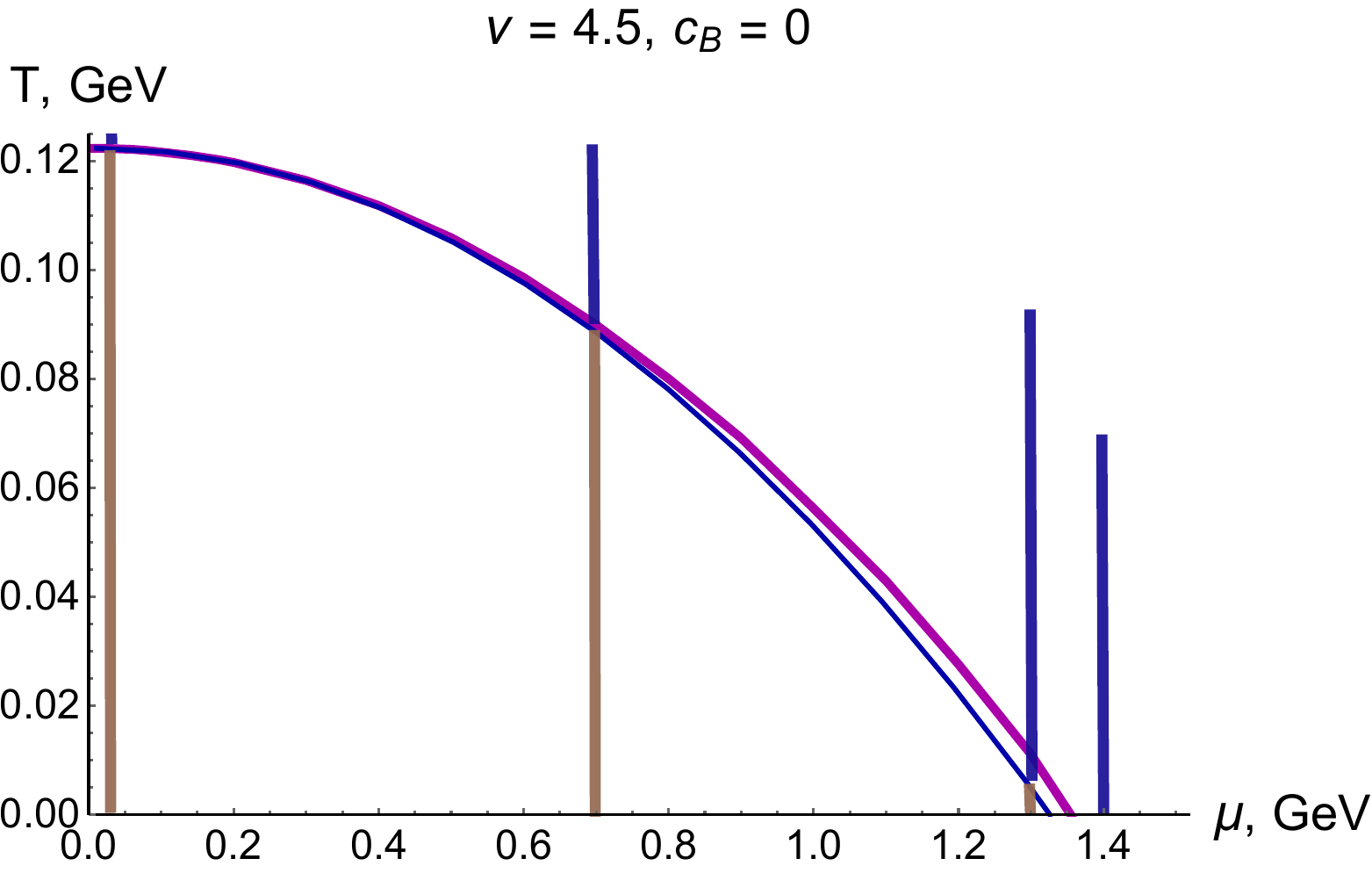}
\\A\\
\includegraphics[scale=0.3]
  {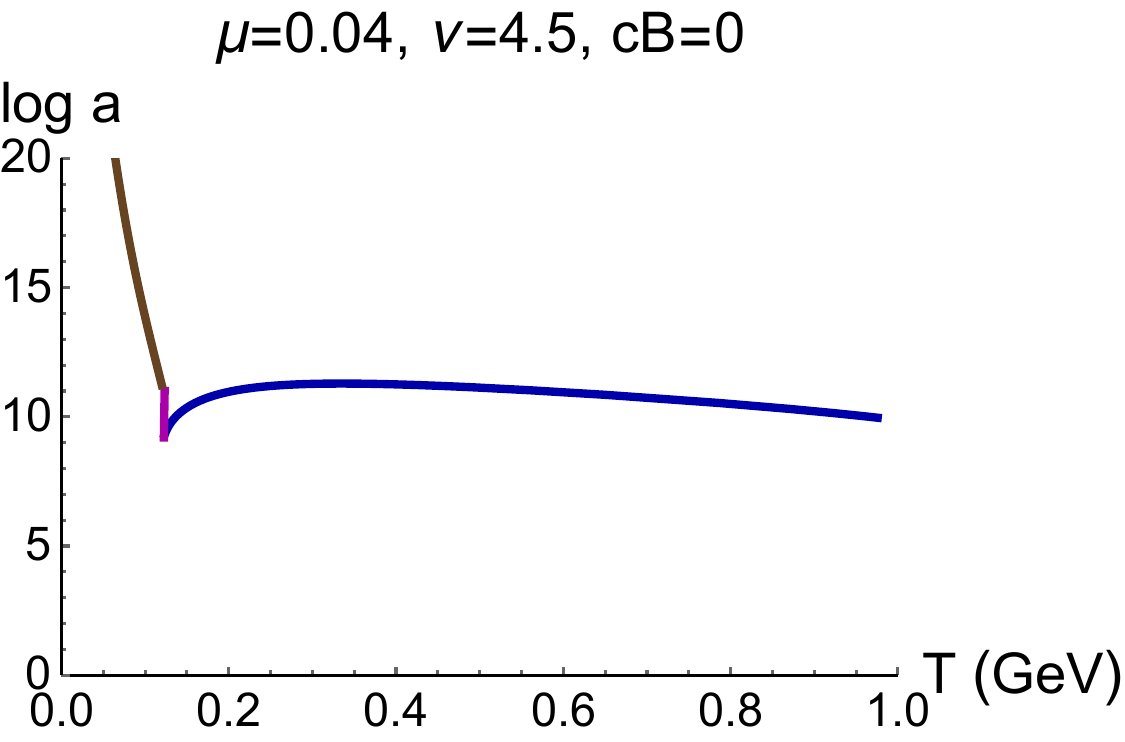}\qquad
  \includegraphics[scale=0.3]
  {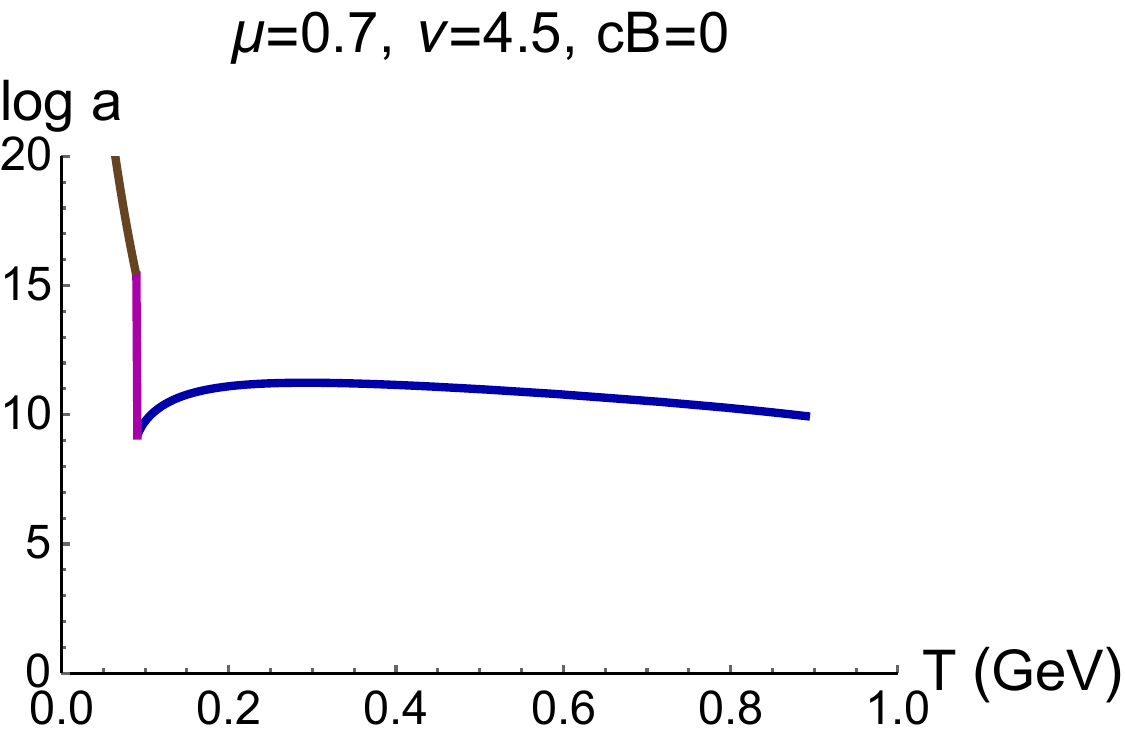}
  \\B\hspace{170pt}C\\$\,$\\
  \includegraphics[scale=0.3]
  {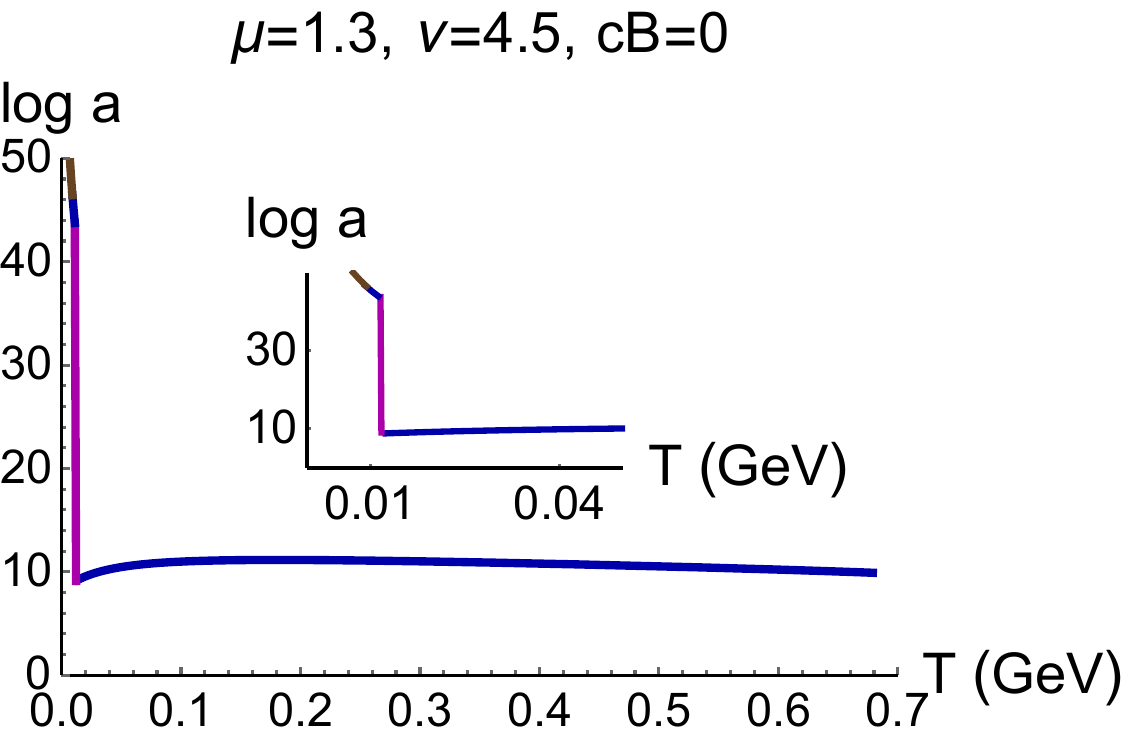}\qquad
  \includegraphics[scale=0.3]
  {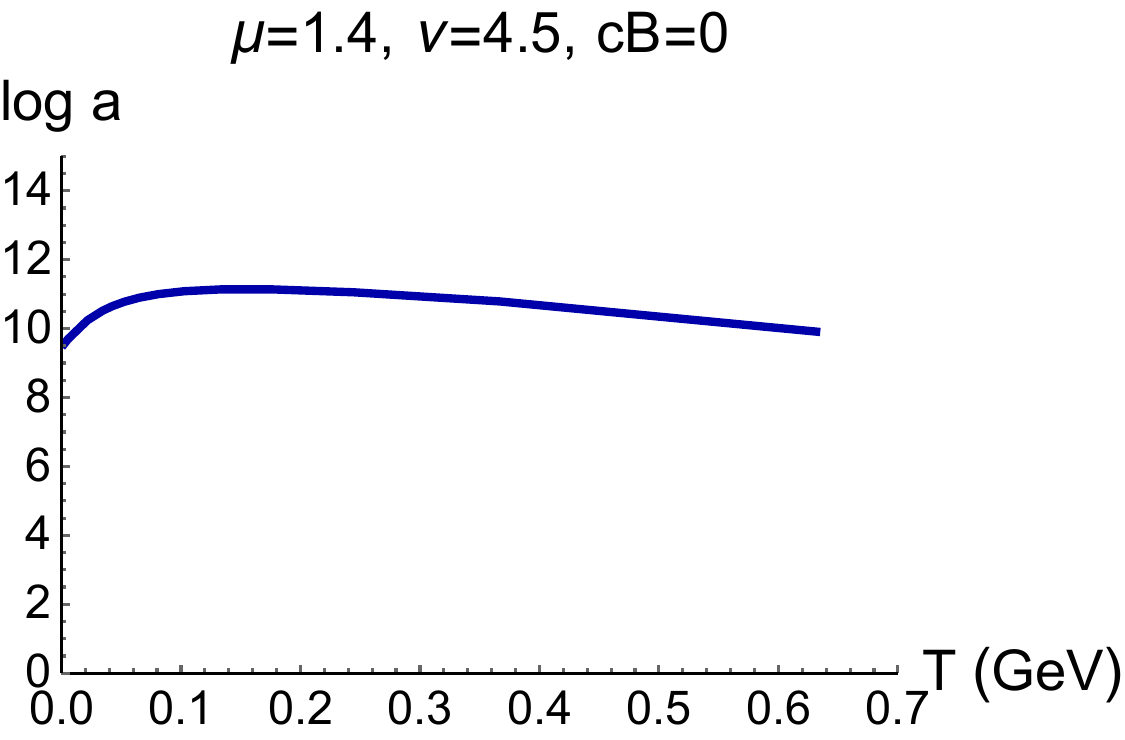}
  \\D\hspace{170pt}E
   \caption{ (A) We calculate the JQ parameter for the LQ model, with   $\nu=4.5$ and $c_B=0$, along vertical lines at fixed $\mu = 0.04$, $0.7$, $1.3$, and $\mu=1.4$ (GeV). Segments of these lines are colored blue, brown, and green corresponding to the QGP, hadronic, and quarkyonic phases they traverse, respectively. The resulting  values for log of the JQ parameter are presented on two bottom panels  (B, C, D, E) are colored using the same scheme.\\
  }
\label{Fig:LQnu45cB0}
\end{figure}

The Phase diagrams in the $(\mu,T)$-plane for $c_B=0$ at $\nu=3$ and $\nu=4.5$ are shown in Figs.\,\ref{Fig:LQnu3cB0}A and \ref{Fig:LQnu45cB0}A, respectively. Both figures employ the same color scheme as Fig.\,\ref{Fig:LQnu15cB0}. While qualitatively similar to Fig.\,\ref{Fig:LQnu15cB0}, these diagrams exhibit quantitative differences: chemical potential values, discontinuity magnitudes in the JQ parameters, and slopes of $\log a$ variations differ.\\

\begin{figure}[htbp]   
\centering
\includegraphics[scale=0.30]{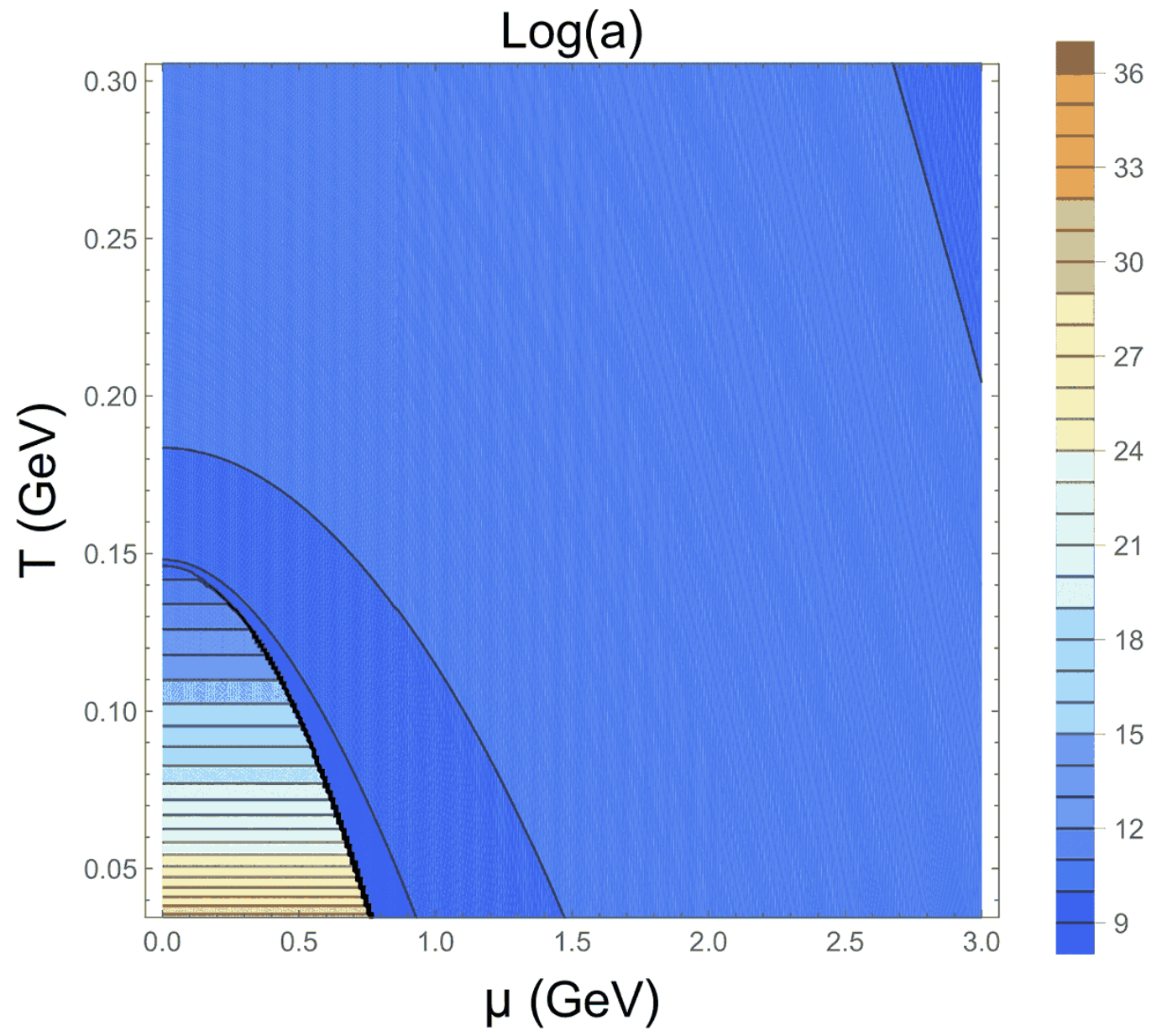}\quad
\includegraphics[scale=0.08]{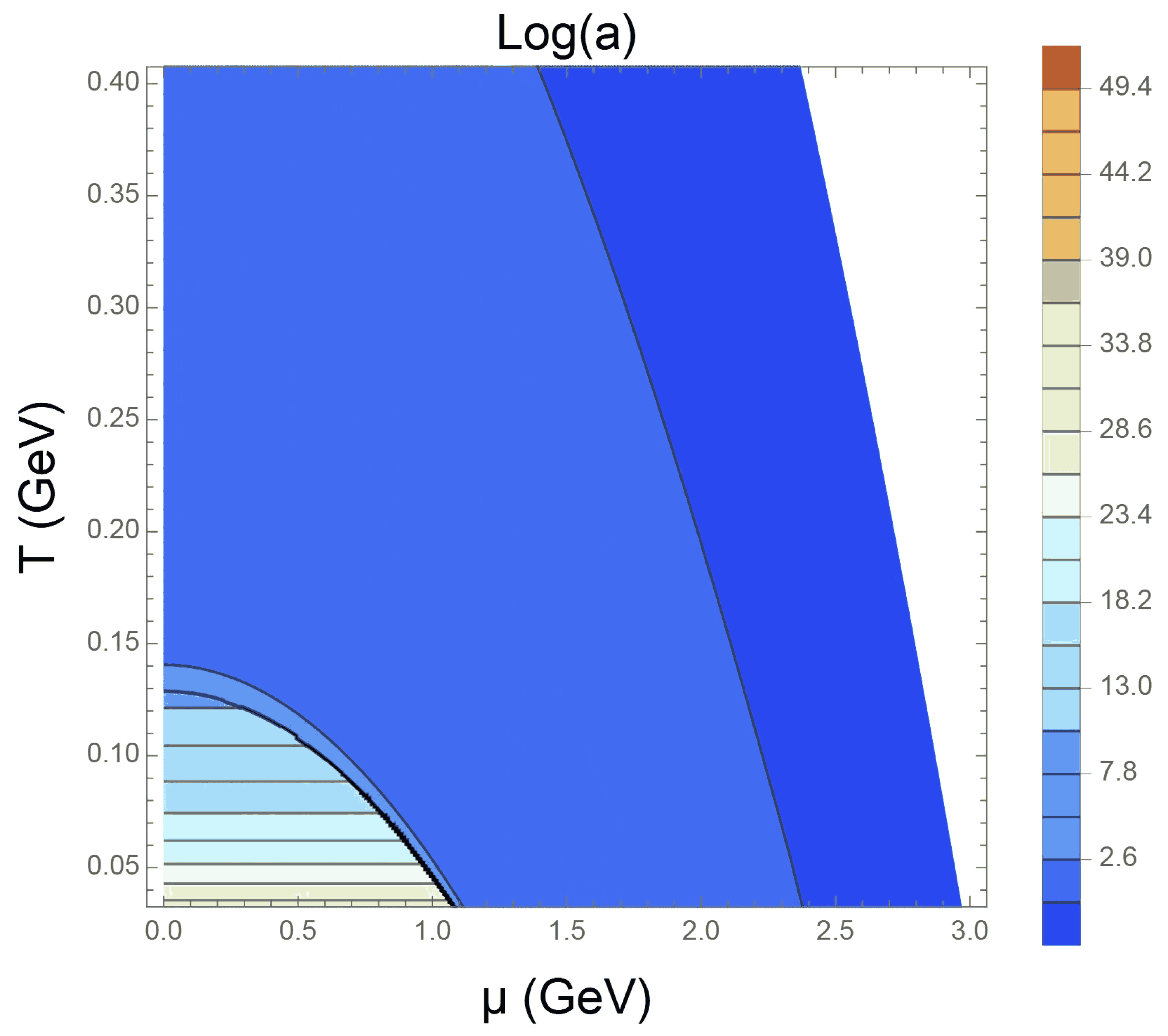} \\
A\hspace{200pt}B\\
\includegraphics[scale=0.085]{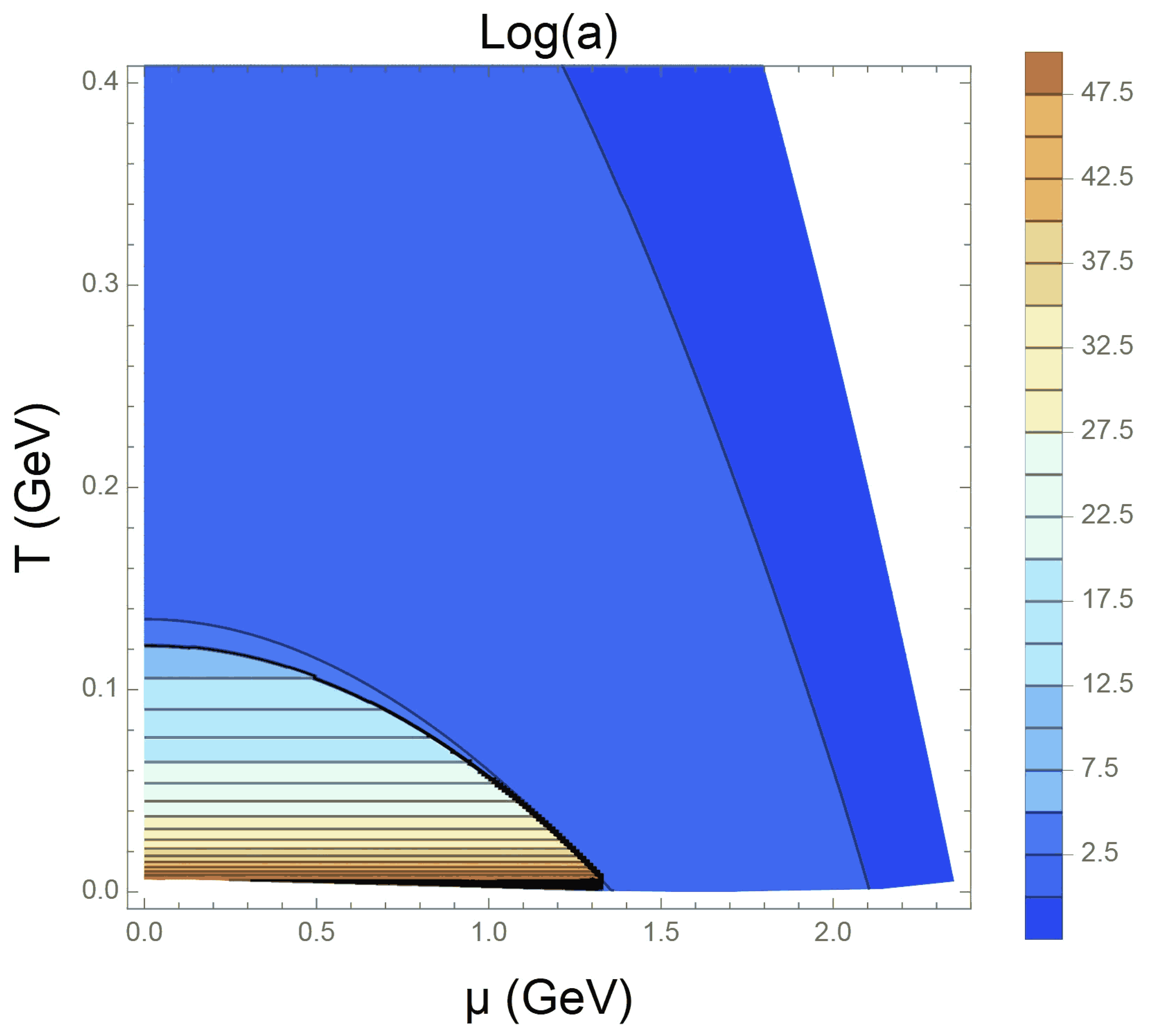}\quad
\includegraphics[scale=0.40]
  {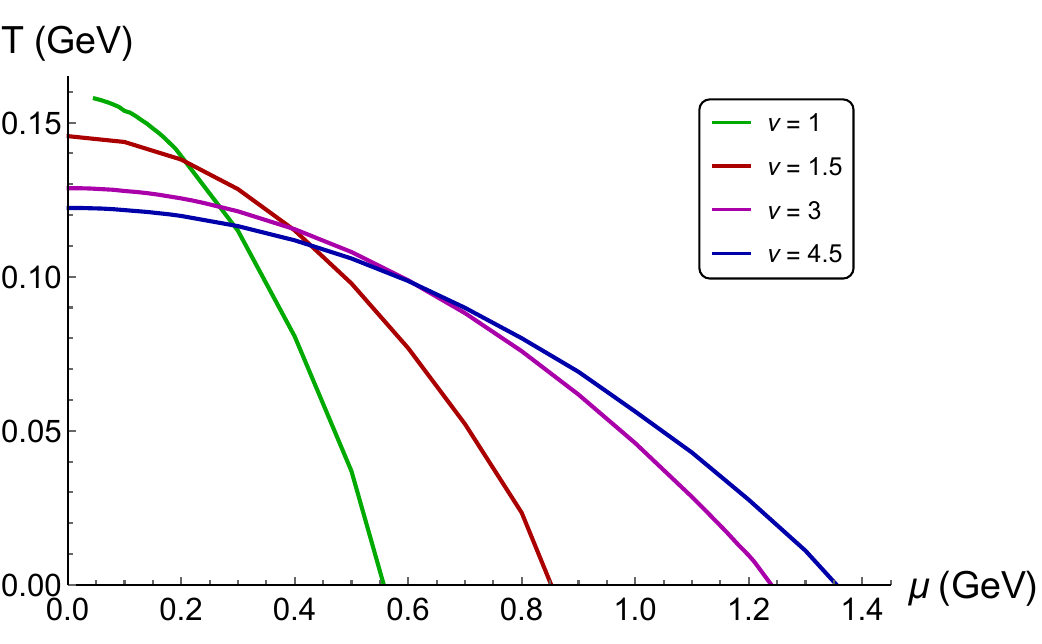}\\
C\hspace{200pt}D 
\caption{ Density plots for the LQ model at zero magnetic field ($c_B = 0$) with spatial anisotropy parameters:  
(A) $\nu=1.5$,  
(B) $\nu=3$,  
(C) $\nu=4.5$.  
(D) First-order phase transition diagrams at $c_B=0$ for the isotropic case ($\nu=1$) and anisotropic cases ($\nu=1.5, 3, 4.5$), from \cite{Arefeva:2022avn}.
}
    \label{Fig:nu15-3-45-cB0}
\end{figure}

\newpage
$\,$
\newpage
$\,$
\newpage
The density plot for the LQ model at zero magnetic field $c_B = 0$ with spatial anisotropies are depicted in 
Figs.\,\ref{Fig:nu15-3-45-cB0}A, \,\ref{Fig:nu15-3-45-cB0}B, and \,\ref{Fig:nu15-3-45-cB0}C corresponding to $\nu=1.5$, $\nu =3$, and $\nu=4.5$, respectively. In Fig.\,\ref{Fig:nu15-3-45-cB0}D the first-order phase transition diagrams at $c_B=0$ are depicted for the isotropic case $\nu = 1$, and the anisotropic cases $\nu = 1.5, 3, 4.5$  \cite{Arefeva:2022avn}.
In the hadronic phase for different spatially anisotropy, the JQ parameter depends mainly on the temperature $T$ and has no significant dependence on the chemical potential $\mu$. However, for QGP phases the JQ parameter depends on both $T$ and $\mu$.

Considering the graphs with density plots at zero magnetic field $c_B=0$ in Fig.\,\ref{Fig:LQnu1cB0} for $\nu=1$, and Fig.\,\ref{Fig:nu15-3-45-cB0} for $\nu=1.5, 3, 4.5$,  we observe that the JQ parameter behavior exhibits discontinuities precisely along the first-order phase transition lines in the $(\mu,T)$-planes. As the anisotropy parameter $\nu$ increases, the phase transition location shifts, causing the transition lines to migrate rightward with a slight downward displacement.

\subsection{Non-zero magnetic field}
\label{sect:NR-LQ-nzero}
Considering a magnetic field oriented along the $x_3$-axis,  in this section  we  mainly study jets propagating along the $x_1$-direction with momentum broadening  perpendicular to it along the $x_2$-direction, i.e. $\hat{q}_2$. In almost all plots in this section we drop the index 2 for the JQ parameter as well as for the IJQ parameter $a$.  
\subsubsection{Non-zero magnetic field, $\nu=1$}
\label{sect:NR-LQ-nzero-nu1}
\begin{figure}[h]
  \centering
\includegraphics[scale=0.23]
{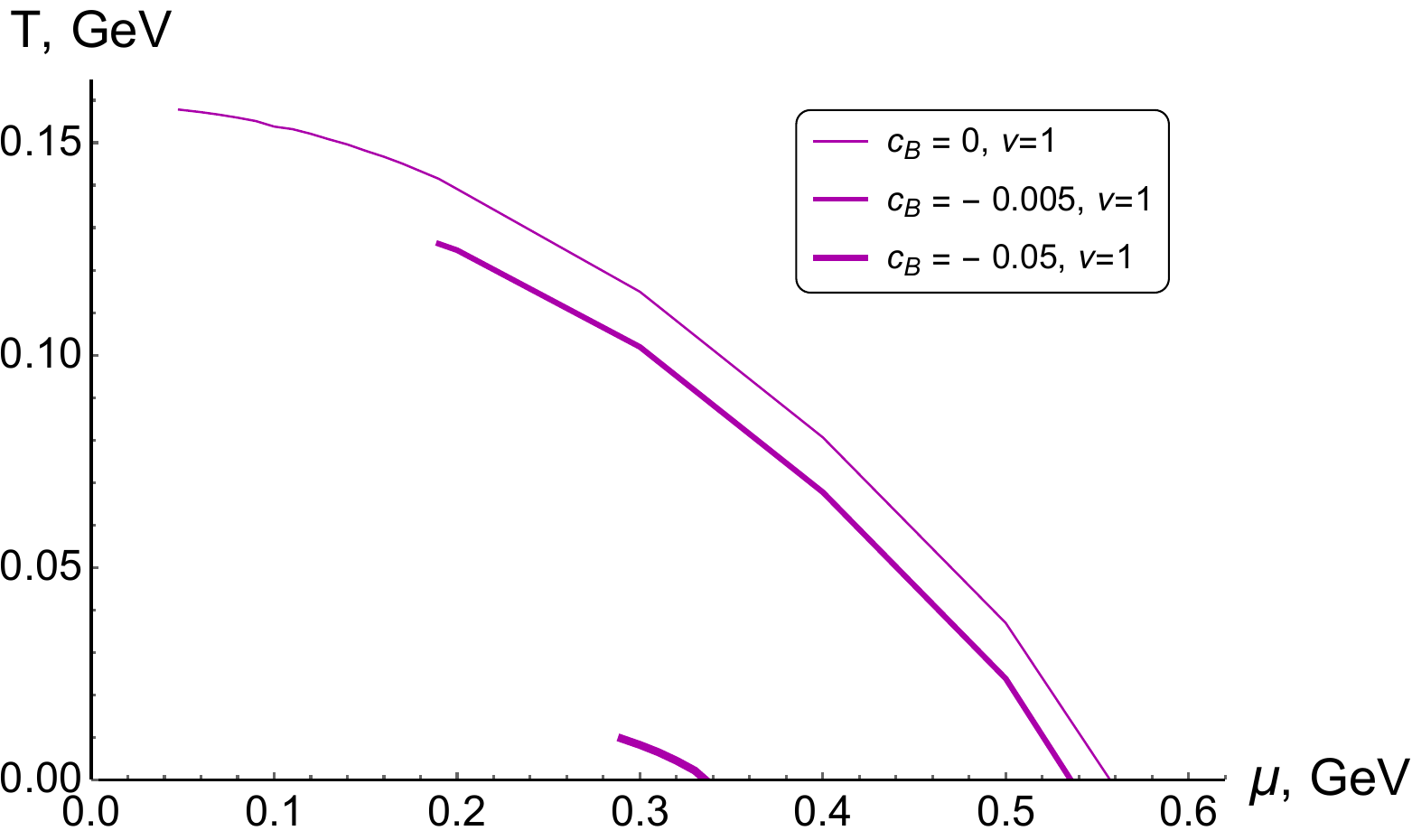}
\quad\includegraphics[scale=0.30]{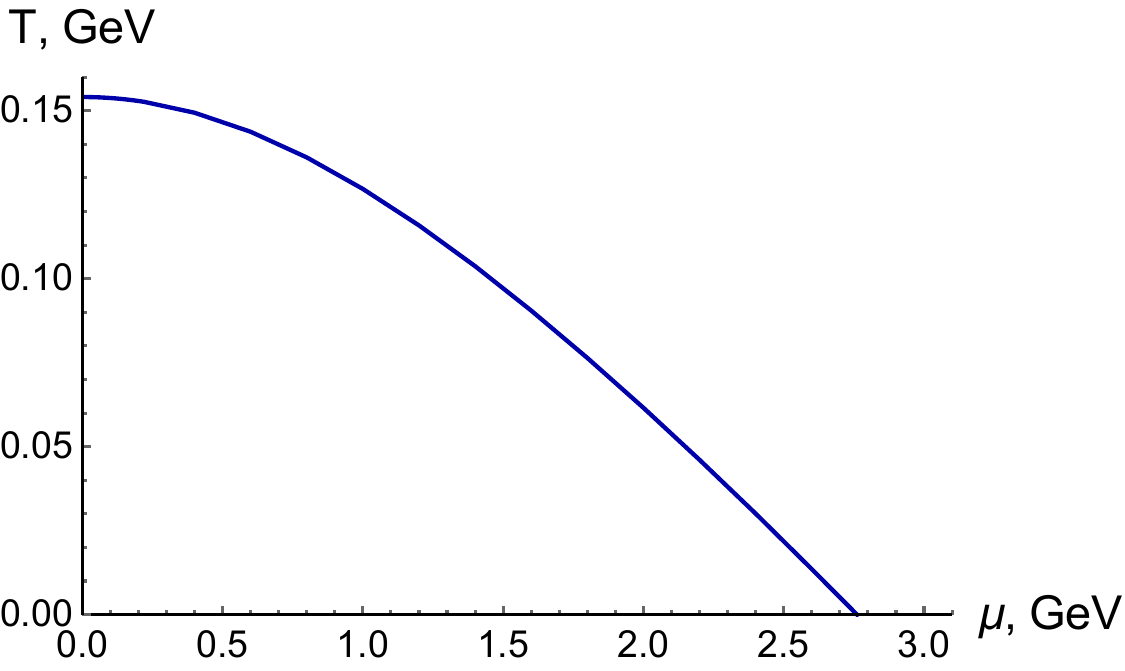}\\
A\hspace{150pt}B\\
\caption{(A) The first-order phase transition lines for the LQ model with $\nu=1$ and different $c_B$. (B) The second-order phase transition line for the LQ model with $\nu=1$ at $c_B = 0$. The confinement/deconfinement phase transition  disappears for $c_B = -0.005$ GeV$^2$, and $c_B =-0.05$ GeV$^2$.
  }
  \label{Fig:LQ-PTnu1cB0-0005-005}
\end{figure}

In this section we examine magnetic field effects on the JQ parameters, beginning with the $\nu=1$ case. Phase diagrams for different magnetic field strengths ($c_B$ values) are shown in Fig.\,\ref{Fig:LQ-PTnu1cB0-0005-005}. panel (A) displays first-order transition lines for the light quark holographic model, while panel (B) demonstrates magnetic field modification of the confinement/deconfinement transition:  increasing $|c_B|$ (for $c_B<0$) suppresses the confinement phase \cite{Arefeva:2022avn}.
This aligns with \cite{Arefeva:2024xmg}, where increasing magnetic fields reduce the running coupling. Furthermore, \cite{Arefeva:2024vom} demonstrates that under physical boundary conditions, the running coupling decreases during hadronic-to-QGP phase transitions. Collectively, these findings indicate magnetic fields restrict systems exclusively to the QGP phase. This suppression of confinement is explicitly shown in Fig.\,\ref{Fig:LQ-PTnu1cB0-0005-005} for $c_B = -0.005$ GeV$^2$, and $c_B =-0.05$ GeV$^2$, where the confinement/deconfinement phase transition is absent.
\\

\begin{figure}[h!]
  \centering
\includegraphics[scale=0.23]
{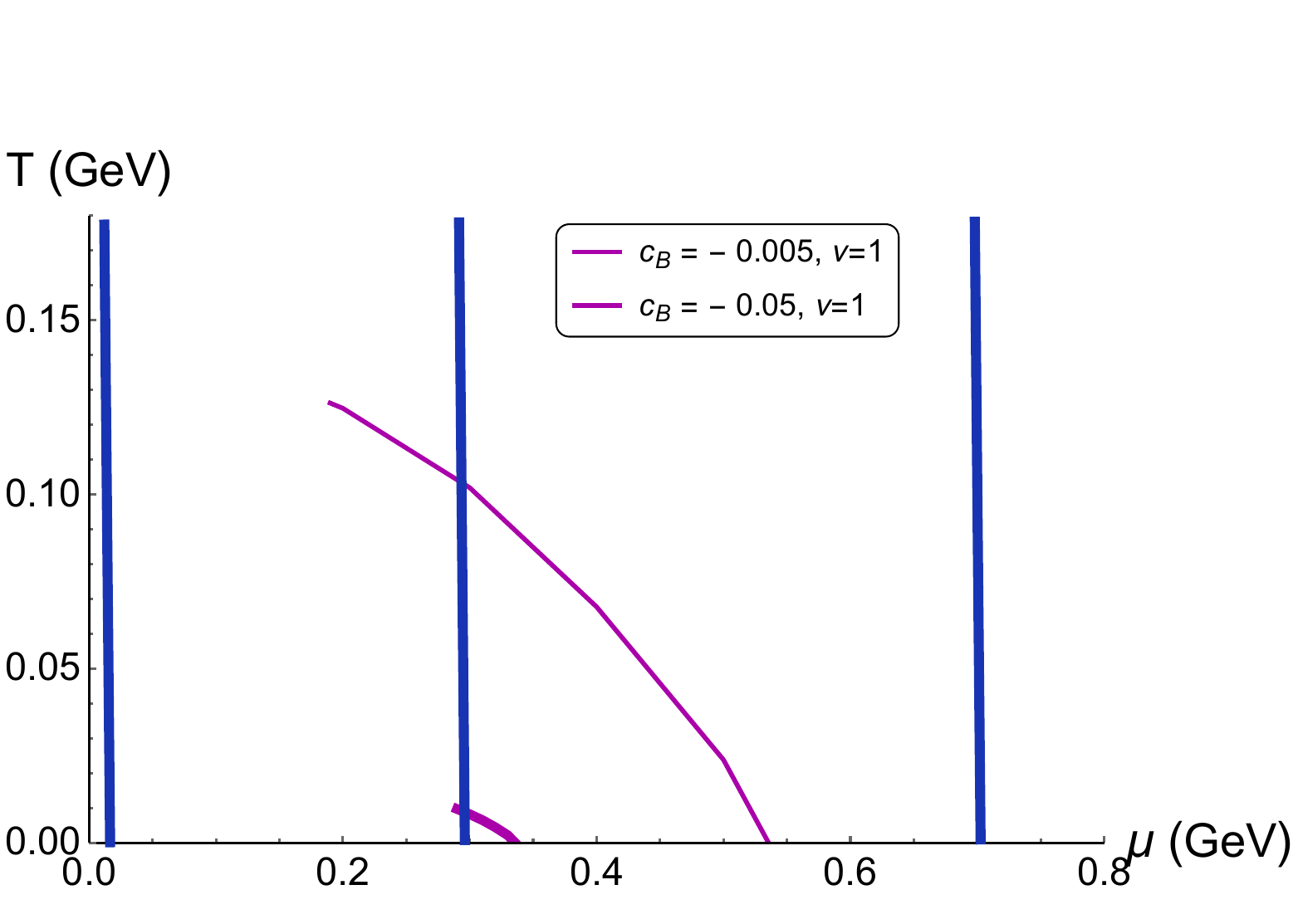}\\
 A\\$\,$\\
\includegraphics[scale=0.2]
   {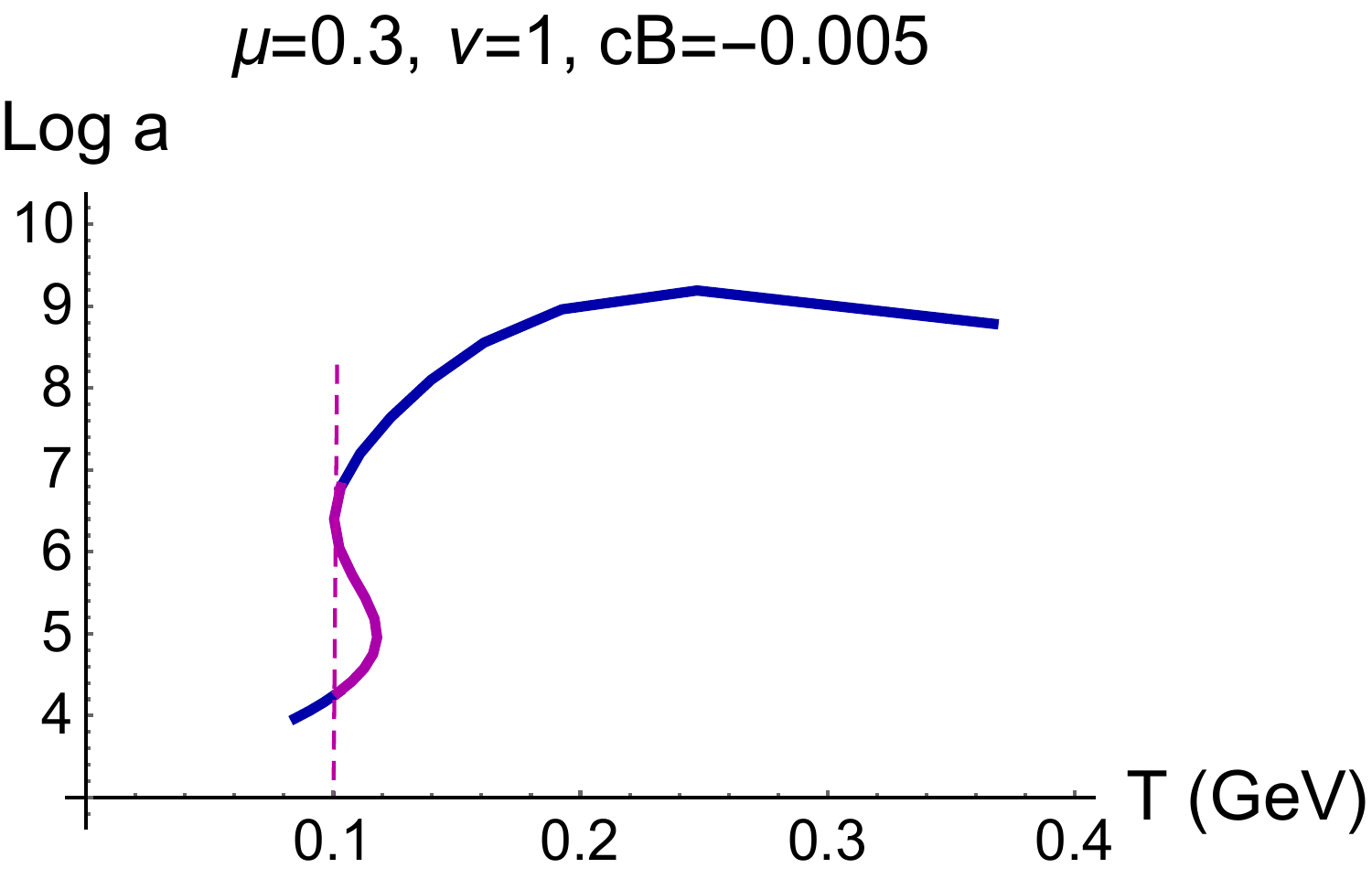} \quad \includegraphics[scale=0.3]  {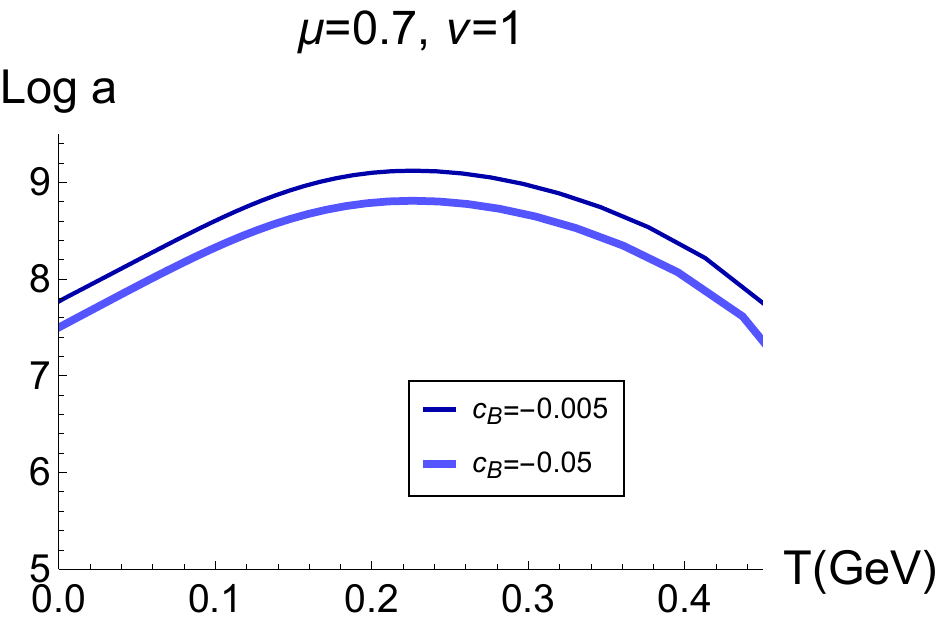}\\
   B\hspace{160pt}C\\
 \includegraphics[scale=0.17]{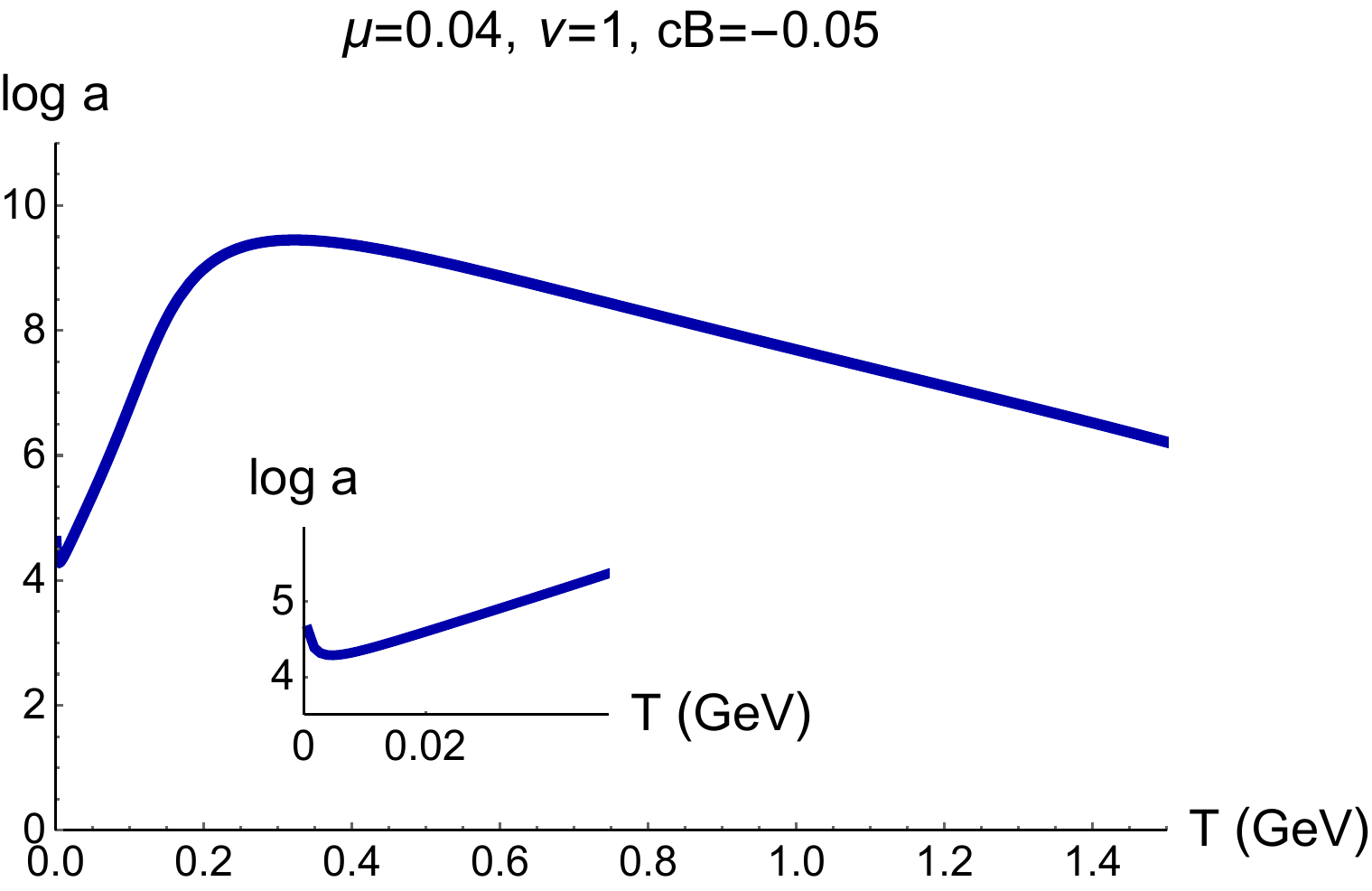}
 \includegraphics[scale=0.25] 
 {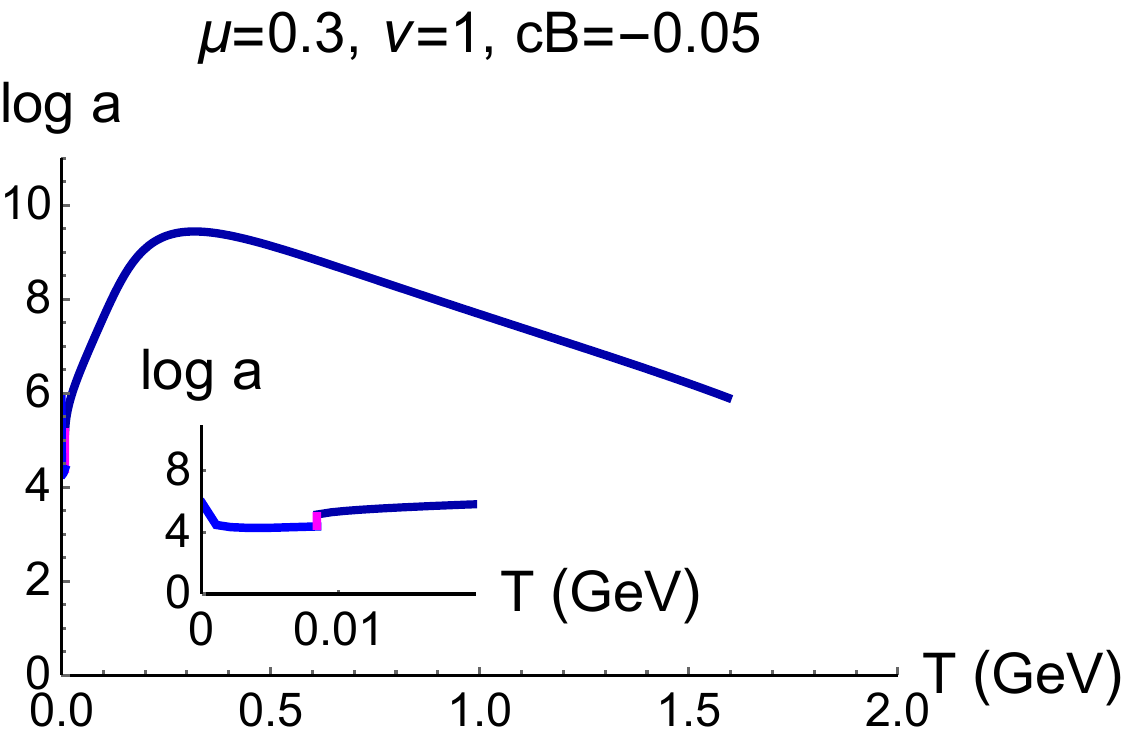}\quad
 \includegraphics[scale=0.21] 
 {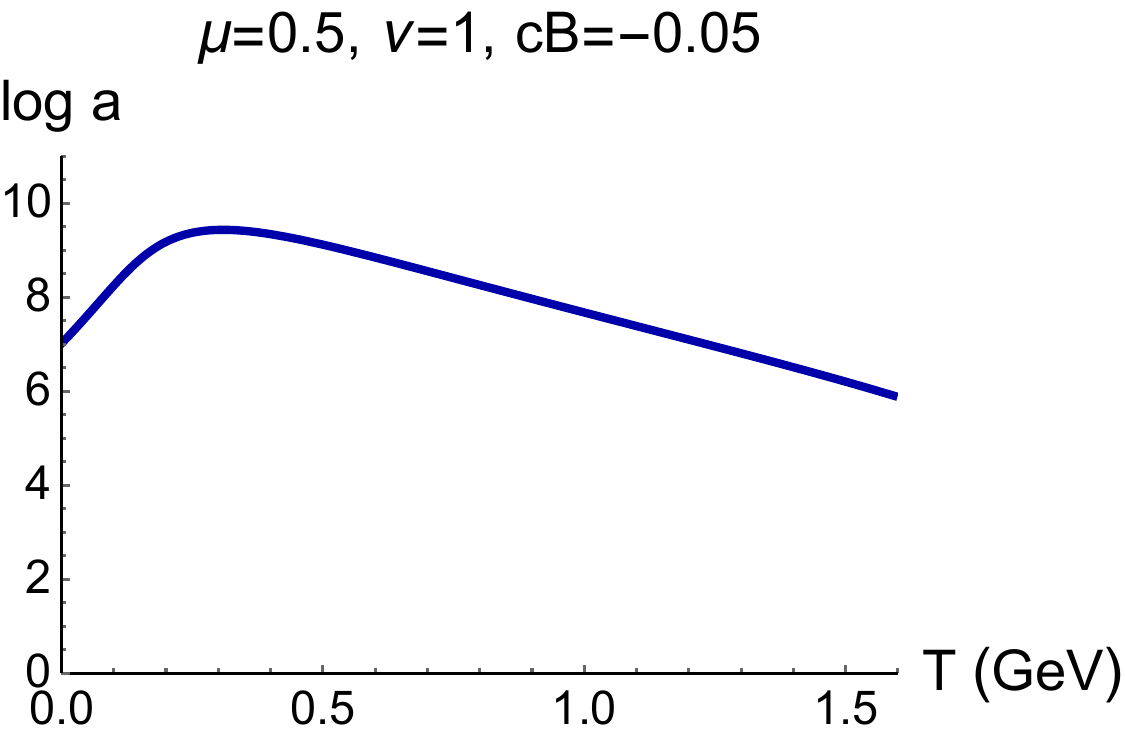}\\
D\hspace{120pt}E\hspace{120pt}F
\caption{
 (A) We calculate the JQ parameter for the LQ model with $\nu=1$ at $c_B=-0.005$ GeV$^2$  and $c_B=-0.05$ GeV$^2$ along vertical lines (constant $\mu$) in the $(\mu,T)$-plane at $\mu = 0.04$, $0.3$, $0.5$ and   $\mu=0.7$ GeV, shown in panel. Segments of these lines are colored blue (QGP), brown (hadronic), and green (quarkyonic) according to the phase traversed. The resulting plots of $\log a$ versus temperature are displayed in the bottom panels (B, C, D, E, F), using the same color scheme.
}
\label{Fig:LQnu1cB0005005}
  \end{figure}

Fig.\,\ref{Fig:LQnu1cB0005005} shows\footnote{Comparing the plots in panels B and E, it is worth noting that panel (B) details the behavior within an unphysical region (which lacks essential physical significance), while panel (E) focuses on a jump occurring in the physical region. In what follows, one of these versions will be presented without further comment. This distinction arises from the different Mathematica codes used to generate the corresponding figures.} discontinuities in the JQ parameter at first-order phase transitions for $c_B=-0.005$  GeV$^2$  and $c_B=-0.05$  GeV$^2$ at $\mu=0.3$ GeV. These discontinuities occur within the QGP phase. After each transition, $\log a$ initially increases with temperature (weakening JQ) until $T \approx 0.25$-$0.3$ GeV (depending on $c_B$ and $\mu$), then decreases (enhancing JQ) at higher temperatures.
Fig.\,\ref{Fig:LQnu1cB0005005} demonstrates that the JQ parameter behavior identifies first-order phase transitions even under external magnetic fields, consistent with the $c_B=0$ results in the previous subsection.

\begin{figure}[h!]
  \centering
\includegraphics[scale=0.1] {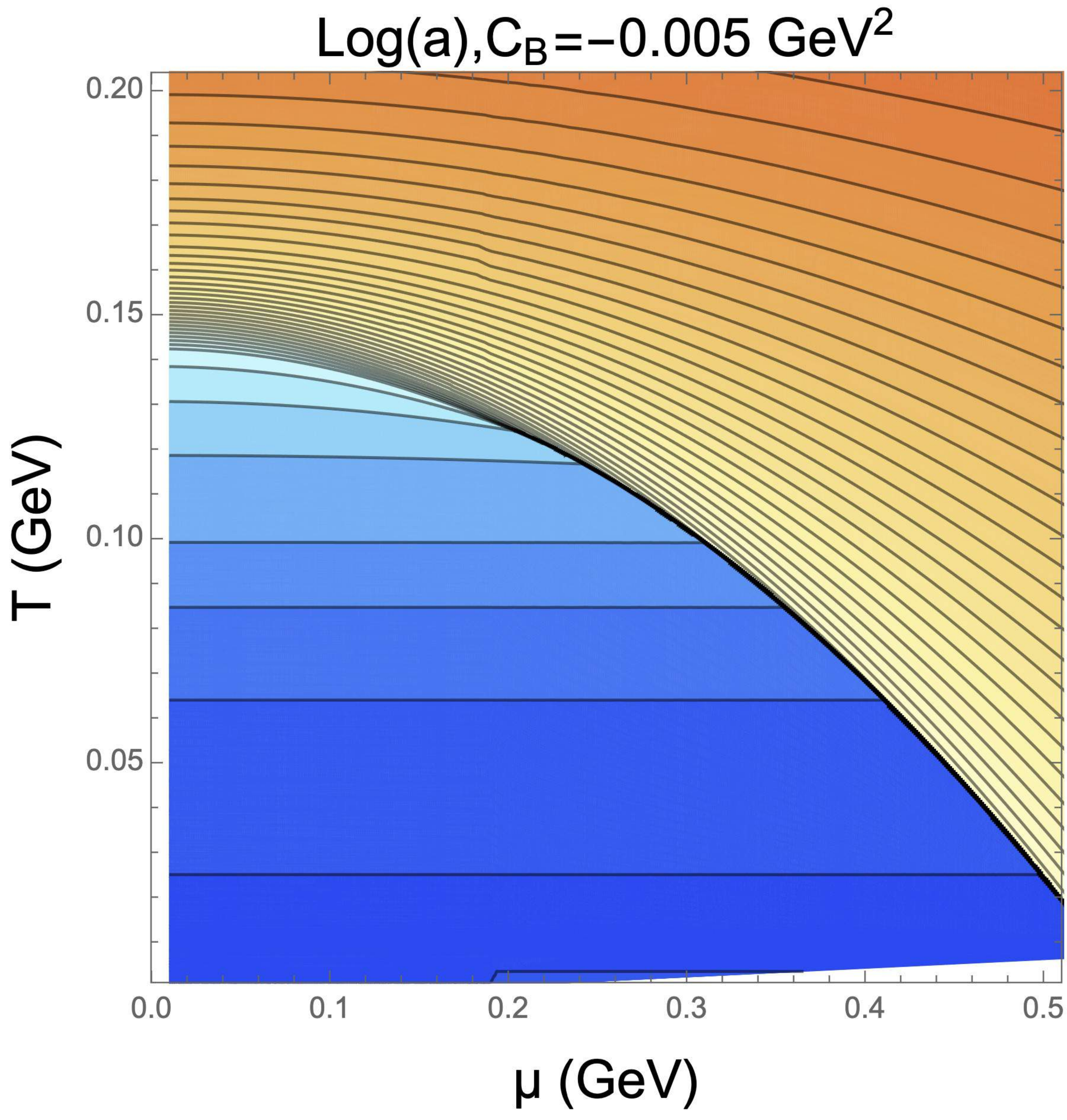}
  \qquad
\includegraphics[scale=0.8]{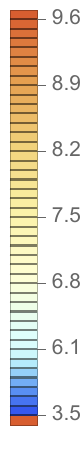}
\includegraphics[scale=0.1]{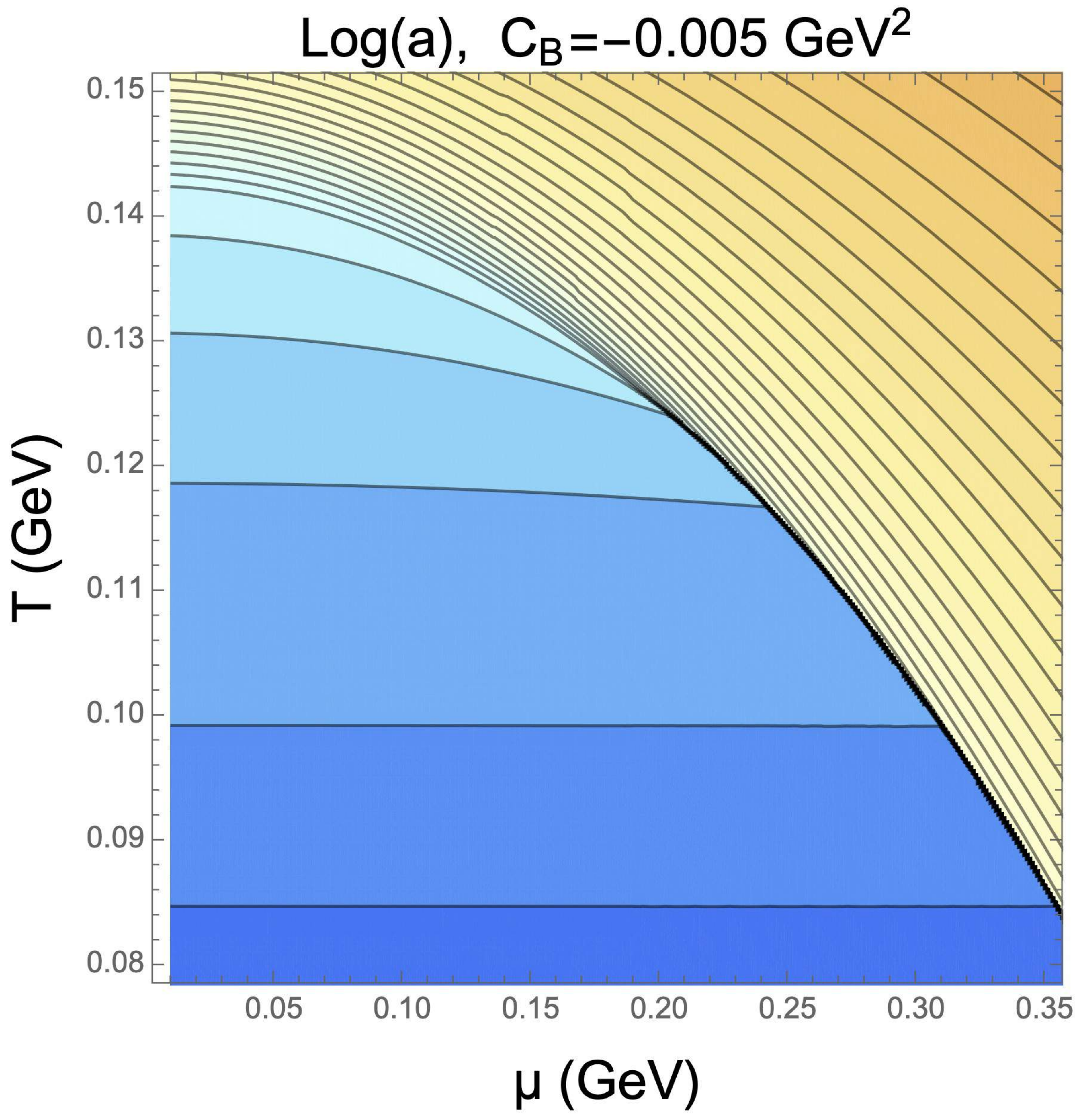} 
   \\
A\hspace{200pt}B
\caption{(A) Density plots of $\log a$ at $\nu=1$ for the LQ model in the presence of the magnetic field $c_B=-0.005$ GeV$^2$, and (B) the zoom view of panel (A) near the CEP at ($\mu \approx 0.19$ GeV , $T \approx0.135$ GeV). }
\label{Fig:LQnu1cB0005}
\end{figure}

\begin{figure}[h]
  \centering
\includegraphics[scale=0.1]
 {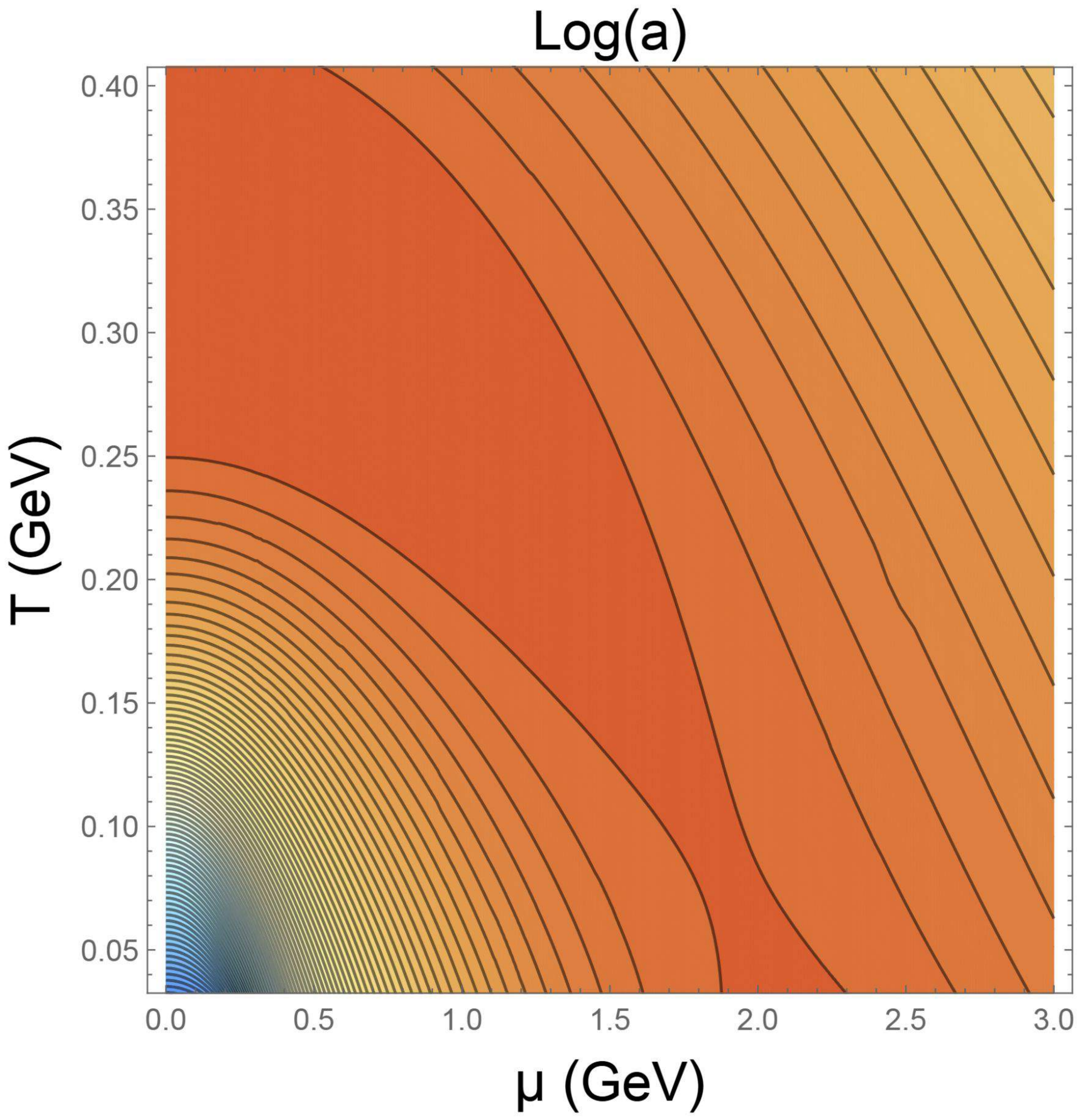}
  \qquad \includegraphics[scale=0.3]{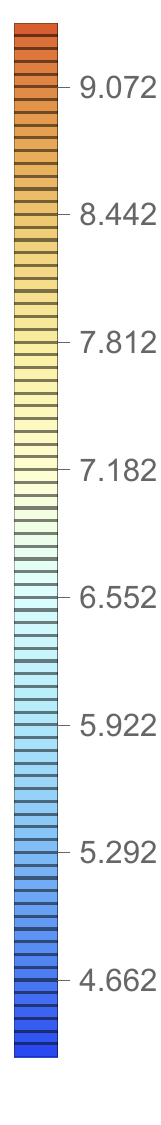}
\includegraphics[scale=0.097]
   {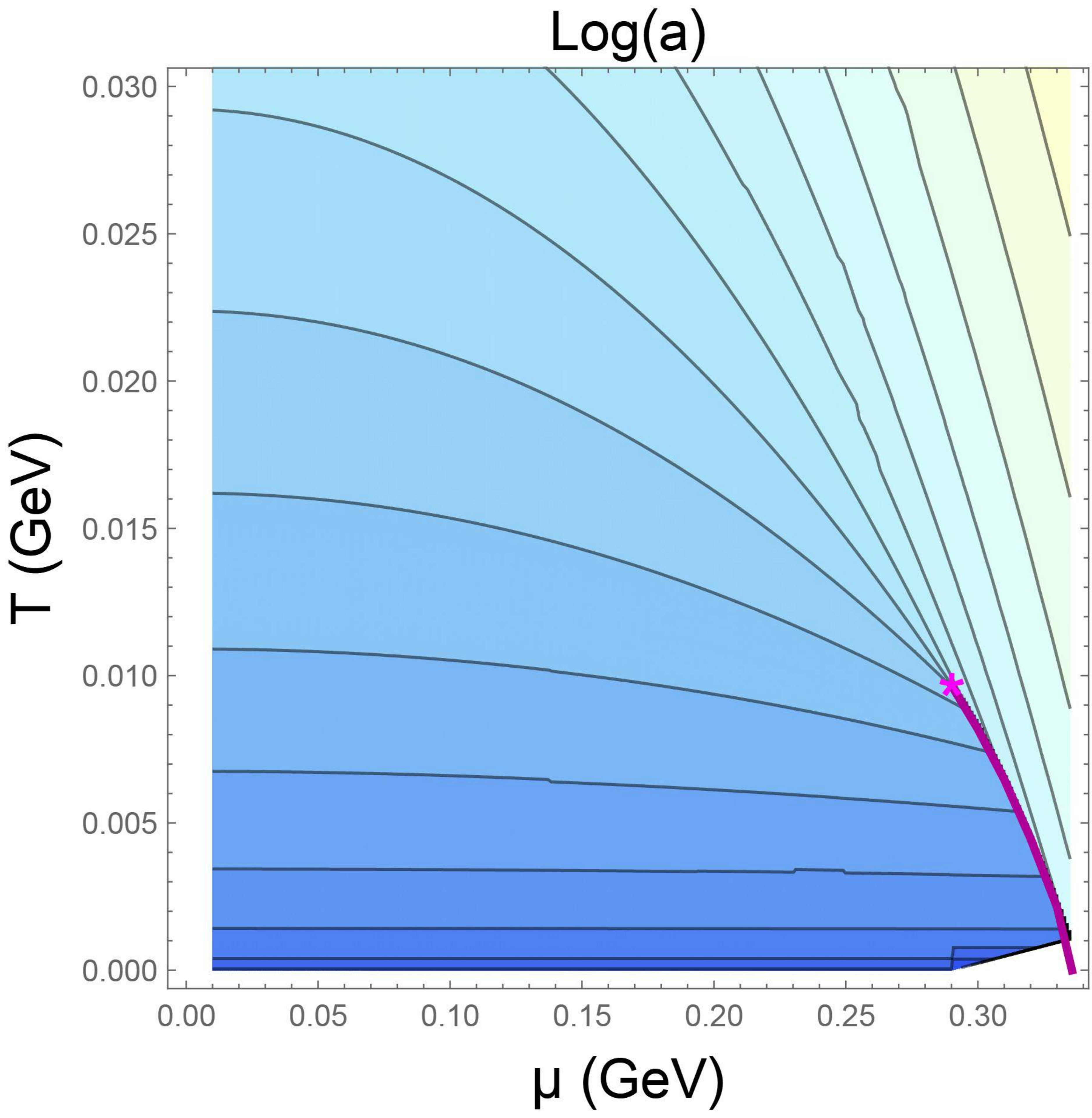} 
   \\ 
   A\hspace{220pt}B\\  
\includegraphics[scale=0.095]
   {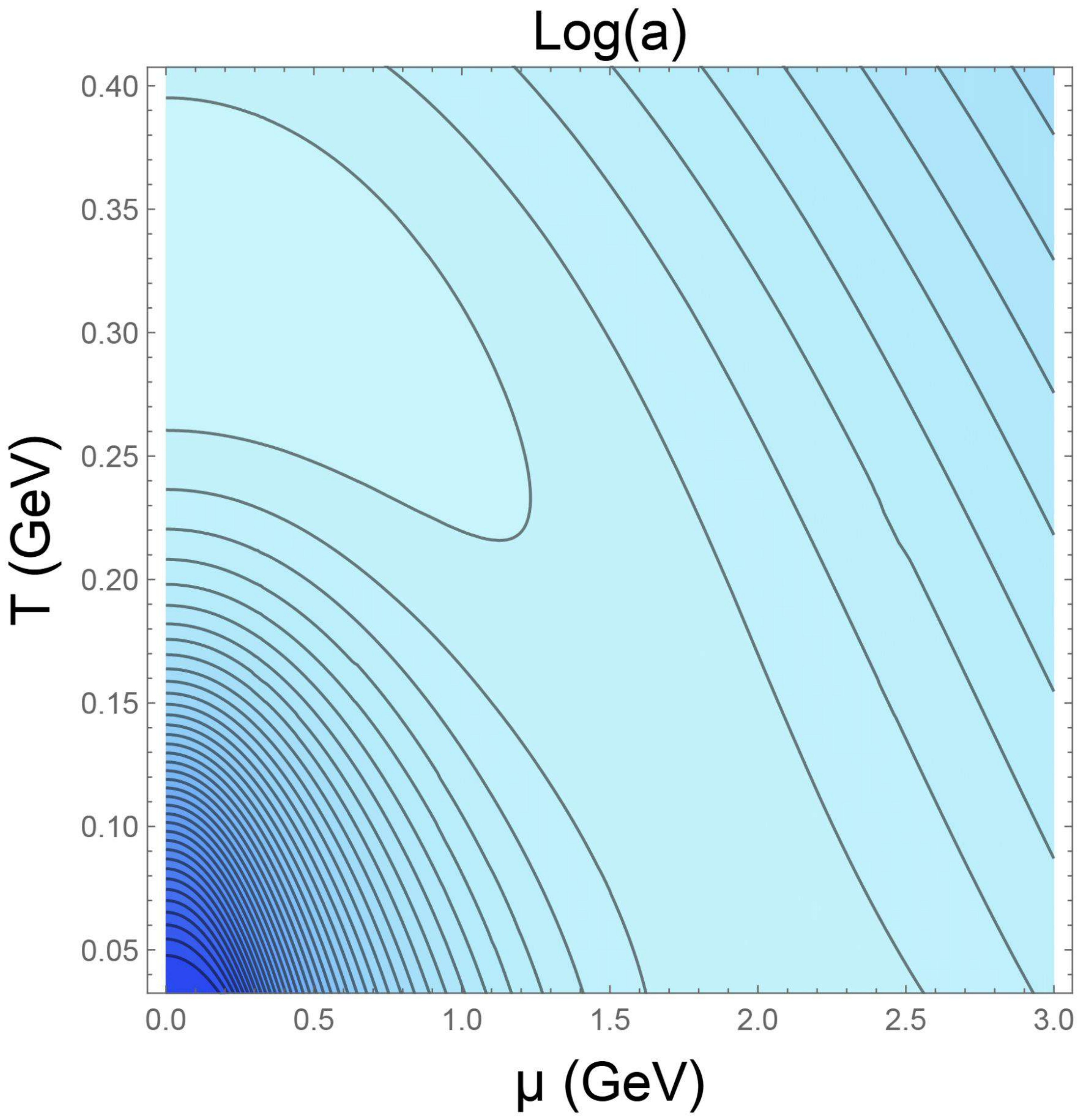}
 \includegraphics[scale=0.45]
  {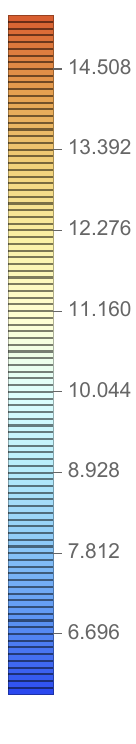}
  \qquad  \includegraphics[scale=0.32]
  {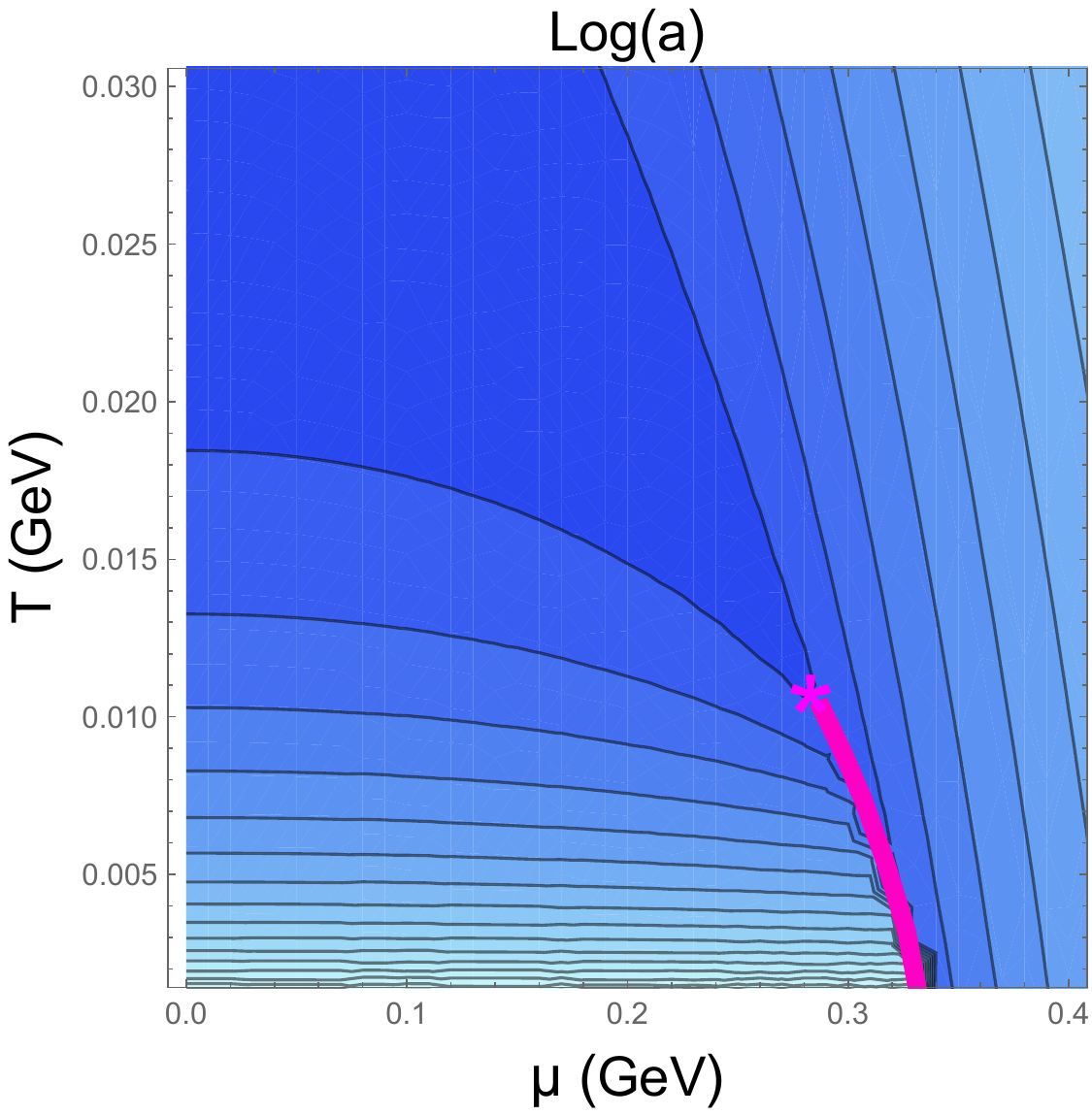} 
   \includegraphics[scale=0.25]
   {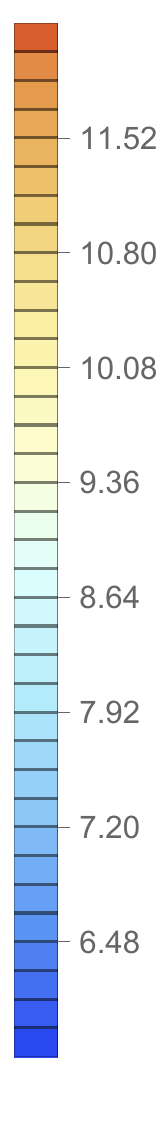} 
   \\
C\hspace{220pt}D
  \caption{(A) Density plots for $\log a_2$ at $\nu=1$ for the LQ model in the presence of the magnetic field $c_B=-0.05$ GeV$^2$, and (B) the zoom of panel (A)  with the phase transition  line. (C) Density plots for $\log a_3$ with the same $\nu$ and $c_B$, and (D) the zoom of panel (C)  with the phase transition  line. The magenta line shows the first-order transition line that starts at CEP with coordinates $(\mu, T) \approx$ (0.29 GeV, 0.01 GeV)  shown by the magenta star. 
 }
  \label{Fig:LQ3nu1cB005}
\end{figure}

Fig.\,\ref{Fig:LQnu1cB0005} and Fig.\,\ref{Fig:LQ3nu1cB005} show $\log a$ density plots at $\nu=1$ for $c_B = -0.005$ \GG and $c_B = -0.05$ \GG , respectively. Fig.\,\ref{Fig:LQnu1cB0005}B zoom in panel (A) near the CEP at $(\mu, T) \approx$ (0.19 GeV, 0.135 GeV). Fig.\,\ref{Fig:LQ3nu1cB005}A and Fig.\,\ref{Fig:LQ3nu1cB005}C depict the density plots for $\log a_2$ and $\log a_3$ experiencing  the hill-like structure located
above the phase transition line.

Figs.\,\ref{Fig:LQ3nu1cB005}B and \ref{Fig:LQ3nu1cB005}D are zooms of  Fig.\,\ref{Fig:LQ3nu1cB005}A and \ref{Fig:LQ3nu1cB005}B with  phase transition markers: magenta curves for first-order transitions and a magenta star for the CEP at $(\mu, T) \approx$ (0.29 , 0.01) (GeV).



\begin{figure}[h]
  \centering
\includegraphics[scale=0.28]
  {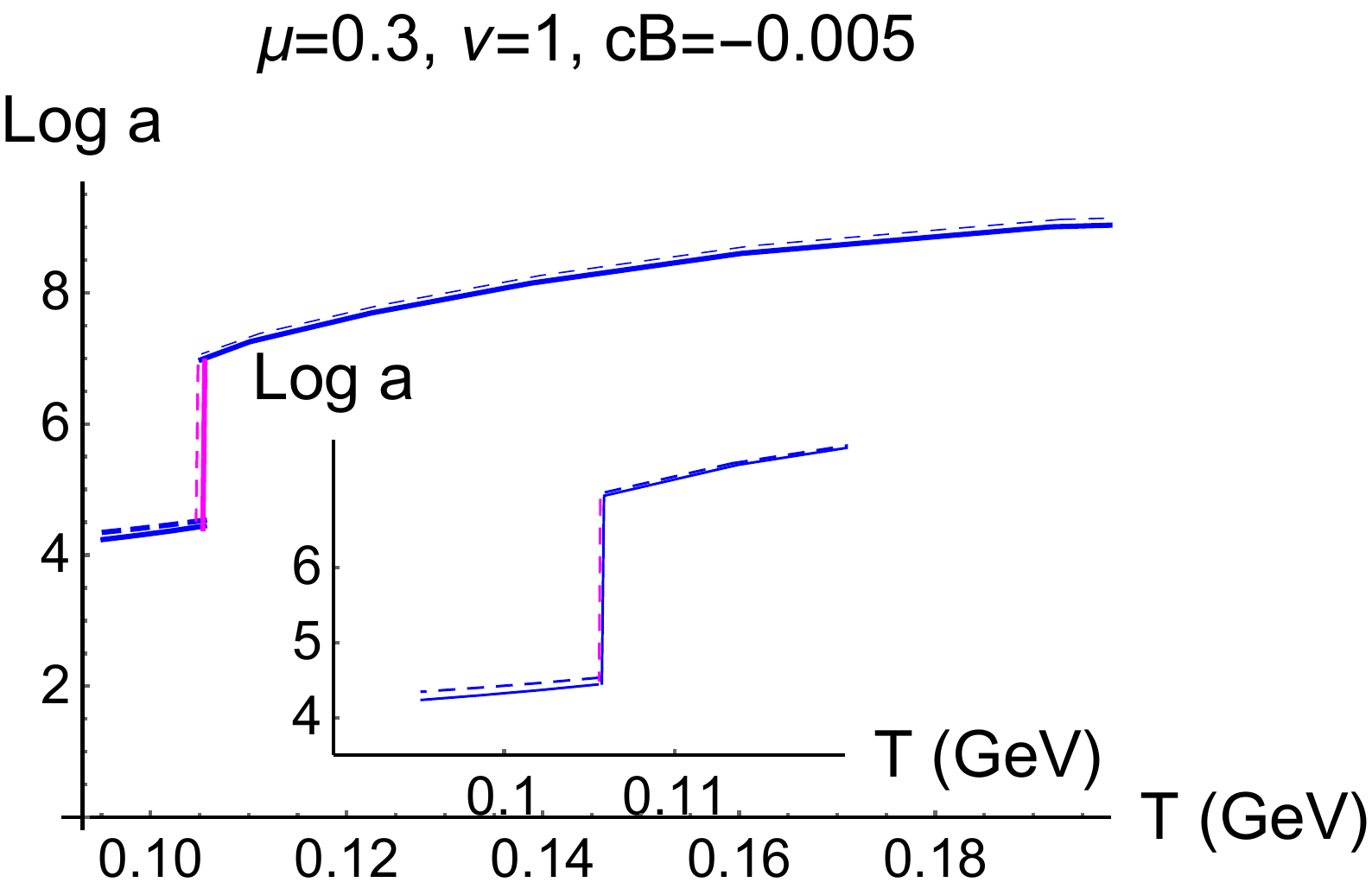}
\caption{ $\log a_2$ (solid lines) and $\log a_3$ (dashed lines) for the LQ model with $\nu = 1$ and $c_B = -0.005$ \GG .
}
\label{Fig:LQnu1q23new}\end{figure}

\begin{figure}[h]
  \centering
\includegraphics[scale=0.15]
  {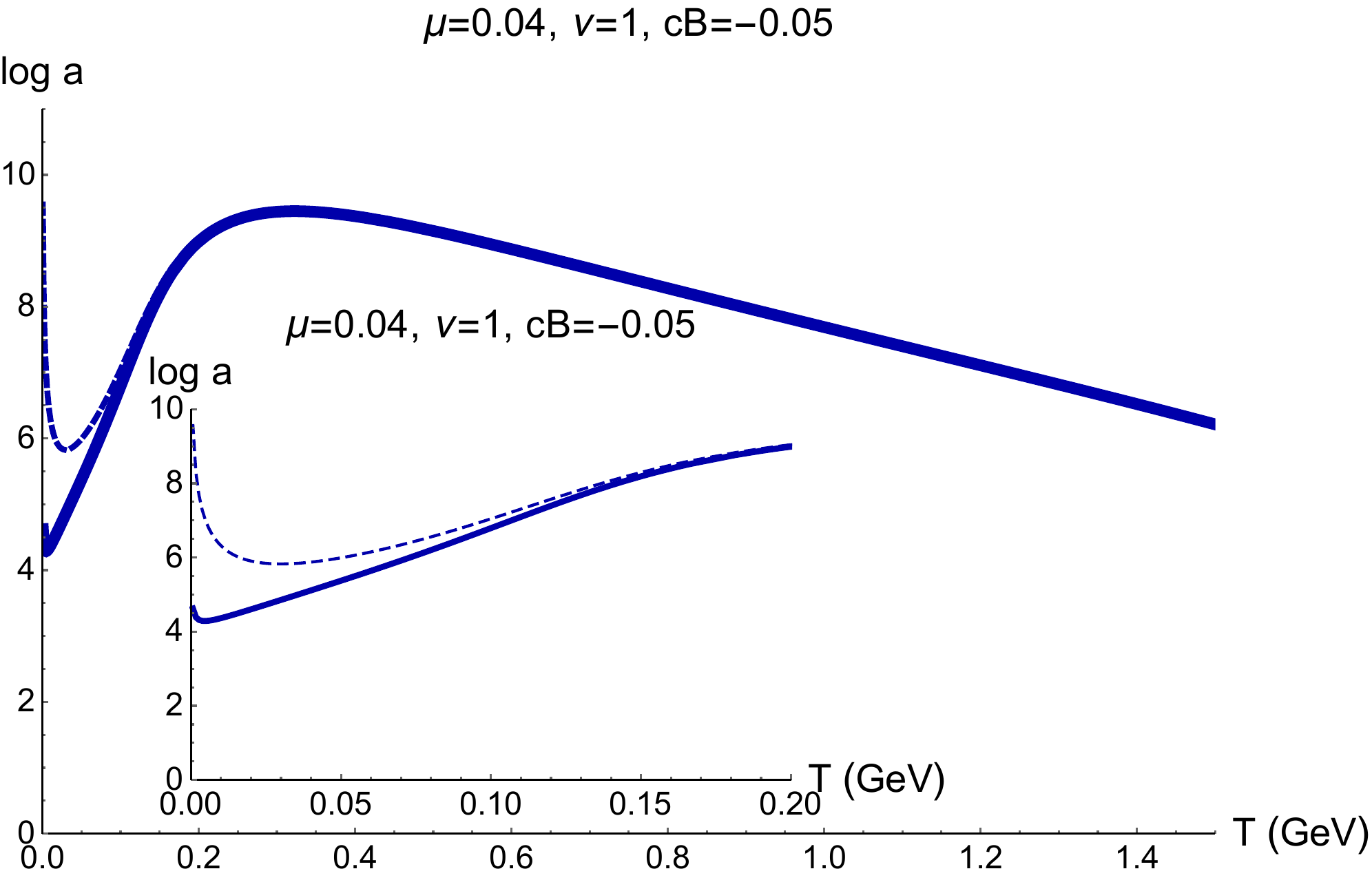}
\includegraphics[scale=0.16]
  {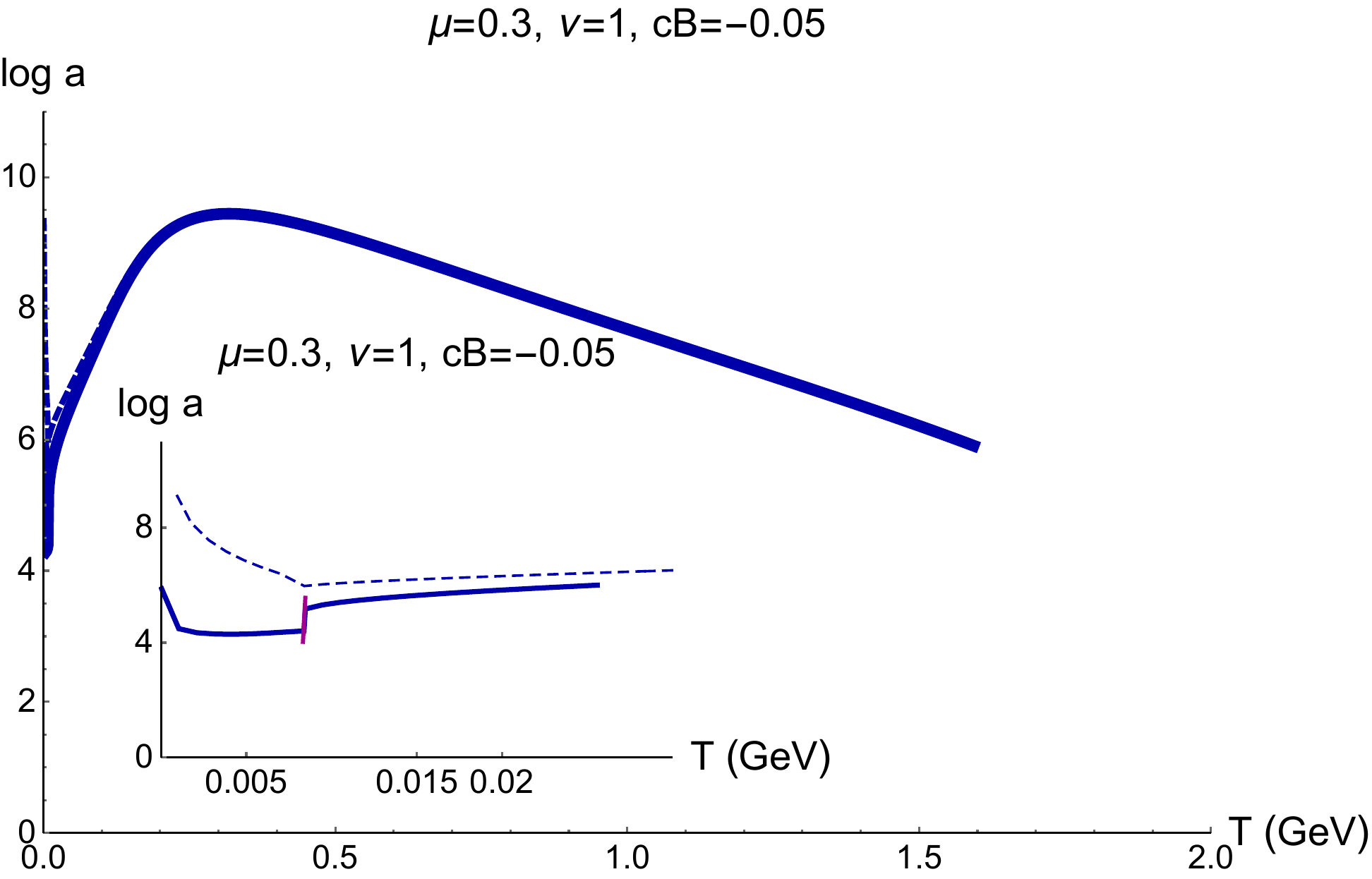
 }
\includegraphics[scale=0.26]
  {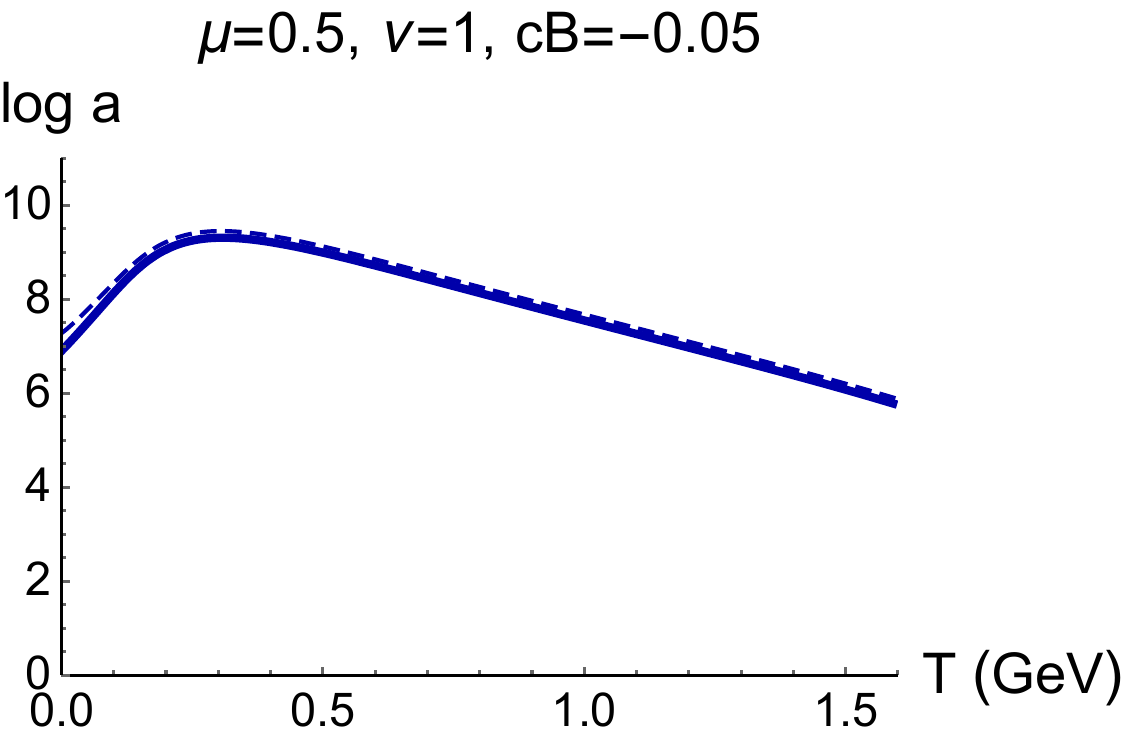}\\A\hspace{150pt}B\hspace{130pt}
  C
\caption{$\log a_2$ (solid lines) and $\log a_3$ (dashed lines) for the LQ model with $\nu = 1$ and $c_B = -0.05$ \GG  versus temperature at fixed chemical potential: (A) $\mu = 0.0.04$ GeV, (B) $\mu = 0.3$ GeV, and (C) $\mu = 0.3$ GeV. In panels (A) and (B) the zoom plots are located inside the basic plots.  
}
\label{Fig:LQnu1q23new2}\end{figure}

We investigated the directional dependence of the JQ parameter by comparing $\hat{q}_3$ (momentum broadening parallel to the magnetic field) and $\hat{q}_2$ (momentum broadening perpendicular to the magnetic field). We observe a subtle but weak orientation dependence:

\begin{itemize}
\item Fig.\,\ref{Fig:LQnu1q23new} and Fig.\,\ref{Fig:LQnu1q23new2} show that for $\nu = 1$, the JQ parameter exhibits anisotropy in the presence of a magnetic field: $\log a_2 \neq \log a_3$ for $c_B = -0.05$ \GG. 

\item 
For all values of the temperature, below and above the first-order phase transition we have $\log a_2 < \log a_3$. 
\item The behavior of JQ parameter in the high-temperature regime is independent of orientations, i.e. increasing $T$ leads to enhancement in JQ parameter.
\end{itemize}

\newpage
$$\,$$
\newpage
$$\,$$
\newpage
$$\,$$
\newpage

\subsubsection{Non-zero magnetic field, $\nu=1.5$}

\begin{figure}[h]
  \centering
\includegraphics[scale=0.18]
{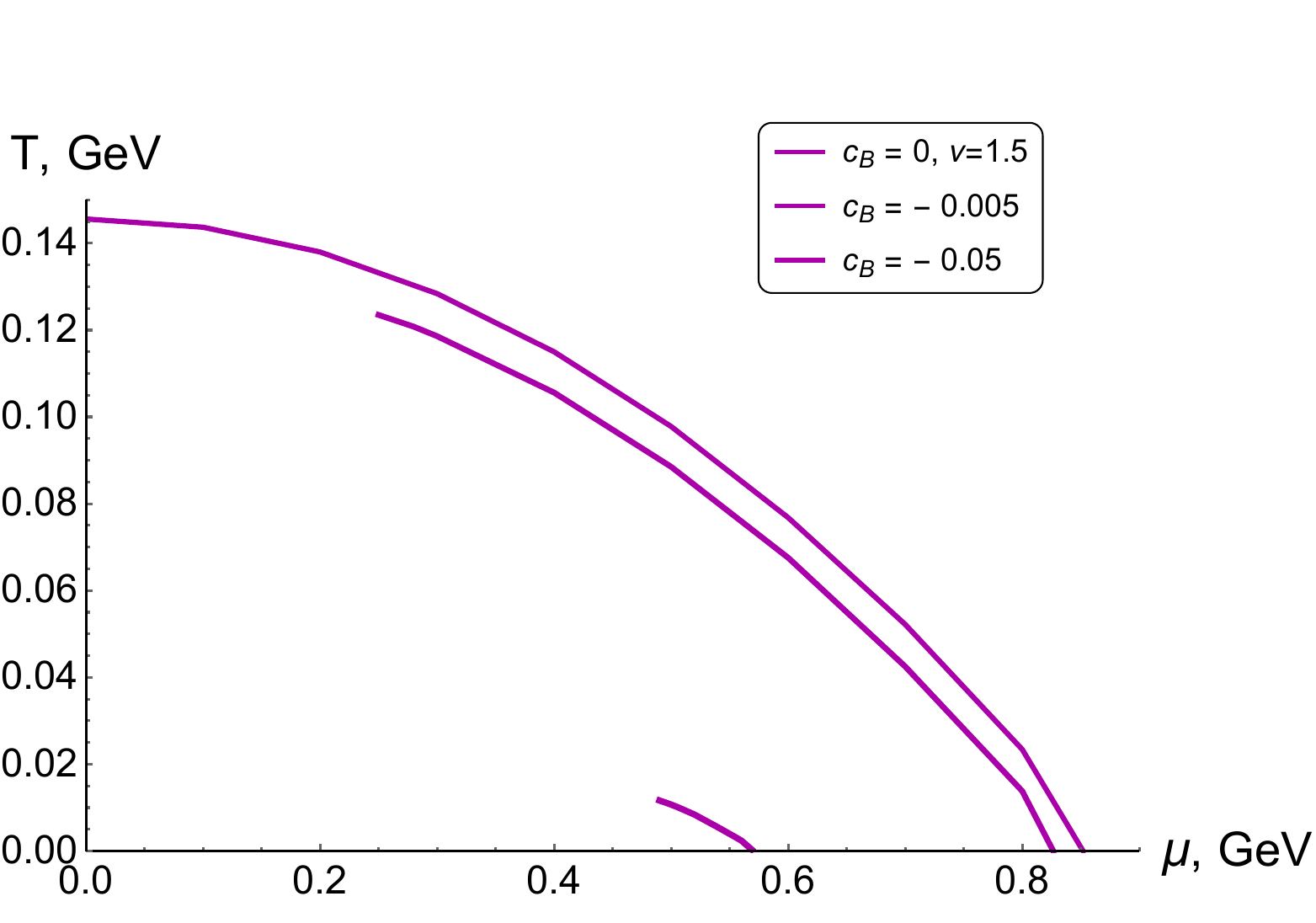}
\quad\includegraphics[scale=0.3]{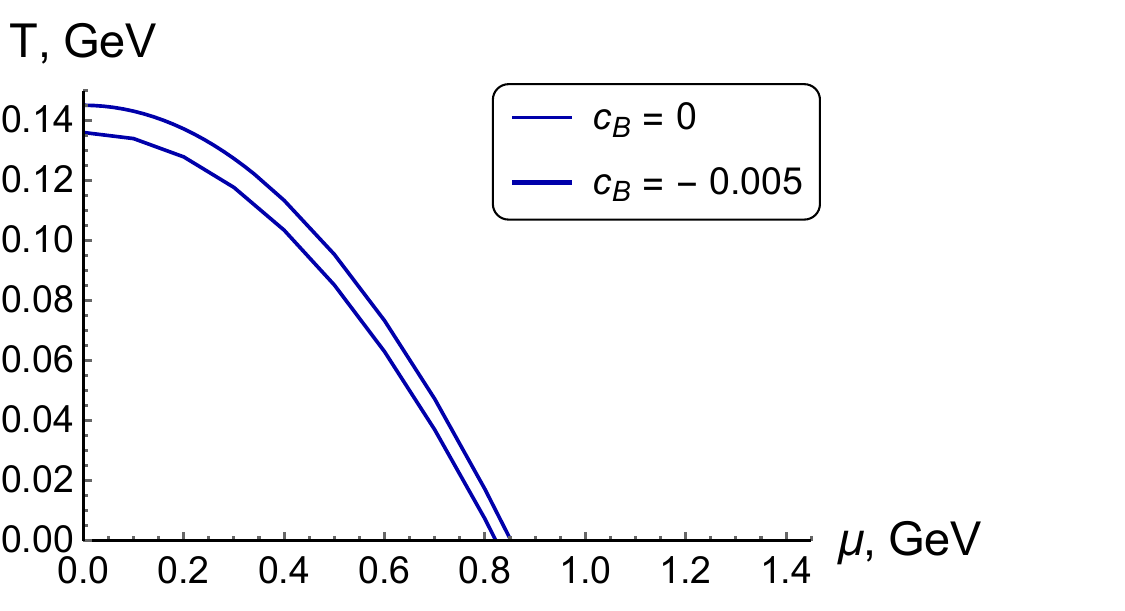}\\
A\hspace{150pt}B
\caption{
(A) The first-order phase transition lines with $c_B=0, -0.005,  -0.05$ \GG, and (B) confinement/deconfinement transition lines with $c_B=0, -0.005$ \GG\, for the LQ model with $\nu=1.5$.
  }
  \label{Fig:LQPTnu15cB0-005-0005}
\end{figure}

\begin{figure}[h!]
  \centering
\includegraphics[scale=0.18]{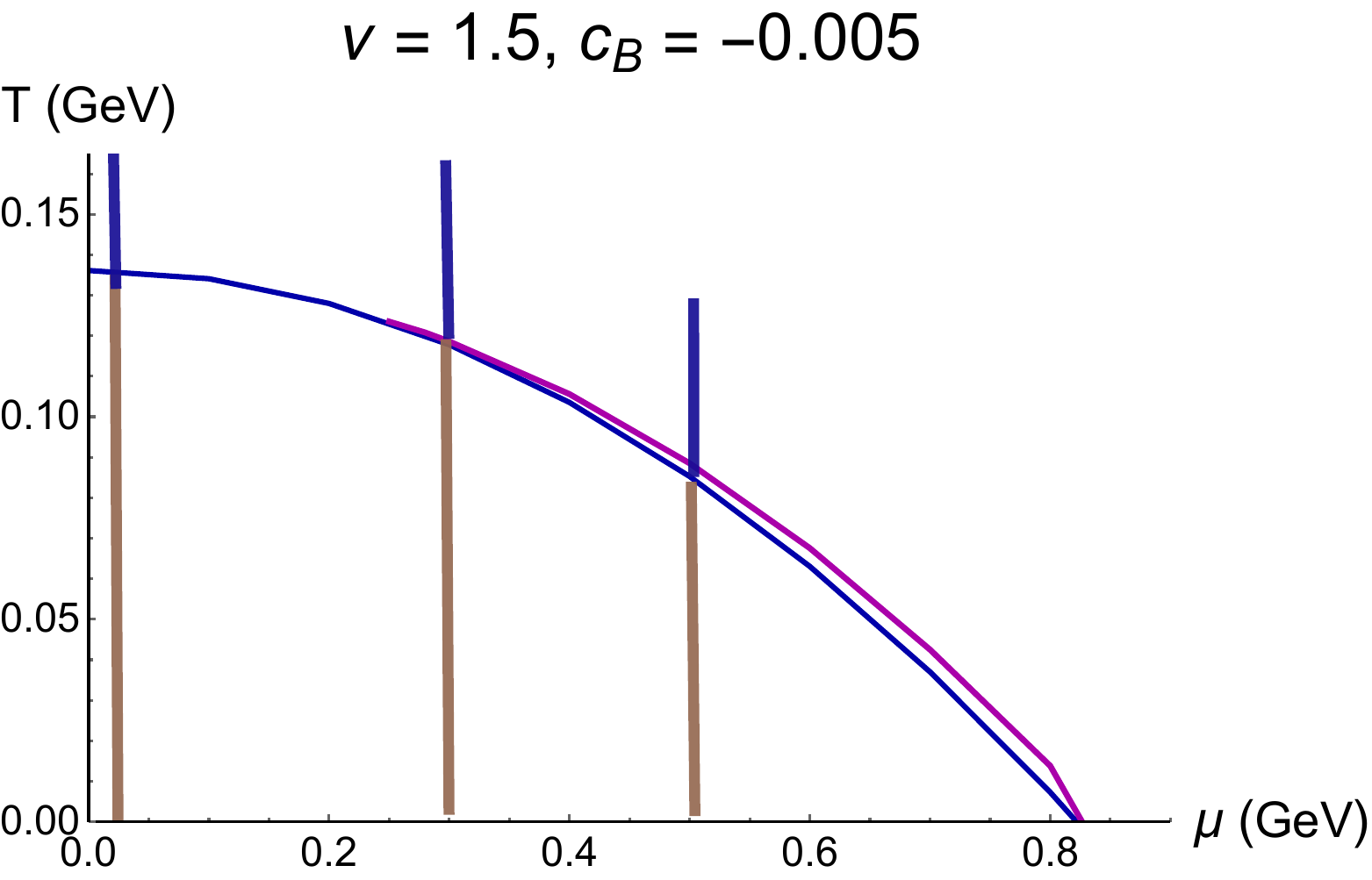}
\qquad
\includegraphics[scale=0.18]
{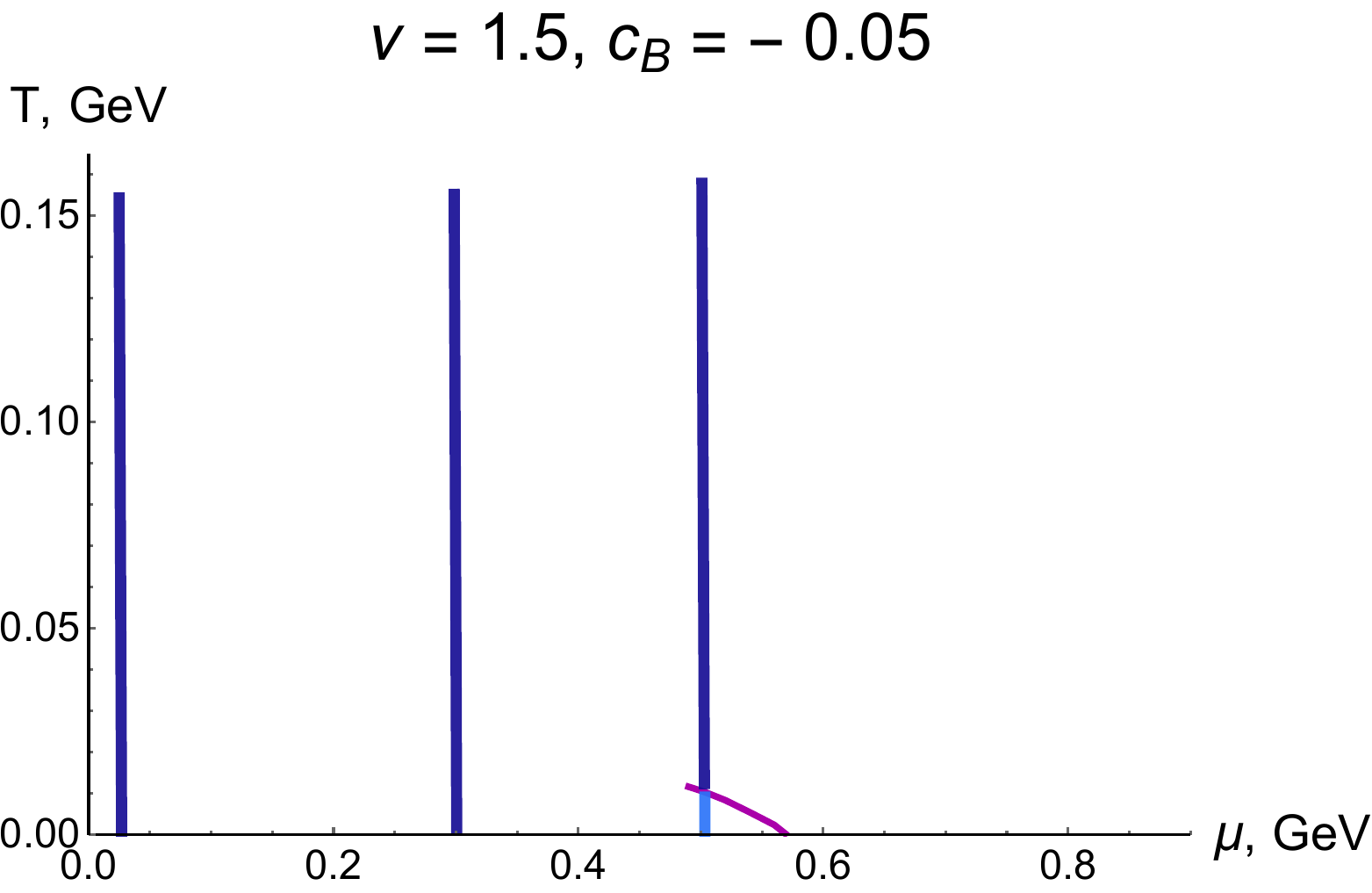}\\
A\hspace{170pt}B\\
\includegraphics[scale=0.33]
   {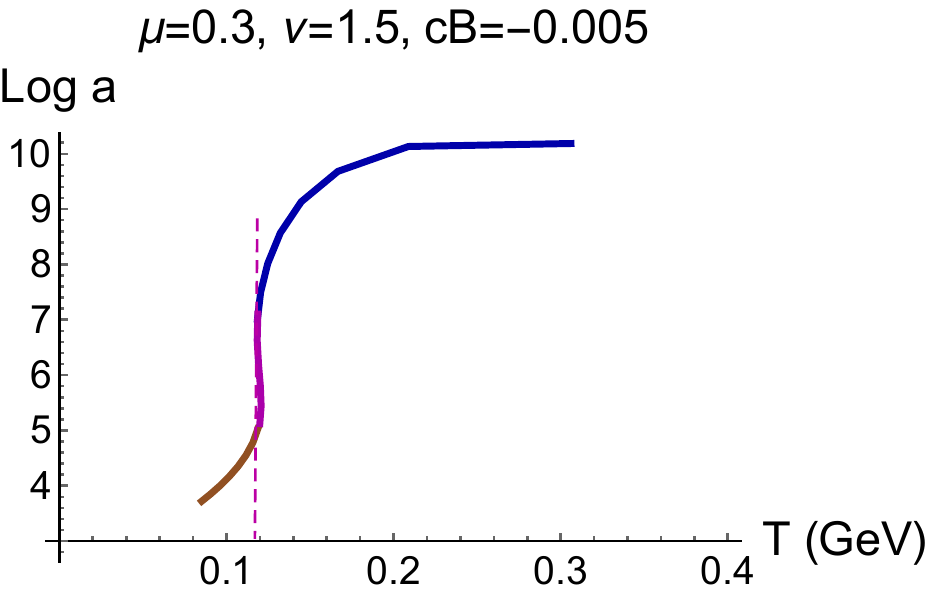}\\C\\$\,$\\
 \includegraphics[scale=0.19]
  {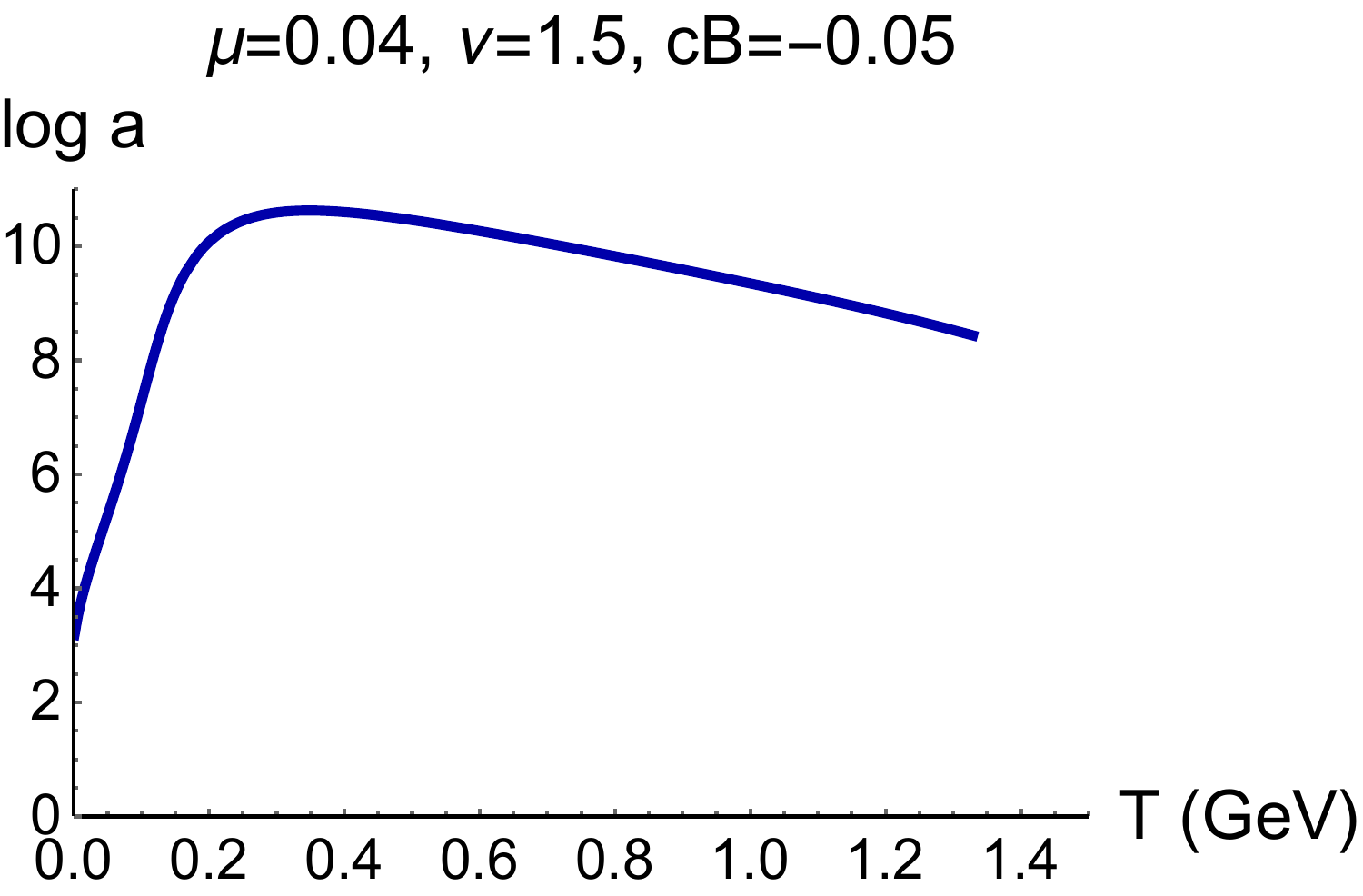}
  \includegraphics[scale=0.19]
   {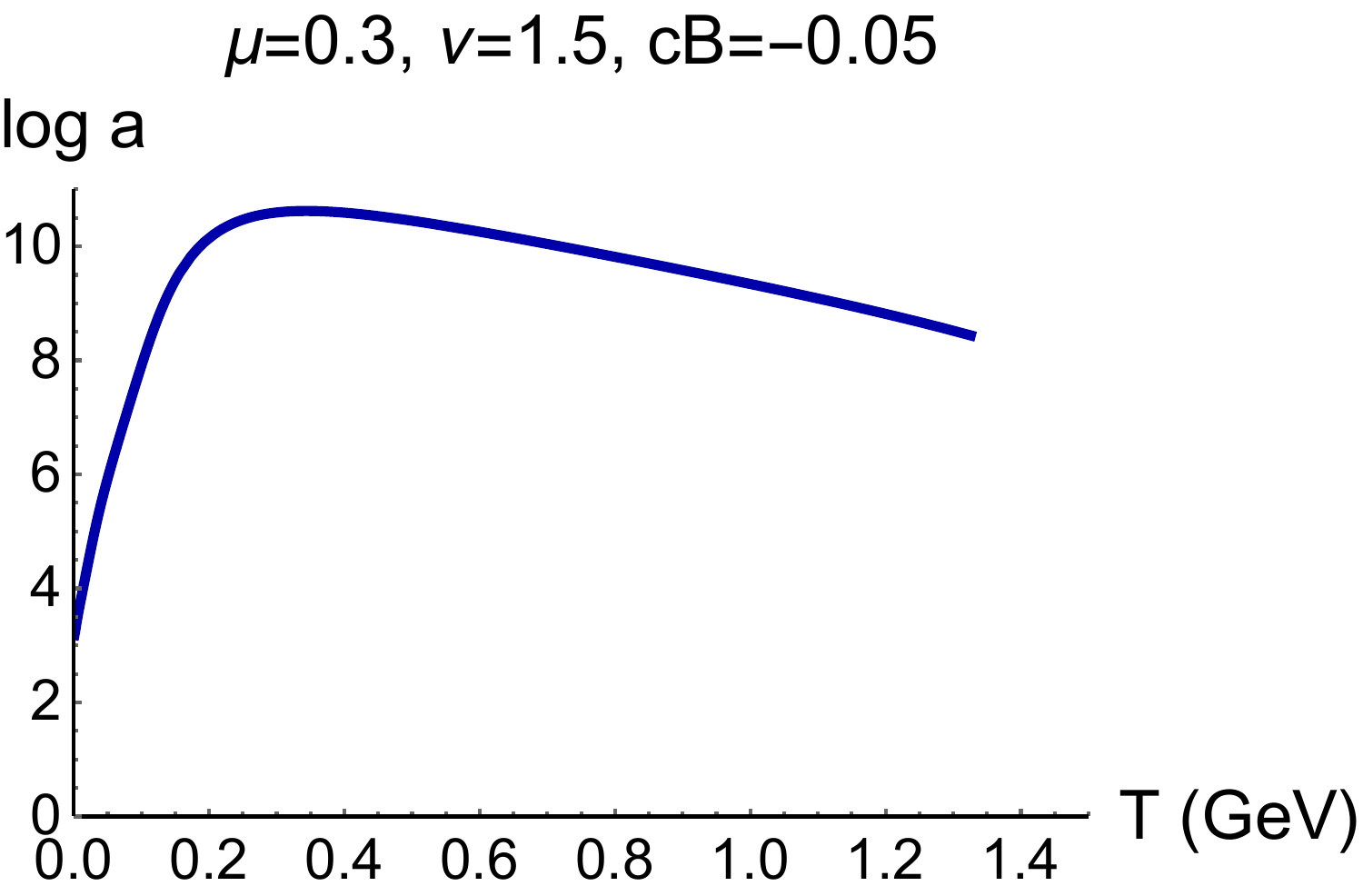}
    \includegraphics[scale=0.18]
   {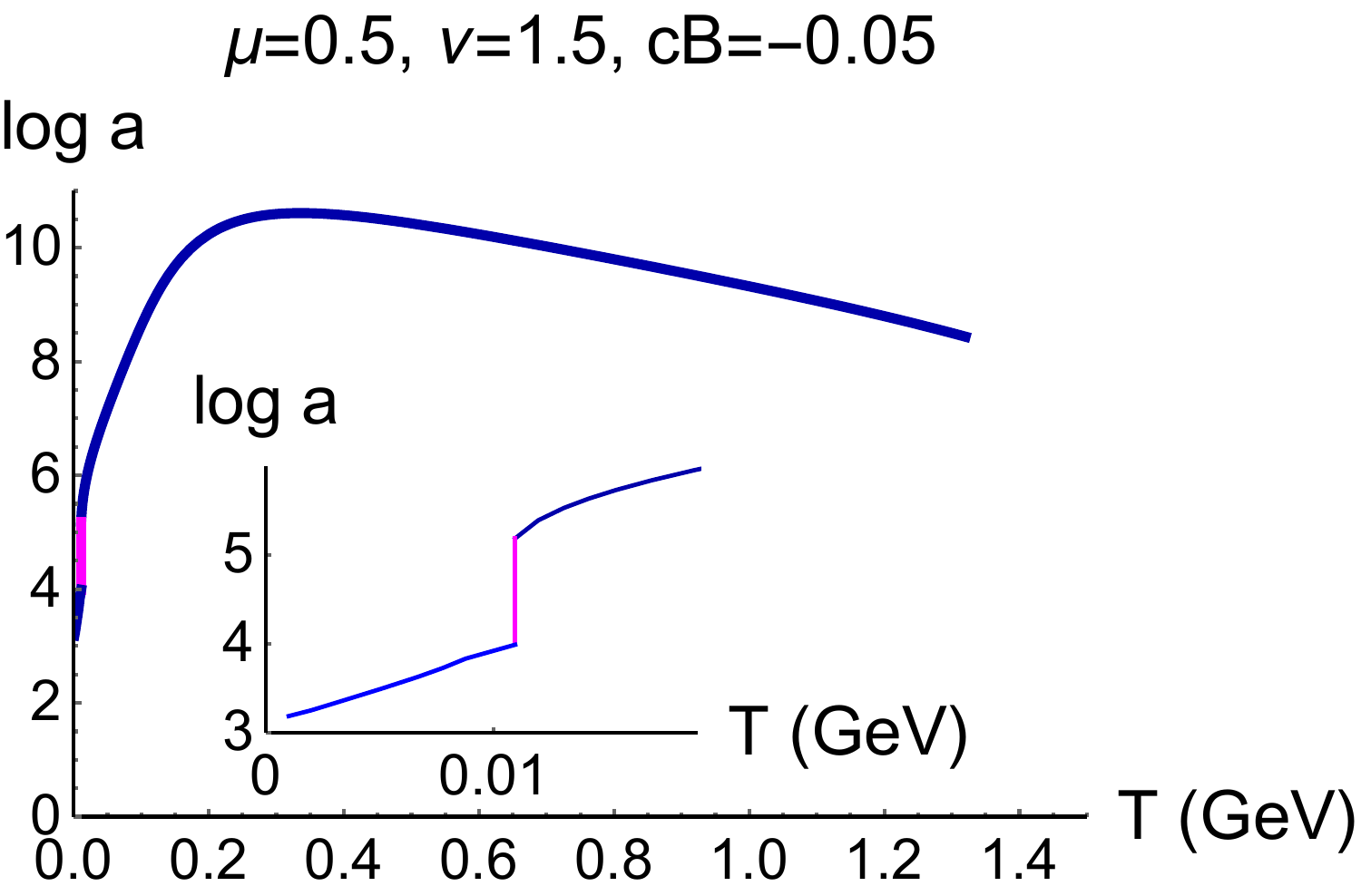}\\
D\hspace{140pt}E\hspace{140pt}F
   \caption{ 
We calculate the JQ parameter for the LQ model with $\nu=1.5$ along vertical lines (constant $\mu$) in the $(\mu,T)$-plane:  
(A) with $c_B = -0.005$ \GG\, at $\mu = 0.04, 0.3, 0.5\,\text{GeV}$,  
(B) with $c_B = -0.05$ \GG\, at $\mu = 0.04, 0.3, 0.5\,\text{GeV}$.
 The segments of these lines are colored blue, brown, and green corresponding to the QGP, hadronic, and quarkyonic phases they traverse. (C) The resulting values for $\log a$ are presented on the bottom panels (for $c_B=-0.005$ \GG), and panels (D, E, F) (for $ c_B=-0.05$ \GG) are colored using the same scheme.
  }
  \label{Fig:LQnu15cB0005005}
\end{figure}

This section demonstrates that for $\nu=1.5$, the behavior of the JQ parameter identifies first-order phase transitions (Fig.\,\ref{Fig:LQPTnu15cB0-005-0005}A) while remaining largely insensitive to confinement/deconfinement transitions (Fig.\,\ref{Fig:LQPTnu15cB0-005-0005}B). We examine: 
$\mu=0.3$ at $c_B=-0.005$ \GG, and 
$\mu=0.04$ , $0.3$, and $\mu=0.5$ GeV at $c_B=-0.05$ \GG.
Among these cases, only the sets with $c_B=-0.005$ \GG, $\mu=0.3$ GeV  and $c_B=-0.05$ \GG, $\mu=0.5$ GeV, cross the first-order transition line; however, only the case $c_B=-0.005$ \GG, $\mu=0.3 $ GeV, crosses the confinement/deconfinement transition line (Figs.\,\ref{Fig:LQnu15cB0005005}A, \ref{Fig:LQnu15cB0005005}B).

$\,$\\

In Fig.\,\ref{Fig:LQnu15cB0005005} we calculate the JQ parameter at $\nu=1.5$ for (A) $c_B=-0.005$ \GG (B), and $c_B=-0.05$ \GG along vertical lines (constant $\mu$) in the $(\mu,T)$-plane at $\mu = 0.04$, $0.3$ (GeV), and   $\mu=0.5$ GeV. The segments of these lines are colored blue (QGP) and  brown (hadronic) according to the phase traversed. Magenta shows the jumps (in the forbidden area).  The resulting plots of $\log a$ versus temperature are displayed in the bottom panels (C, D, E, F), using the same color scheme. Fig.\,\ref{Fig:LQnu15cB0005005}C shows in hadronic phase for $c_B=-0.005$ \GG \, the JQ parameter decreases until the first-order phase transition at $T\sim 0.12$ GeV. Also, in the QGP phase the JQ parameter decreases to $T\sim 0.3$ GeV.
Fig.\,\ref{Fig:LQnu15cB0005005}D shows in QGP phase for $c_B=-0.05$ \GG \,the JQ parameter decreases up to the minimum at $T\sim 0.3$ GeV and for $T\gtrsim 0.3$ GeV the JQ parameter increases. The same behavior is seen at $\mu=0.3$ GeV in Fig.\,\ref{Fig:LQnu15cB0005005}E. In addition, the JQ parameter at Fig.\,\ref{Fig:LQnu15cB0005005}F decreases  till jump at $T\sim 0.01$ GeV and keeps decreasing behavior till minimum at $T\sim 0.3$ GeV and then increases for $T\gtrsim 0.3$ GeV in the  QGP phase.

\newpage

\begin{figure}[h]
  \centering
\includegraphics[scale=0.29]{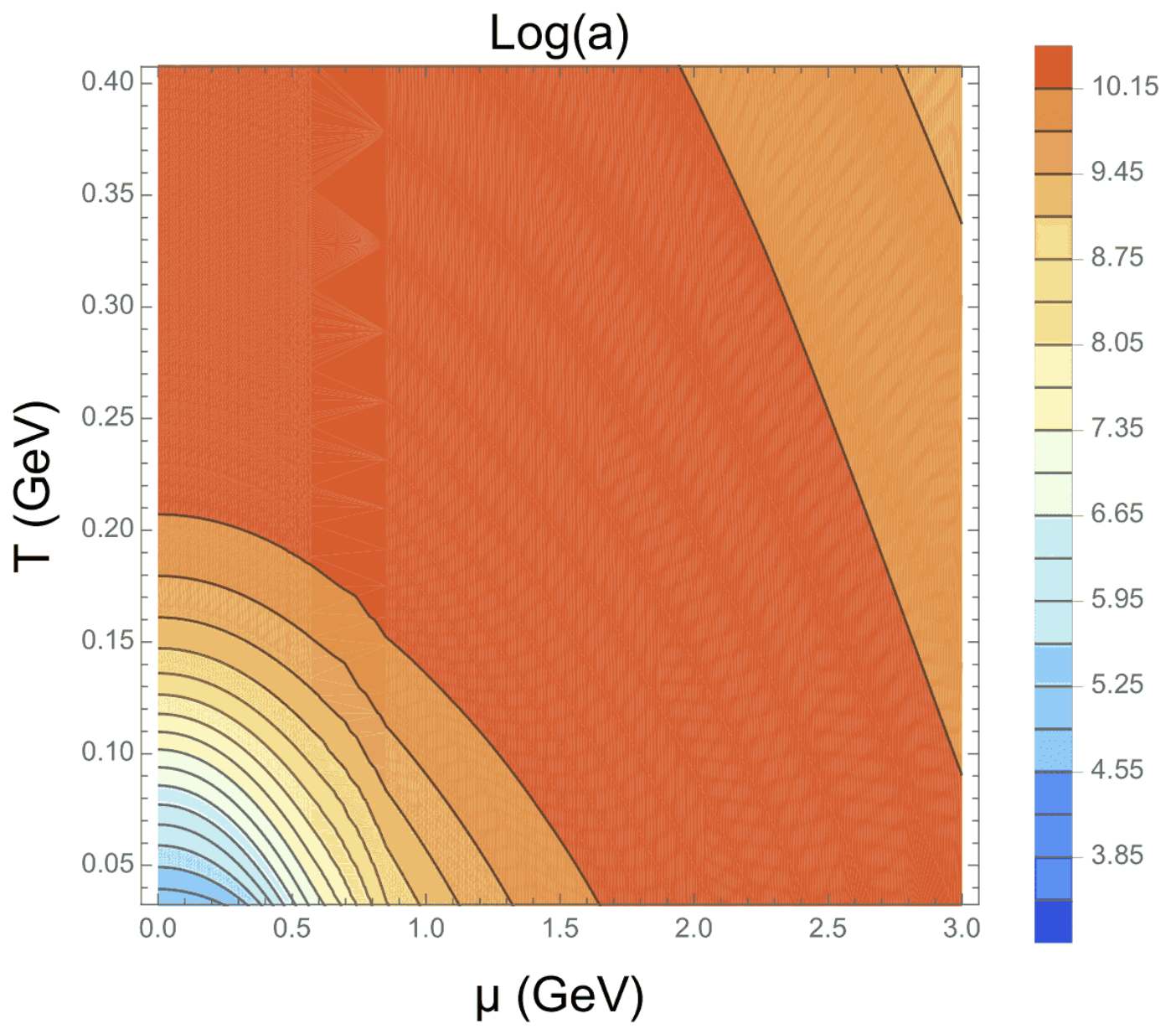}\quad
\includegraphics[scale=0.46]{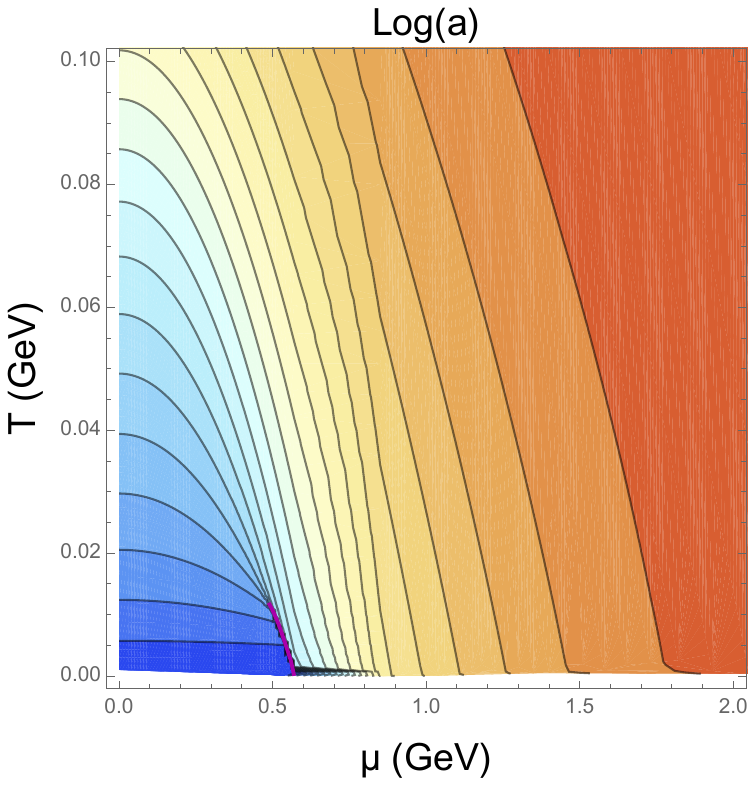}
  \\
A\hspace{200pt}B\\
$\,$\\
 \includegraphics[scale=0.30]{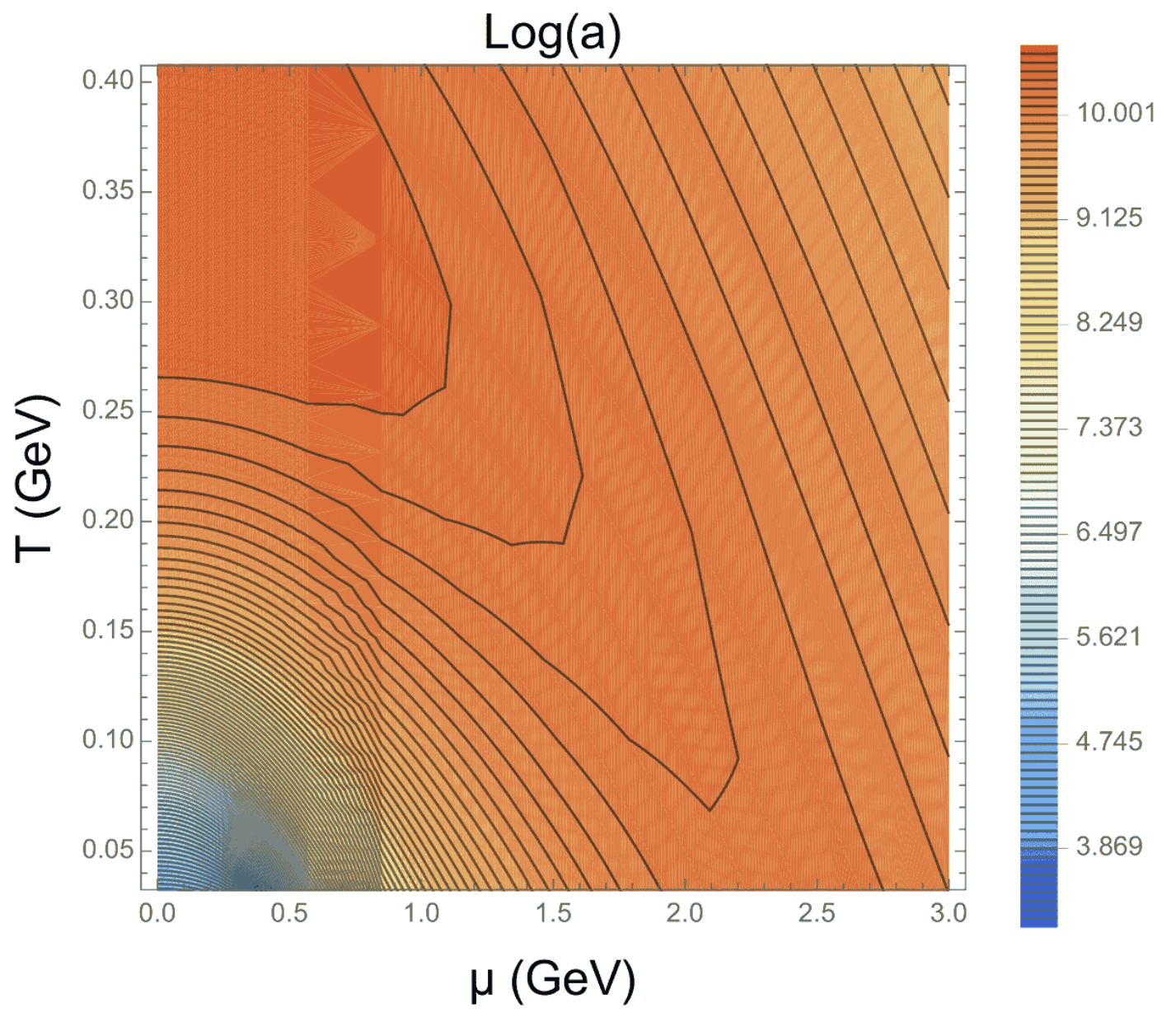}\\
C
  \caption{ (A) Density plots for $\log a$ at $\nu=1.5$ for 
  the LQ model  in the presence of the magnetic field, $c_B=-0.05$ \GG. Note that for these values of the magnetic field and anisotropy parameter, no hadronic phase exists; consequently, the second-order confinement/deconfinement phase transition is absent, and the system remains entirely in the QGP phase. 
  (B) Zoom of panel (A)  with  the magenta line showing the first-order transition line.
  The CEP  is located approximately at $(\mu_{CEP},T_{CEP}) \approx (0.49\, GeV, \,0.012 \, GeV)$. 
 (C) Zoom of panel (A) with increased contour density (100 contours), revealing the hill-like structure located above the phase transition line. 
}
  \label{Fig:LQnu15cB005}
\end{figure}

Fig.\,\ref{Fig:LQnu15cB005} displays the density plots of $\log a$ at $\nu=1.5$ for $c_B = -0.05$ \GG. panel (B) zooms in panel (A) near the critical endpoint at $(\mu_{CEP},T_{CEP}) \approx (0.49\, , \,0.012)$ (GeV), with the magenta line marking the first-order transition. panel (C) provides a zoom of panel (A) using increased contour density (100 contours), revealing a hill-like structure above the phase transition line. it is important to note that for these values of the magnetic field and anisotropy parameter, no hadronic phase exists; consequently, the second-order confinement/deconfinement phase transition is absent, and the system remains entirely in the QGP phase.

$\,$
\\

Detailed examination of the density plot in Fig.\,\ref{Fig:LQnu15cB005} and 2D plots in Fig.\,\ref{Fig:LQnu15cB0005005} (panels D, E, and F) reveals for $c_B=-0.05$ \GG:
\begin{itemize}
\item A discontinuity in $\log a$ within 0.52 GeV $< \mu <$ 0.59 GeV.
\item For 0.52 GeV $< \mu <$\,   0.59 GeV:
\begin{itemize}
\item $\log a$ increases before the discontinuity.
\item continues increasing in $\log a$ immediately after the discontinuity.
\item log a decreases above $T \gtrsim 0.4$ GeV
 due to a hill feature. For any fixed 
$\mu$, increasing 
$T$ beyond this threshold reduces $\log a$, indicating 
$\mu$-independent behavior.

\end{itemize}
\item For $\mu < 0.52$ GeV:
\begin{itemize}
\item $\log a$ increases until $T \approx 0.4$ GeV.
\item $\log a$ decreases at higher temperatures.
\end{itemize}
\item No discontinuity occurs at $\mu < 0.52$ GeV, though $\log a$ shifts from increasing to decreasing near $T \approx 0.4$ GeV.
\item At high temperatures ($T \gg 0.4$ GeV), $\log a$ universally decreases with increasing $T$.
\end{itemize}


\newpage
\subsubsection{Non-zero magnetic field, $\nu=3$}

\begin{figure}[h]
  \centering
\includegraphics[scale=0.18]
{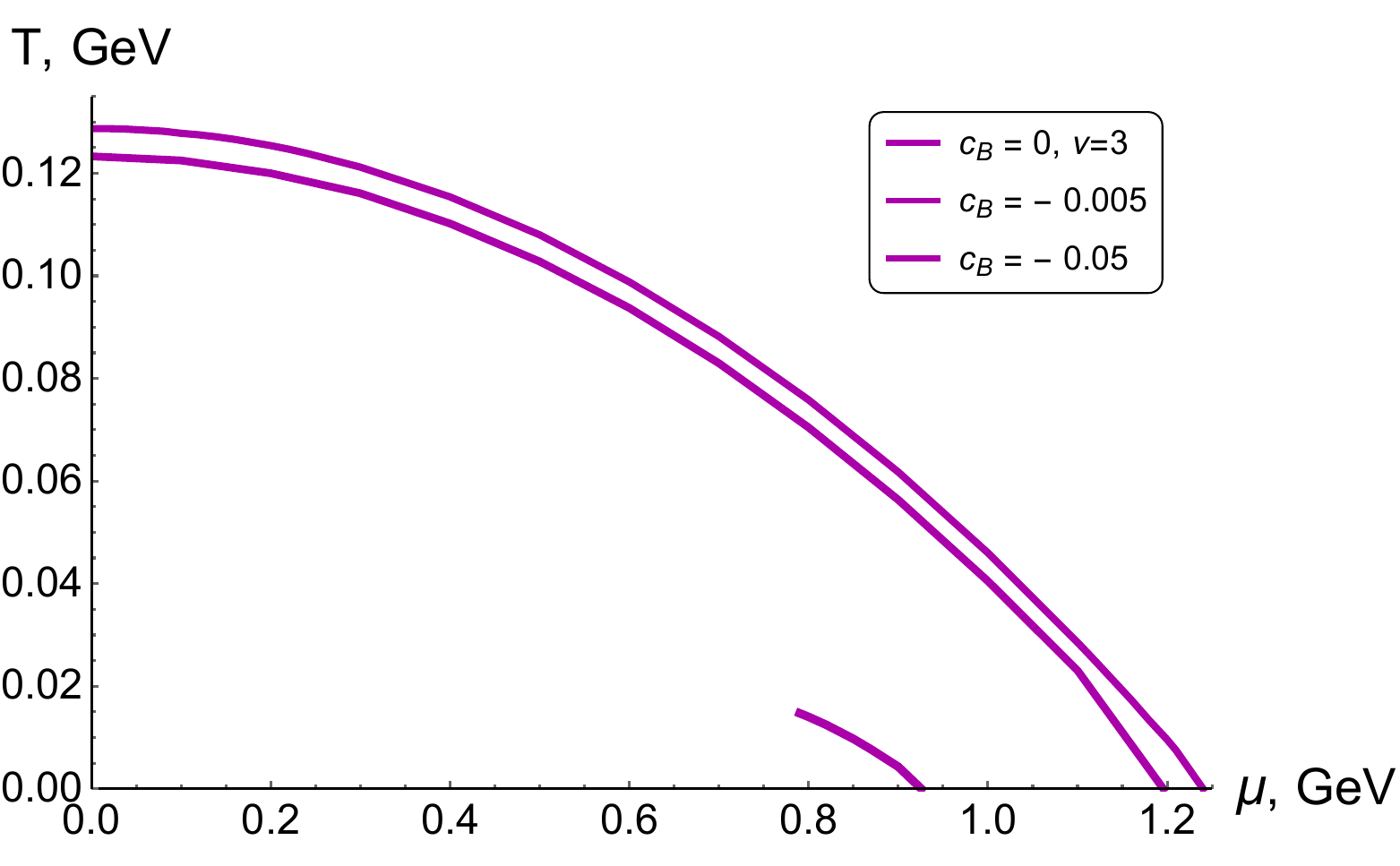}
\quad\includegraphics[scale=0.18]{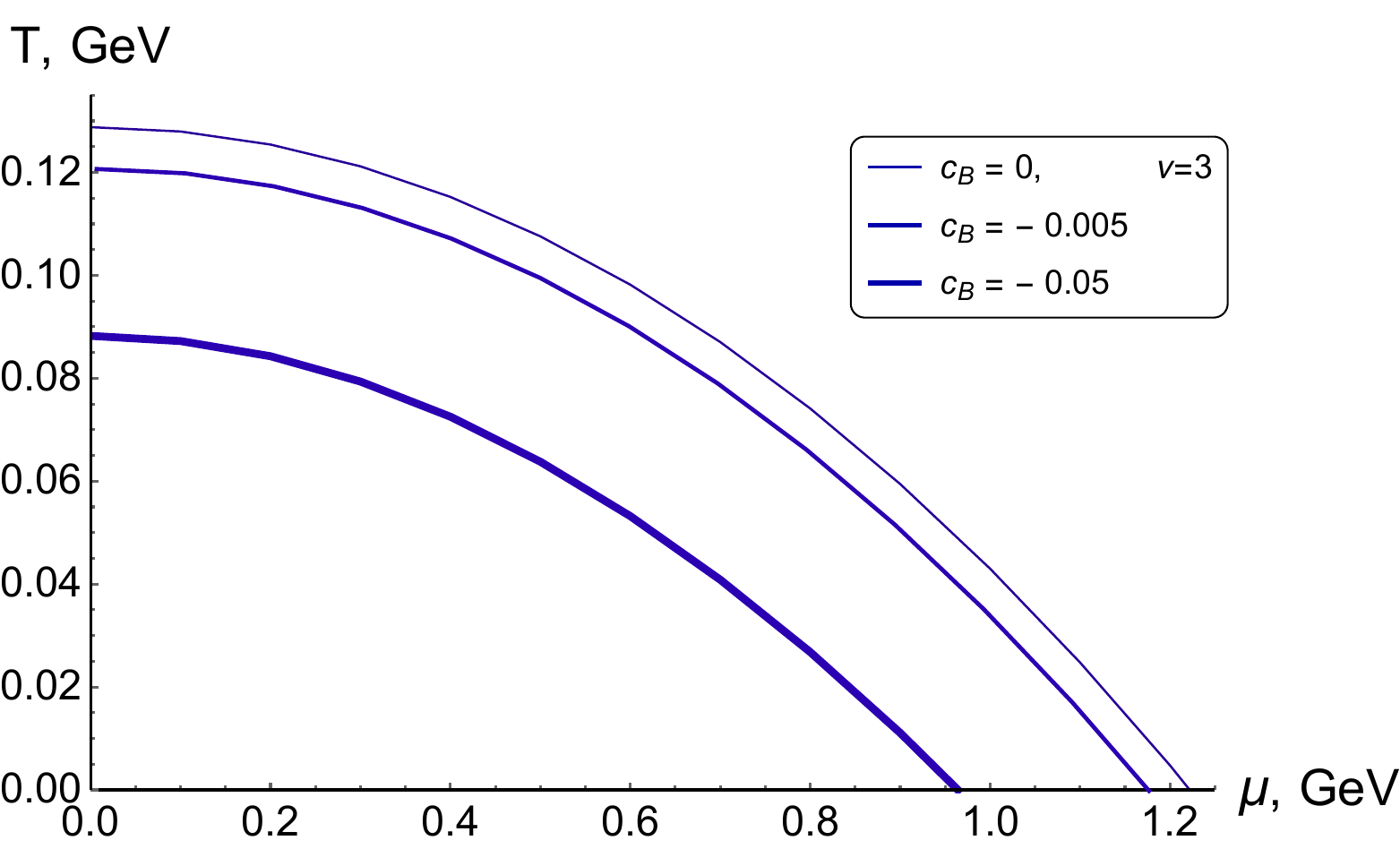}\\
A\hspace{150pt}B
\caption{
(A) The first-order phase transition lines (B) and confinement/deconfinement transition lines for the LQ model with $\nu=3$ and $c_B=0, -0.005,  -0.05$ \GG.
  }
  \label{Fig:LQ-firstTLnu3}
\end{figure}

In this section, we demonstrate that for $\nu=3$ and varying $\mu$, the behavior of the JQ parameter detects the first-order phase transitions shown in Fig.\,\ref{Fig:LQ-firstTLnu3}A, while remaining practically insensitive to the confinement/deconfinement transition in Fig.\,\ref{Fig:LQ-firstTLnu3}B.

\begin{figure}[h]
  \centering
\includegraphics[scale=0.18]
{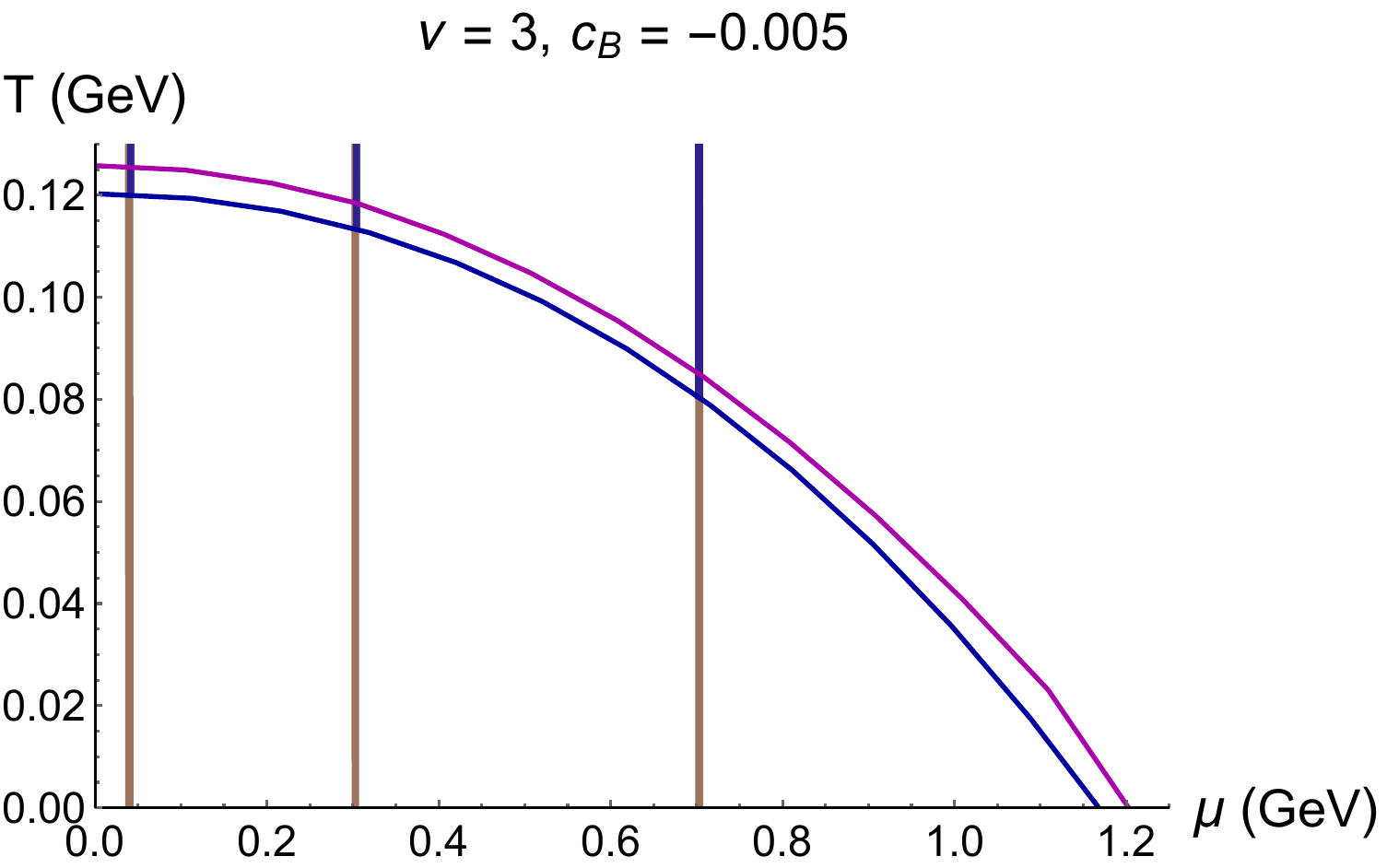}\quad
\includegraphics[scale=0.15]
{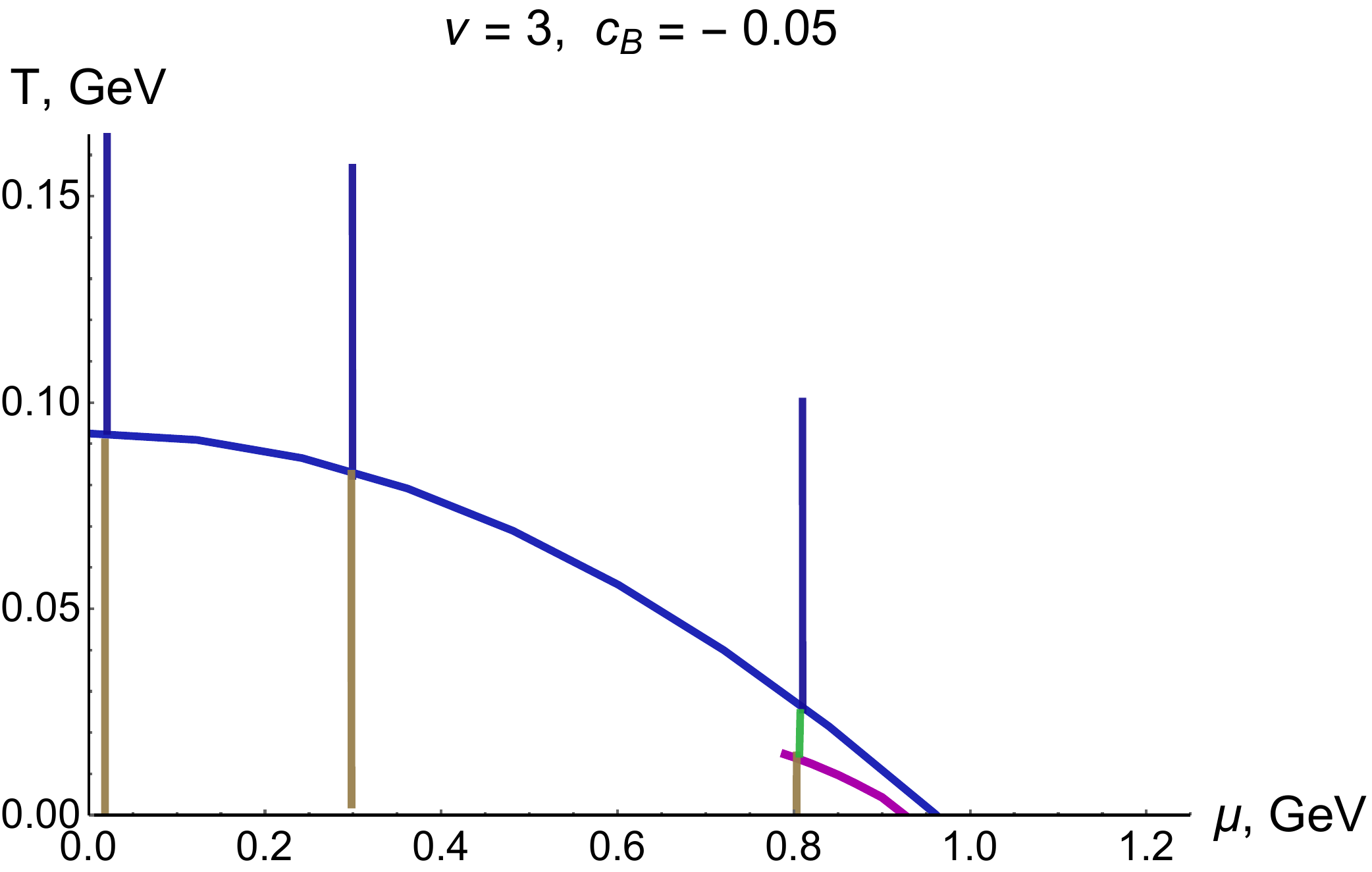}\\
 A\hspace{150pt}B\\ 
    \includegraphics[scale=0.32]
    {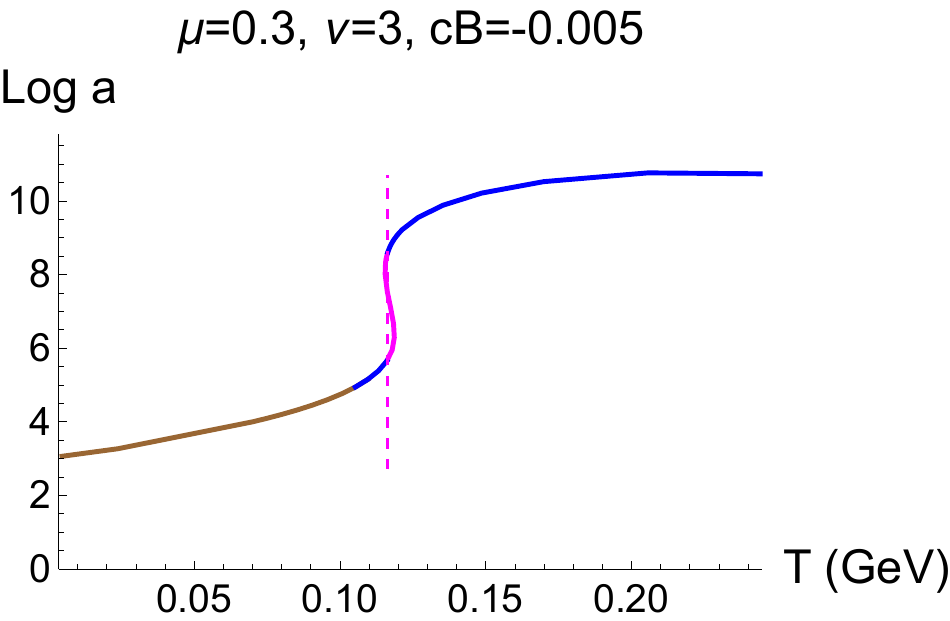}\\
    C\\$\,$\\
\includegraphics[scale=0.18]
   {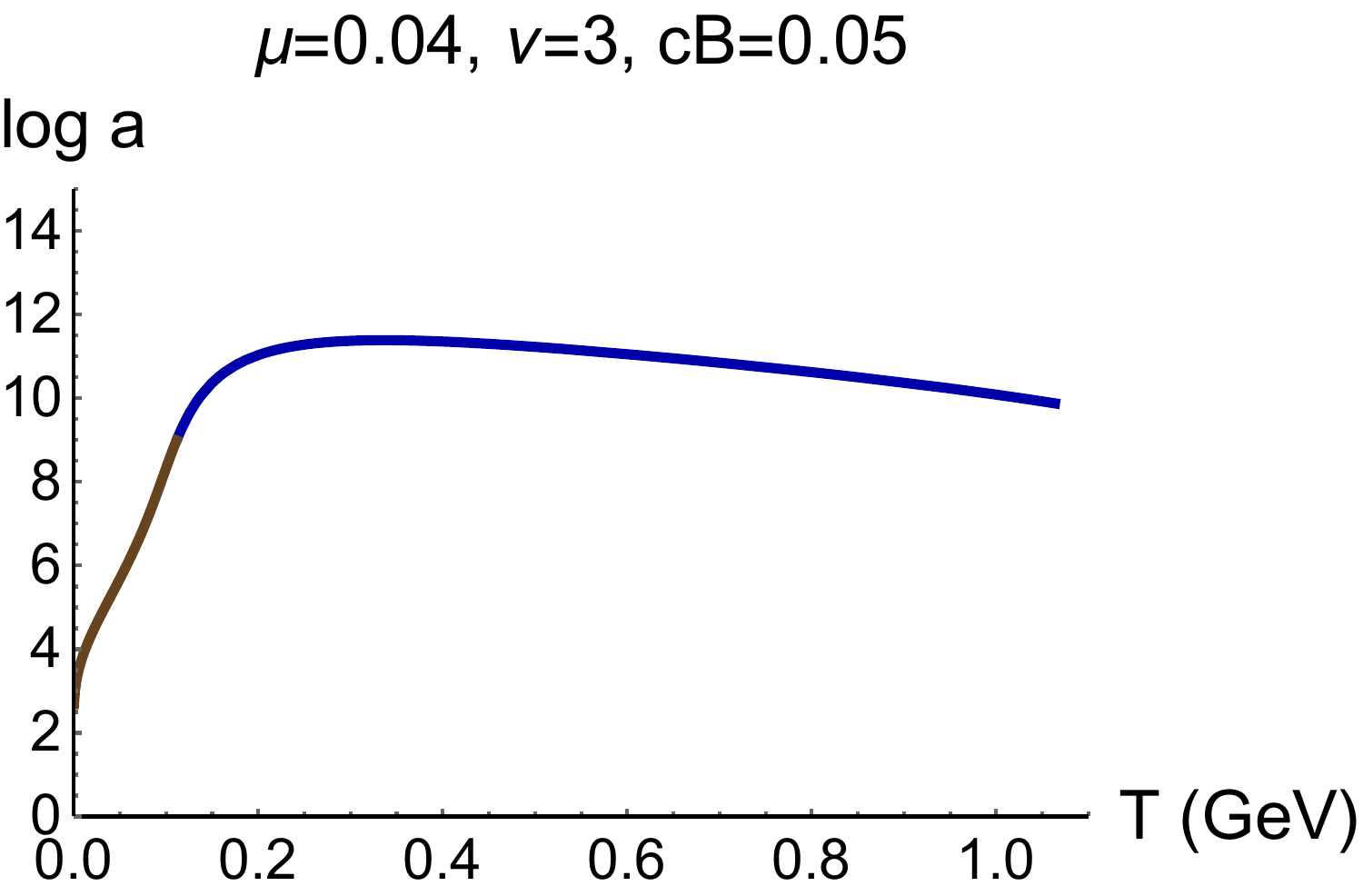}\quad\includegraphics[scale=0.18]
     {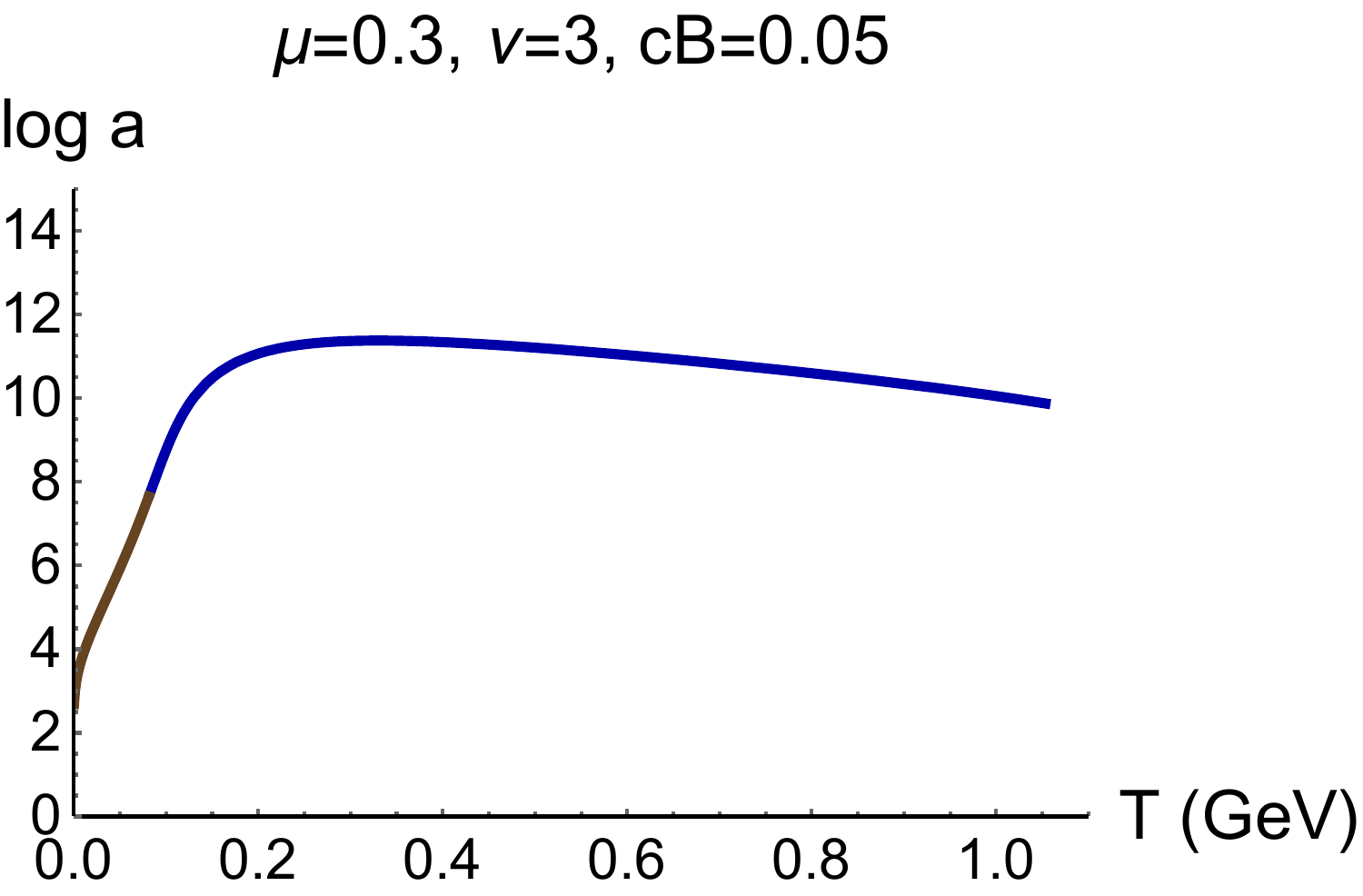}
   \quad \includegraphics[scale=0.14]
   {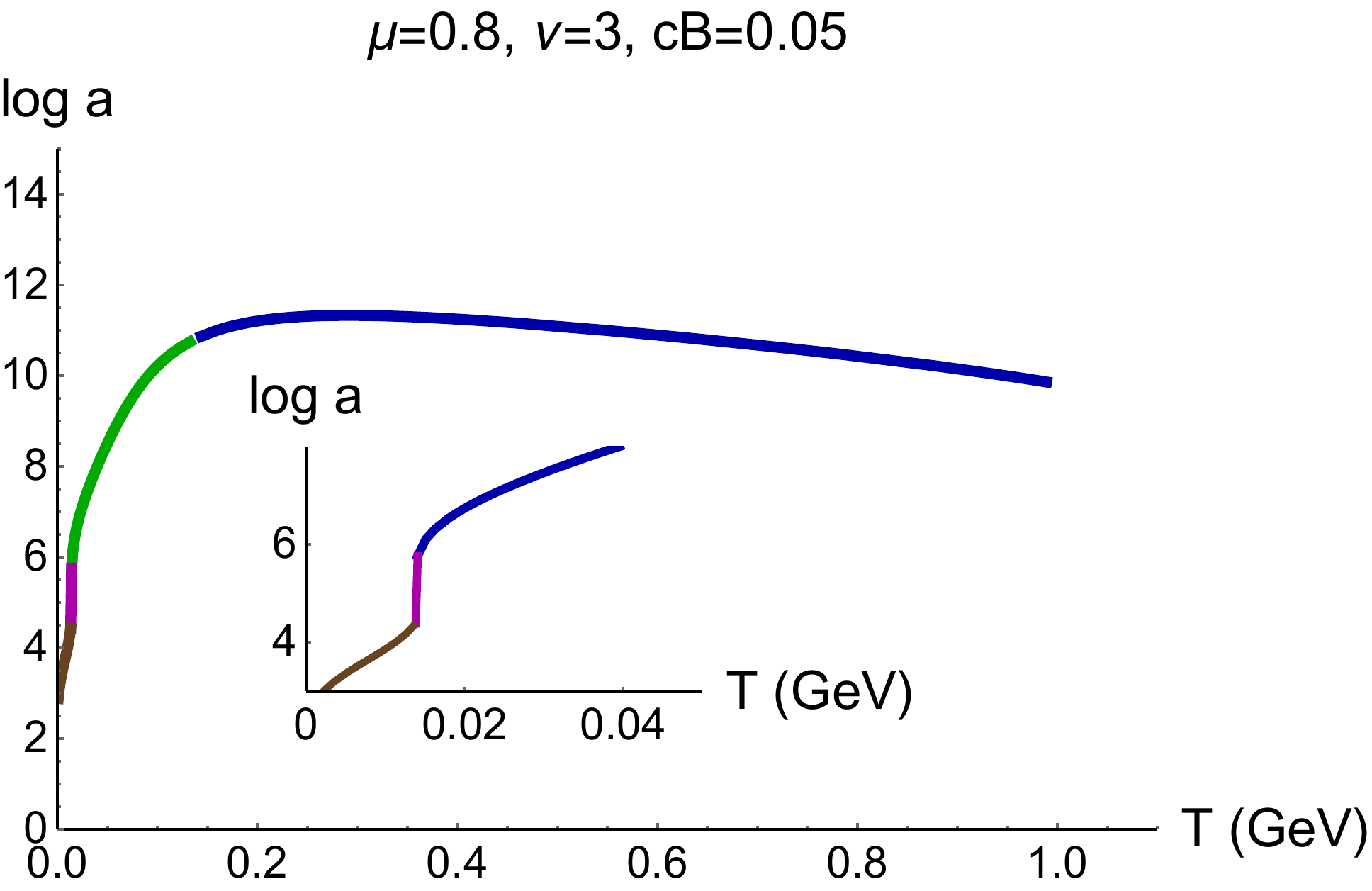}
   \\ D\hspace{140pt}E\hspace{140pt}F
\caption{We calculate the JQ parameter for the LQ model with $\nu=3$ along vertical lines (constant $\mu$) in the $(\mu,T)$-plane:  
(A) with $c_B = -0.005$ \GG\, at $\mu = 0.04, 0.3, 0.7\,\text{GeV}$,  
(B) with $c_B = -0.05$ \GG\, at $\mu = 0.04, 0.3, 0.8\,\text{GeV}$.
 The segments of these lines are colored blue, brown, and green corresponding to the QGP, hadronic, and quarkyonic phases they traverse. (C) The resulting values for $\log a$ are presented on the bottom panels (for $c_B=-0.005$ \GG), and panels (D, E, F) (for $ c_B=-0.05$ \GG) are colored using the same scheme.}
  \label{Fig:LQnu3cB0005005}
\end{figure}

We perform calculations for two sets of parameters:

(i) $c_B = -0.005$ \GG, $\mu = 0.3$ GeV.

(ii) $c_B = -0.05$ \GG, $\mu = 0.04$, $0.3$, $0.8$ GeV.
\\
All cases include the confinement/deconfinement transition, but only the following exhibit first-order phase transitions:

\begin{itemize}
   \item Case (i) (Figs.\,\ref{Fig:LQnu3cB0005005}A, \ref{Fig:LQnu3cB0005005}C). 
    \item The subcase of (ii) with  $\mu = 0.8$ GeV (Figs.\,\ref{Fig:LQnu3cB0005005}B, \ref{Fig:LQnu3cB0005005}F).
\end{itemize}

Results for $c_B = -0.005$ \GG\, at $\mu = 0.004, 0.7$ (GeV) are omitted, as they closely resemble the behavior shown in Fig.\,\ref{Fig:LQnu3cB0005005}C.
\\

Our calculations show:
\begin{itemize}
    \item A jump in $\log a$ occurs at $\mu=0.3$ GeV with $c_B=-0.005$ \GG\,, illustrated  in Fig.\,\ref{Fig:LQnu3cB0005005}C.  This picture is similar to the plot presented in  Fig.\,\ref{Fig:LQnu15cB0005005}C.  We observe that $\log a$ increases rapidly before the jump and just after the jump, and becomes slowly decrease  afterward making the JQ parameter increasing at high temperature regime. 
    \item 
    Although no discontinuity  is observed in Figs.\,\ref{Fig:LQnu3cB0005005}D and \ref{Fig:LQnu3cB0005005}E, a distinct change in slope occurs near the confinement/deconfinement transition. The color transition from brown (hadronic) to blue (QGP) occurs near the hills in these plots.

  However, a similar change in slopes occurs even in cases without a confinement/deconfinement phase transition, as shown, for example, in panels D and E of Fig.\,\ref{Fig:LQnu15cB0005005}. Naively, this suggests that for fixed parameters (while varying only the spatial anisotropy $\nu$), the JQ parameter exhibits identical temperature dependence. We expect universal asymptotic behavior in the JQ parameter at high temperature regime.

    \item In all cases, $\log a$ decreases indicating enhancement in the JQ parameter with increasing temperature in the high-temperature regime.
\end{itemize}


\subsubsection{Non-zero magnetic field, $\nu=4.5$}

\begin{figure}[h]
  \centering
  $$\nu=4.5$$\\
\includegraphics[scale=0.21]
{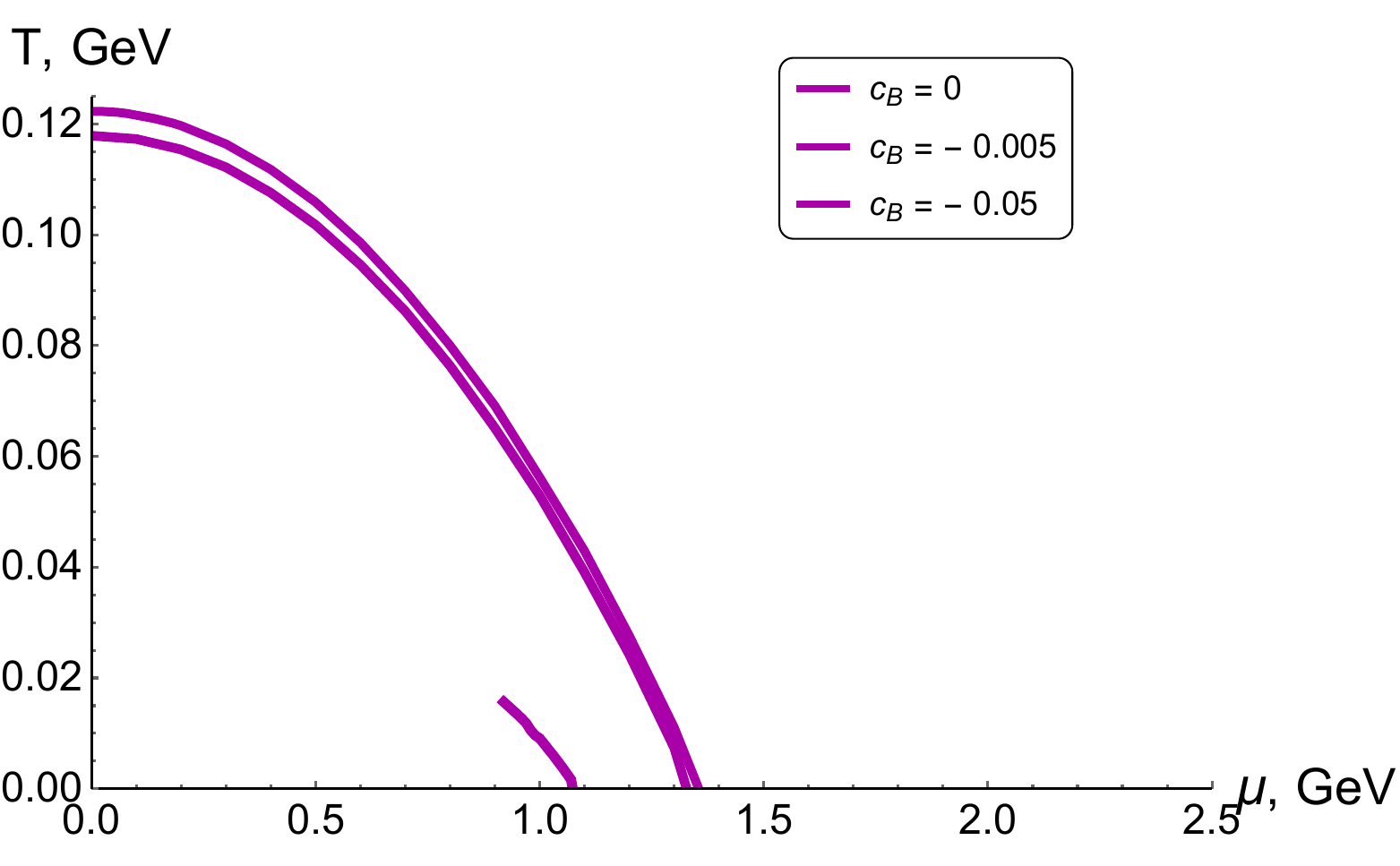}
\quad\includegraphics[scale=0.29]{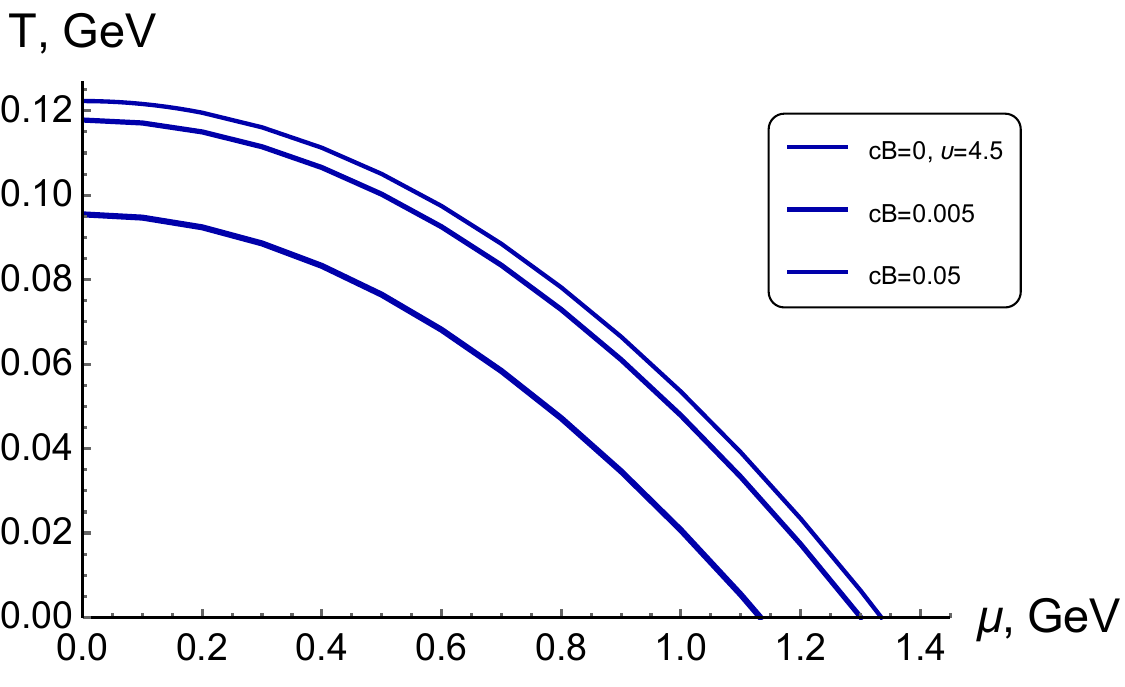}\\
A\hspace{150pt}B
\caption{(A) The first-order phase transition lines,  and  (B) confinement/deconfinement transition lines for the LQ model with $\nu=4.5$ and $c_B = 0, -0.005$ \GG , $-0.05$ \GG.}
  \label{Fig:LQ-first TLnu45}
\end{figure}

In this section, we study $\nu=4.5$ for various $\mu$. Figure~\ref{Fig:LQ-first TLnu45} presents $(\mu,T)$-phase diagrams at $\nu=4.5$ and $c_B = 0$, $-0.005$, $-0.05$ (\GG), with first-order phase transitions in (A) and confinement/deconfinement transitions in (B).
The behavior of the JQ parameter closely resembles that observed for $\nu=3$, differing primarily in the chemical potential value where the first-order phase transition occurs.

\begin{figure}[h!]
  \centering
  \includegraphics[scale=0.3]
 {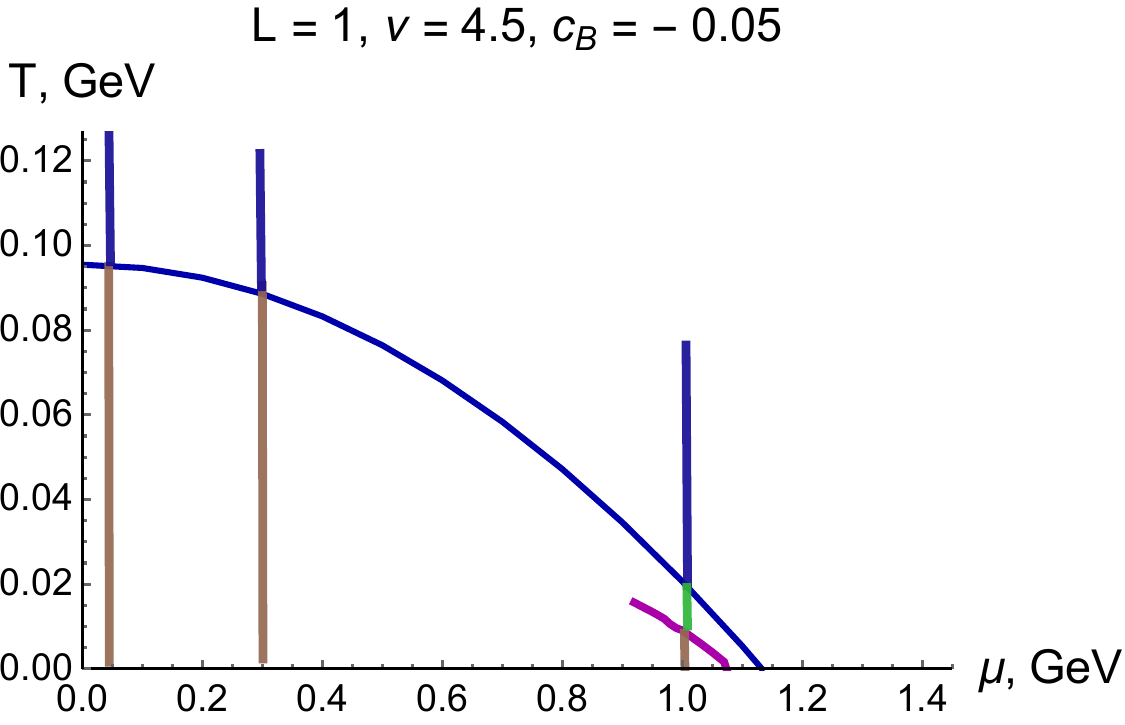}\\
 A
 \\
\includegraphics[scale=0.25]
   {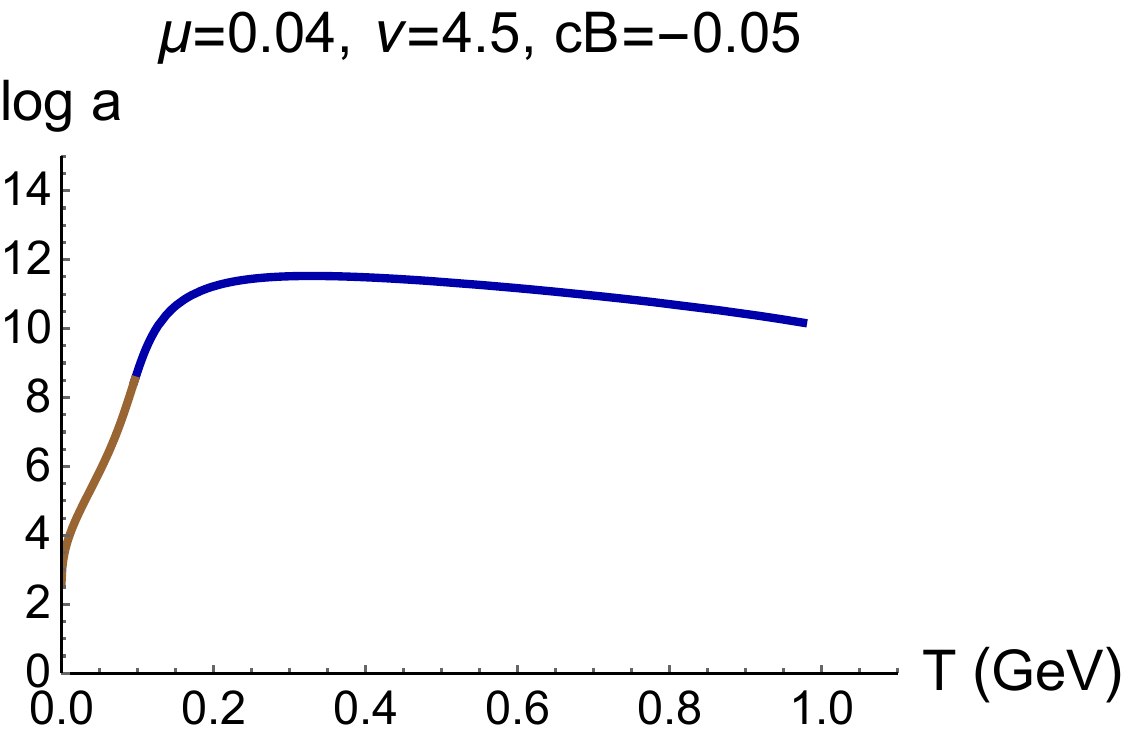}
  \includegraphics[scale=0.25]
   {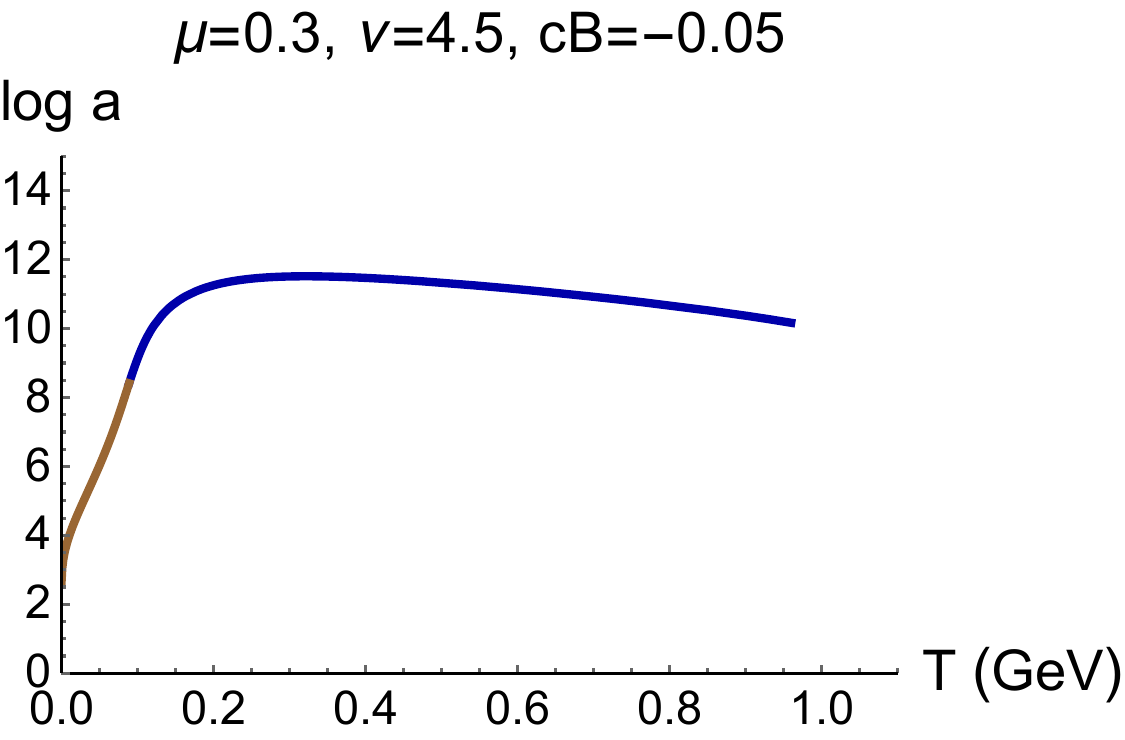}
  \includegraphics[scale=0.23]
   {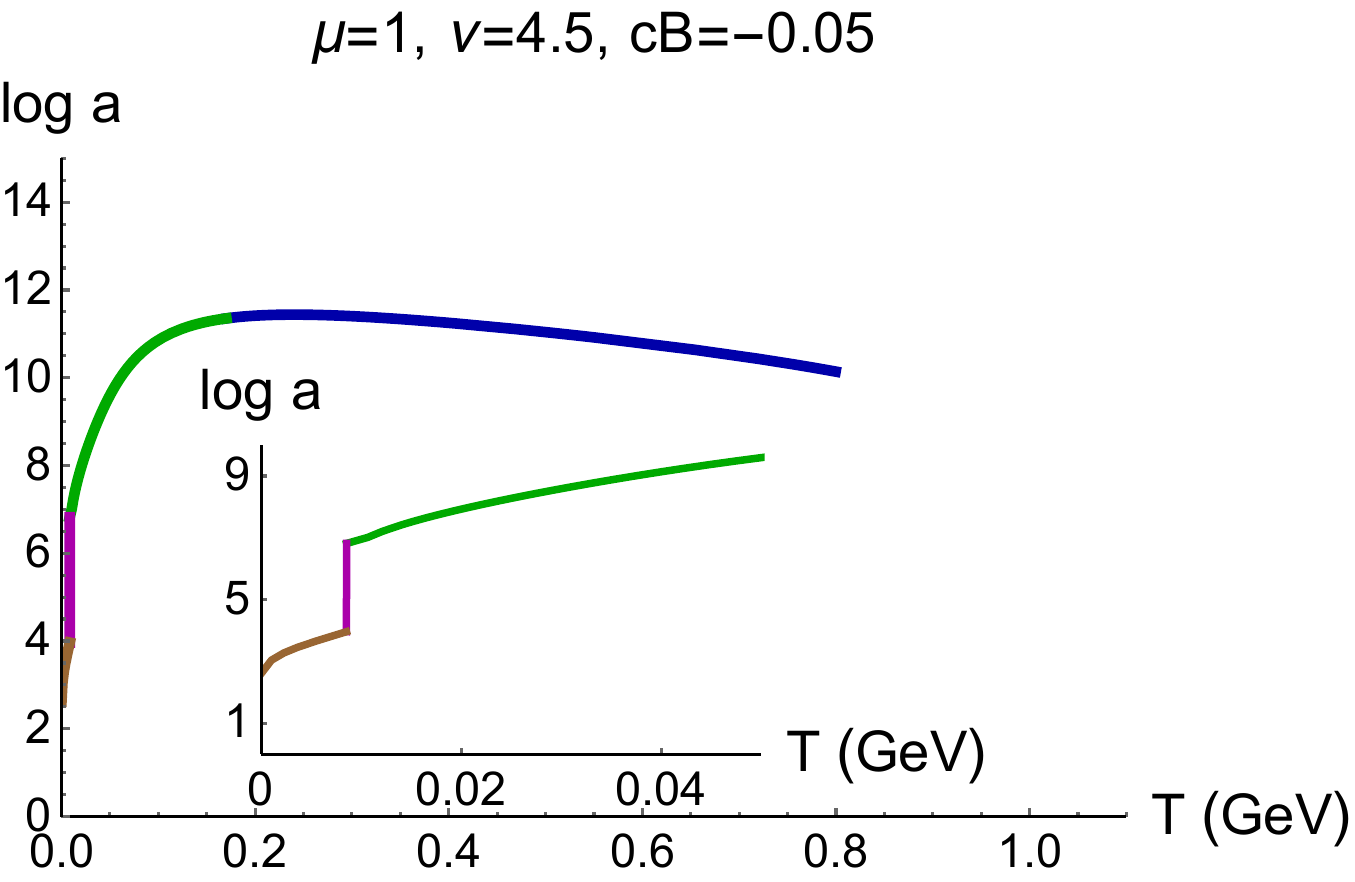}
   \\B\hspace{140pt}C\hspace{140pt}
  D
\caption{ Considering $\nu=4.5$ and $c_B=-0.05$ \GG,  (A) we calculate the JQ parameter for the LQ model along vertical lines (constant $\mu$) in the $(\mu,T)$-plane shown in panel. The calculations are performed at fixed chemical potentials:  
(B) $\mu = 0.04\,\text{GeV}$,  
(C) $\mu = 0.3\,\text{GeV}$,  
(D) $\mu = 1\,\text{GeV}$.
 The segments of these lines are colored blue, brown, and green corresponding to the QGP, hadronic, and quarkyonic phases they traverse, respectively (A). The resulting  values for log of the JQ parameter are presented on the bottom panels (B, C, D) are colored using the same scheme.
   }
  \label{Fig:Qmu004031nu45cb005}
\end{figure}

\begin{figure}[h]
  \centering
\includegraphics[scale=0.21]{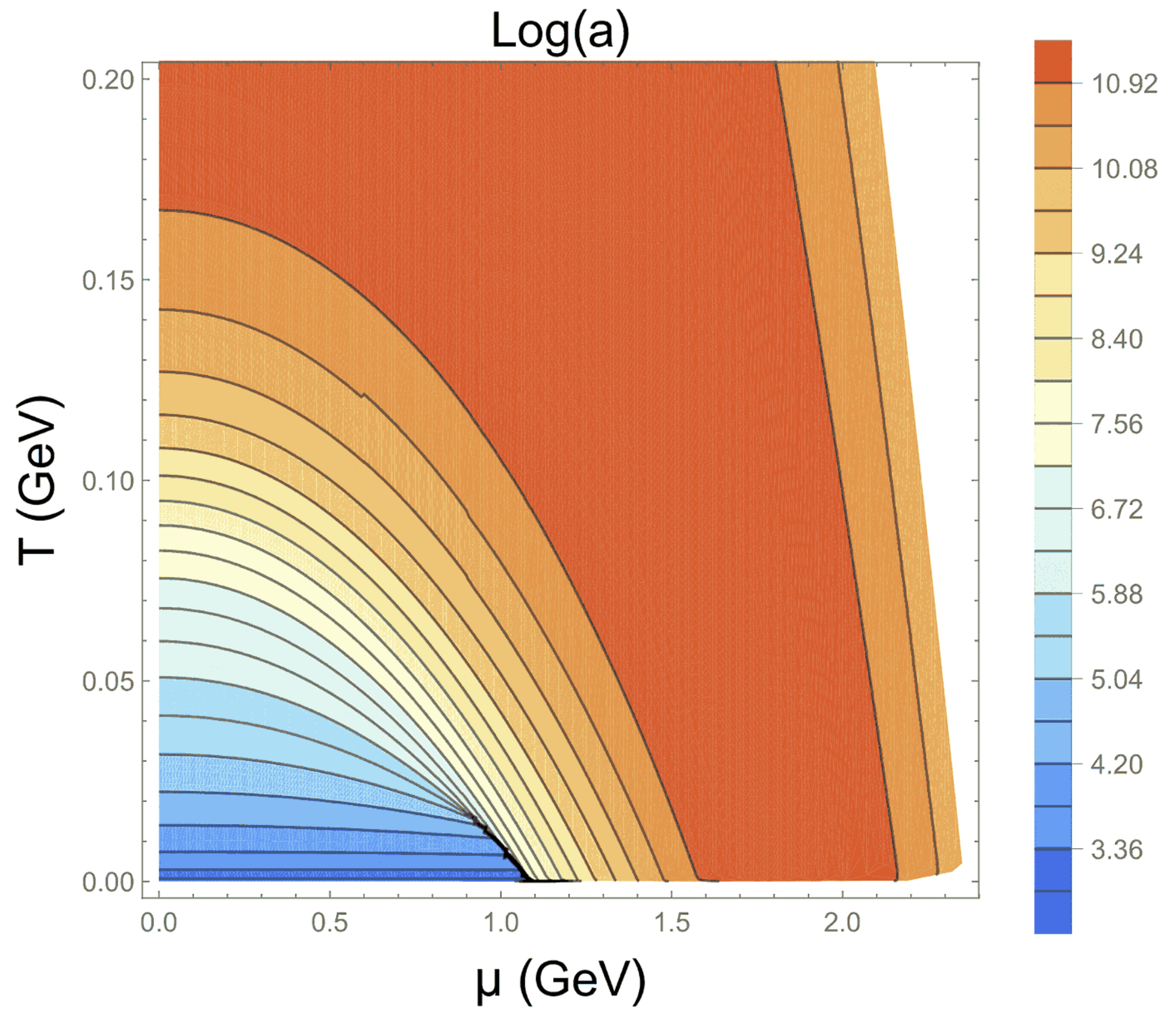}\quad
\includegraphics[scale=0.104]{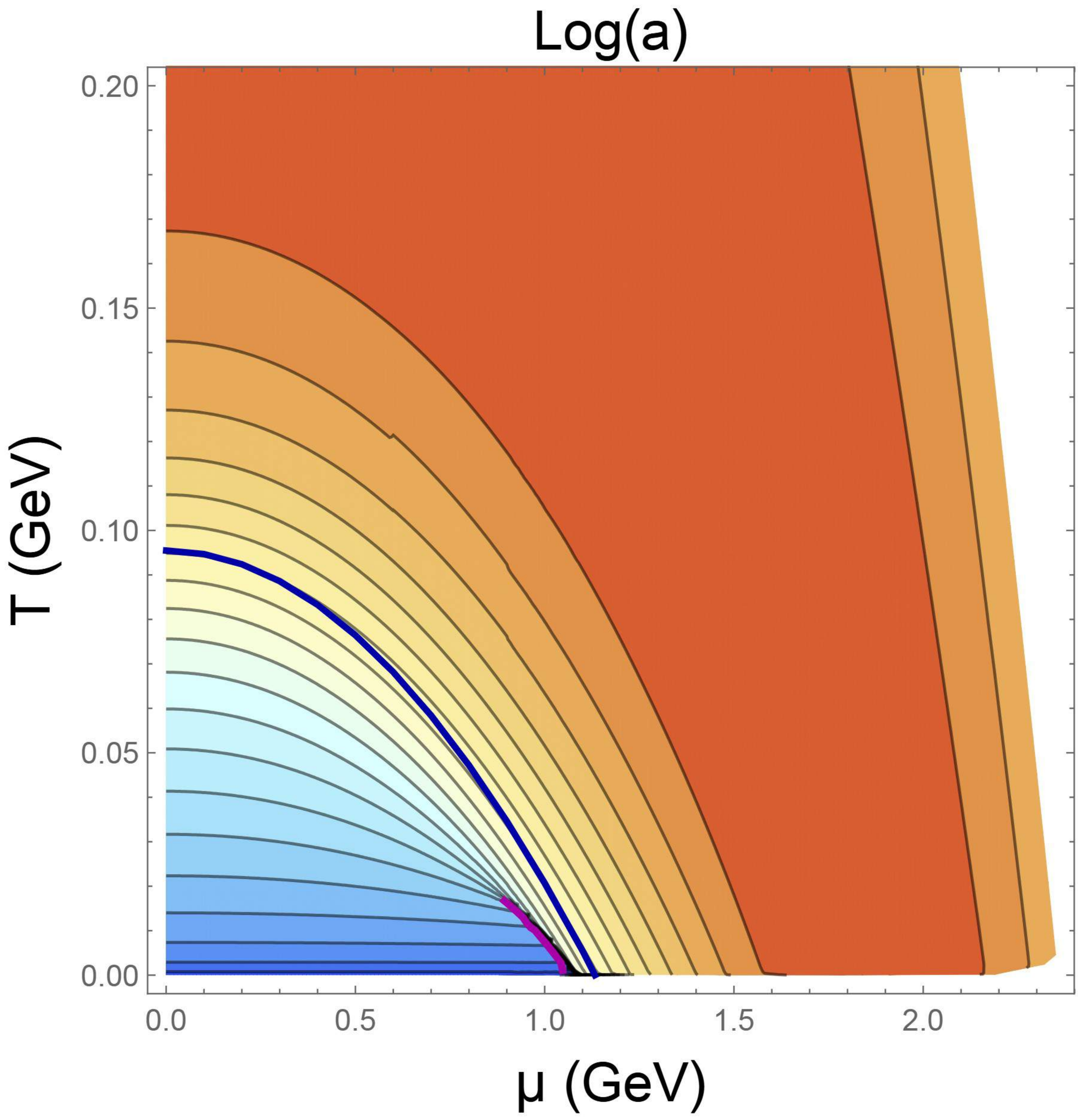}
\\
A\hspace{220pt}B\\
\includegraphics[scale=0.12]{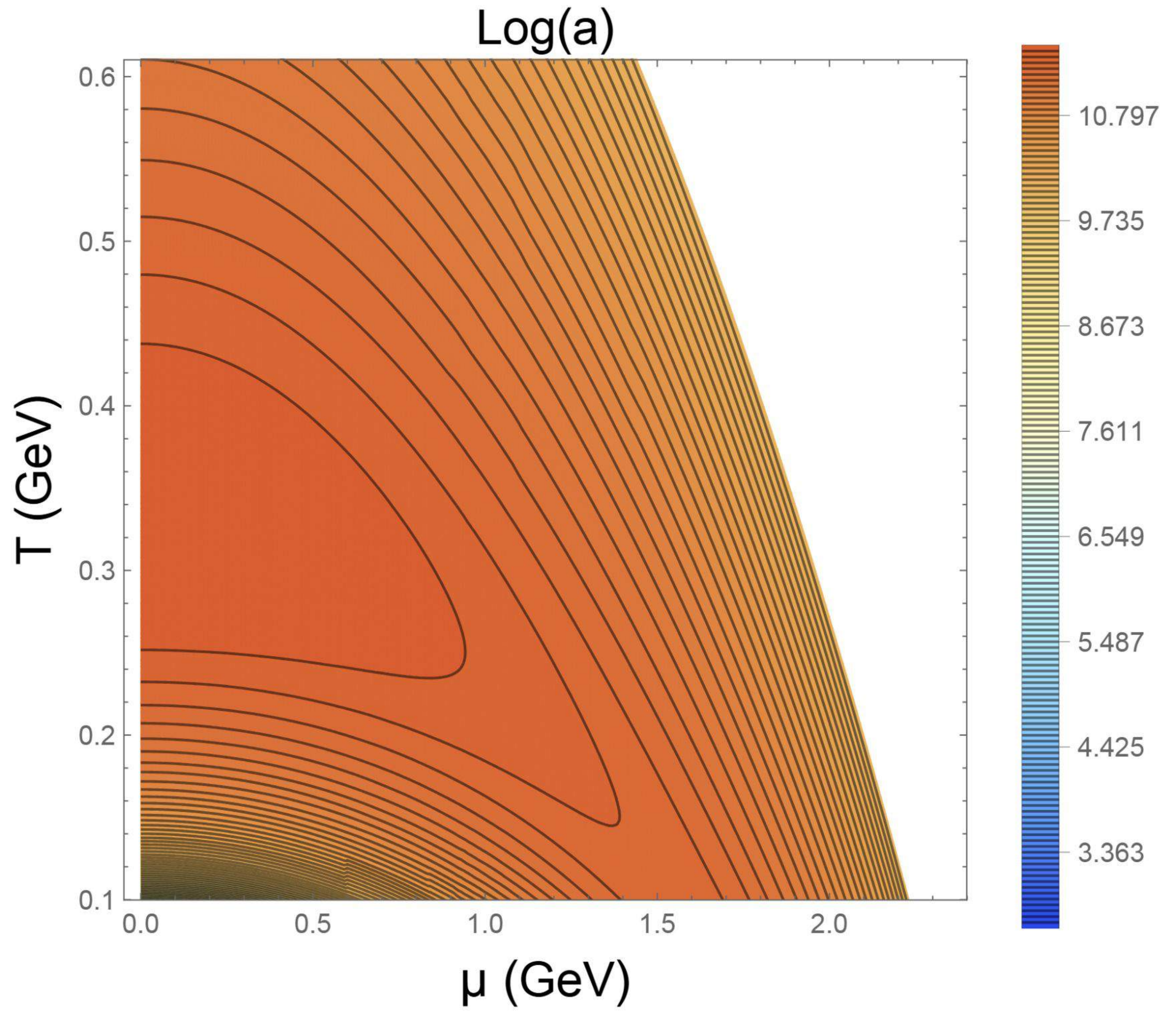}
  \\  C
\caption{ 
(A) Density plots of $\log a$  for the LQ model, with $c_B=-0.05$ \GG\, and $\nu = 4.5$; including 20 contours. (B) The same with phase transition lines: i) the magenta line presents the first-order transition line, ii), and the blue one presents the confinement/deconfinement transition line. (C) Zoom of panel (A) with increased contour density (80 contours), revealing the hill-like structure located above the phase transition lines. 
 }
  \label{Fig:LQnu45cB005}
\end{figure}

Our calculations show:
\begin{itemize}
\item No discontinuity appears in Figs.\,\ref{Fig:Qmu004031nu45cb005}B and \ref{Fig:Qmu004031nu45cb005}C, but a distinct change in slope occurs near the confinement/deconfinement transition. This change is clearly visible in the density plot of Fig.\,\ref{Fig:LQnu45cB005}C.

\item For $c_B=-0.05$ \GG \, at $\mu=1$ GeV (Fig.\,\ref{Fig:Qmu004031nu45cb005}D), $\log a$ exhibits:
\begin{itemize}
\item rapid increase before the discontinuity.
\item a jump at $T=0.00757$ GeV.
\item continued increase in the quarkyonic phase.
\item slow decrease in the QGP phase.
\end{itemize}

\item In all cases, $\log a$ decreases indicating enhancement in the JQ parameter with increasing temperature in the high-temperature regime.
\end{itemize}

\newpage

\newpage
\subsection{Summary of  jet quenching results for the LQ model
}\label{sect-slq}

We have calculated the JQ parameter $\hat{q}$ for quarks propagating along light-like trajectories in a QCD medium with light quarks at finite temperature ($T$) and chemical potential ($\mu$). Our results reveal a non-trivial dependence on both $T$ and $\mu$, particularly near the phase transition temperature $T_c$.
\\

From previous studies \cite{Yang:2015aia}, we know that in the isotropic case, both first-order and second-order phase transitions occur. The corresponding phase structure is shown in Figs.\,\ref{Fig:LQ-PTnu1cB0-0005-005}A and \ref{Fig:LQ-PTnu1cB0-0005-005}B. When varying parameters $c_B$ and $\nu$, the following behavior emerges \cite{Arefeva:2022avn}:
\begin{itemize}
\item As the magnetic field increases (controlled by $c_B$):
  \begin{itemize}
  \item the second-order phase transition disappears (Fig.\,\ref{Fig:LQ-PTnu1cB0-0005-005}A).
  \item the first-order phase transition shifts toward lower temperatures and chemical potentials (inverse magnetic catalysis), becoming negligible at extreme values (Fig.\,\ref{Fig:LQ-PTnu1cB0-0005-005}B).
  \end{itemize}
  
\item As $\nu$ increases:
  \begin{itemize}
  \item the confinement/deconfinement transition reappears for large negative $c_B$ (high magnetic fields).
  \item the first-order transition and critical endpoint shift rightward in the $(\mu,T)$-plane (Fig.\,\ref{Fig:LQ-first TLnu45}).
  \end{itemize}
\end{itemize}
 
 Examining the JQ parameter for the LQ model across different $c_B$ and $\nu$ values, we observed:
\begin{itemize}
    \item A discontinuity  at the first-order phase transition line.
    \item A change in the slope of $\log a$ curve near second-order phase transition lines.
\end{itemize}

The corresponding plots are shown in the figures referenced in Tables~\ref{table:LQ1} and \ref{tab:LQ-density}.

\begin{table}[h]
\centering
\begin{tabular}{|c|l|c|c|c|}
\hline
\multicolumn{1}{|l|}{\diagbox{$|c_B|$}{$\nu$}} & \multicolumn{1}{c|}{1}                     & 1.5                                                                 & 3                                                                   & 4.5                                                                 \\ \hline
0                               & Fig.\,\ref{Fig:LQnu1cB0}   & \multicolumn{1}{c|}{Fig.\,\ref{Fig:LQnu15cB0} } & \multicolumn{1}{c|}{Fig.\,\ref{Fig:LQnu3cB0}} & \multicolumn{1}{c|}{Fig.\,\ref{Fig:LQnu45cB0}} \\ \hline
0.005                           & Fig.\,\ref{Fig:LQnu1cB0005005} & 
Fig.\,\ref{Fig:LQnu15cB0005005}                                                  & Fig.
\ref{Fig:LQnu3cB0005005}                                                                 & -                                                            \\ \hline
0.05                            & Fig.\,\ref{Fig:LQnu1cB0005005}  &  Fig.\,\ref{Fig:LQnu15cB0005005}                                                          & Fig.\,\ref{Fig:LQnu3cB0005005}                                                                    & Fig.\,\ref{Fig:Qmu004031nu45cb005}                                                                  \\ \hline
\end{tabular}
\caption{Schematic layout of 2D plots of dependence of $\log a$ of the JQ parameter for light quarks on $T$ for fixed values of $\mu$ at different $c_B$ and $\nu$. }
\label{table:LQ1}
\end{table}

\begin{table}[h]
\centering
\begin{tabular}{|c|l|c|c|c|}
\hline
\multicolumn{1}{|l|}{\diagbox{$|c_B|$}{$\nu$}} & \multicolumn{1}{c|}{1}                     & 1.5                                                                 & 3                                                                   & 4.5                                                                 \\ \hline
0                               & Fig.\,\ref{Fig:DensityLQnu1cB0}    & \multicolumn{1}{l|}{Fig.\,\ref{Fig:nu15-3-45-cB0}A} & \multicolumn{1}{c|}{Fig.\,\ref{Fig:nu15-3-45-cB0}B} & \multicolumn{1}{l|}{Fig.\,\ref{Fig:nu15-3-45-cB0}C} \\ \hline
0.005                           & Fig.\,\ref{Fig:LQnu1cB0005} & -                                                                   & -                                                                   & -                                                                   \\ \hline
0.05                            & Fig.\,\ref{Fig:LQ3nu1cB005}  & Fig.\,\ref{Fig:LQnu15cB005}                                                                   & -                                                                   & Fig.\,\ref{Fig:LQnu45cB005}                                                                  \\ \hline
\end{tabular}
\caption{Density plots of $\log a$ of the the JQ parameter in $(\mu,T)$-plane for the LQ model  at different $c_B$ and $\nu$.}
\label{tab:LQ-density} 
\end{table}

\newpage
$$\,$$
\section{JQ Parameter for heavy quarks: numerical results}\label{sect:HQ-NR}

In this section, we study the JQ parameter within a specialized holographic model for heavy quarks exhibiting magnetic catalysis. Experimentally, heavy-quark QCD displays quantum catalysis, where an external magnetic field increases the critical temperature of the first-order phase transition. This model was constructed in \cite{Arefeva:2023jjh}. Our results for the JQ parameter for HQ models are different from other  holographic HQ models just due to the difference of the models themselves.
\\

The phase diagram of the HQ model in the $(\mu,T)$-plane describes two different types of phase transitions, i.e. the first-order phase transition and confinement/deconfinement phase transition; see Fig.\,\ref{Fig:HQnu1cB0}C. These phase transitions can also be presented in the $(\mu,z_h)$-plane. As noted in Sect.\,\ref{sect:LQ-NR}, the phase structure in the $(\mu,z_h)$-plane provides a convenient basis for examining how physical quantities depend on thermodynamic parameters $T$ and $\mu$.   
 $ (\mu,z_h)$-phase diagrams for our model are presented for two cases:
 \begin{itemize}
 \item
 $\nu=1$ with zero magnetic field, Fig.\,\ref{Fig:HQPT-mu-zh}.
\item
$\nu \neq 1$ with non-zero magnetic field,  Fig.\,\ref{Fig:HQnu15} and Fig.\,\ref{Fig:HQnu45}.
\end{itemize}
The corresponding phase diagrams in the $(\mu,T)$-planes are presented  in Fig.\,\ref{Fig:HQnu1cB0}C, Fig.\,\ref{Fig:HQ-PD-muT-nu},   Fig.\,\ref{Fig:HQnu15}, and   Fig.\,\ref{Fig:HQnu45}; see \cite{Arefeva:2023jjh}, \cite{Arefeva:2024xmg}, and \cite{Arefeva:2025okg}.
 \\

Our goal in this section is to show that a detailed examination of density plots of the JQ parameter on the $(\mu,T)$-plane  reveals that the JQ parameter responds only at the first-order phase transition line, exhibiting a distinct change there. In contrast, the change of the JQ parameter  across the second-order phase transition line is smooth.

\subsection{Jet quenching for zero magnetic field}\label{sect:HQ-NR-zero}

\begin{figure}[t!]
  \centering
  \includegraphics[scale=0.082]{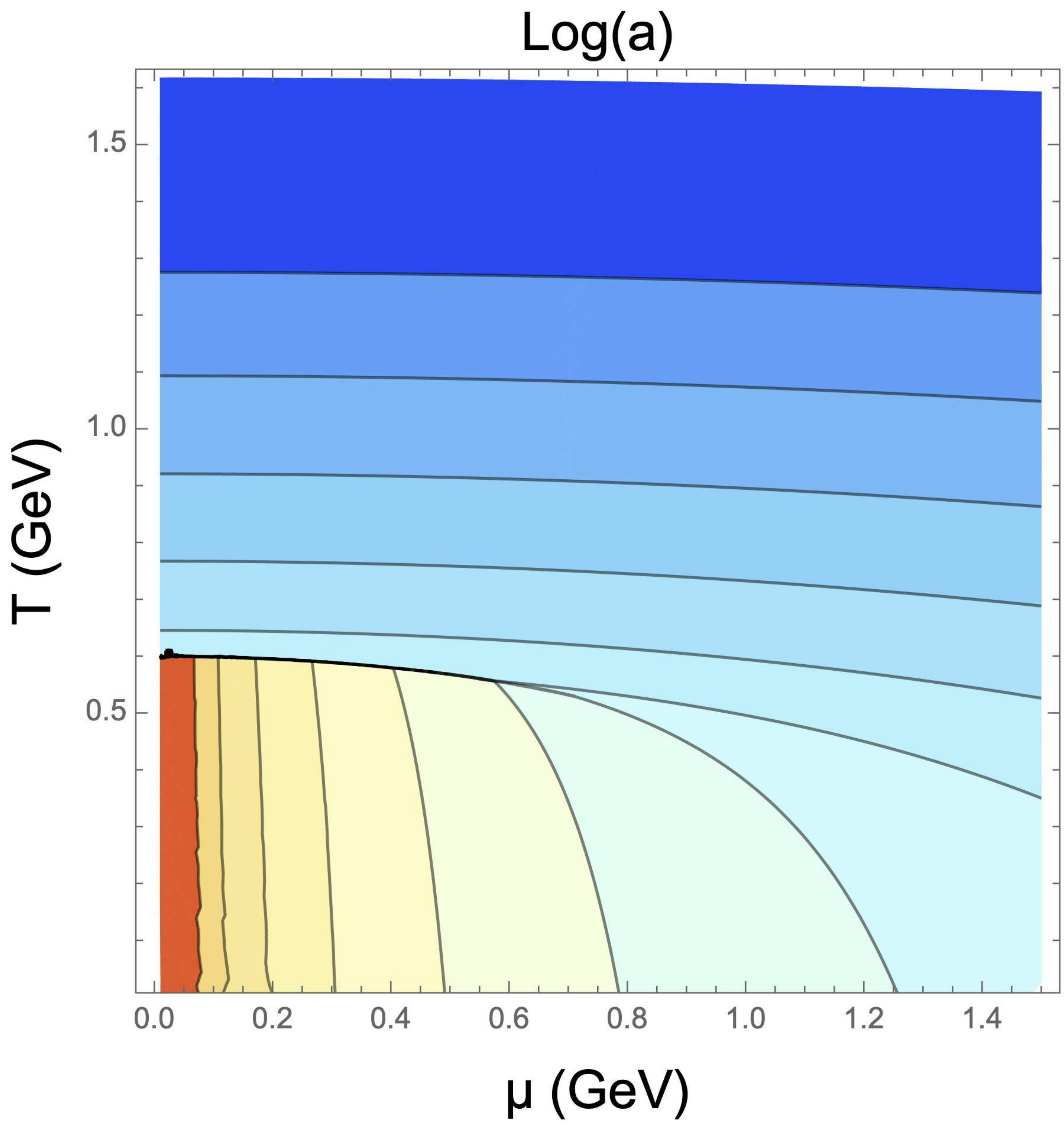}
  \includegraphics[scale=0.41] {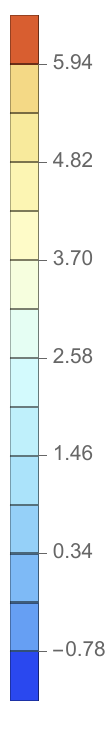}\qquad\quad
\includegraphics[scale=0.081]{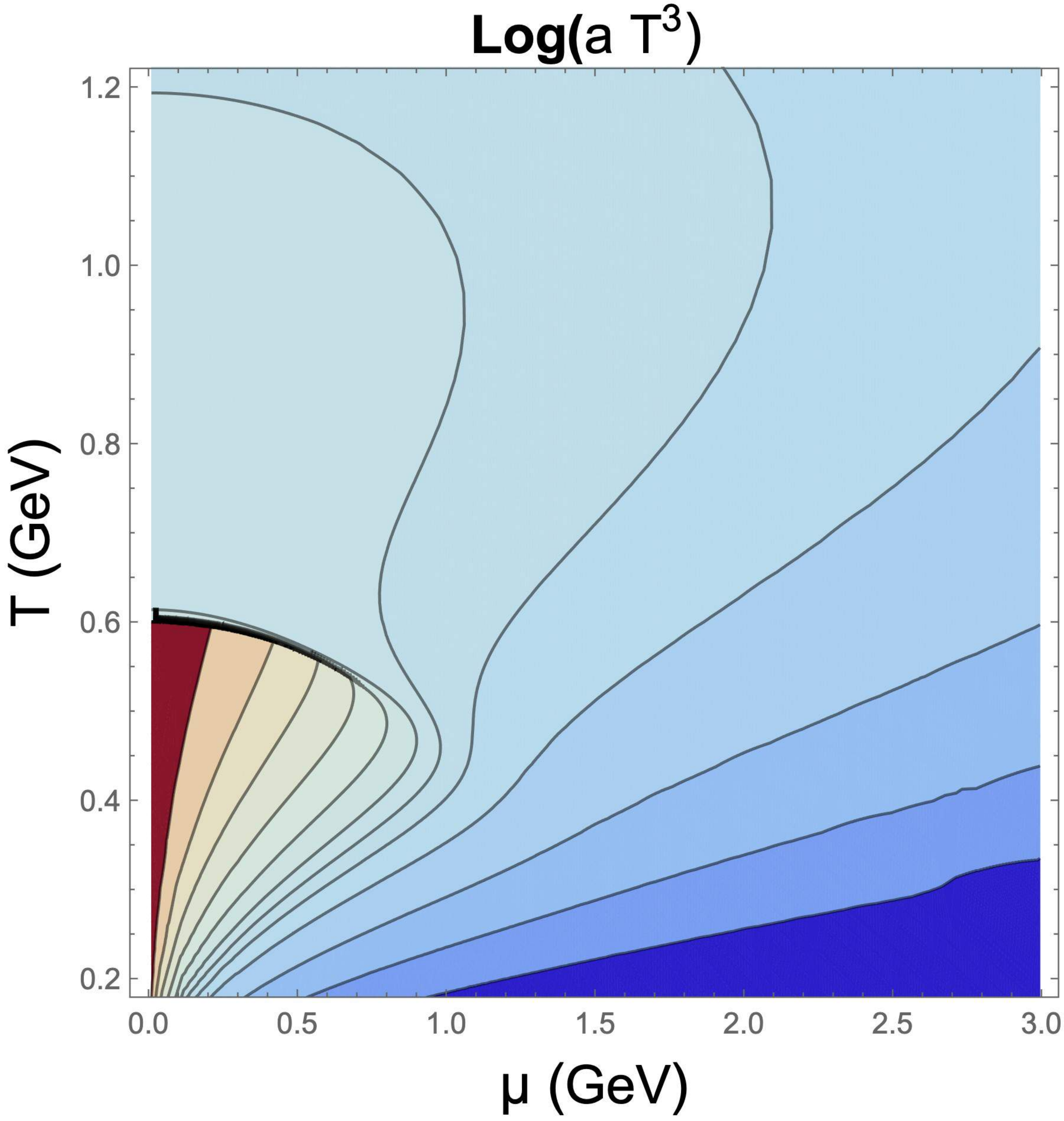}\qquad
\includegraphics[scale=0.31]
  {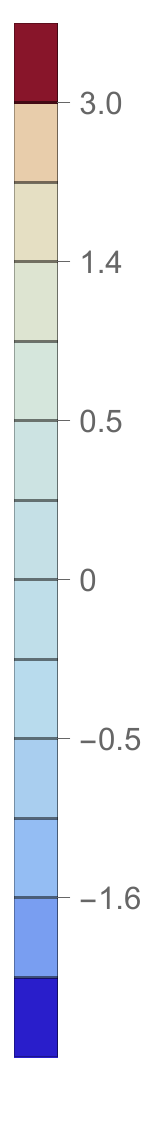}\\
  A \hspace{170pt} B\\ $\,$\\
 \includegraphics[scale=0.15]{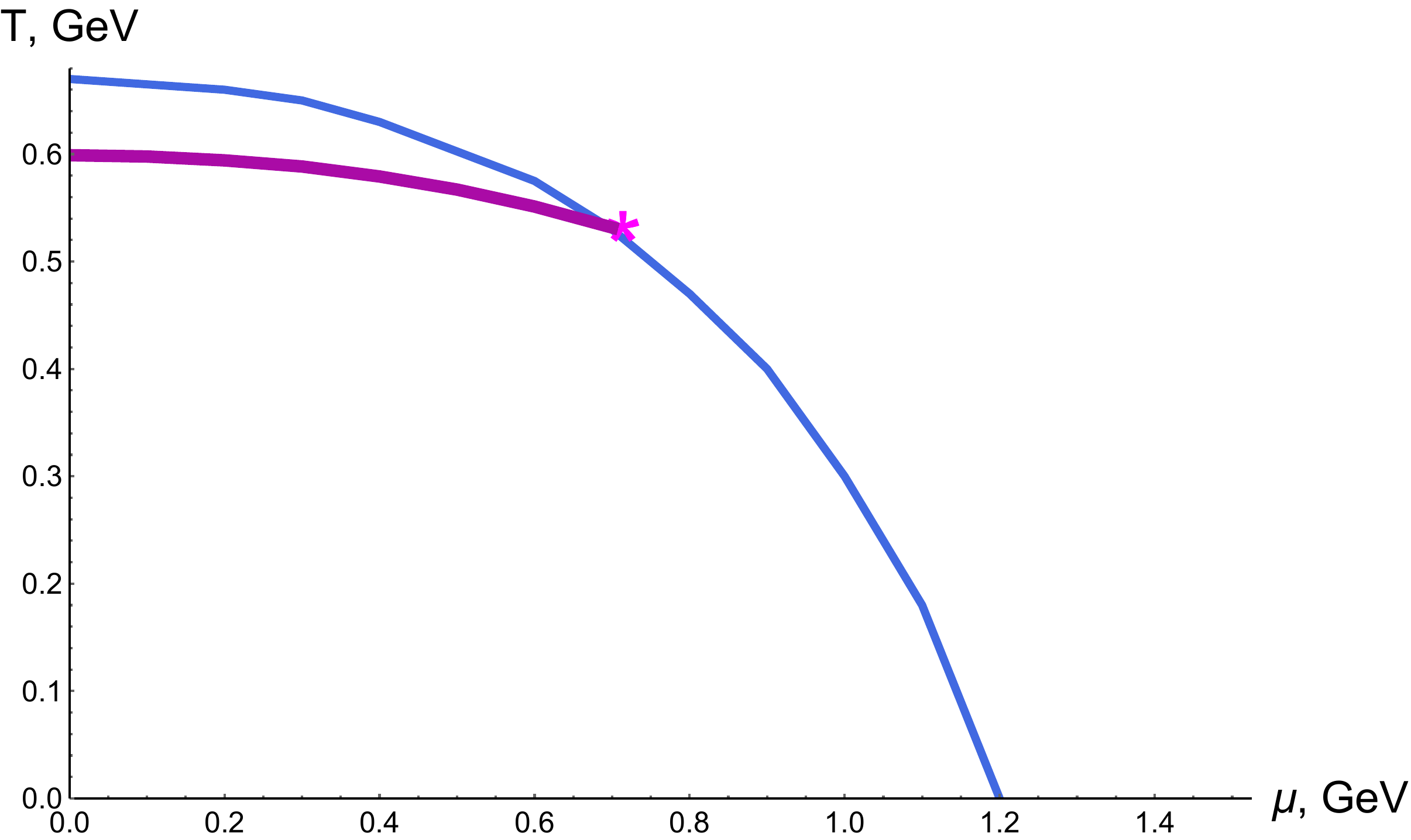} 
\includegraphics[scale=0.08]{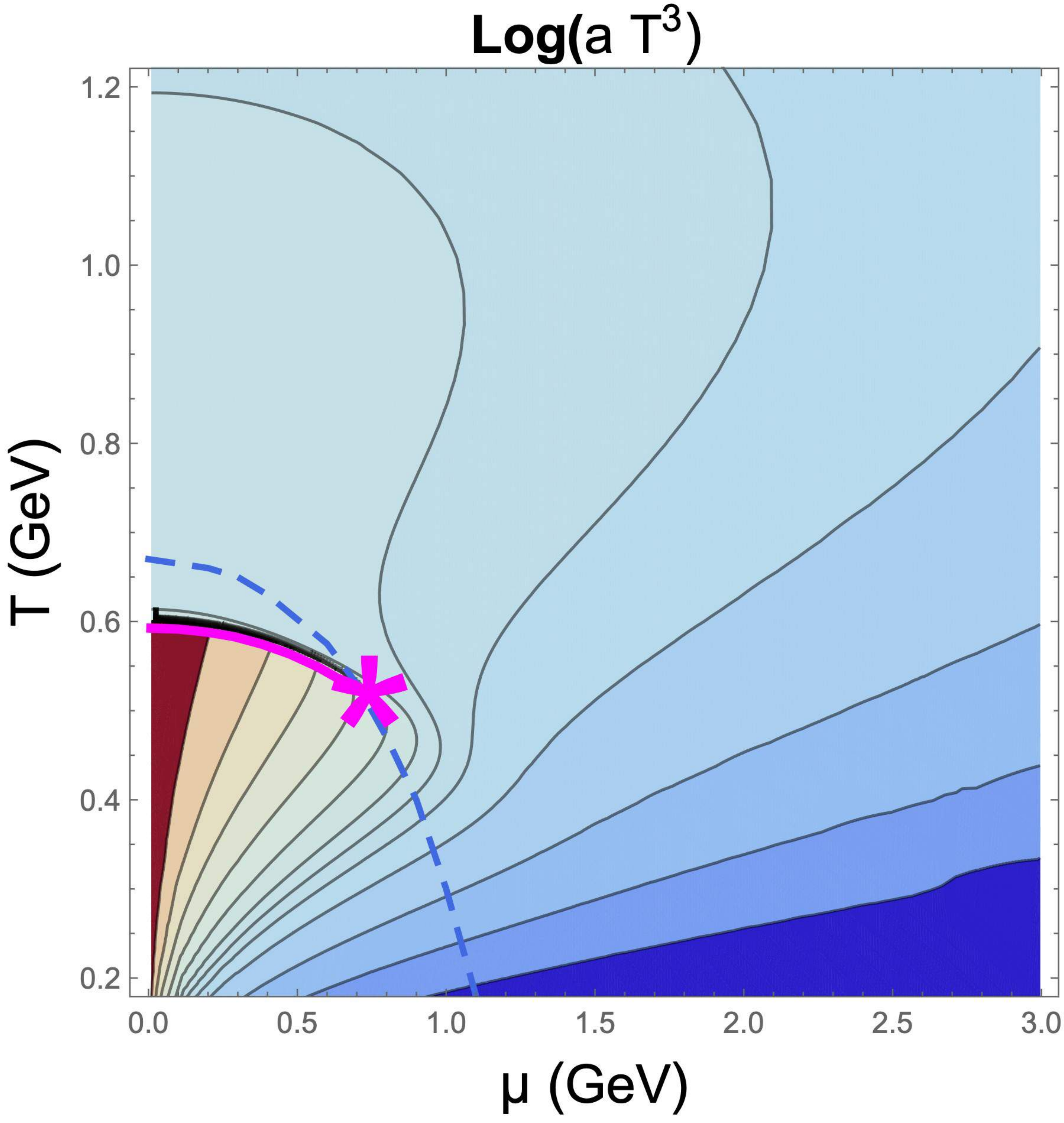}\\
 C \hspace{170pt} D 
\caption{
(A) Density plot with contours  for $\log a$ of the HQ model, $c_B=0$ and $\nu=1$, with  boundary condition \eqref{QH-nbc}. (B) Density plots with contours  for $\log (a T^3)$ for the same parameters as for the panel (A). (C) The phase diagram for the HQ model, $c_B=0$ and $\nu=1$. (D) The same as panel (B) with lines indicating phase transition lines: magenta line denotes first-order, dashed blue line denotes second-order and the magenta star presents the CEP.
  }
  \label{Fig:HQnu1cB0}
\end{figure}

The plot in Fig.\,\ref{Fig:HQnu1cB0}A  shows the density plot for  $\log a$ of the isotropic HQ model (for $\nu=1$ and $c_B=0$) with  boundary condition \eqref{QH-nbc}. Here are also some contours.  Each contour corresponds to a fixed value of $\log a$.
Fig.\,\ref{Fig:HQnu1cB0}B displays the density plot for $\log (a T^3)$ with the same parameters as Fig.\,\ref{Fig:HQnu1cB0}A.
Fig.\,\ref{Fig:HQnu1cB0}D is the same as Fig.\,\ref{Fig:HQnu1cB0}B, but includes additional curves and stars marking phase transition lines: magenta for first-order and dashed blue for second-order and the magenta star presents the CEP. The arrangement of constant $\log(a T^3)$ contours allows us to determine the location of the first-order phase transition in the $(\mu,T)$-plane shown in Fig.\,\ref{Fig:HQnu1cB0}C.

\newpage
$$\,$$

\begin{figure}[h]
    \centering
  \includegraphics[width=0.4\textwidth]{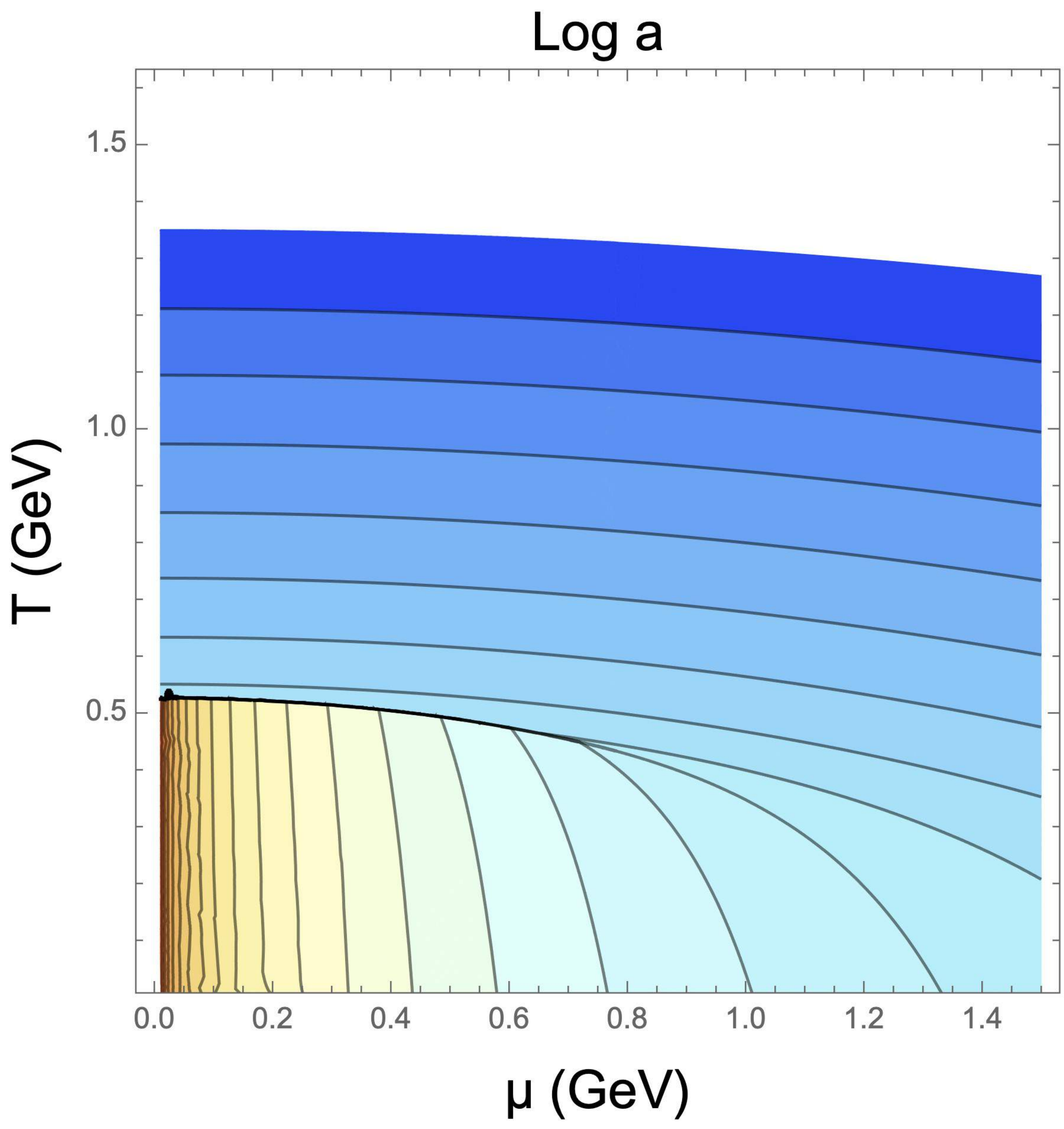}
\includegraphics[width=0.05\textwidth]{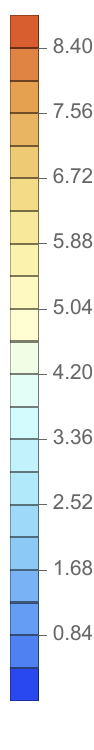}
\quad
\includegraphics[width=0.4\textwidth]{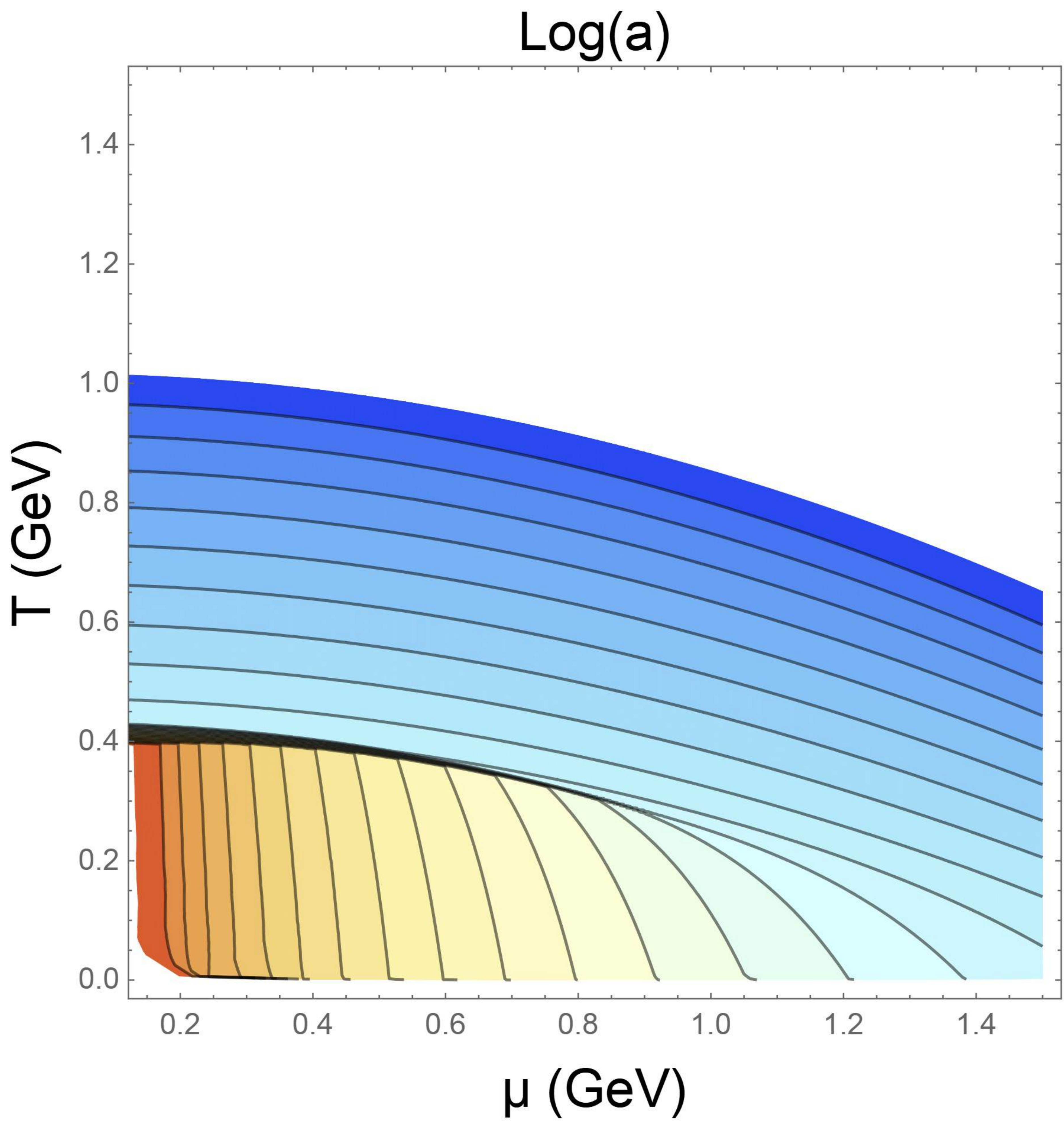}
\includegraphics[width=0.05\textwidth]{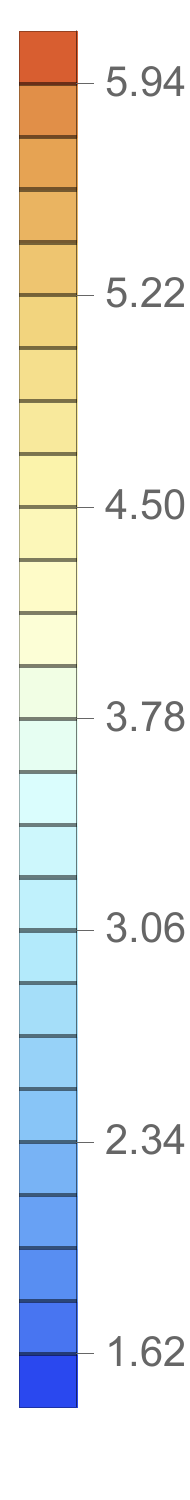}
\\A\hspace{150pt}B
\caption{ The density plots for the HQ model in the ($\mu, T$)-plane with contours for $\log a$   at $c_B = 0$ with (A)  $\nu = 1.5$, and (B) $\nu=4.5$.
}
    \label{Fig:HQnu1545-cB0}
\end{figure}

\begin{figure}[h!]
  \centering
  \includegraphics[scale=0.43]
 {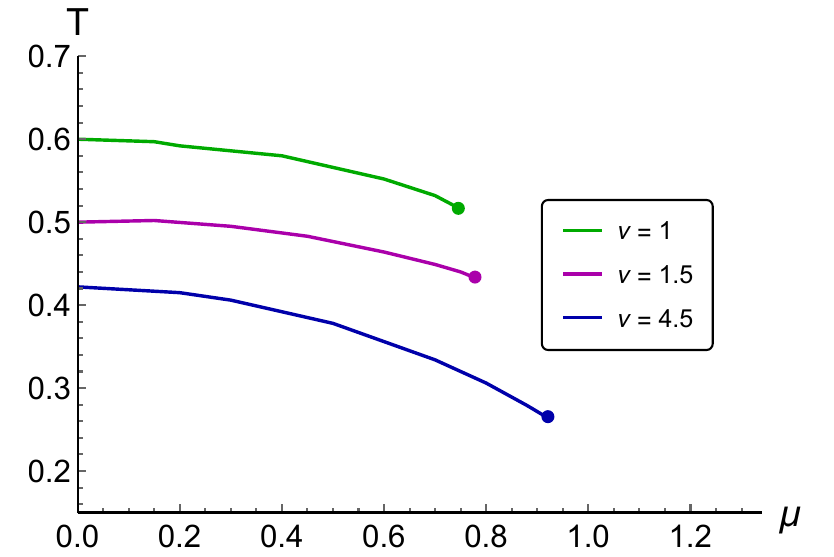}
  \caption{The first-order phase transition lines for the HQ model at $c_B=0$ considering spatial anisotropies $\nu=1,1.5$, and $\nu=4.5$. 
}\label{Fig:HQ-PD-muT-nu}
\end{figure}

The plots in Fig.\,\ref{Fig:HQnu1545-cB0} show density plots with contours for the HQ model with $c_B = 0$ for   $\nu = 1.5$ and $\nu=4.5$.  The orientation of the contours differs between the regions below and above $T \approx 0.5$ GeV (Fig.\,\ref{Fig:HQnu1545-cB0} A) and $T \approx 0.4$ GeV (Fig.\,\ref{Fig:HQnu1545-cB0} B). Furthermore, the density appears smoother for $\mu > 0.81$ GeV in Fig.\,\ref{Fig:HQnu1545-cB0} A and $\mu > 0.95$ GeV in Fig.\,\ref{Fig:HQnu1545-cB0} B. Fig.\,\ref{Fig:HQnu1545-cB0} shows that in the HQ model in hadronic phase the JQ parameter mostly depends on $\mu$ and slightly on $T$ that is completely  in opposite to the LQ model. These pictures reveal the first-order phase transition in $(\mu,T)$-plane shown in Fig.\,\ref{Fig:HQ-PD-muT-nu} and corresponding locations of the CEP.

\newpage

\subsection{Non-zero magnetic field}
\label{sect:HQ-NR-nzero}

In this subsection, with a magnetic field oriented along the $x_3$-axis, we focus on jets propagating along the $x_1$-direction and their momentum broadening primarily parallel to the magnetic field (along $x_3$), denoted $\hat{q}_3$. In all plots, we omit the subscript 3 for both the jet quenching parameter $\hat{q}$ (Eq. \eqref{qi}) and the parameter $a$ (Eq. \eqref{a3}). When comparing $\hat{q}$ values across different directions, we restore the corresponding indices (see Figs.\,\ref{Fig:HQnu1q23}, \ref{Fig:HQ23nu15cB005}, and \ref{Fig:HQ23nu45cB005}).

\subsubsection{Non-zero magnetic field, $\nu=1$}
\label{NR-HQ-nzero-nu1}

\begin{figure}[h!]
  \centering
\includegraphics[scale=0.27]
{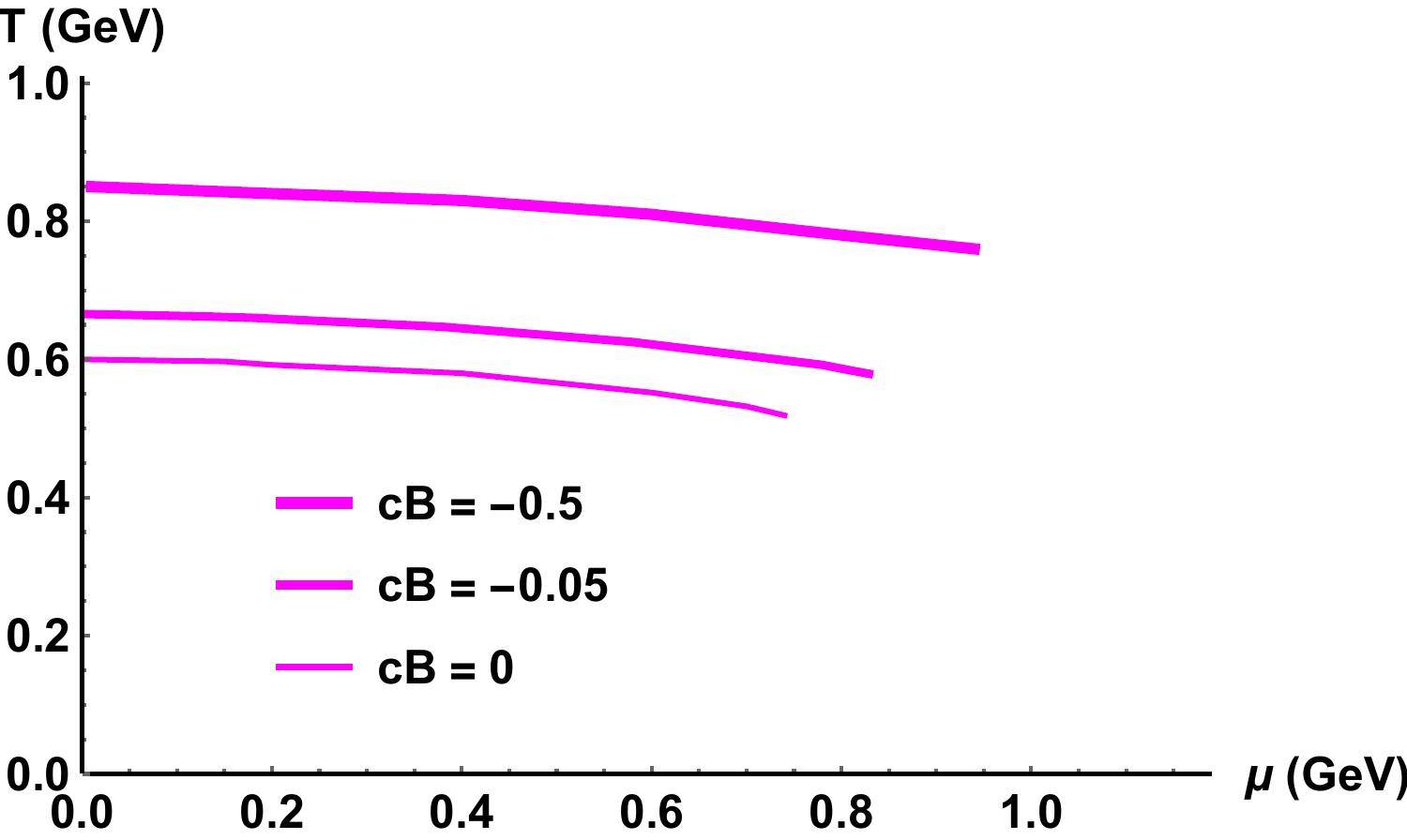}
\quad\includegraphics[scale=0.28]{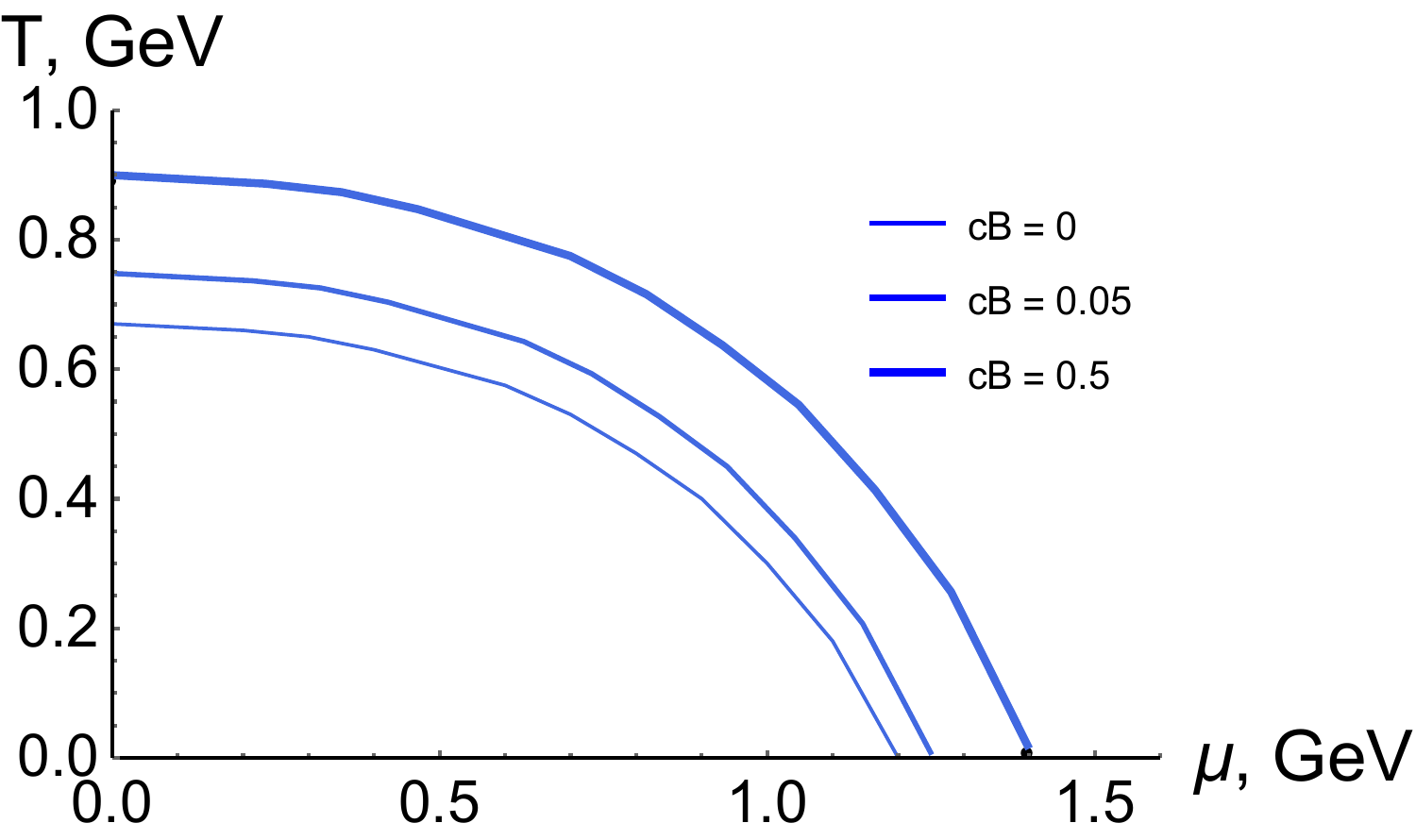}\\
A\hspace{150pt}B\\
\caption{(A) The first-order phase transition lines for the HQ model with $\nu=1$ and $c_B = 0, -0.05$ \GG , $-0.5$ \GG, and  (B) the second-order phase transition lines for the HQ model with the same $\nu$ and $c_B$.
  }
  \label{Fig:HQ-PTnu1cB0-005-05}
\end{figure}

The plots in Fig.\,\ref{Fig:HQ-PTnu1cB0-005-05} show the phase transitions in the HQ model for $\nu=1$ and different magnetic fields.

\begin{figure}[h!]
  \begin{center}
\includegraphics[scale=0.07]{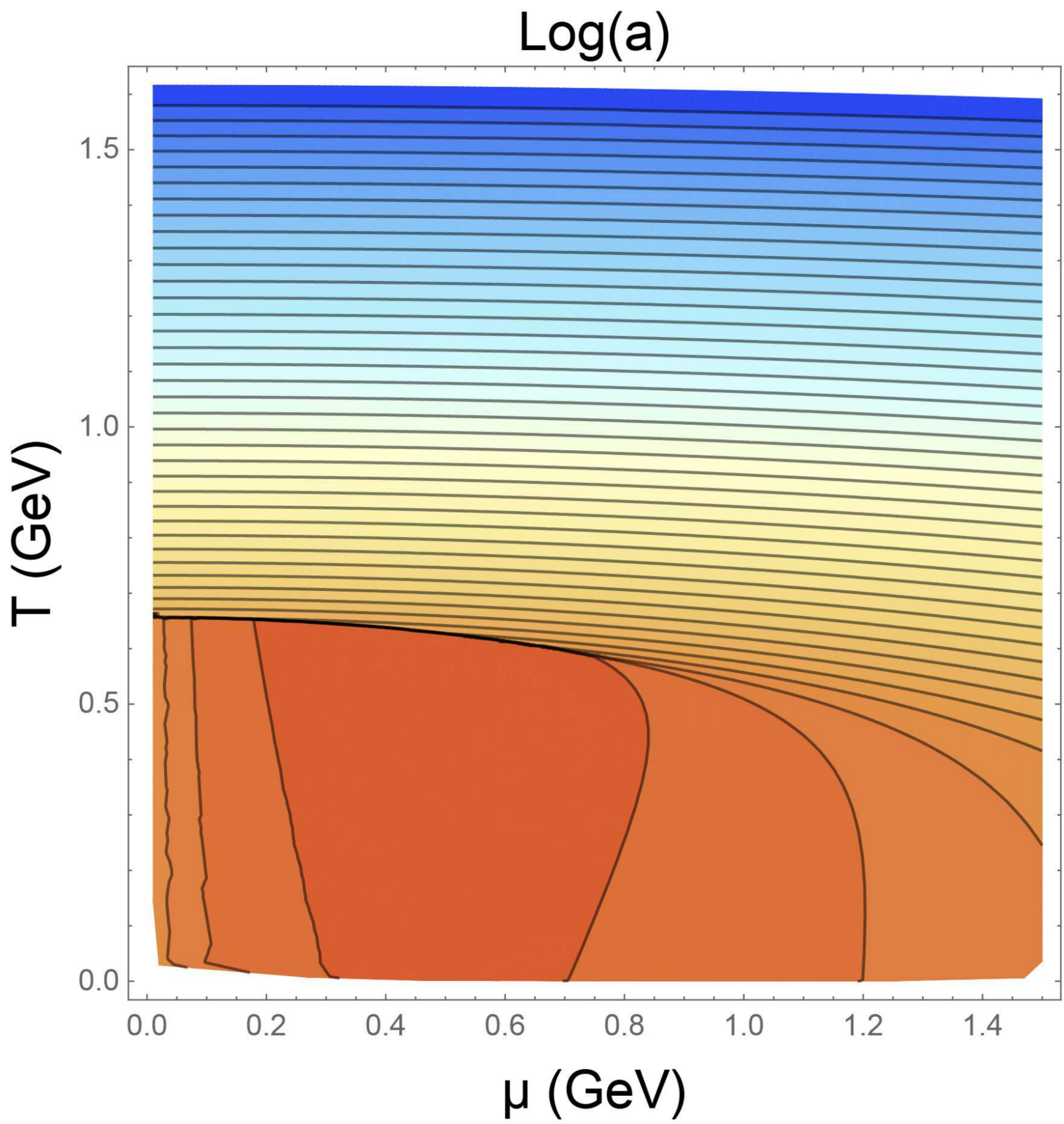}
\includegraphics[scale=0.18]{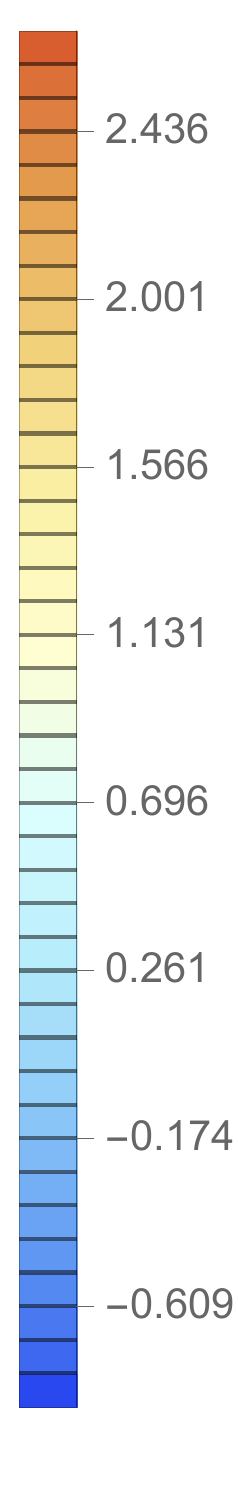}\quad
 \includegraphics[scale=0.171]
{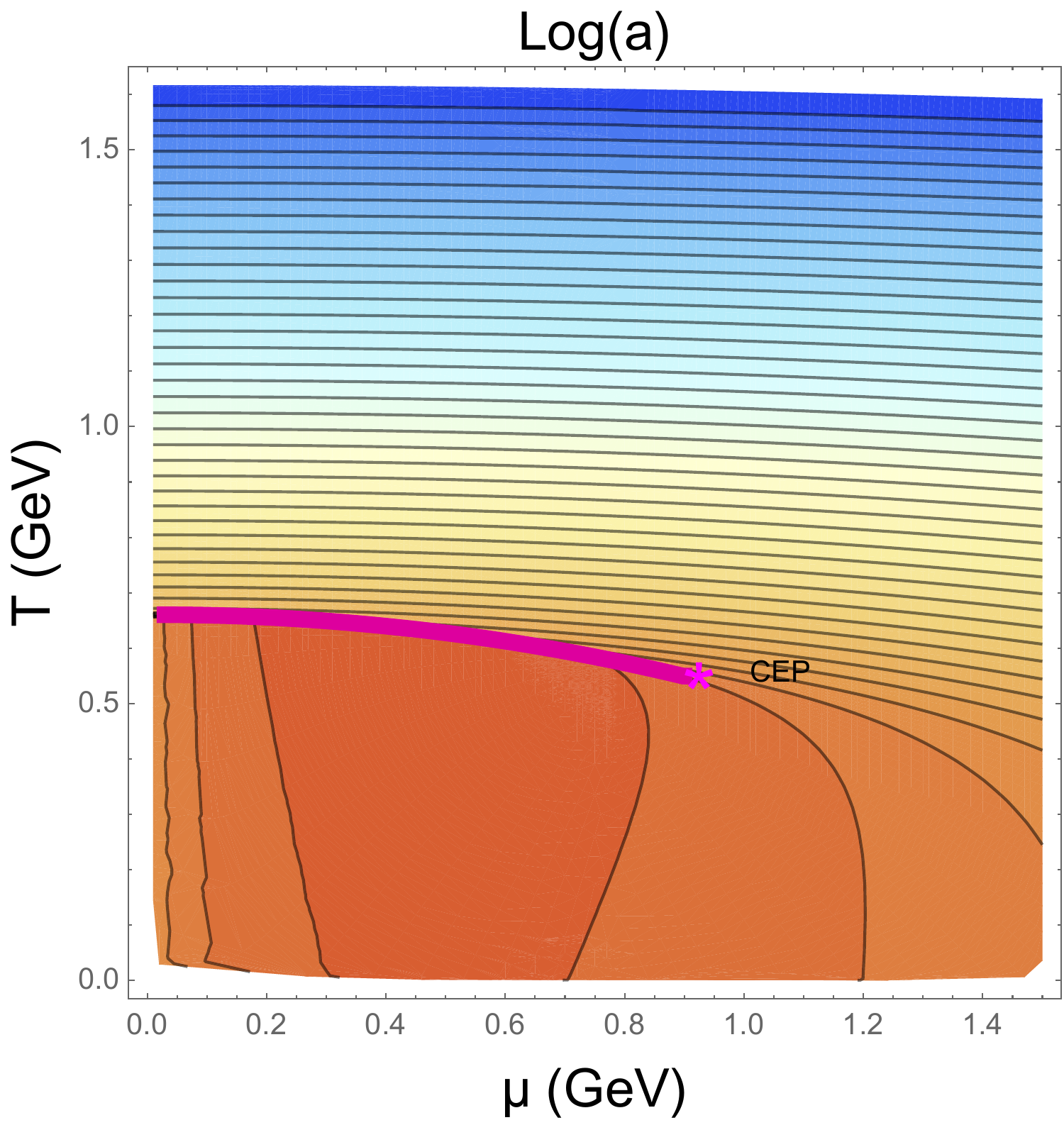}
\includegraphics[scale=0.17] {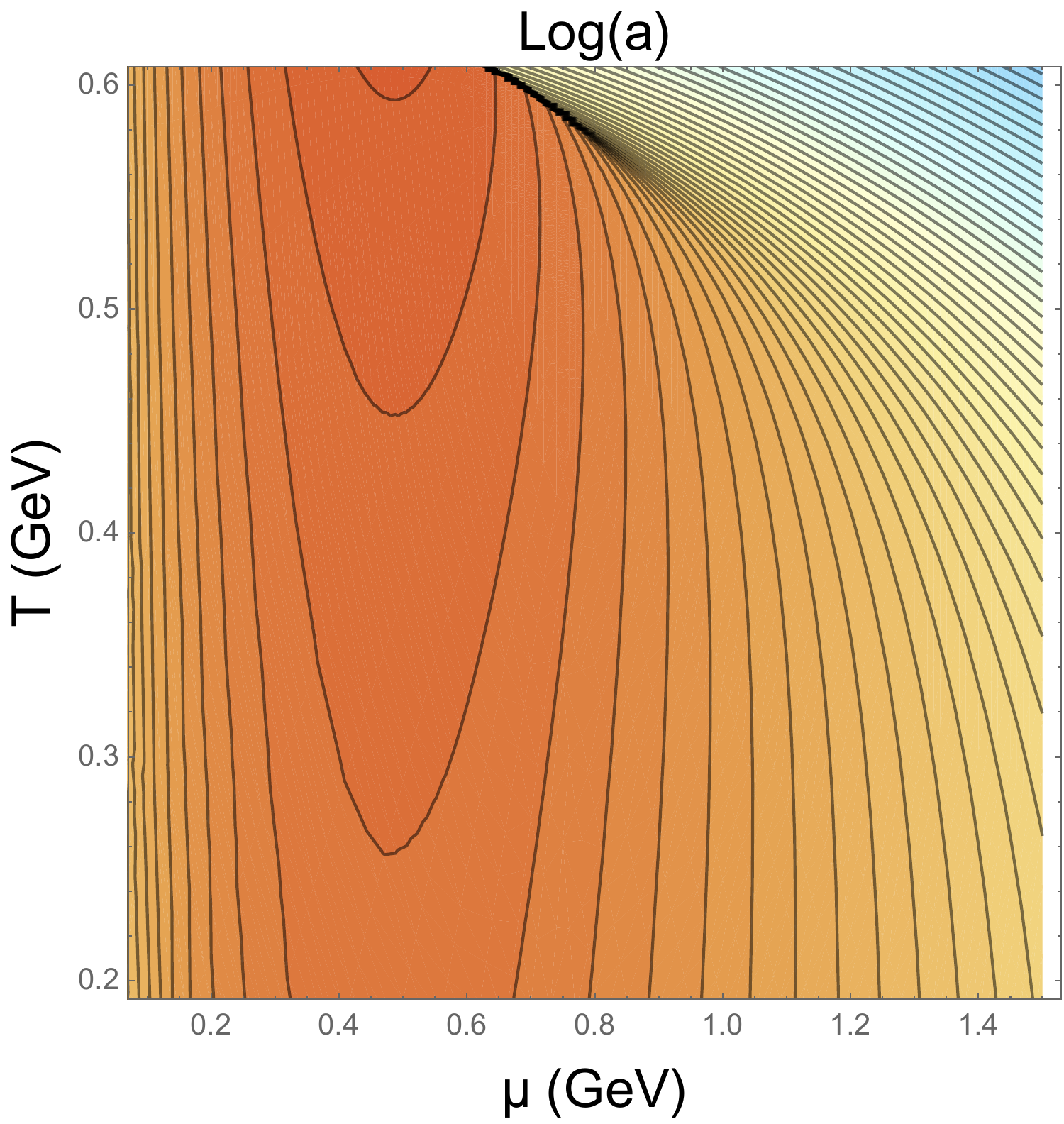}
\includegraphics[scale=0.17] {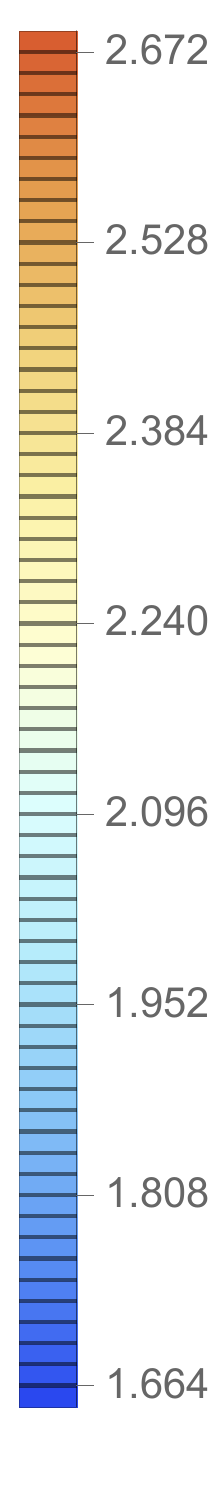}\\
A\hspace{140pt}B\hspace{140pt}C
\\
\includegraphics[scale=0.072] {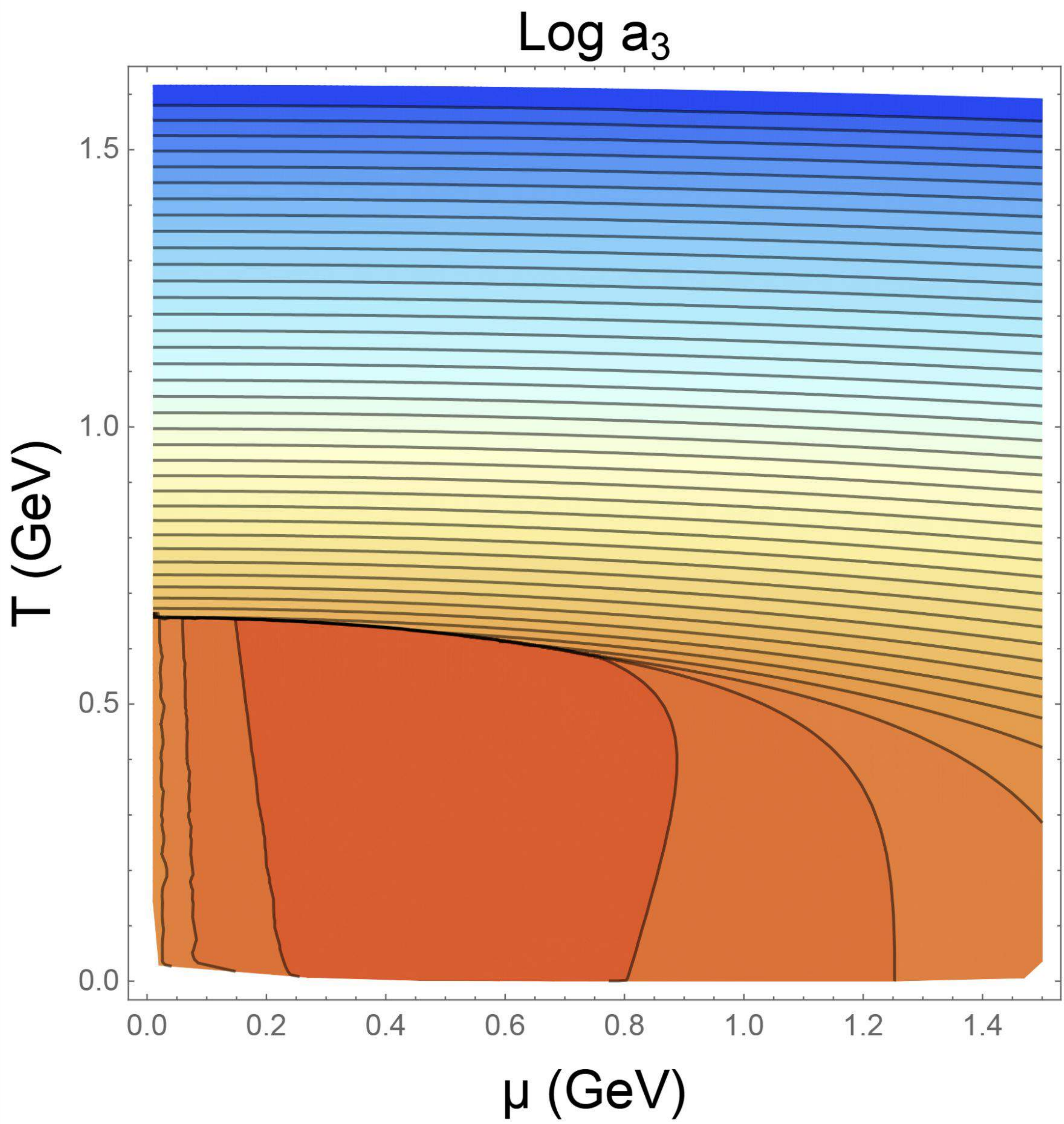}
\includegraphics[scale=0.18] 
{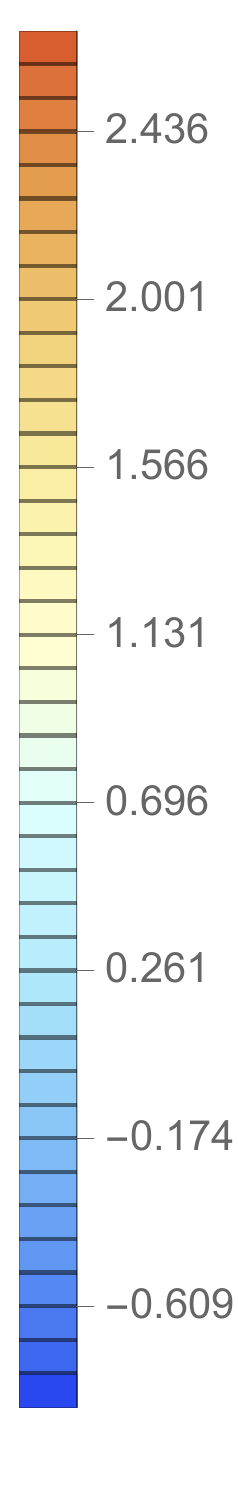}
\includegraphics[scale=0.173] {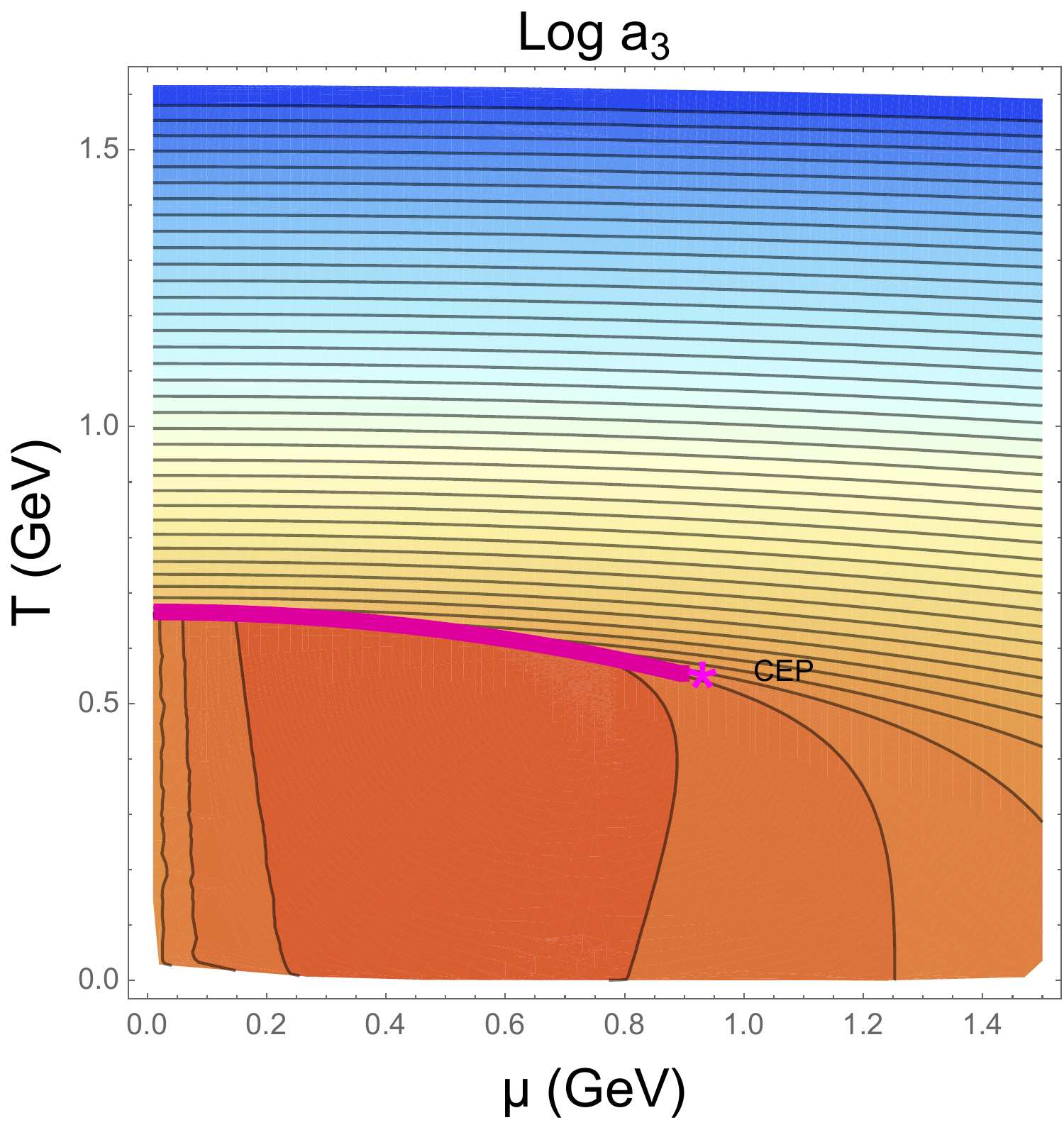}
\includegraphics[scale=0.173] {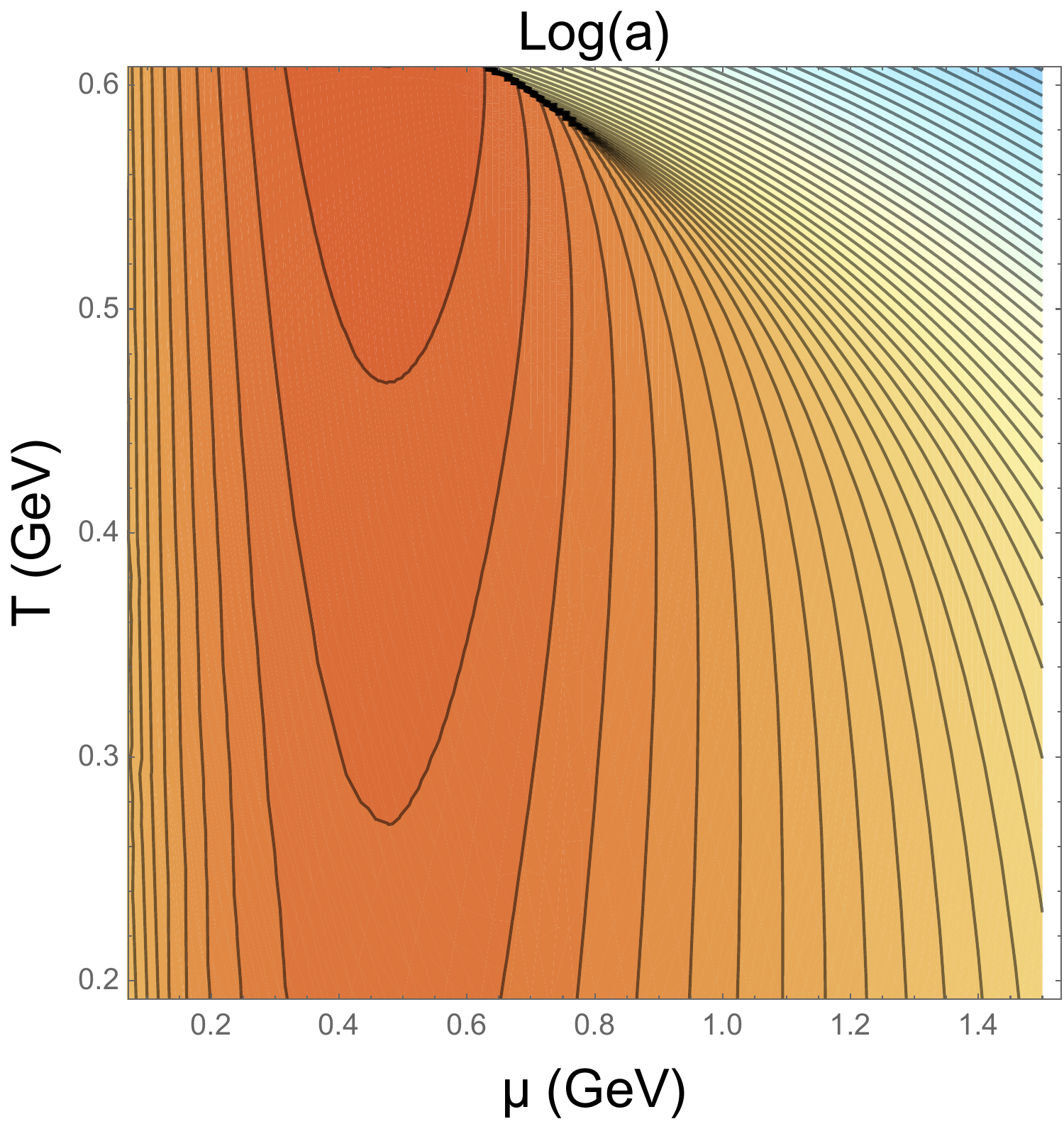}
\includegraphics[scale=0.18] {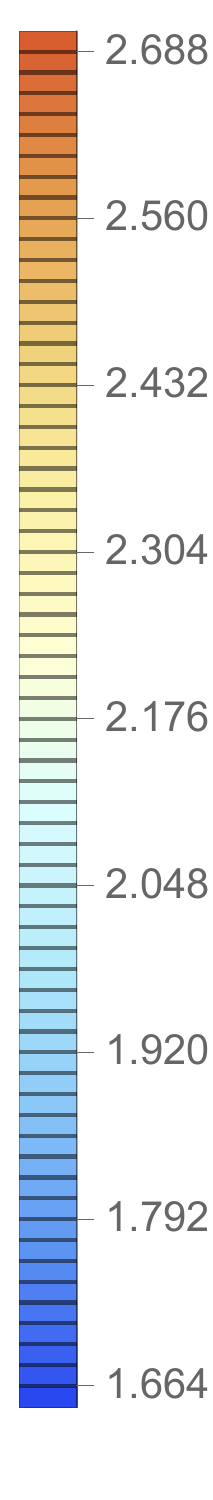}\\
D\hspace{140pt}E\hspace{140pt}F
\end{center}
\caption{Contour plots of (A) $\log a_2$ and (D) $\log a_3$ for the HQ model in the $(\mu, T)$-plane with non-zero magnetic field $c_B = -0.05$ \GG and $\nu = 1$, both displaying 20 contours. Panels (B) and (E) replicate panels A and D respectively, with first-order phase transition lines in magenta and magenta stars marking the critical endpoints (CEPs). Panels (C) and (F) show zoomed regions of panels A and D with higher contour density (200 contours), revealing hill-like structures beneath the phase transition lines. The first-order phase transition begins at $(\mu, T) = (0, 0.65$ GeV). Increased contour density in (C) and (F) shows no significant differences between $\log a_2$ (top row) and $\log a_3$ (bottom row).
  }
  \label{Fig:HQnu1cB005}
\end{figure}

\begin{figure}[h!]
  \begin{center}
 \includegraphics[scale=0.1]{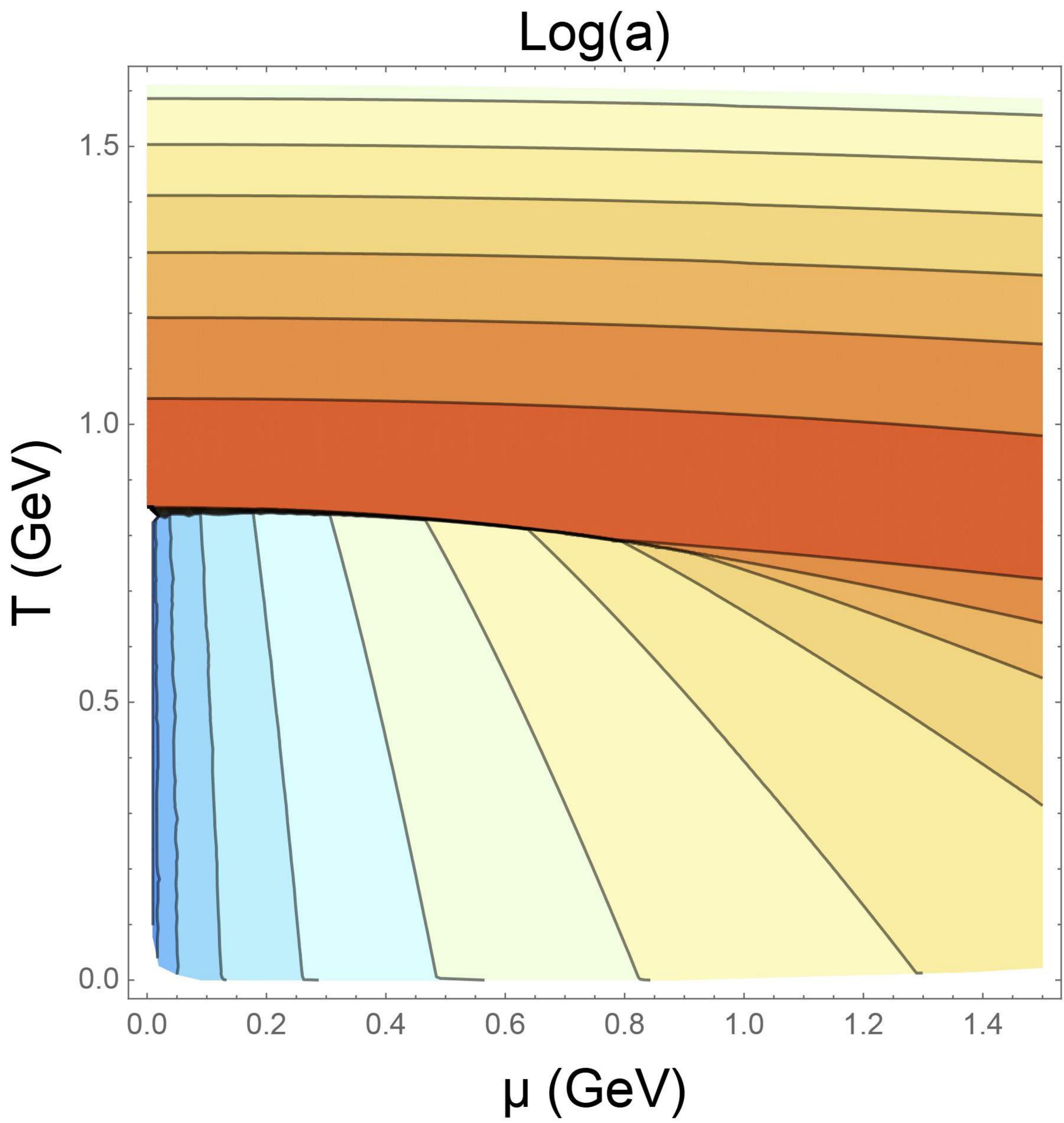}
 \includegraphics[scale=0.25]{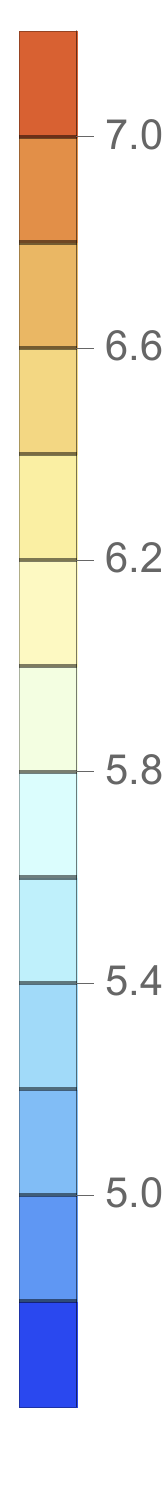}\qquad
\includegraphics[scale=0.1]{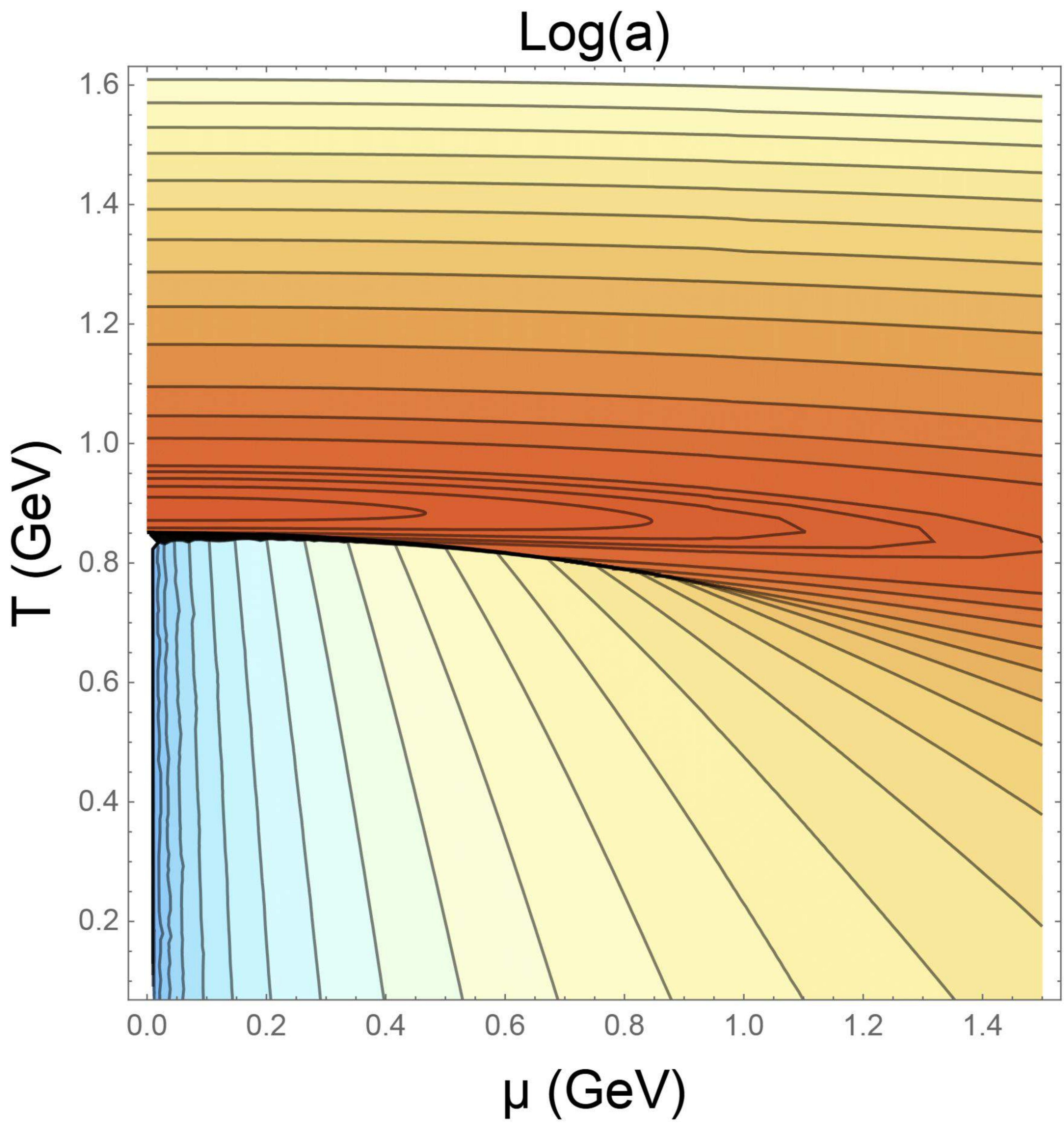}
  \includegraphics[scale=0.25]{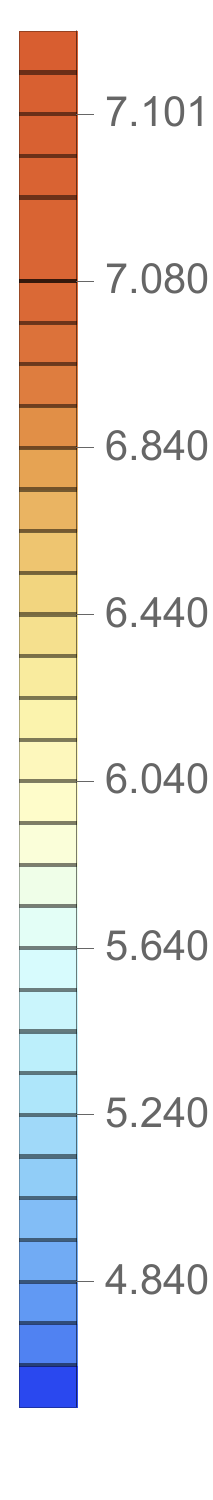}\\
 A\hspace{220pt}B\\
\,\ \\
\includegraphics[scale=0.1]{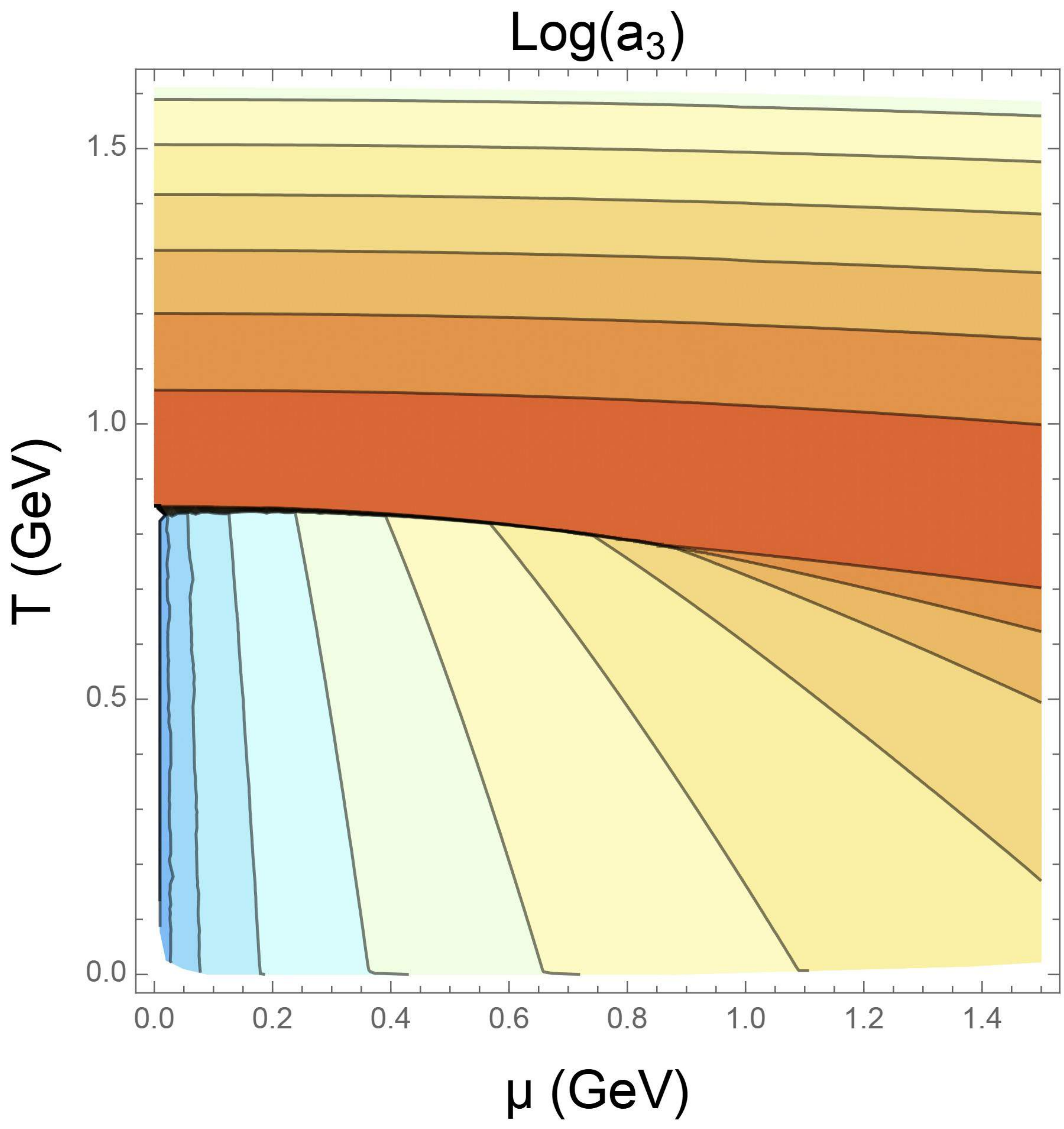}
 \includegraphics[scale=0.5]{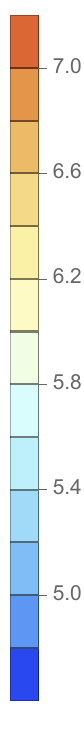}\qquad
\includegraphics[scale=0.5]{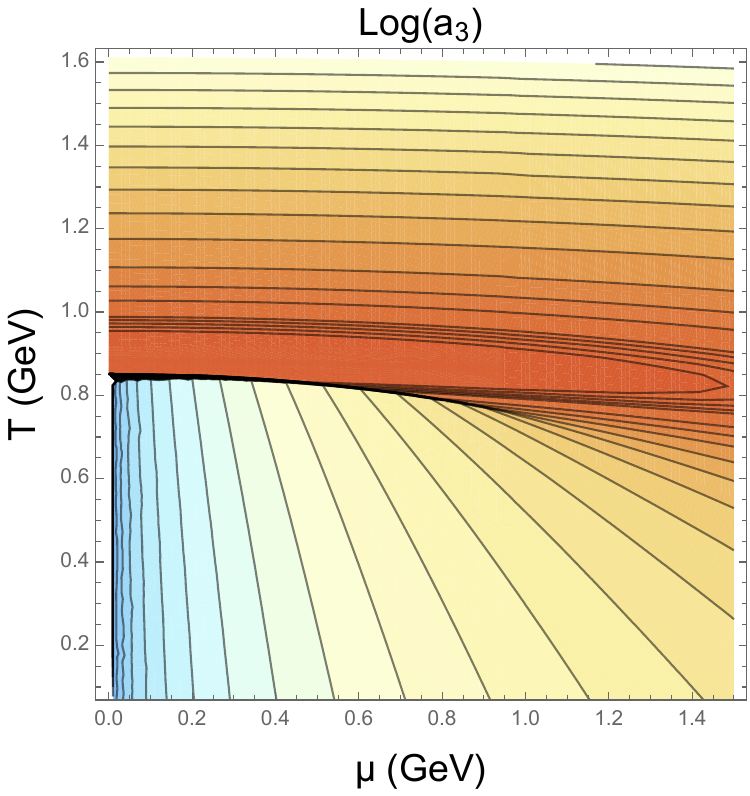}
  \includegraphics[scale=0.5]{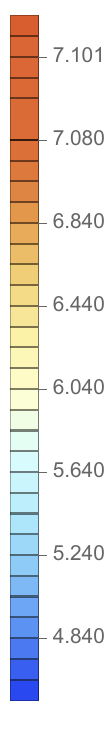}\\
 C\hspace{220pt}D
 \end{center}
  \caption{Contour plots for the HQ model in the $(\mu, T)$-plane with $c_B = -0.5$ GeV and $\nu = 1$:
(A) $\log a_2$ and (C) $\log a_3$ at standard contour density;
(B) and (D) show higher contour density versions of (A) and (C) respectively.
Differences between $\log a_2$ (top row) and $\log a_3$ (bottom row) are only apparent at higher contour density, as seen in panels (B) and (D).
  }
  \label{Fig:HQnu1cB05}
  \end{figure}

\begin{figure}[h!]
  \begin{center}

\includegraphics[scale=0.25]
  {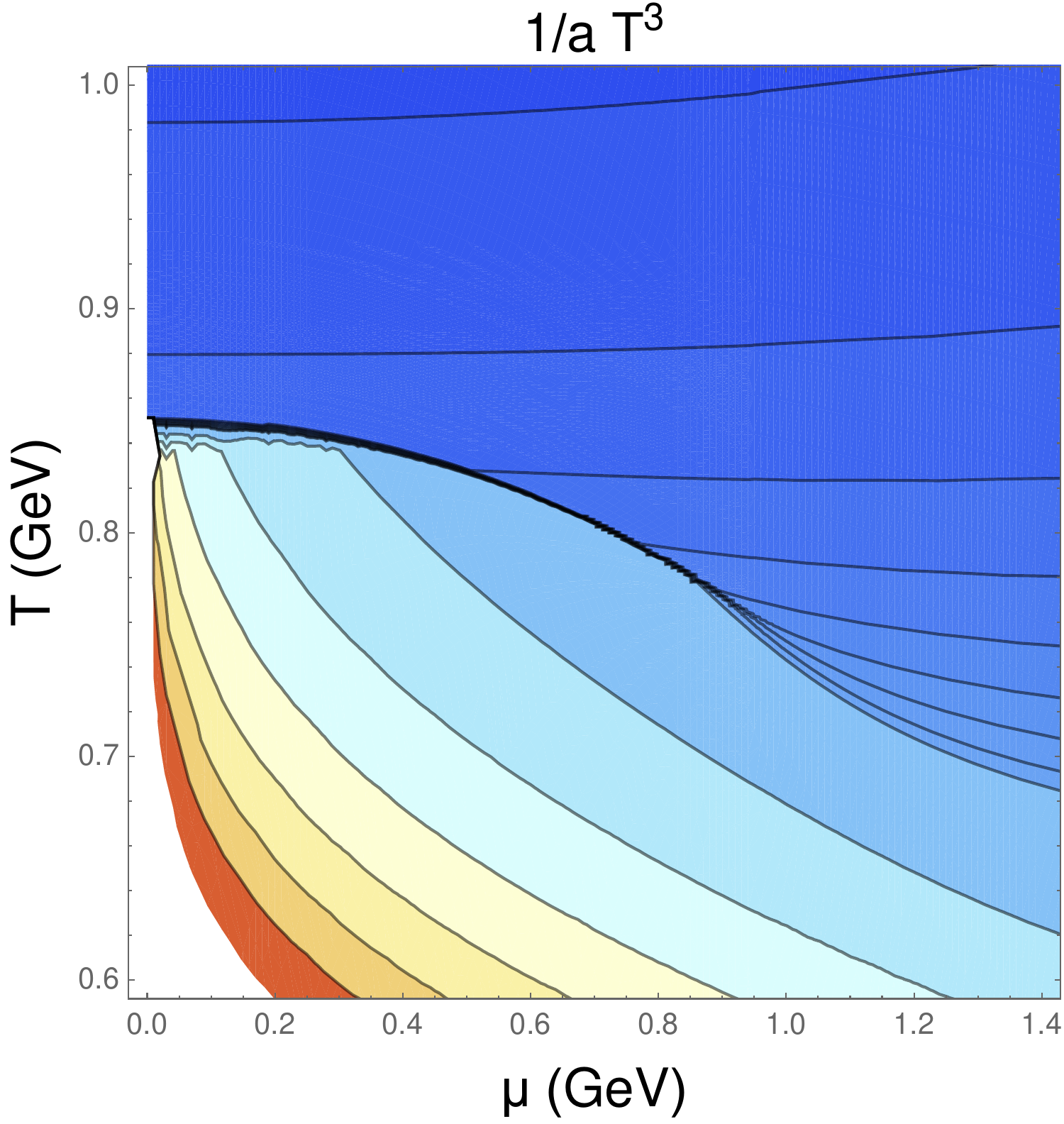}\quad
\includegraphics[scale=0.25]
  {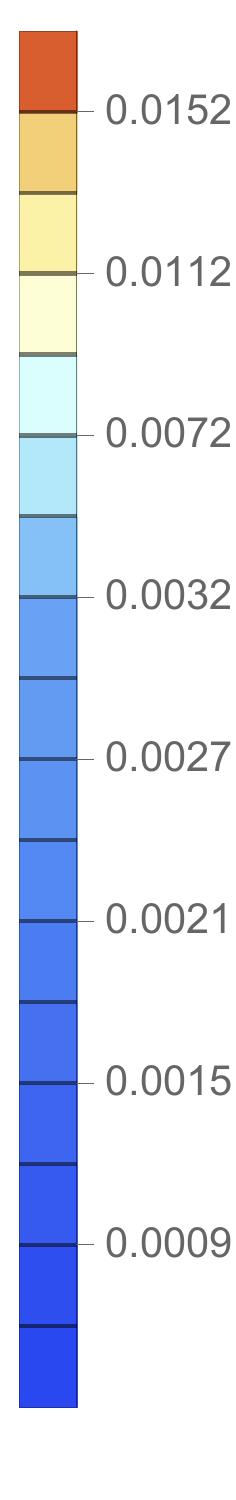}\quad
  \includegraphics[scale=0.25]
  {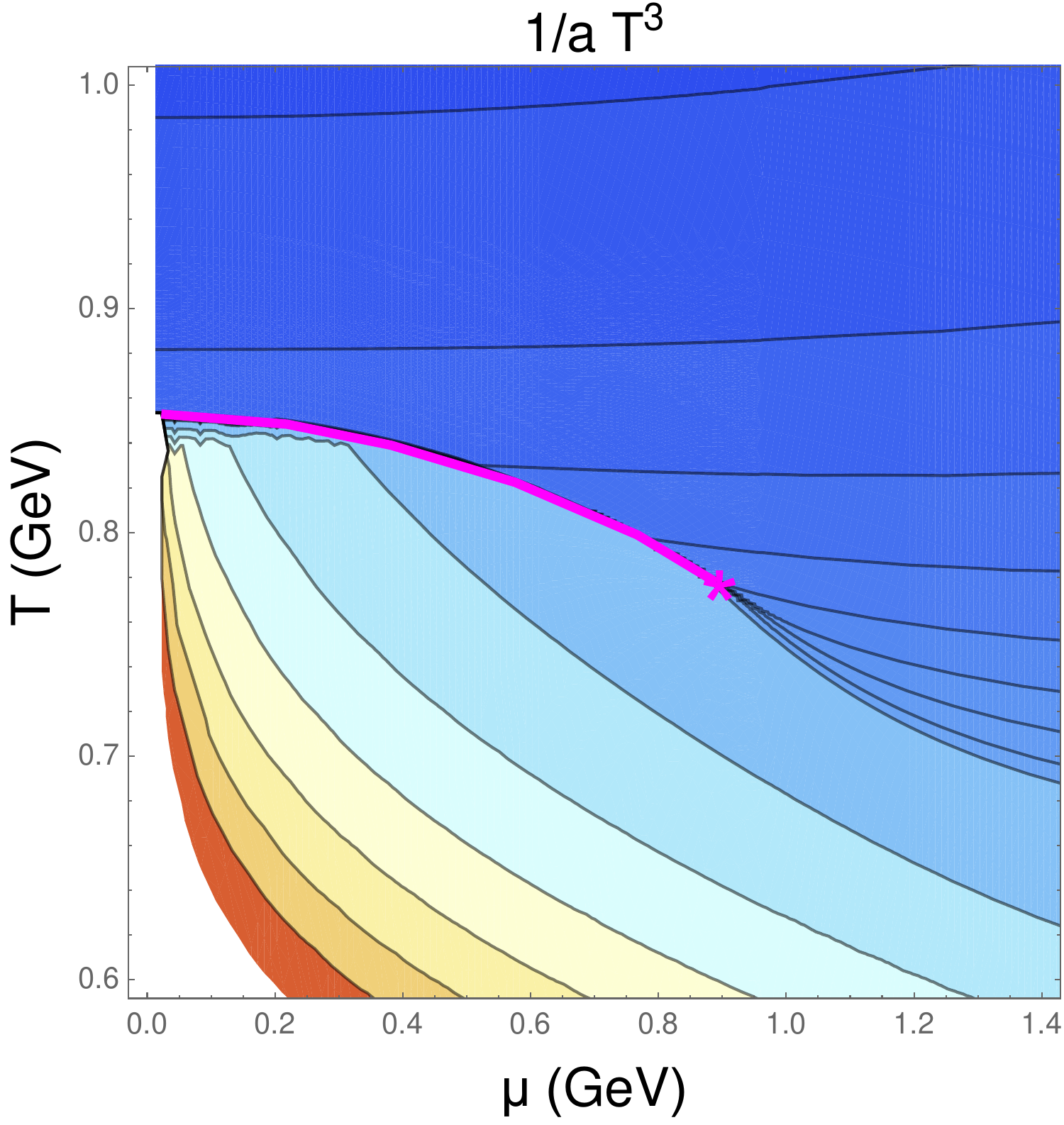}
\\
   A\hspace{220pt}B
 \end{center}
  \caption{ (A) Density plot with contours for $\log a_2 T^3$  for the HQ model with  non-zero magnetic field   $c_B=-0.5$ \GG and $\nu = 1$.
  (B) The same as in panel (A), with the first-order transition indicated by a magenta line.  Here, the magenta line starts at $\mu=0$, $T=0.85$ GeV. 
  }
  \label{Fig:HQnu1cB05T3}
  \end{figure}

\begin{figure}[h!]
  \begin{center}
\includegraphics[scale=0.14]
  {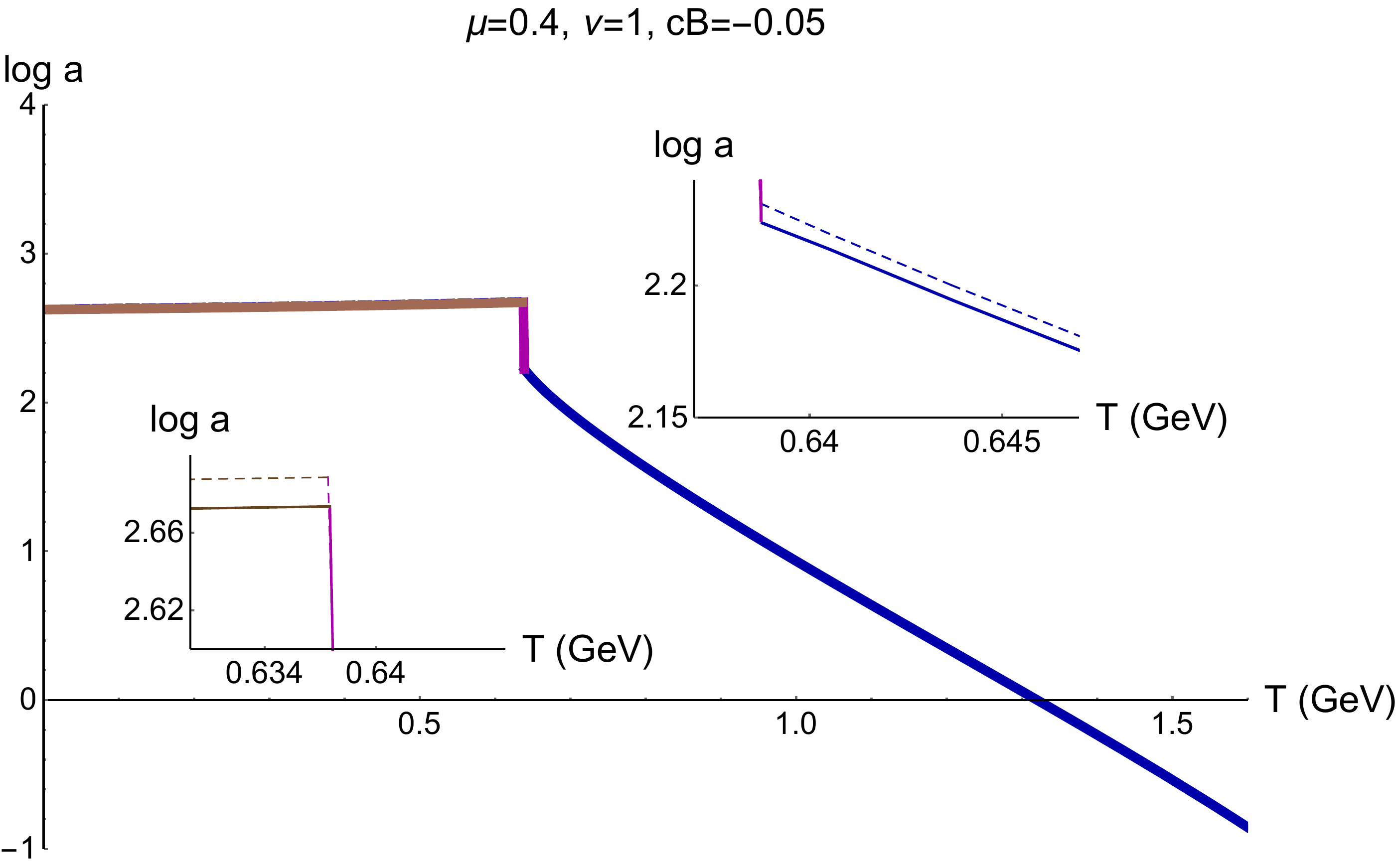}\qquad
   \includegraphics[scale=0.16]
  {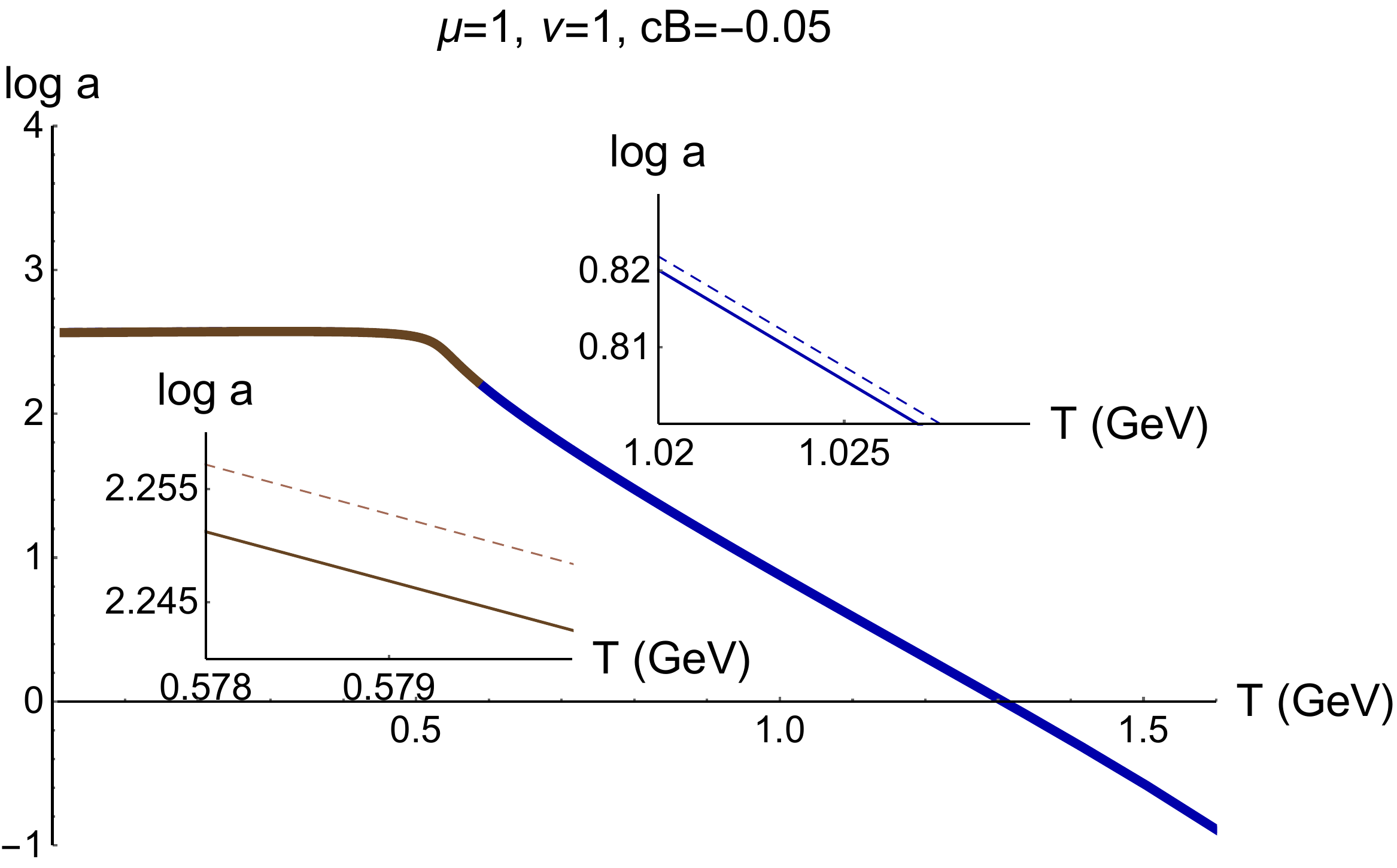}\\A\hspace{190pt} B 
  \end{center}
\caption{ $\log a_2$ (solid lines) and $\log a_3$ (dashed lines) for the HQ model, with  $\nu = 1$ and $c_B=-0.05$ \GG, versus temperature at fixed chemical potentials:  (A) $\mu = 0.4$ GeV and (B)  $\mu = 1$ GeV. The solid  and dashed blue lines correspond to the QGP, and solid brown and dashed gray lines correspond to the hadronic phases. The magenta lines (solid and dashed) correspond to unstable regions. We see  jumps at panel (A)  and  changes of the slopes near the second-order phase transitions at panel (B). 
  }
  \label{Fig:HQnu1q23}
\end{figure}

The plots in Fig.\,\ref{Fig:HQnu1cB005},  Fig.\,\ref{Fig:HQnu1cB05}, and Fig.\,\ref{Fig:HQnu1cB05T3} demonstrate that the behavior of the JQ parameter enables identification of the first-order phase transition position even in the presence of an external magnetic field. This conclusion is further supported by the two-dimensional graphs shown in Fig.\,\ref{Fig:HQnu1q23}.

Comparing the density plots of the JQ parameter in Fig.\,\ref{Fig:HQnu1cB0} (zero magnetic field) and Fig.\,\ref{Fig:HQnu1cB005} (non-zero magnetic field), significant differences emerge in the behavior of constant-value contours. Specifically, as noted in Sect.\,\ref{sect:HQ-NR-zero}, the contours in Fig.\,\ref{Fig:HQnu1cB0} are nearly vertical, while those in Fig.\,\ref{Fig:HQnu1cB005} are predominantly horizontal. This contrast concerns hadronic phases.

\begin{itemize}
  \item The plots in Fig.\,\ref{Fig:HQnu1q23} show that for $\nu = 1$, the JQ parameter exhibits anisotropy in the presence of a magnetic field: $\log a_2 \neq \log a_3$ for $c_B = -0.05$ \GG.
  
  \item Both $\log a_2$ and $\log a_3$ are nearly temperature-independent in the hadronic phase but decrease with temperature in the quark-gluon phase. Each exhibit jumps at first-order phase transitions with approximately equal magnitudes.
  
  \item For large chemical potentials, the slope of constant-$\mu$ curves changes near second-order phase transitions. These slopes are almost orientation-independent.
  \item Comparison of Figs.~\ref{Fig:HQnu1cB05} and \ref{Fig:HQnu1cB05T3} shows that replacing $\log (a T)$ with $\log (a T^3)$ eliminates the hill-like characteristic.
\end{itemize}

\subsubsection{Non-zero magnetic field, $\nu=1.5$}\label{NR-HQ-nzero-nu15}

\begin{figure}[h!]
  \centering
  \includegraphics[scale=0.25]
  {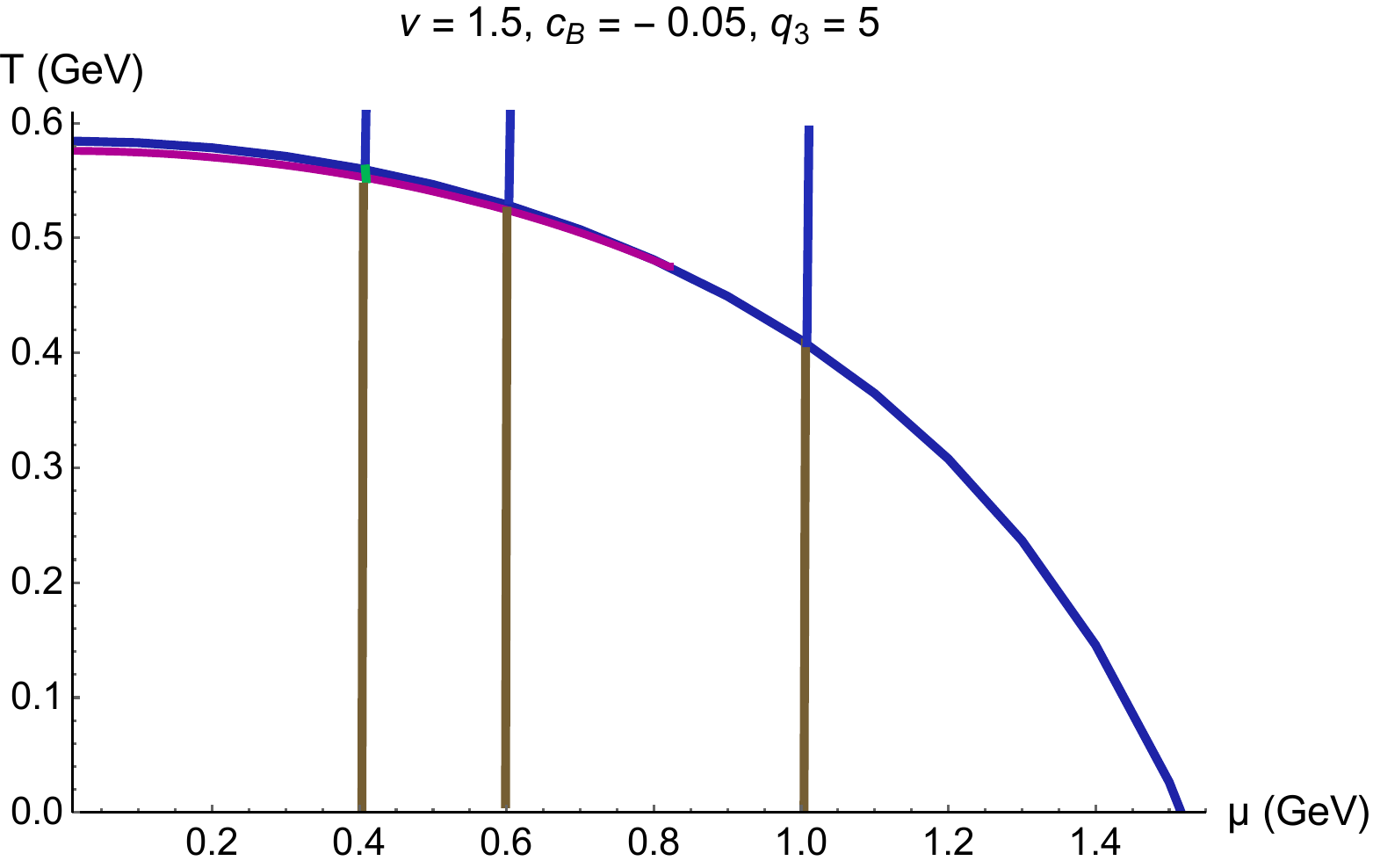}
\caption{The phase diagram for the HQ model in the ($\mu, T$)-plane for $c_B = -0.05$ \GG \,and $\nu = 1.5$. Magenta and blue lines indicate the first-order and the second-order phase transitions, respectively. The vertical lines at fixed $\mu=0.4, 0.6, 1$ (GeV) show the paths we calculate the JQ parameter.
}
\label{Fig:HQTmunu15cB005}
\end{figure}

 The phase diagram in the ($\mu, T$)-plane including the first-order and second-order phase transitions for the HQ model denoted by magenta and blue lines, in $c_B = -0.05$ \GG \,and $\nu = 1.5$ is depicted in Fig.\,\ref{Fig:HQTmunu15cB005}.

\begin{figure}[t!]
  \centering
\includegraphics[scale=0.2]
  {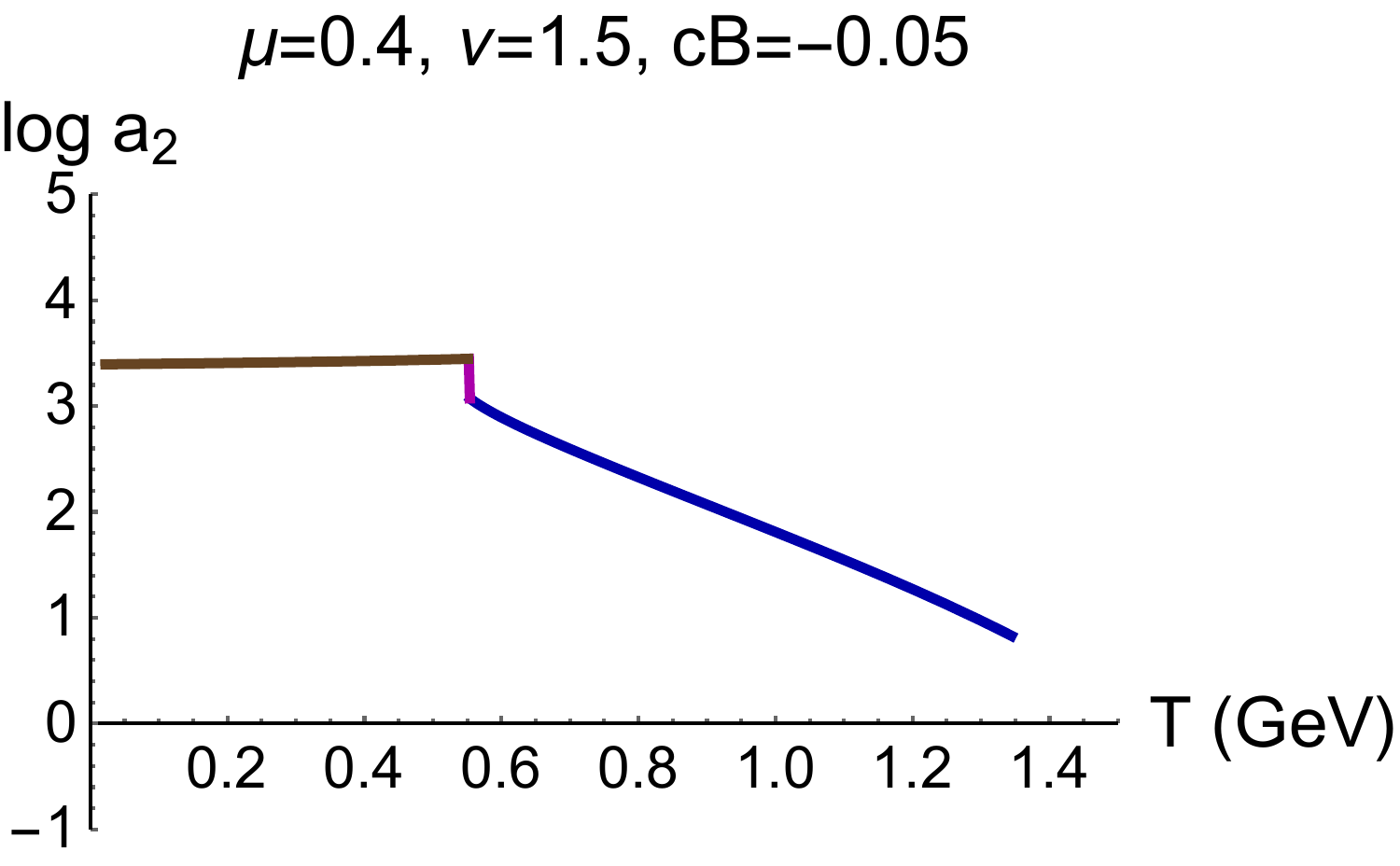}
  \includegraphics[scale=0.19]
{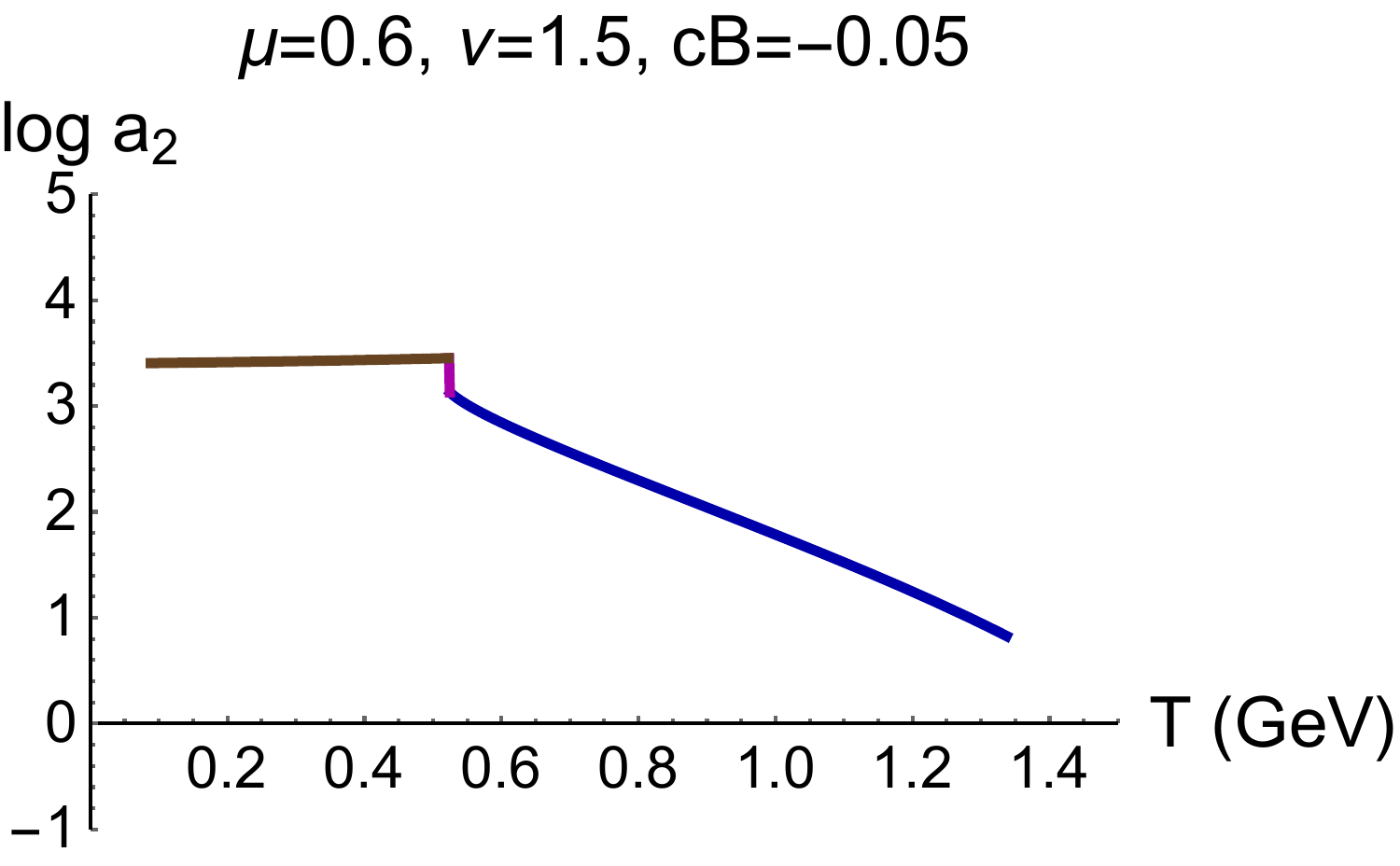}
\includegraphics[scale=0.19]
{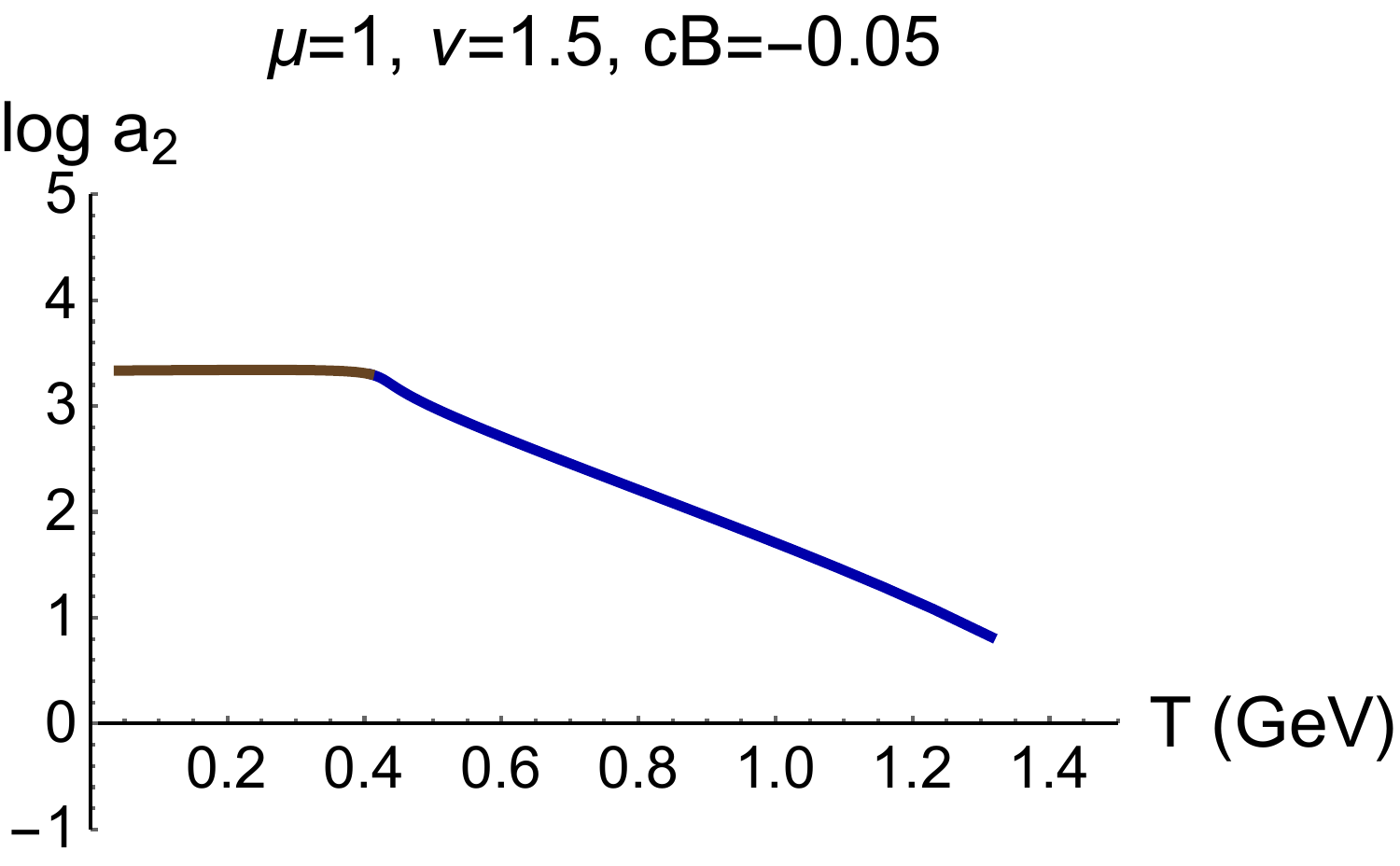}
  \\
A\hspace{130pt}B\hspace{130pt}C\\
\caption{ $\log a_2$  for the HQ model versus temperature for $c_B = -0.05$ \GG\, and $\nu=1.5$ at fixed chemical potentials (A) $\mu = 0.4$ GeV , (B) $\mu = 0.6$ GeV, and (C) $\mu = 1$ GeV. The blue and  brown  lines correspond to the QGP and  hadronic  phases, respectively. The magenta lines  correspond to unstable regions. We see  jumps in panels (A) and (B), and a smooth  change of the slopes  near the second-order phase transition in panel (C).
}
  \label{Fig:HQnu15cB005}
\end{figure}

\newpage
\begin{itemize}
 \item The plots in Fig.\,\ref{Fig:HQnu15cB005} show that for $\nu = 1.5$,  $\log a_2$ is nearly temperature-independent in the hadronic phase but decreases with temperature in the quark-gluon phase. It exhibits jumps at first-order phase transitions and the values of these jumps decrease with increasing the chemical potential until it reaches a critical value. Indeed, the jump value in Fig.\,\ref{Fig:HQnu15cB005}A is larger than the jump in  Fig.\,\ref{Fig:HQnu15cB005}B and the jump is absent in Fig.\,\ref{Fig:HQnu15cB005}C.
 \end{itemize}

\begin{figure}[t!]
  \centering
 \includegraphics[scale=0.13]
{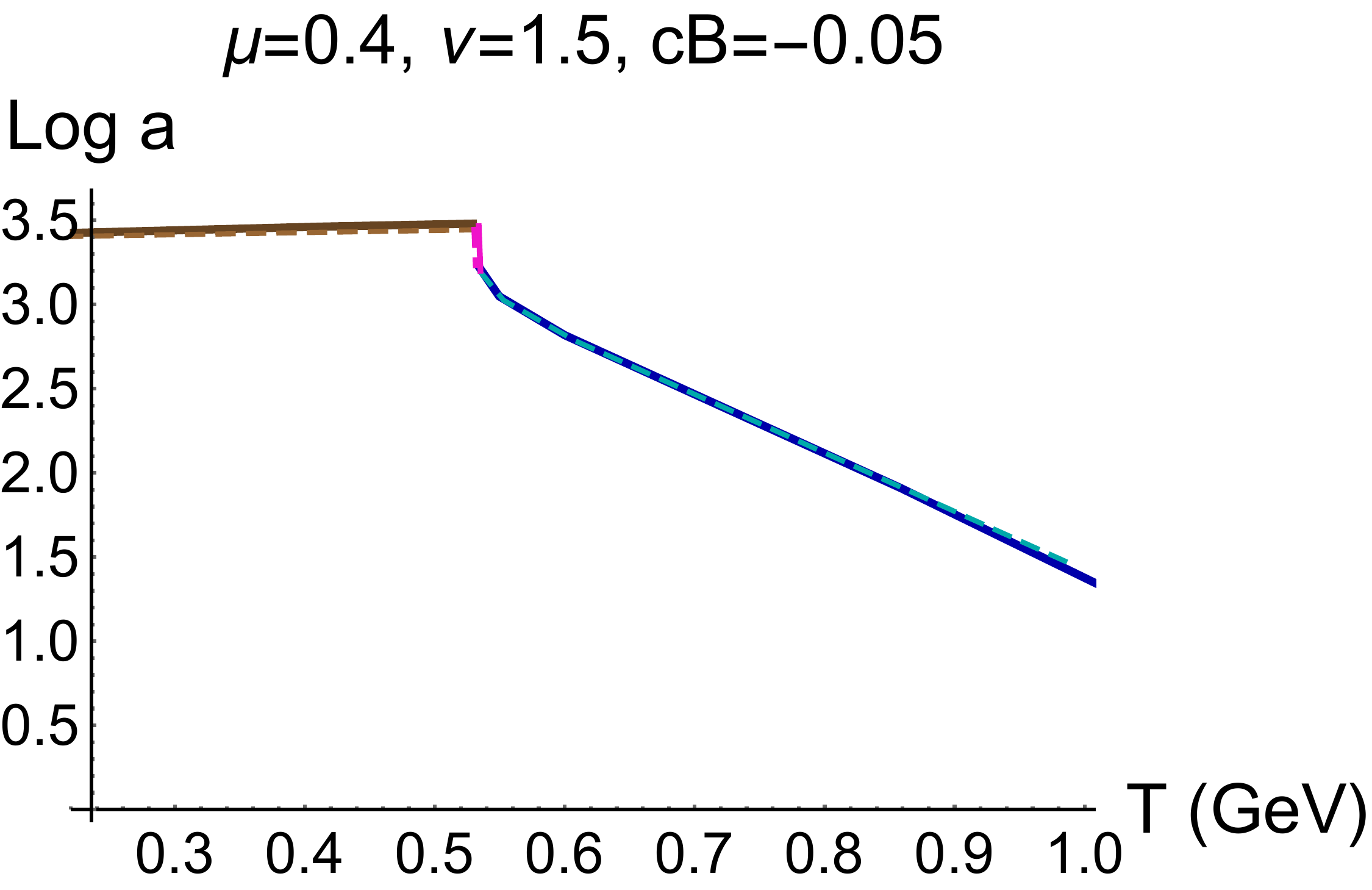}\qquad
\includegraphics[scale=0.13]{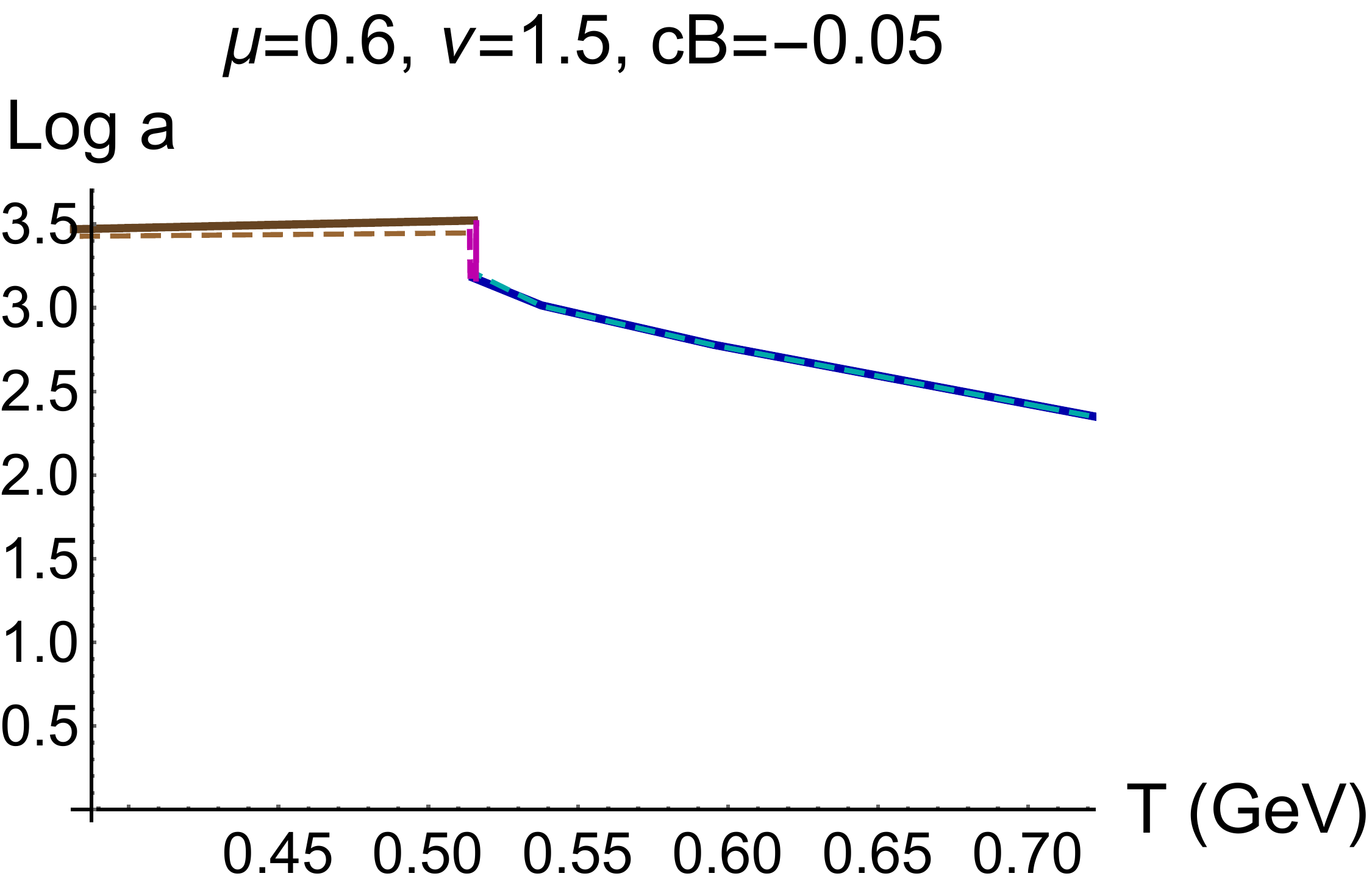} 
  \includegraphics[scale=0.18]
  {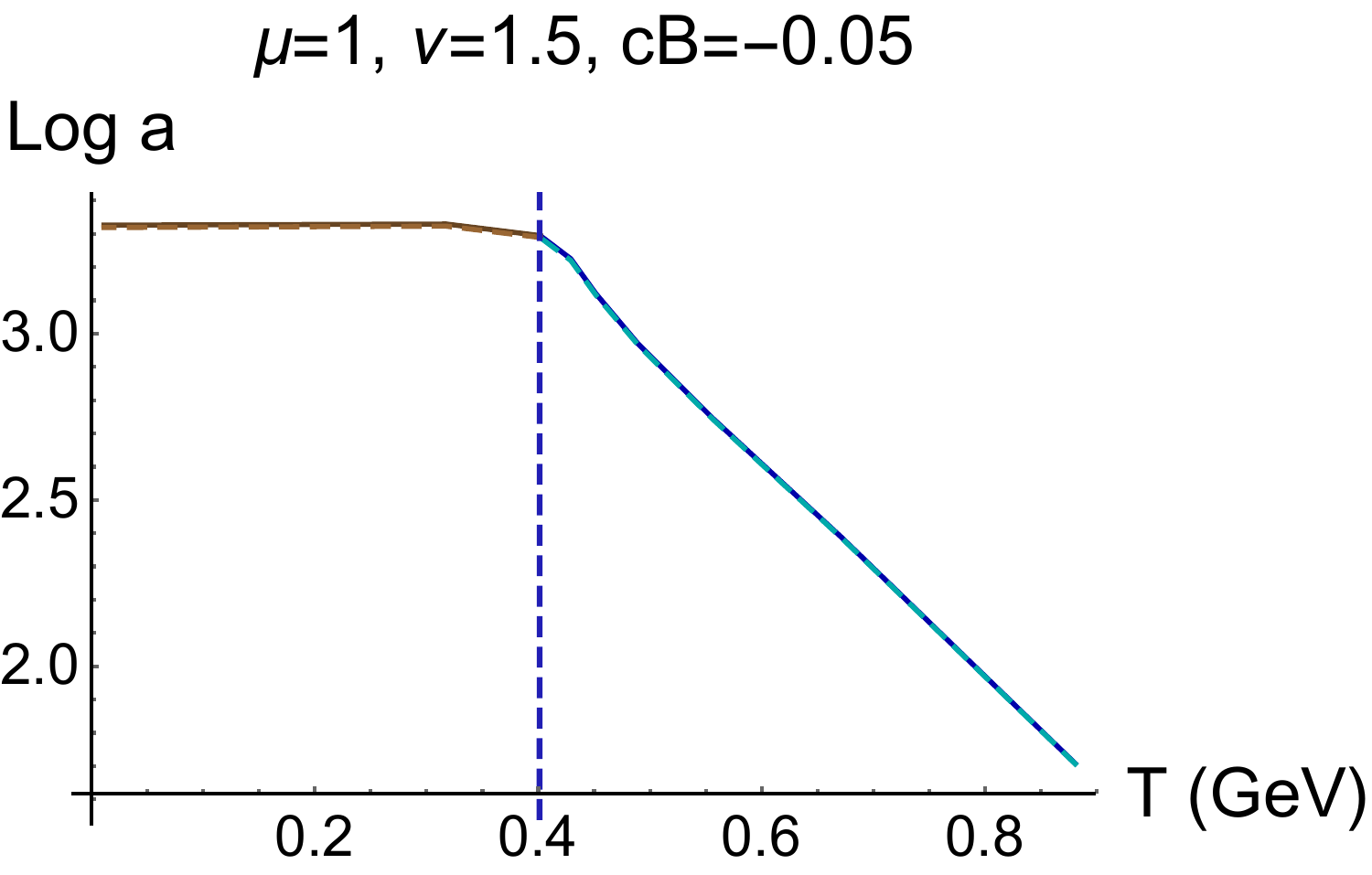}\\
A\hspace{140pt}B\hspace{140pt}C
\caption{$\log a_2$ (solid lines) and $\log a_3$ (dashed lines) are presented for the HQ model as a function of temperature for $c_B = -0.05$ \GG\, and $\nu=1.5$ at fixed chemical potentials (A) $\mu = 0.4$ GeV, (B) $\mu = 0.6$ GeV, and (C) $\mu = 1$ GeV. The blue and cyan correspond to the QGP and brown and gray lines correspond to the hadronic phases. The magenta lines (solid and dashed) correspond to unstable regions. 
We see  jumps in panel (A)  and a smooth  change of the slopes  near the second-order phase transition in panel (B). 
 }
  \label{Fig:HQ23nu15cB005}
\end{figure}

 \begin{itemize}
\item Plots in Fig.\,\ref{Fig:HQ23nu15cB005} show that for $\nu = 1.5$, the JQ parameter exhibits anisotropy in the presence of a magnetic field: $\log a_2 \neq \log a_3$ for $c_B = -0.05$ \GG, but the anisotropy  for $c_B\neq 0$ is very small, 
\be\label{a2a3}
\Big |\log a_2-\log a_3\Big |_{c_B=-0.05}<<\Big |\log a_2-\log a_3\Big |_{c_B=0}.\ee
  
  \item Both $\log a_2$ and $\log a_3$ are nearly temperature-independent in the hadronic phase but decrease with temperature in the quark-gluon phase. Each exhibit jumps at first-order phase transitions with approximately equal magnitudes.
  
  \item For large chemical potentials the slope of constant-$\mu$ curves changes near second-order phase transitions. These slopes are almost orientation-independent.
\end{itemize}

\begin{figure}[h!]
  \centering
  \includegraphics[scale=0.23]
  {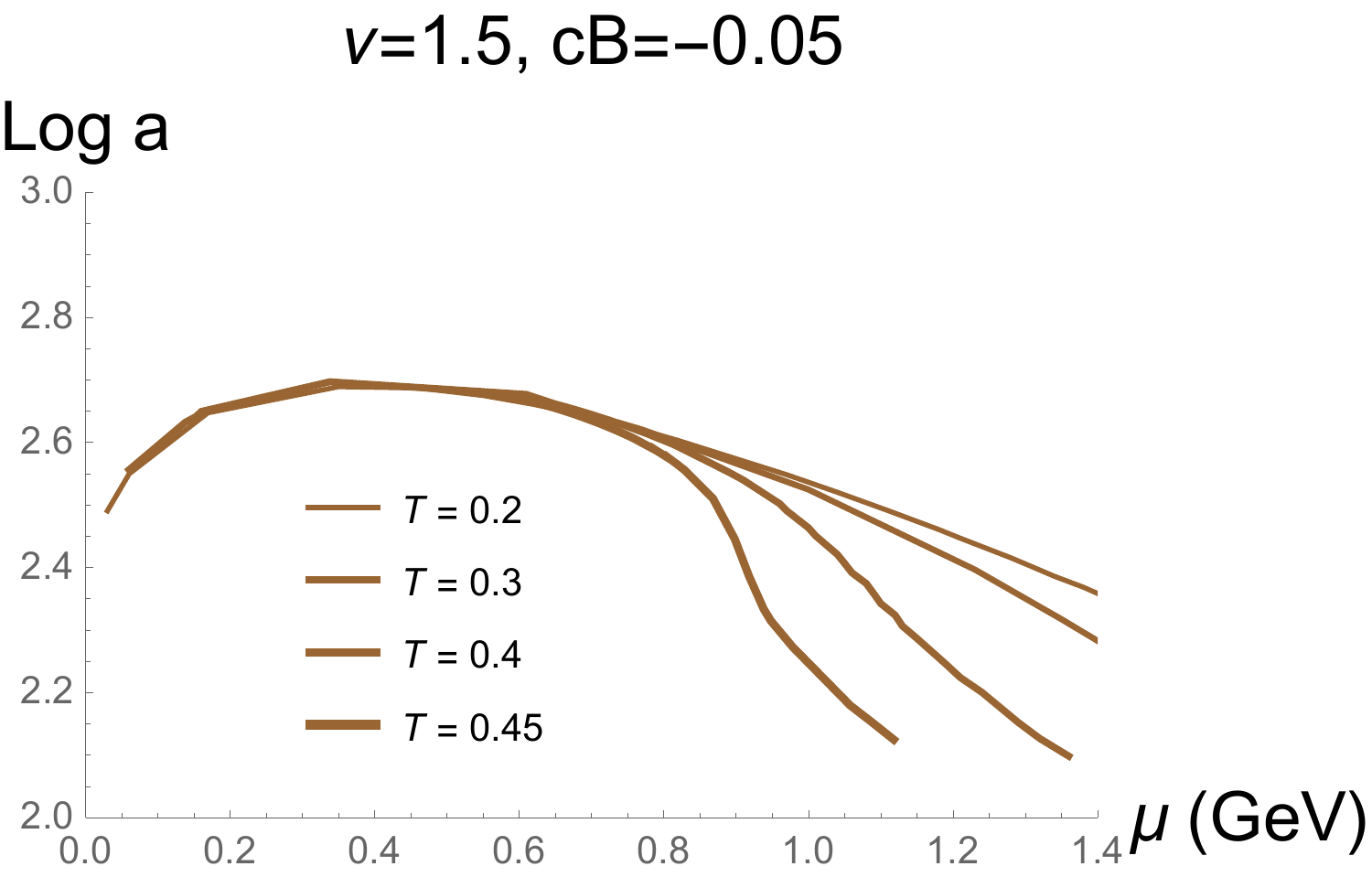}\qquad
\includegraphics[scale=0.22]
{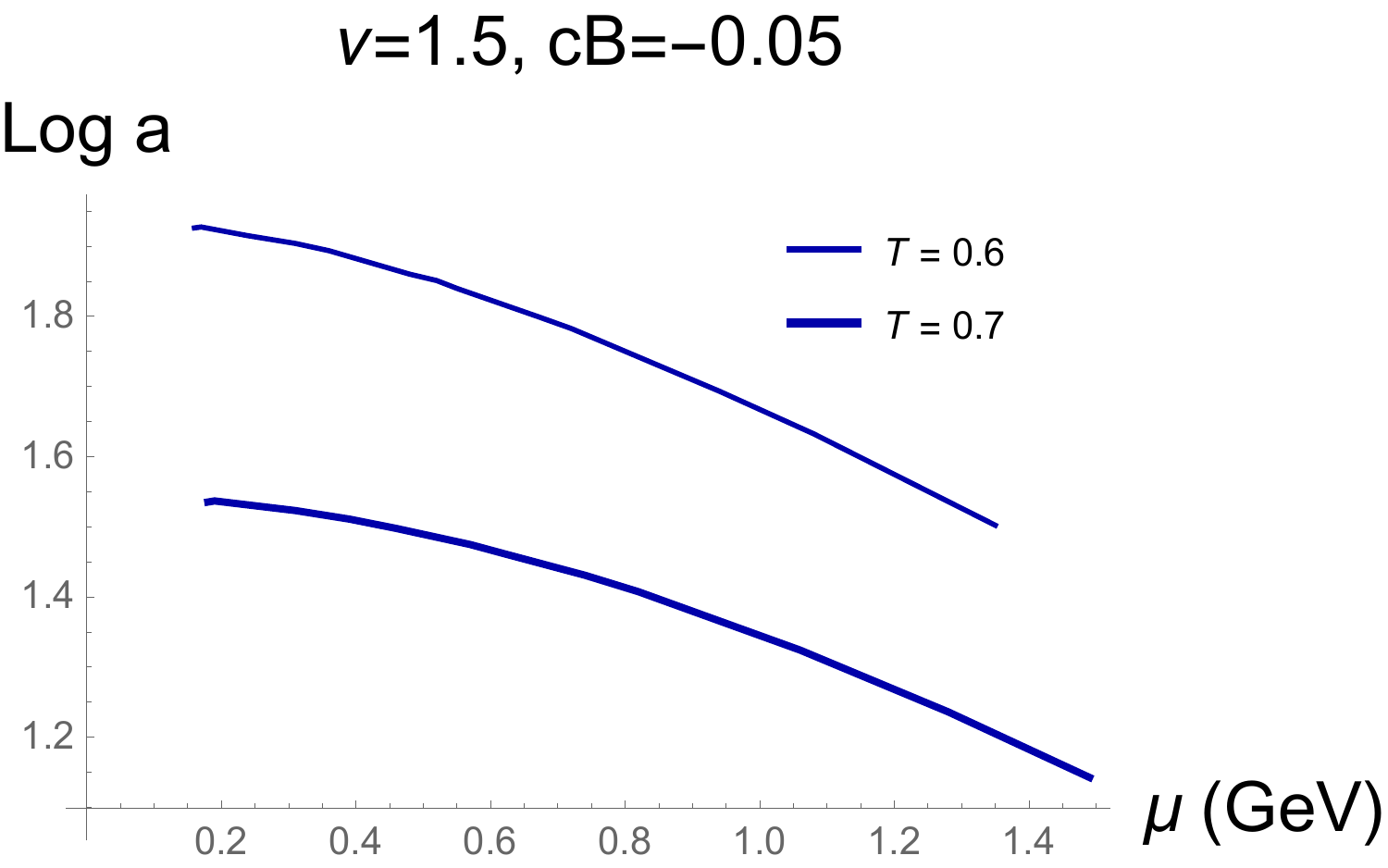} 
 \\  A\hspace{170pt}B
  \caption{ $\log a$ for the HQ model as a function of the chemical potential $\mu$ for $c_B = -0.05$ \GG\, and $\nu=1.5$ at fixed temperatures: (A) $T=0.2,0.3, 0.4$ (GeV) and $T=0.45$ GeV at below the first-order phase transition line, and (B) above the phase transition line at  $T=0.6$  GeV and $T=0.7$ GeV.  We see  hills near $\mu_{hill}\approx 0.5$ GeV.
  }
  \label{Fig:HQ-hill}
 \end{figure}

The Plots in Fig.\,\ref{Fig:HQ-hill} show $\log a$ for $c_B = -0.05$ \GG\, and $\nu=1.5$ as a function of the chemical potential $\mu$ at fixed temperatures below and above the first-order phase transition line:

\begin{itemize}
    \item \textit{Below the first-order phase transition}: Fig.\,\ref{Fig:HQ-hill}A shows $\log a$ versus $\mu$ at fixed temperatures: $T = 0.2$, 0.3, 0.4 (GeV), and $T=0.45$ GeV. With increasing chemical potential, $\log a$ first increases, reaches a maximum near $\mu_{\text{hill}} \approx 0.5$ GeV, then decreases. Below $\mu_{\text{hill}}$, the growth is nearly temperature-independent. Above $\mu_{\text{hill}}$, the decrease shows a clear temperature dependence, occurring faster at lower temperatures.
    
    \item \textit{Above the first-order phase transition}: Fig.\,\ref{Fig:HQ-hill}B displays $\log a$ versus $\mu$ at fixed temperatures: $T = 0.6$ GeV and $T =0.75$ GeV. Here, $\log a$ decreases monotonically, with reduced magnitude at higher temperatures.
\end{itemize}

\begin{figure}[h!]
    \centering
 \includegraphics[width =0.34\linewidth]
{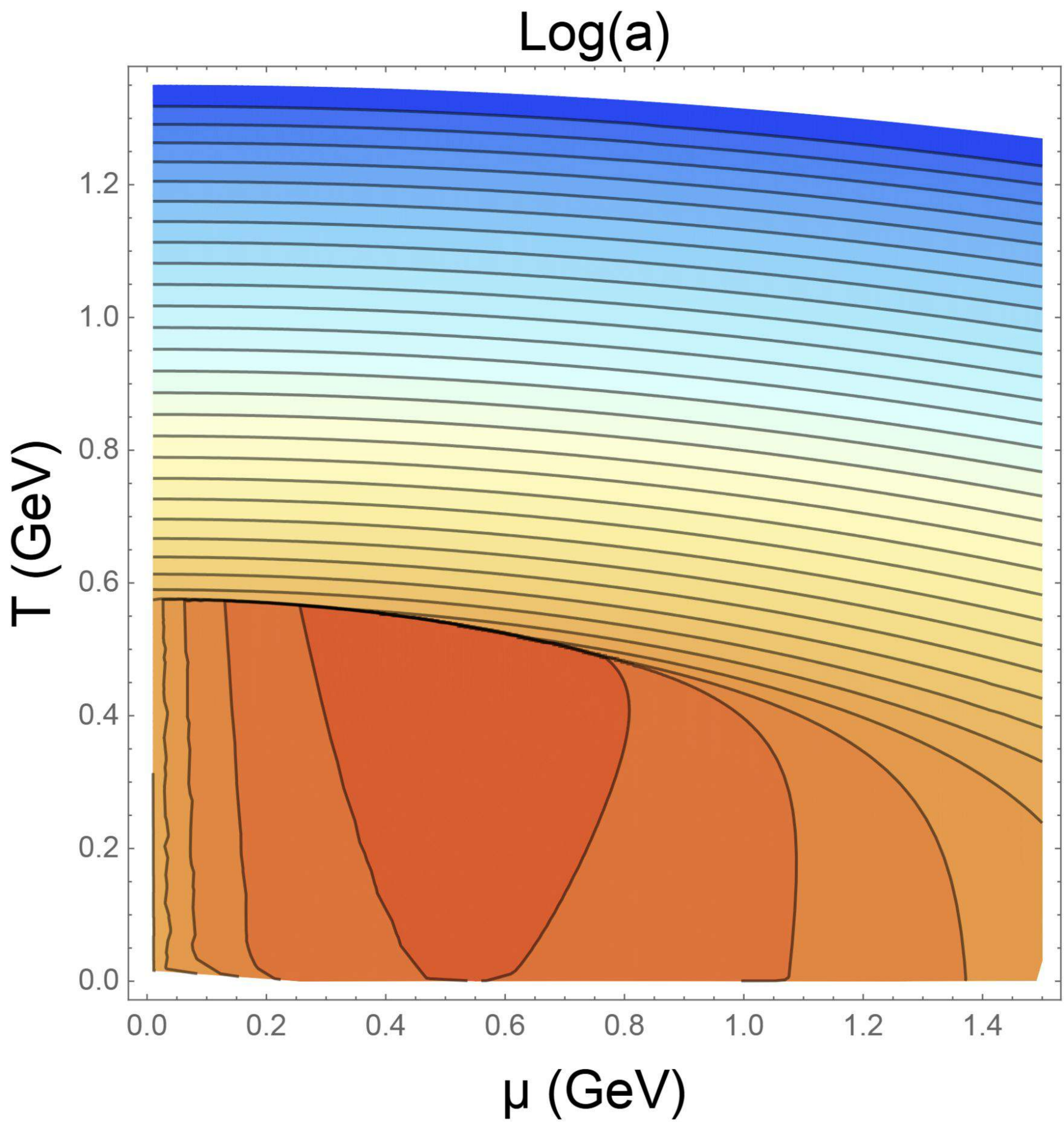} 
\includegraphics[width =0.045\linewidth]
 {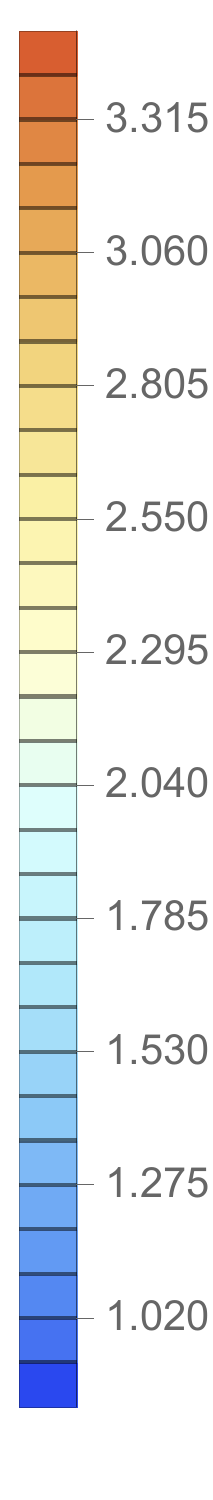}\quad
 \includegraphics[width =0.34\linewidth]{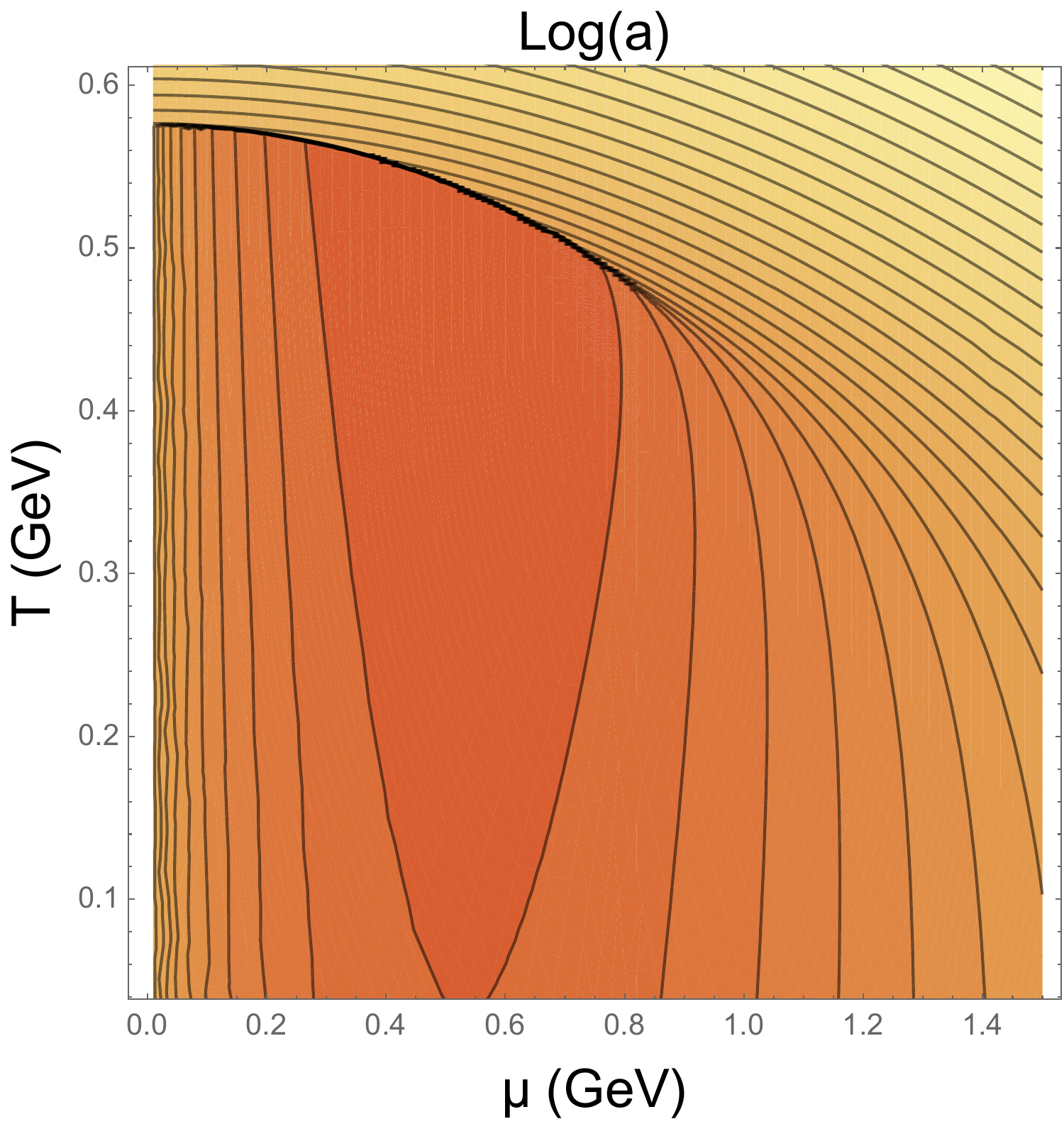}
  \includegraphics[width =0.045\linewidth]{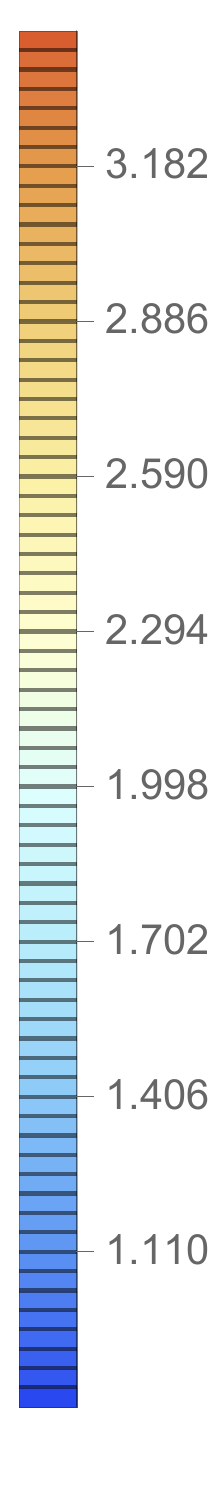}
 \\
A\hspace{180pt}B\\
 \includegraphics[width =0.34\linewidth]
{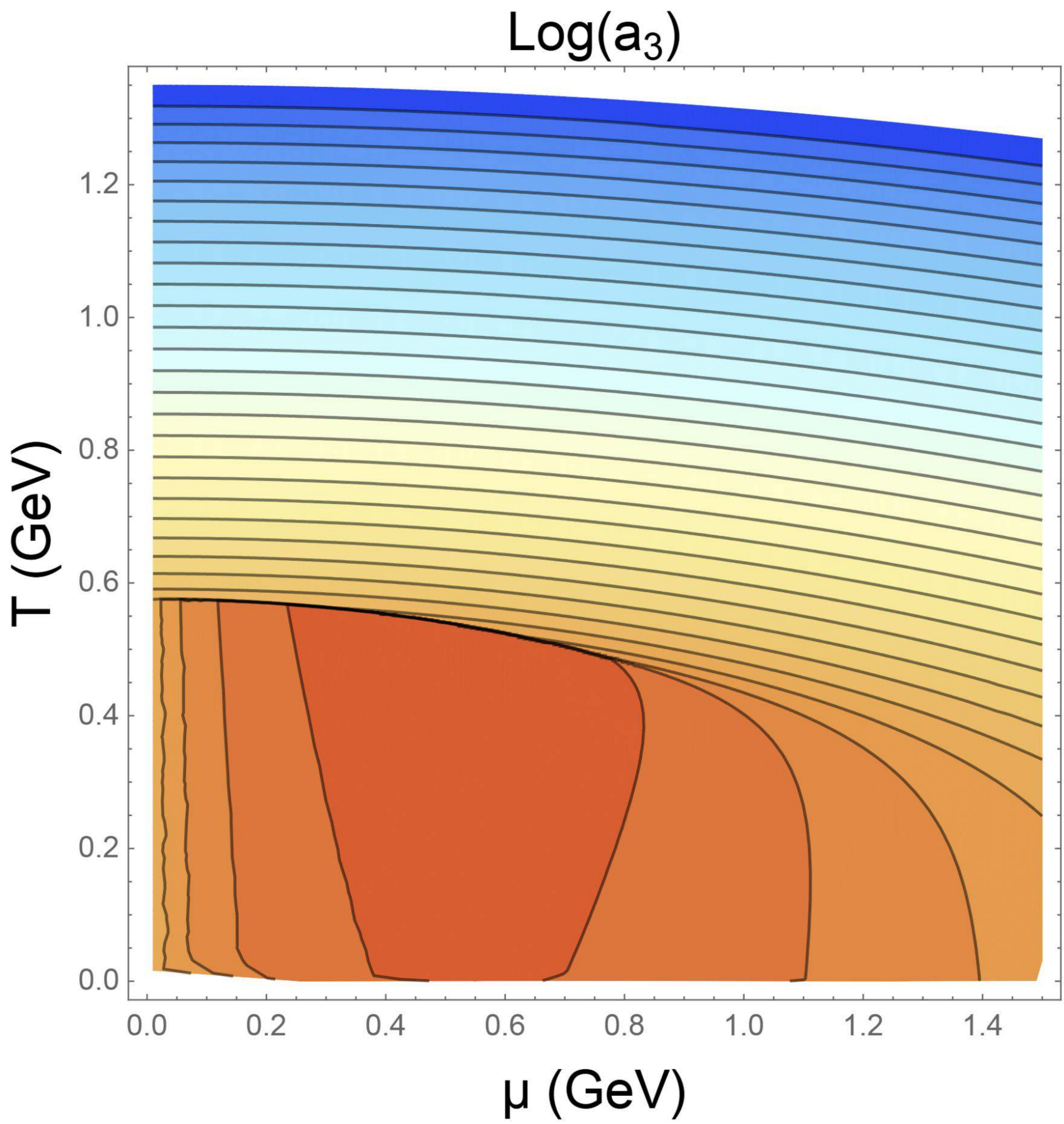} 
\includegraphics[width =0.045\linewidth]
 {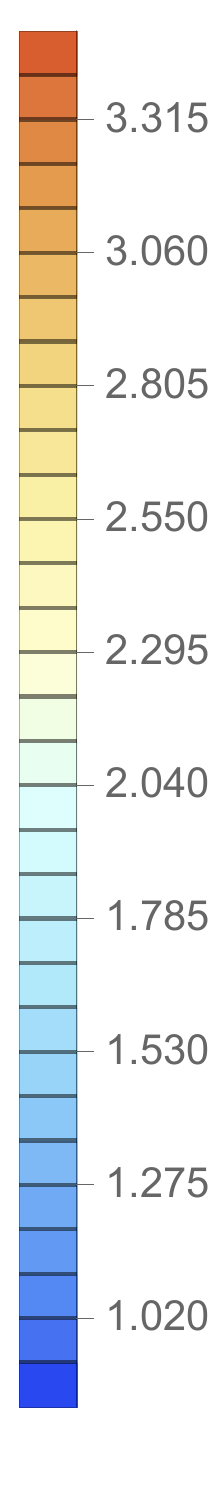}\quad
 \includegraphics[width =0.34\linewidth]{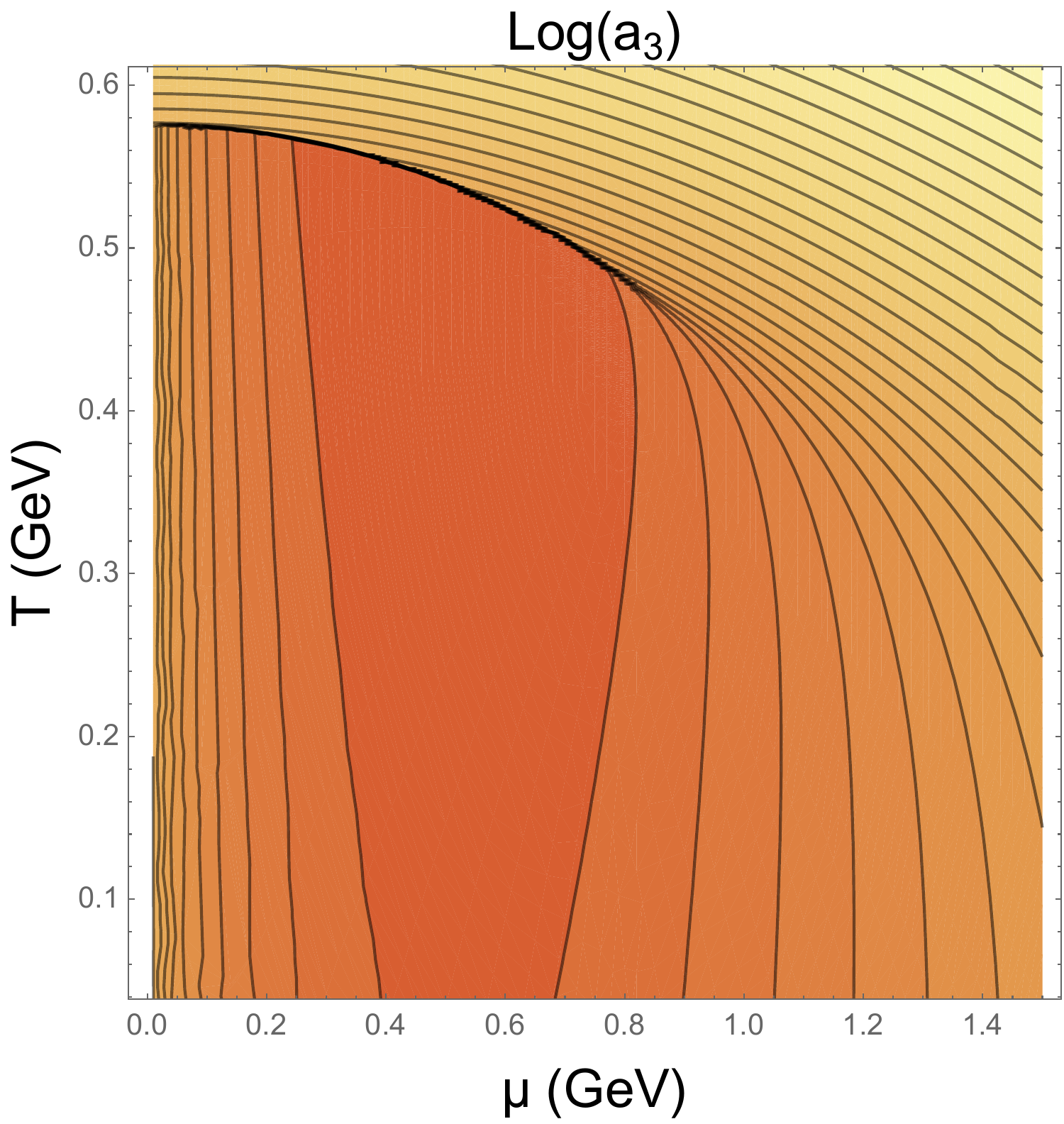}
  \includegraphics[width =0.045\linewidth]{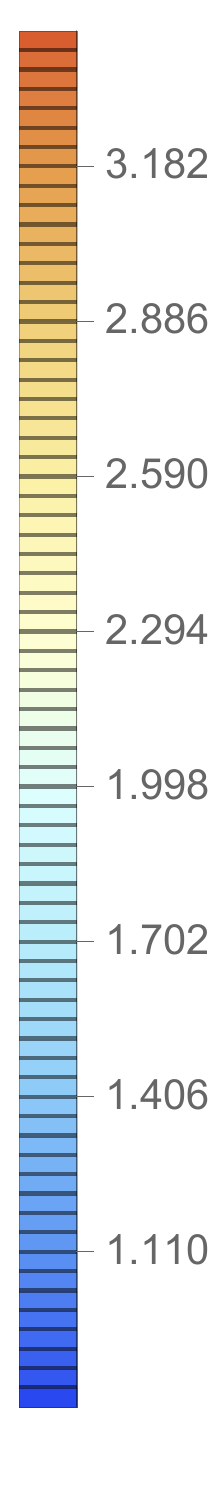}
 \\ C\hspace{180pt}D
\caption{ Contour plots of (A) $\log a_2$ and (C) $\log a_3$ for the HQ model in the $(\mu, T)$-plane with non-zero magnetic field $c_B = -0.05$ \GG and $\nu = 1.5$. Panels (B) and (D) are zooms of panels (A) and (C) under the first-order phase transition line, respectively. The difference between the top and bottom panels is only visible in their zoomed versions (B and D).
    }
    \label{Fig:HQnu15cb-005}
\end{figure}

Fig.\,\ref{Fig:HQnu15cb-005} shows contour plots for the HQ model in the $(\mu, T)$-plane with $c_B = -0.05$ \GG\, and $\nu = 1.5$:
(A) $\log a_2$; (C) $\log a_3$;
(B) and (D) show zoomed regions of (A) and (C) respectively, below the first-order phase transition line.
Differences between $\log a_2$ (top row) and $\log a_3$ (bottom row) are only apparent in these zoomed regions (panels B and D).

\newpage
$$\,$$

\subsubsection{Non-zero magnetic field, $\nu=4.5$}\label{NR-HQ-nzero-nu45}

\begin{figure}[h!]
  \centering
  \includegraphics[scale=0.25]
  {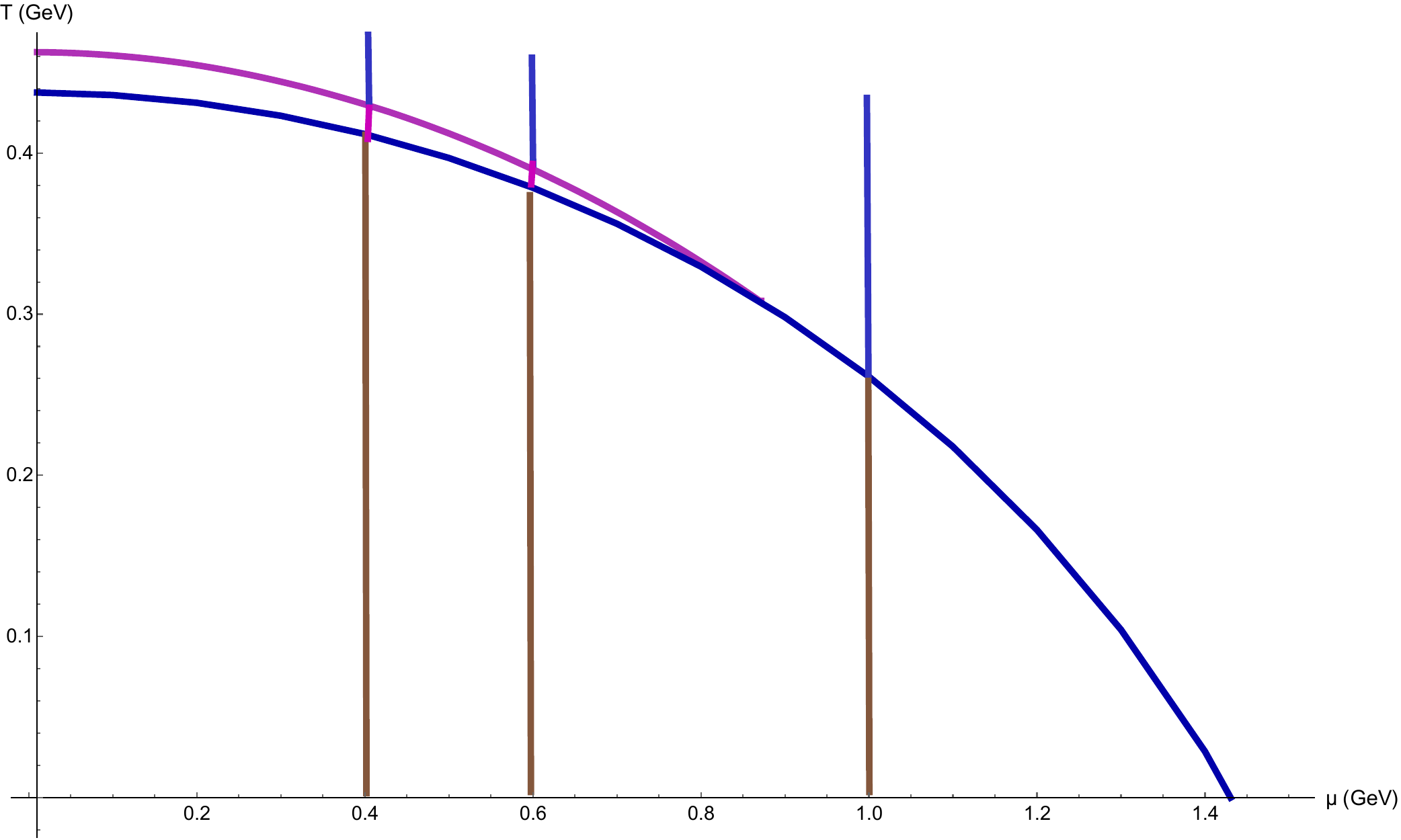}
\caption{ The first-order and second-order phase transitions for the HQ model denoted by magenta and blue lines, in the ($\mu, T$)-plane for $c_B = -0.05$ \GG \, and $\nu = 4.5$. The vertical lines with brown and blue colors indicate the hadronic and the QGP phases, respectively.
}
\label{Fig:HQTmunu15cB005m}
\end{figure}

The phase diagram in the ($\mu, T$)-plane including the first-order and second-order phase transitions for the HQ model denoted by magenta and blue lines, at $c_B = -0.05$ \GG \,and $\nu = 4.5$ is depicted in Fig.\,\ref{Fig:HQTmunu15cB005m}.

\begin{figure}[h!]
  \centering
   \includegraphics[scale=0.13]
  {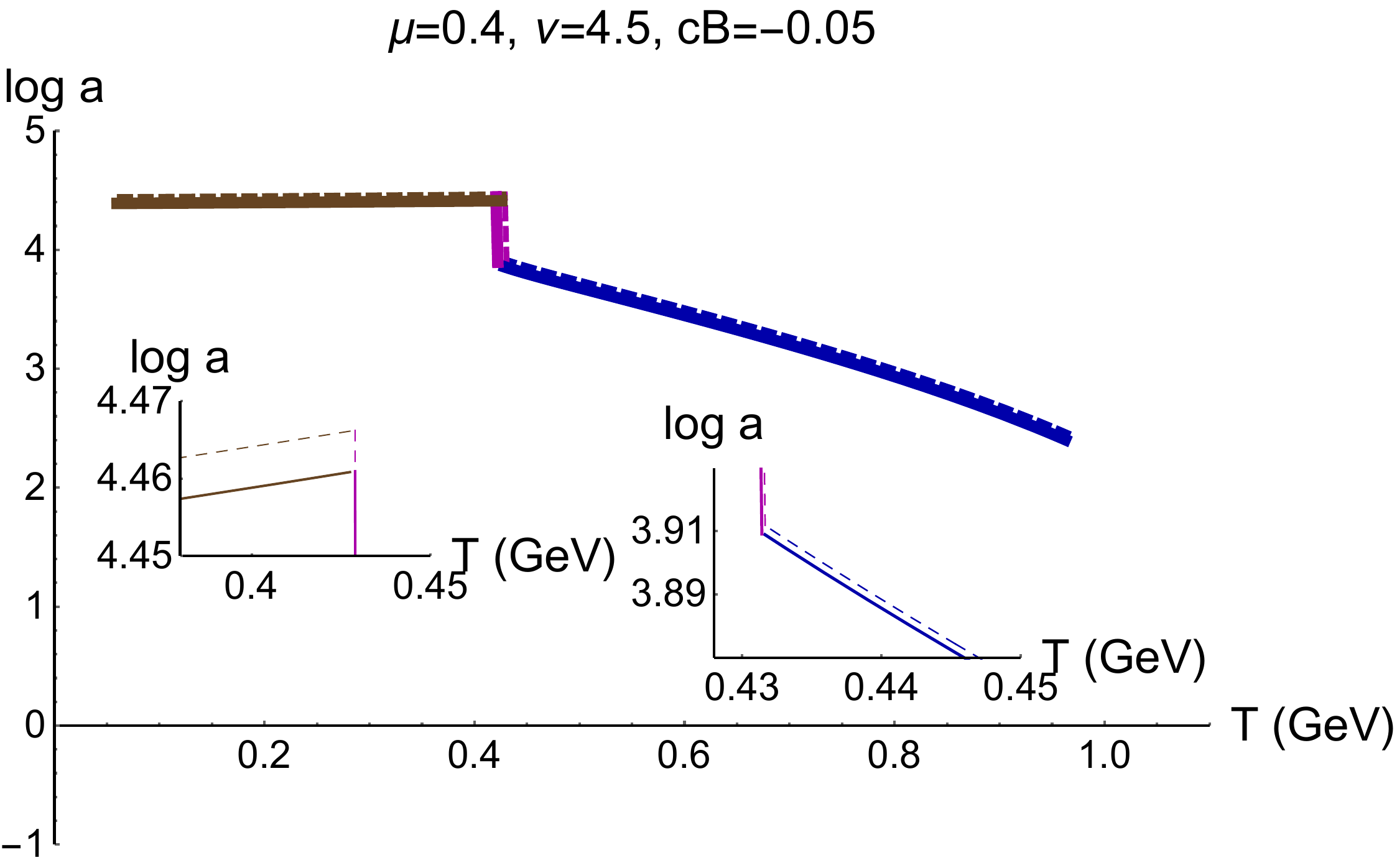}
  \includegraphics[scale=0.18]
{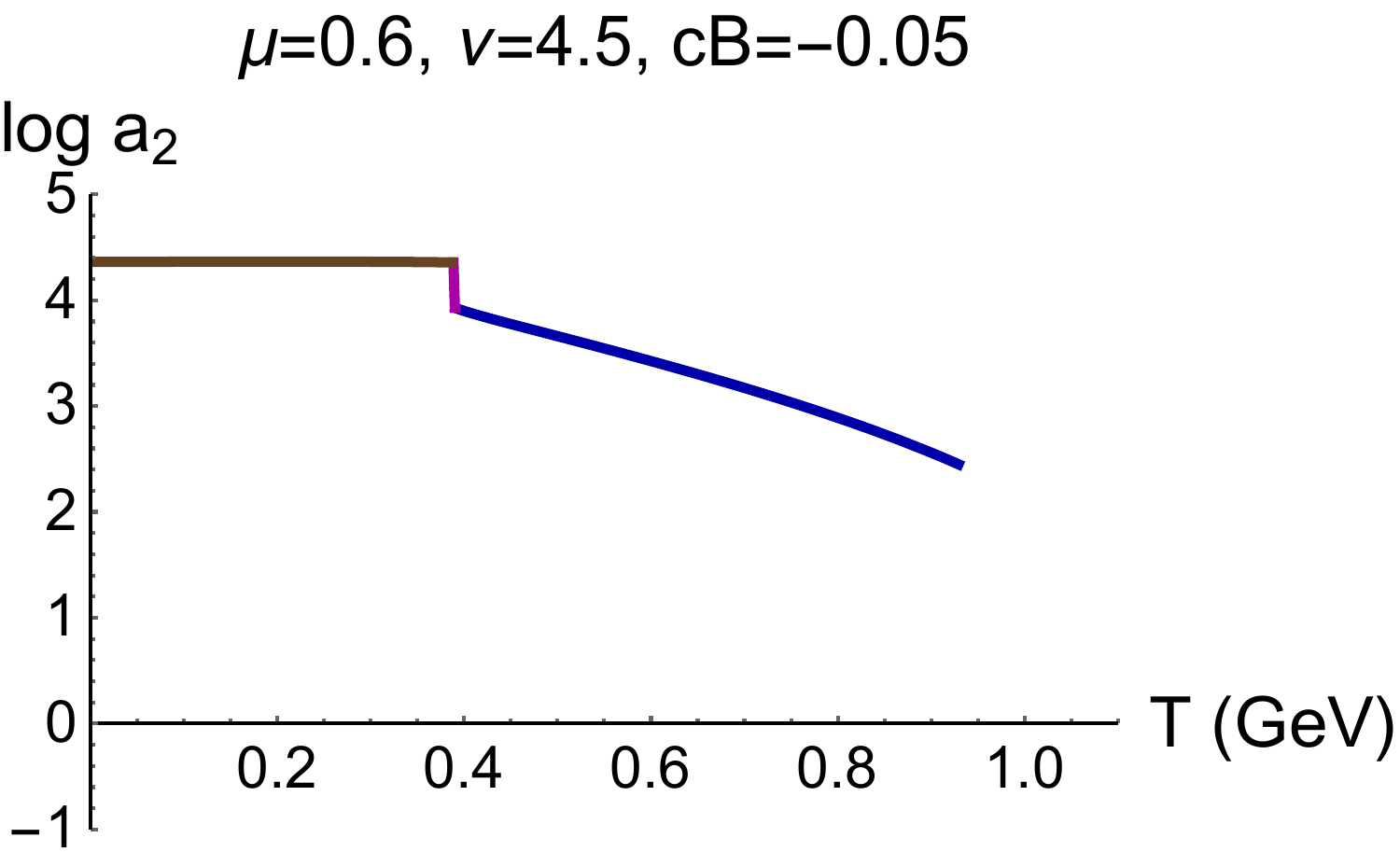}
  \includegraphics[scale=0.16]
  {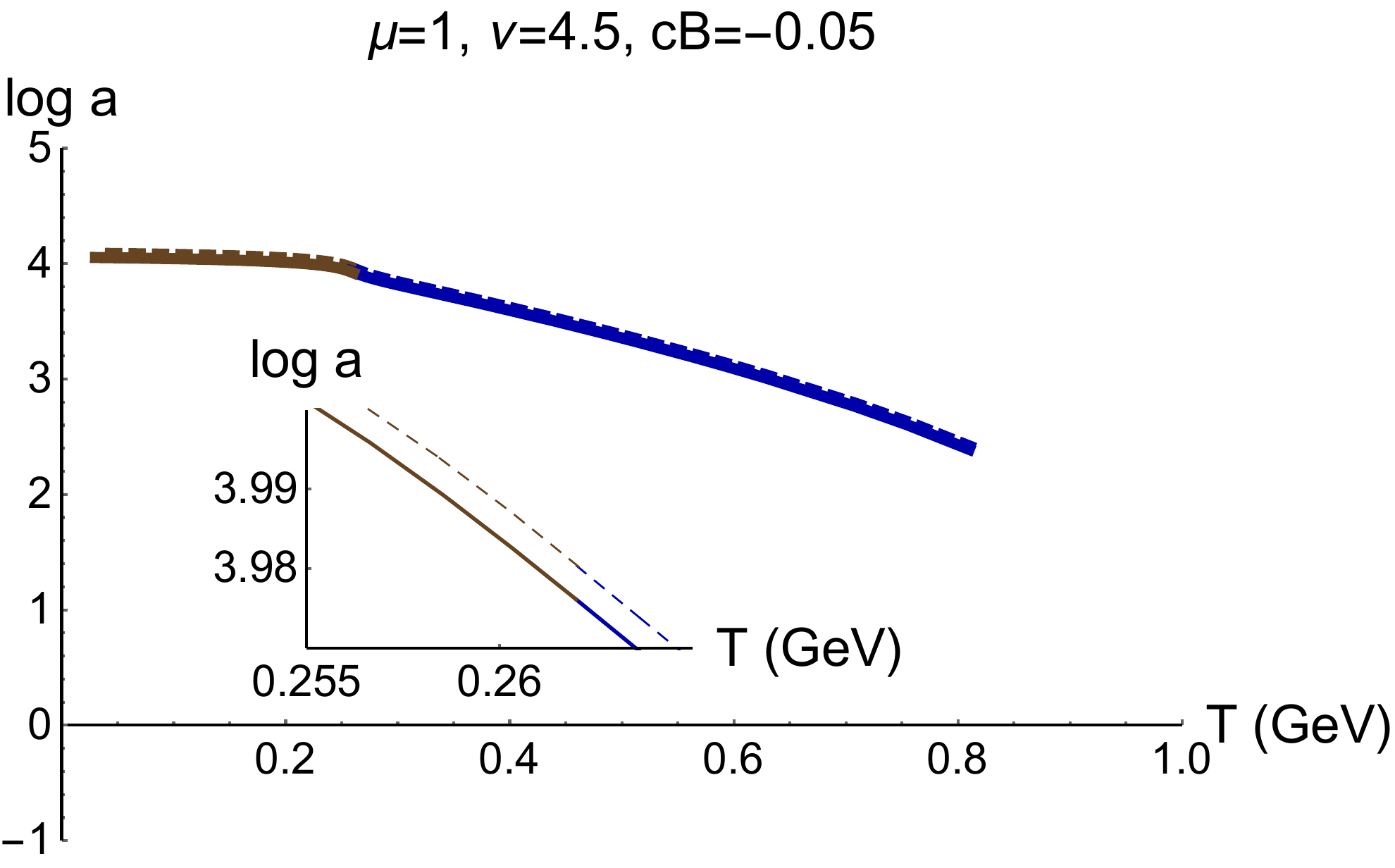} \\
 A\hspace{150pt}B\hspace{150pt}C\\
\caption{The dependence of $\log a_2$ (solid lines) and $\log a_3$ (dashed lines) on $T$ for the HQ model, with $\nu=4.5$ and $c_B=-0.05$ \GG\,, at fixed chemical potentials (A) $\mu = 0.4$ GeV, (B) $\mu = 0.6$ GeV (only $a_2$; $a_3$ is not included ), and (C) $\mu = 1$ GeV. The blue and cyan correspond to the QGP and brown and gray lines correspond to the hadronic phases. The magenta lines (solid and dashed) correspond to unstable regions. 
We see  jumps in panel (A)  and a smooth  change of the slopes  near the second-order phase transition at panel (B). 
}
\label{Fig:HQ23nu45cB005}
\end{figure}

\begin{itemize}
\item The plots in Fig.\,\ref{Fig:HQ23nu45cB005} show that for $\nu = 4.5$, the JQ parameter exhibits anisotropy in the presence of a magnetic field: $\log a_2 \neq \log a_3$ for $c_B = -0.05$ \GG, but the anisotropy  for $c_B\neq 0$ is very small; see \eqref{a2a3}.

\item Both $\log a_2$ and $\log a_3$ are nearly temperature-independent in the hadronic phase but decrease with temperature in the quark-gluon phase. Each exhibit jumps at first-order phase transitions with approximately equal magnitudes.
  
\item For large chemical potentials, the slope of constant-$\mu$ curves changes near second-order phase transitions. These slopes are almost orientation-independent.
\end{itemize}

\begin{figure}[h!]
  \centering
  \includegraphics[scale=0.25]
  {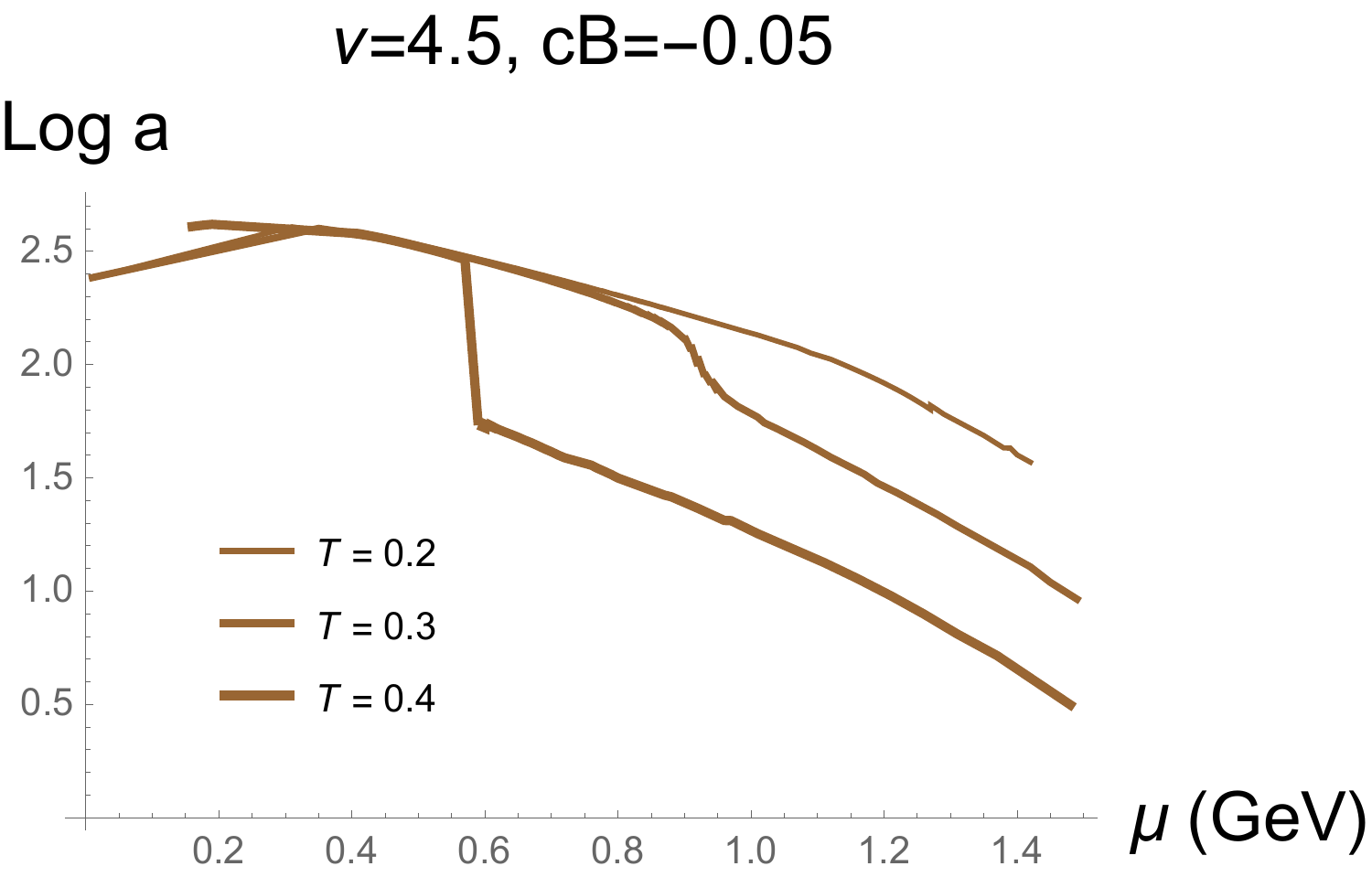}\qquad
\includegraphics[scale=0.22]
{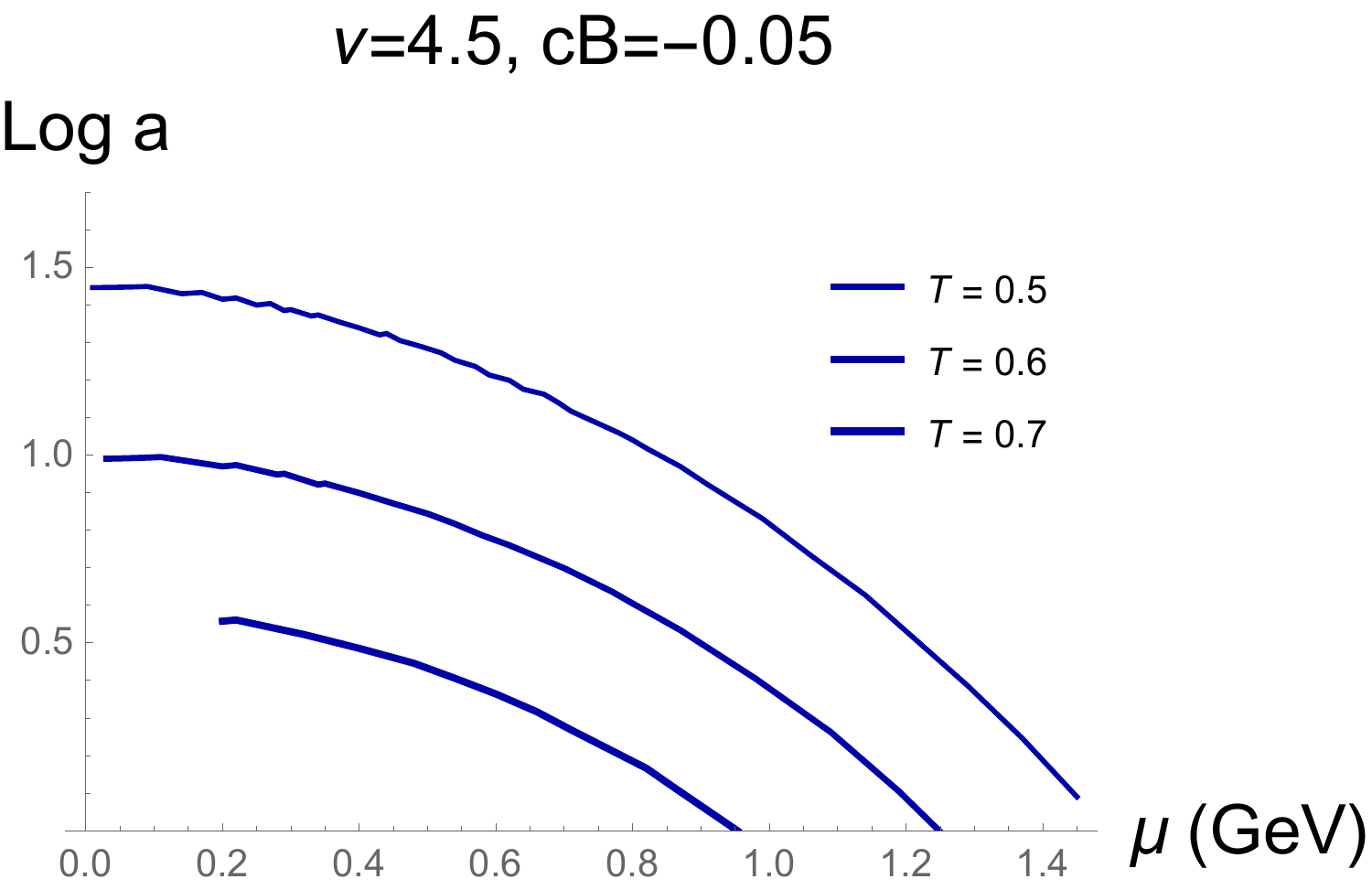} 
 \\
  A\hspace{150pt}B
  \caption{ 
  (A) The dependence of $\log a$ on $\mu$ for the HQ model, with $\nu=4.5$ and $c_B=-0.05$ \GG\,, at fixed temperatures (A)  $T=0.2,0.3$ and $T=0.5$ (GeV) below the first-order phase transition line (B), and above the phase transition line for $T=0.5,0.6$ and $T=0.7$ (GeV). We see  hills near $\mu\approx 0.4$ GeV.
  }
  \label{Fig:HQn-A-F}
\end{figure}

The Plots in Fig.\,\ref{Fig:HQn-A-F} show $\log a$ for $c_B = -0.05$ \GG \, and $\nu = 4.5$ as a function of the chemical potential $\mu$ at fixed temperatures below and above the first-order phase transition line:

\begin{itemize}
    \item \textit{Below the first-order phase transition}: Fig.\,\ref{Fig:HQn-A-F}A shows $\log a$ versus $\mu$ at fixed temperatures: $T = 0.2$, 0.3 and $T =0.4$ (GeV). For $T = 0.2, 0.3$ (GeV), with increasing chemical potential, $\log a$ first increases, reaches a maximum near $\mu_{\text{hill}} \approx 4$ GeV, then decreases. Below $\mu_{\text{hill}}$, the growth is nearly temperature-independent. Above $\mu_{\text{hill}}$, the decrease shows a clear temperature dependence, occurring faster at lower temperatures.
    For $T = 0.4$ GeV, no clearly defined hill is observed; however, near $\mu \approx 0.6$ GeV, a sharp drop occurs, followed by a gradual decrease.
 
    \item \textit{Above the first-order phase transition}: Fig.\,\ref{Fig:HQn-A-F}B displays $\log a$ versus $\mu$ at fixed temperatures: $T = 0.5,0.6$ and $T =0.7$ (GeV). Here, $\log a$ decreases monotonically, with a reduced magnitude at higher temperatures.
\end{itemize}

\begin{figure}[t!]
    \centering
 \includegraphics[width =0.35\linewidth]
{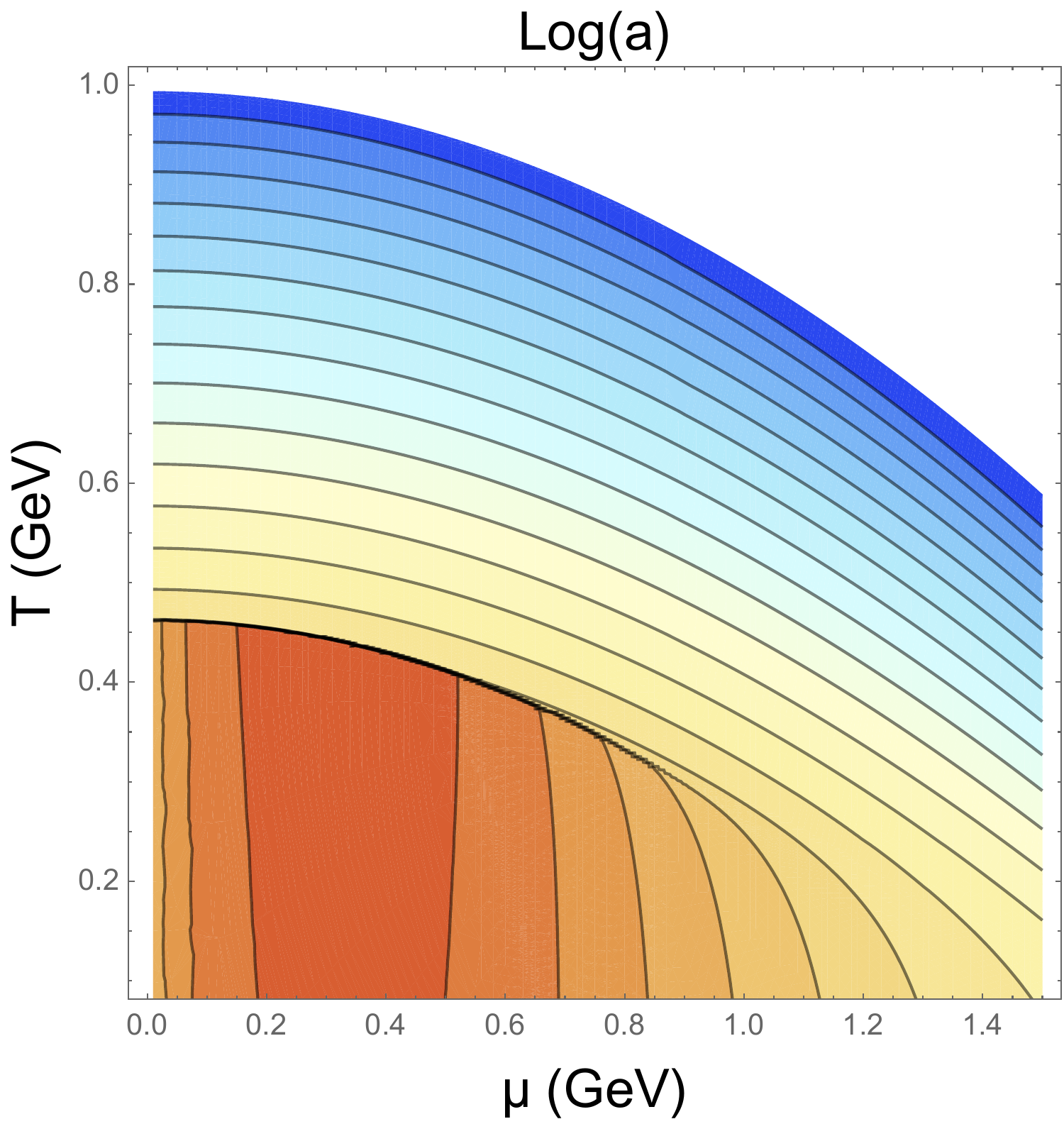} 
\includegraphics[width =0.045\linewidth]{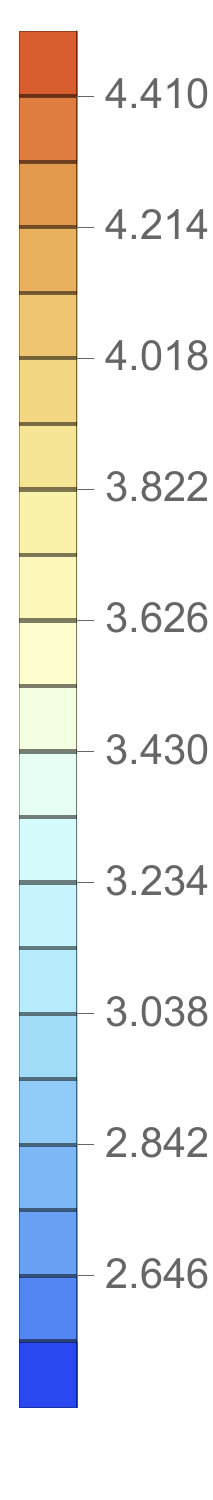}\qquad
\includegraphics[width =0.35\linewidth]
{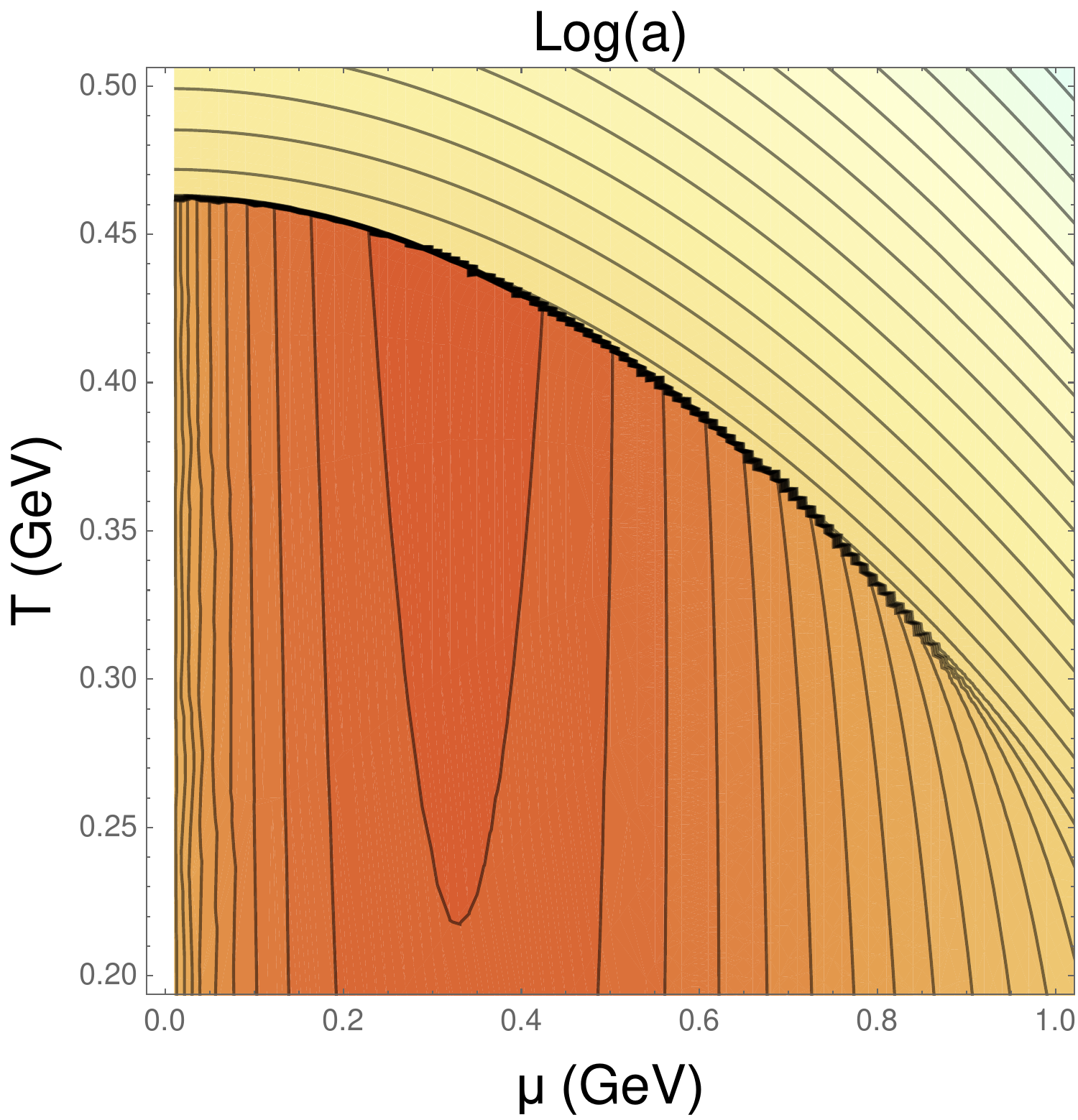}
\includegraphics[width =0.05\linewidth]{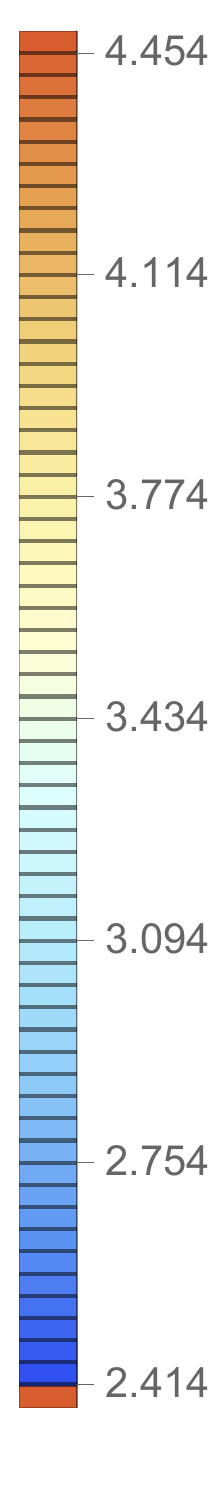}\\
A\hspace{170pt}B\\
\includegraphics[width =0.35\linewidth]
{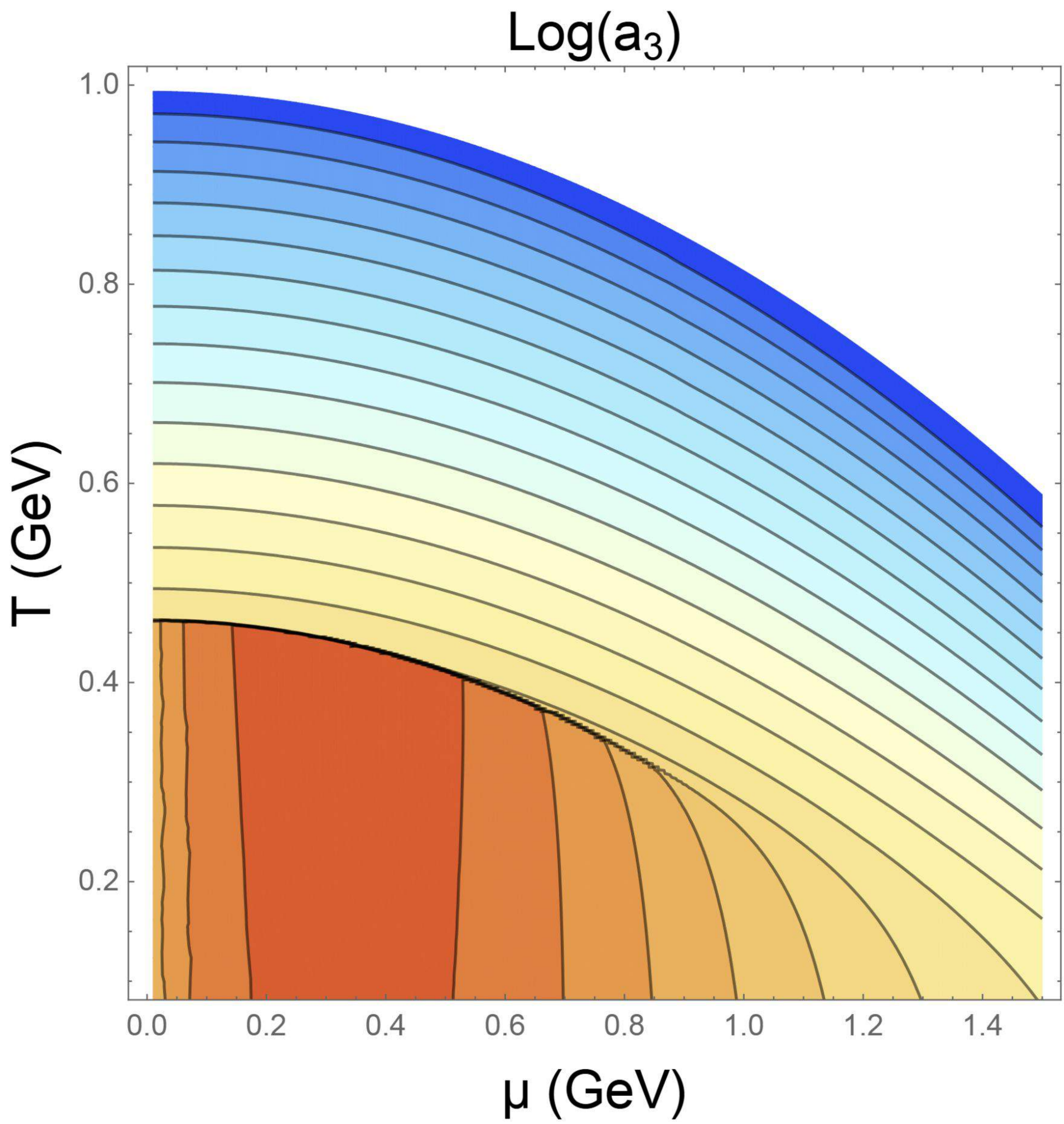} 
\includegraphics[width =0.045\linewidth]{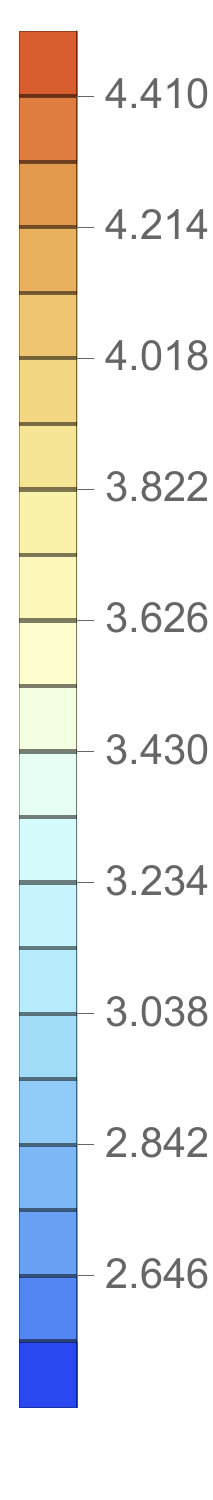}\qquad
\includegraphics[width =0.35\linewidth]
{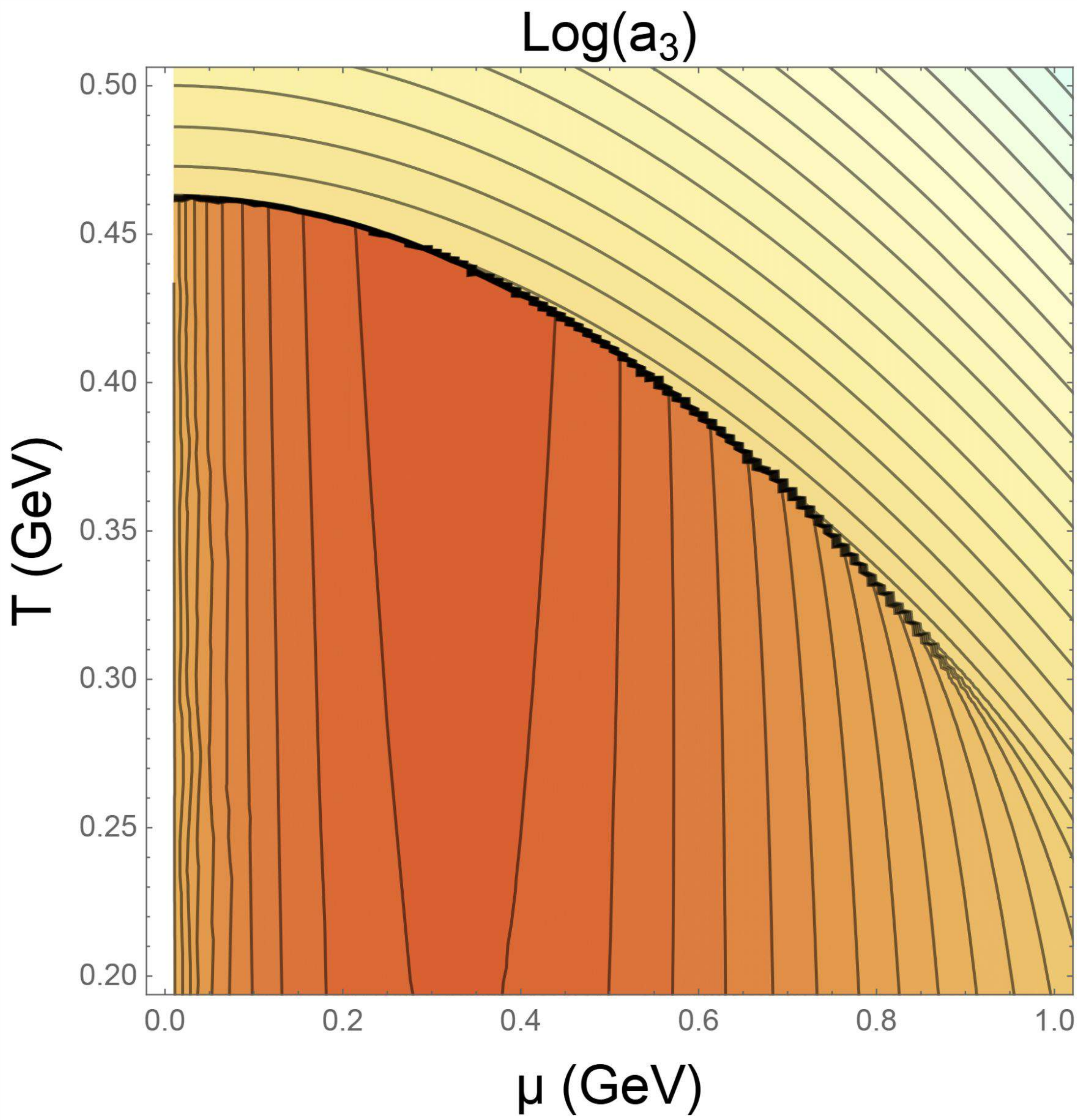}
\includegraphics[width =0.05\linewidth]{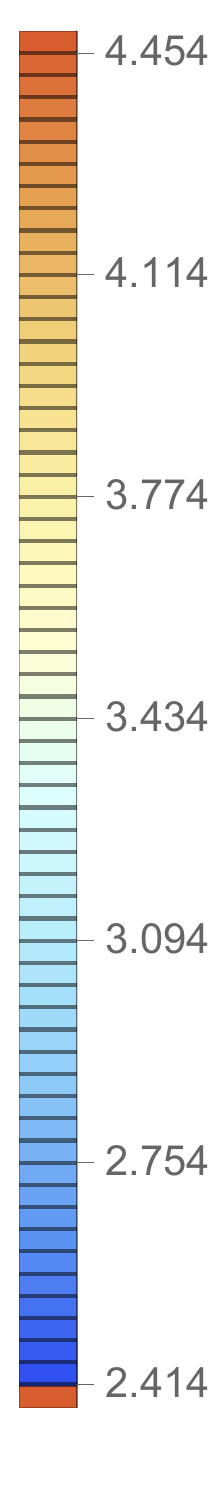}\\
C\hspace{170pt}D
  \caption{ Contour plots of (A) $\log a_2$ and (C) $\log a_3$ for the HQ model in the $(\mu, T)$-plane with non-zero magnetic field $c_B = -0.05$ \GG and $\nu = 4.5$. Panels (B) and (D) are zooms of panel panels (A) and (c)  under the first-order phase transition line, respectively. The difference between the top and bottom panels is only visible in zoomed versions (B and D). 
}
    \label{Fig:HQnu45cb-005}
\end{figure}

The results presented in Fig.\,\ref{Fig:HQnu45cb-005} summarize our discussions in Sections~\ref{NR-HQ-nzero-nu15} and~\ref{NR-HQ-nzero-nu45}, showing the density distribution of $\log a_2$ in the $(\mu,T)$-plane. 

\begin{figure}[t!]
    \centering
 \includegraphics[width =0.45\linewidth]
 {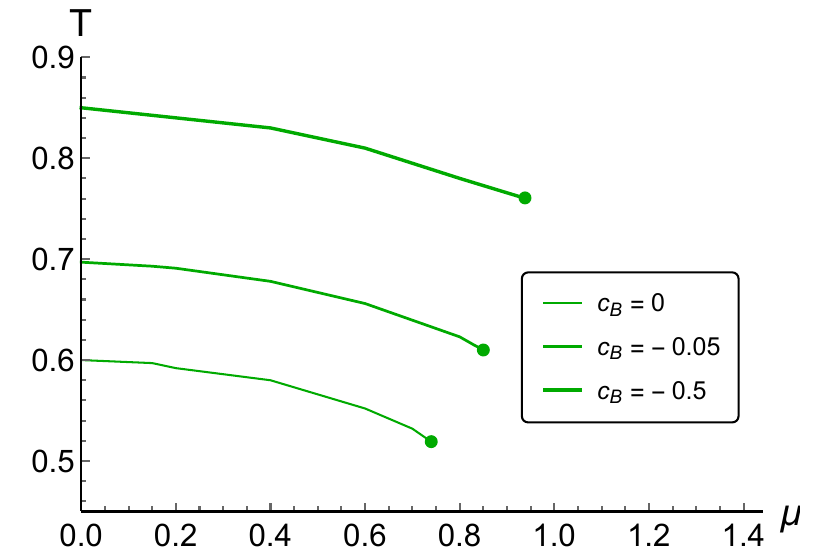}\quad
\includegraphics[width =0.45\linewidth]{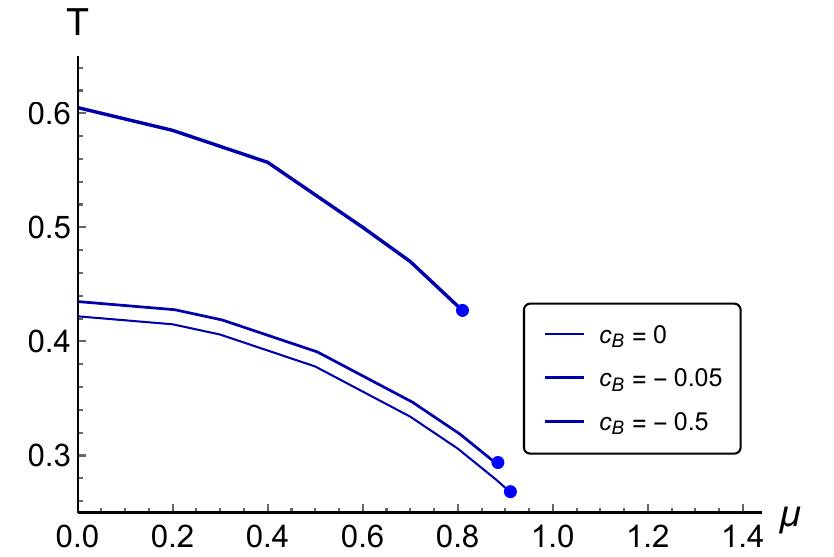}\\
A\hspace{150pt}B
 \caption{The first-order phase transition lines for the HQ model in the $(\mu,T)$-plane, with $c_B=0, -0.05$, and $c_B=-0.5$ (\GG)\, at fixed values of anisotropies (A) $\nu=1$, and (B) $\nu=4.5$. 
    }
    \label{Fig:HQPTnu11545cb0-005}
\end{figure}

\begin{itemize}
    \item These plots clearly reveal a first-order phase transition. Notably, the transition temperature for $\nu=4.5$ is lower than for $\nu=1.5$. For comparison, Fig.\,\ref{Fig:HQPTnu11545cb0-005} displays phase transition lines in the $(\mu,T)$-plane obtained from free energy analyses in our previous papers. Specifically, the plots in Fig.\,\ref{Fig:HQPTnu11545cb0-005} correspond to those in Fig.\,21A and Fig.\,21B of \cite{Arefeva:2023jjh} for $c_B = 0$, $0.05$, and $0.5$ (\GG).
    
    \item Fig.\,\ref{Fig:HQnu15cb-005}  clearly demonstrates a prominent hill for $\nu=1.5$. In contrast, Fig.\,\ref{Fig:HQnu45cb-005}  shows a less distinct hill but clearly exhibits a decrease in $\log a$ with increasing chemical potential: for $\mu \gtrsim 0.4~\text{GeV}$ in Fig.\,\ref{Fig:HQnu15cb-005}B and $\mu \gtrsim 0.42~\text{GeV}$ in Fig.\,\ref{Fig:HQnu45cb-005}B.
\end{itemize}

\begin{figure}[h!]
  \centering
\includegraphics[scale=0.27]
  {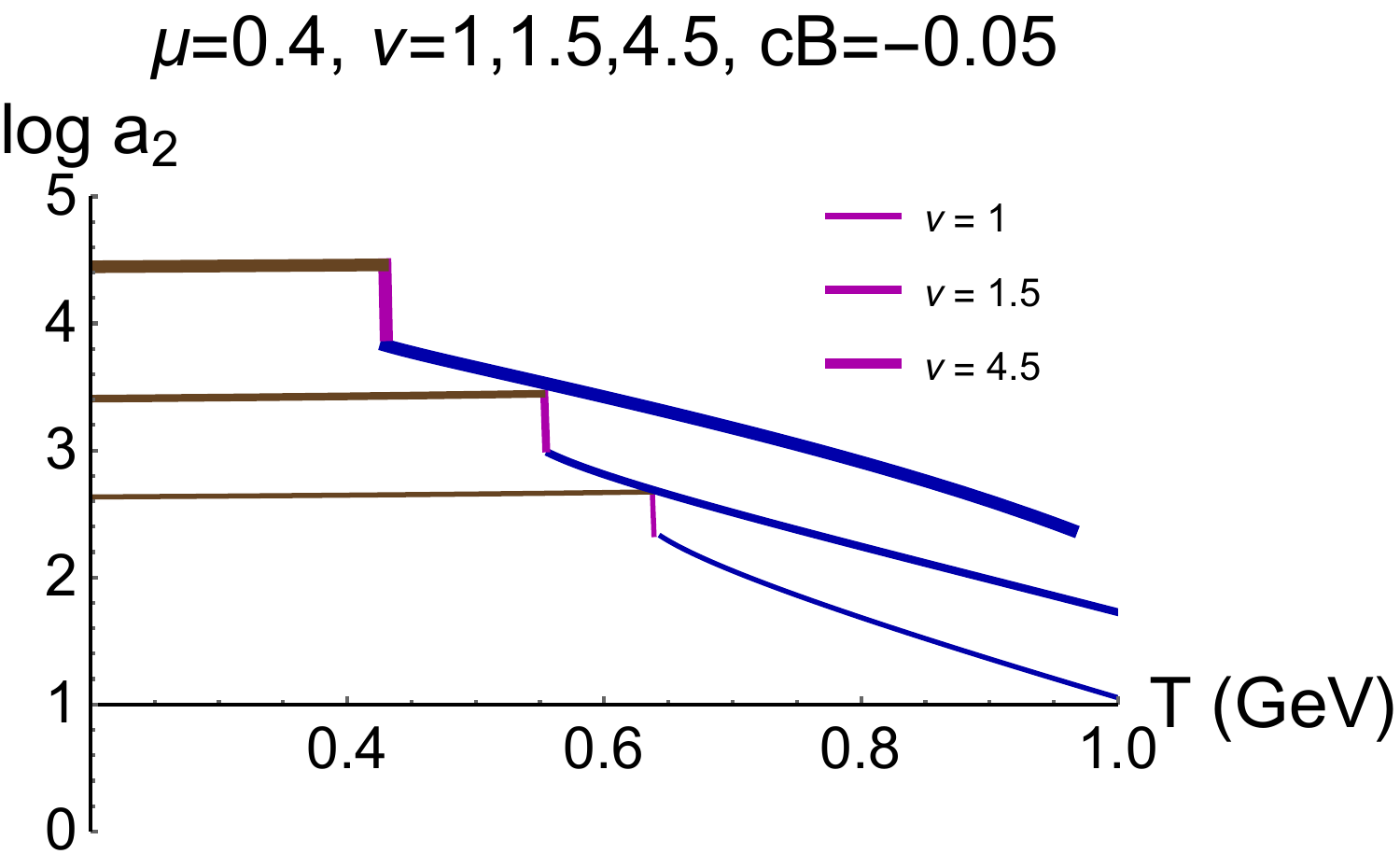}
 \caption{The dependence of $\log a_2$ on $T$ for the HQ model, with $\mu=0.4$ and $c_B=-0.05$ \GG\,, at fixed values of anisotropies: $\nu=1$, $1.5$, and $\nu=4.5$. The magnitude of the jumps in magenta color increases with the anisotropy parameter $\nu$.
 }
  \label{Fig:HQ-nu11545cB005-jump}
\end{figure}

It is instructive to examine the dependence of discontinuity magnitudes on anisotropy. For this purpose, Fig.\,\ref{Fig:HQ-nu11545cB005-jump} plots $\log a$ versus temperature $T$ at fixed chemical potential $\mu = 0.4~\text{GeV}$ for anisotropy parameters $\nu = 1$, $1.5$, and $\nu=4.5$. Line thickness increases with $\nu$ to visually distinguish the curves. Magenta segments indicate unstable regions and should be ignored. The observed discontinuities increase slightly with increasing anisotropy parameter $\nu$.

\newpage

\subsection{Summary of  jet quenching results for the HQ model}\label{sect-shq}

In this section, we have calculated the JQ parameter for quarks moving along light-like trajectories in a QCD-like medium with heavy quarks at finite temperature 
$T$, chemical potential $\mu$, magnetic field, and the anisotropy parameter $\nu$. Our results are presented through multiple plots, where:
\begin{itemize}
\item 
Table\,\ref{tab:HQ1} summarizes the schematic layouts of 2D plots showing 
the JQ parameter ($\log a$, in fact) of heavy quarks versus temperature 
$T$ at fixed $\mu$ values.

\item Table\,\ref{tab:HQ2}  presents schematic layouts of density plots for the JQ parameter ($\log a$, in fact).
\end{itemize}

These results reveal nontrivial dependencies of the JQ parameter on both the temperature and the chemical potential. Specifically, we observe:
\begin{itemize}
\item A discontinuous jump at the first-order phase transition line.
\item A change in the slope of 
$\log a$ versus temperature $T$
(at fixed $\mu$) at the second-order phase transition line.
\end{itemize}

The JQ parameter increases with $T$ and $\mu$ in the QGP phase. Conversely, the hadron phase displays complex behavior, featuring a hill-like structure in the $(\mu,T)$-plane below the first-order phase transition line at non-zero magnetic field. We further analyzed the effects of magnetic field strength and anisotropy parameter $\nu$ on these behaviors.

$$\,$$

\begin{table}[h!] 
\centering
\begin{tabular}{|c|l|c|c|c|}
\hline
\multicolumn{1}{|l|}{\diagbox{$|c_B|$}{$\nu$}} & \multicolumn{1}{c|}{1}                     & 1.5      & 3                                                                   & 4.5                                                                 \\ \hline
0                               & Fig.\,\ref{Fig:HQnu1mucB0}   & 
\multicolumn{1}{c|}{-} & \multicolumn{1}{c|}{-} & \multicolumn{1}{c|}{-} \\ \hline
0.05                           & Fig.\,\ref{Fig:HQnu1q23}  & Fig.\,\ref{Fig:HQnu15cB005}            & -  & Fig.\,\ref{Fig:HQ23nu45cB005}    \\ \hline
\end{tabular}
\caption{Schematic layout of 2D plots of dependence of $\log a$ of the JQ parameter for heavy quarks on $T$ for fixed values of $\mu$ at different $c_B$ and $\nu$. }
\label{tab:HQ1}
\end{table}

\begin{table}[h!] 
\centering
\begin{tabular}{|c|l|c|c|c|}
\hline
\multicolumn{1}{|l|}{\diagbox{$|c_B|$}{$\nu$}} & \multicolumn{1}{c|}{1}       & 1.5  & 3 & 4.5\\
\hline
0 & Fig.\,\ref{Fig:HQnu1cB0}& 
\multicolumn{1}{c|}{Fig.\,\ref{Fig:HQnu1545-cB0}A} & \multicolumn{1}{c|}{-} & \multicolumn{1}{c|}{Fig.\,\ref{Fig:HQnu1545-cB0}B} \\ \hline
0.05 & Fig.\,\ref{Fig:HQnu1cB005}& 
\multicolumn{1}{c|}{Fig.\,\ref{Fig:HQnu15cb-005}} & \multicolumn{1}{c|}{-} & \multicolumn{1}{c|}{Fig.\,\ref{Fig:HQnu45cb-005}} \\ \hline
0.5                           & Fig.\,\ref{Fig:HQnu1cB05}  & -            & -  & -      \\ 
\hline
\end{tabular}
\caption{Schematic layout of density plots for $\log a$, showing the JQ parameter for heavy quarks at different $c_B$ and $\nu$. }
\label{tab:HQ2}
\end{table}

\newpage
\section{ Conclusion} \label{Sect:Conc}

In this paper, we investigate the behavior of the JQ parameter  in an anisotropic holographic model under a strong magnetic field for light and heavy quark cases \cite{Arefeva:2022avn,Arefeva:2023jjh}.  Both models are constructed using the Einstein-dilaton-three-Maxwell action and incorporate a 5-dimensional metric with a warp factor, extending earlier isotropic and partially anisotropic frameworks, specifically the HQ of \cite{Andreev:2006ct,He:2010ye,Arefeva:2018hyo,Bohra:2019ebj,Arefeva:2020vae,Yang:2015aia,Arefeva:2023jjh,Arefeva:2024vom,Arefeva:2025okg} and the LQ model in \cite{Li:2017tdz,Arefeva:2022avn, Arefeva:2024vom,Arefeva:2020byn}.
\\

The LQ and HQ models have both similarities and significant differences. 
First, both have first-order and second-order phase transitions. But locations of phase transitions in $(\mu,T)$-planes are quite different; see plots in  Fig.\,\ref{Fig:DensityLQnu1cB0}C for light quarks and \ref{Fig:HQnu1cB0}C
for heavy quarks. The locations of phase transition depend on the anisotropy and the value of the magnetic field. 

\begin{itemize}
\item i)
For light quarks, increasing the anisotropy parameter $\nu$ reduces the temperature of the CEP  and shifts it towards lower values of temperature and chemical potential down to zero; see 
Fig.\,\ref{Fig:nu15-3-45-cB0}D, 
Fig.\,\ref{Fig:LQ-PTnu1cB0-0005-005}A, Fig.\,\ref{Fig:LQnu15cB0005005}A,
Fig.\,\ref{Fig:LQnu3cB0005005}A, and  Fig.\,\ref{Fig:LQ-first TLnu45}A.

\item ii) For heavy quarks, increasing  the anisotropy parameter $\nu$ reduces the temperature of the CEP  and shifts it towards lower values of temperature and the chemical potential down to zero, Fig.\,\ref{Fig:HQPTnu11545cb0-005}.
\item iii)
The differences between the LQ and HQ models are exacerbated when magnetic fields are included.
They respond differently to the magnetic field: the HQ model exhibits magnetic catalysis, whereas the LQ model demonstrates inverse magnetic catalysis; see  Fig.\,\ref{Fig:LQ-PTnu1cB0-0005-005}A,
Fig.\,\ref{Fig:LQnu3cB0005005}A and Fig.\,\ref{Fig:LQ-first TLnu45} for light quarks and Fig.\,\ref{Fig:HQPTnu11545cb0-005} for heavy quarks.
\end{itemize}

Studies of the JQ parameter allow us to accurately, at least theoretically, localize the first-order phase transition. For this purpose, we studied the density  plots of $\log a$ and its dependence on the parameter $\nu$ and the magnetic field parameter $c_B$.

\begin{itemize}
    \item Table\,\ref{tab:LQ-density} shows the map for the density plots of $\log a$ of the JQ parameter for the LQ model.
 We see:  
    \begin{itemize}
    \item that the location of the first-order phase transition, which we can read from the plot for the isotropic case in Fig.\,\ref{Fig:DensityLQnu1cB0}A, is in accordance with Fig.\,\ref{Fig:DensityLQnu1cB0}C.
   
    \item when we increase $\nu$ we get a map for the first-order phase transition that is in accordance with expected features mentioned in (i) above.
      \end{itemize}
     \item The Table\,\ref{tab:HQ2} shows the map for the density plots of $\log a$ of  the JQ parameter for the HQ model.
 We see:  
 \begin{itemize}
 \item that location of the first-order phase transition  that we can read from the plot for isotropic case  in Fig.\,\ref{Fig:HQnu1cB0}A   is in accordance with Fig.\,\ref{Fig:HQnu1cB0}C.
    \item when we increase $\nu$ we get a map for the first-order phase transition that is in accordance with expected features mentioned in (ii) above.
    \end{itemize}

\item The distinct positions of the first-order phase transition lines in the two models mentioned above in (iii) lead to contrasting behaviors in the discontinuity of 
$\log a$
 across the first-order phase transition:
 \begin{itemize}
\item  for the LQ model, the jump in of $\log a$ 
increases with increasing  $\mu$. We see this clearly for totally isotropic case (Fig.\,\ref{Fig:DensityLQnu1cB0}A, zero magnetic field for different $\nu$ (Fig.\,\ref{Fig:nu15-3-45-cB0}A, \ref{Fig:nu15-3-45-cB0}B and \ref{Fig:nu15-3-45-cB0}C), and for non-zero magnetic field in the cases where the fist-order phase transition take place, for example in the cases presented in Fig.\,\ref{Fig:LQnu1cB0005} and Fig.\,\ref{Fig:LQ3nu1cB005}.

\item  in the HQ model, the jump in $\log a$ decreases with increasing $\mu$ for the isotropic case (Fig.\,\ref{Fig:HQnu1cB0}A) and for various $\nu$ (Fig.\,\ref{Fig:HQnu1545-cB0}). However, under non-zero magnetic fields, the behavior becomes more complex. As shown in Fig.\,\ref{Fig:HQnu1cB005} ($c_B=-0.05$ \GG, $\nu=1$), the jump initially increases until $\mu=0.45$ GeV, then decreases to zero at the CEP. Similar behavior occurs in Fig.\,\ref{Fig:HQnu1cB05}B ($c_B=-0.5$ \GG, $\nu=1$), where the jump increases until $\mu=0.75$ GeV before decreasing to zero at the CEP. This pattern also holds for $c_B=-0.05$ \GG, $\nu=1.5$ (Fig.\,\ref{Fig:HQnu15cb-005}) and $c_B=-0.05$ \GG, $\nu=4.5$ (Fig.\,\ref{Fig:HQnu45cb-005}). 

\end{itemize}
\item 
Moreover, in large $T$, the isotropic LQ model exhibits the behavior $a T^3 \sim const$, as shown in Fig.\,\ref{Fig:LQT3nu1cB0}. This  behavior is also exhibited by the LQ model in \cite{Zhu:2023aaq}. It is consistent with the predictions of the conformal invariant model \cite{Liu:2006ug}, and shows good agreement with the experimental data from RHIC and LHC. The magnetic field breaks the conformal symmetry, as shown in Fig.\,\ref{Fig:LQnu1cB005m}.

\item Our isotropic HQ model at $\mu=0$ does not exhibit the behavior predicted by conformal invariance, where $a T^3 \sim const$ at large $T$ (see Fig.\,\ref{Fig:HQnu1cB0mu06081}). This finding contrasts with the behavior of the JQ parameter for the HQ model in \cite{Heshmatian:2023yzz},  which approaches conformality as $T \to \infty$, while ours does not. The deviation is expected because our HQ model lies far from conformality. Interestingly, the asymptotic behavior of $\log (a T^3)$ in large $T$ remains independent of the chemical potential $\mu$, as shown in Fig.\,\ref{Fig:HQnu1cB0mu06081}D.

\item  However, there are also some similarities: for both light and heavy quarks, $\log a$ decreases linearly at large temperatures, as shown in Fig.\,\ref{Fig:LQnu1cB0005005} and Fig.\,\ref{Fig:LQnu15cB0005005}.

\item Considering a magnetic field oriented along the $x_3$-axis,  we have studied jets propagating along the $x_1$-direction with momentum broadening both parallel to the magnetic field ($\hat{q}_3$) and perpendicular to it along the $x_2$-direction ($\hat{q}_2$). In Sect.\,\ref{sect:LQ-NR}, our analysis focused specifically on the transverse broadening component $\hat{q}_2$ for the LQ model, while in Sect.\,\ref{sect:HQ-NR}, we examined the parallel broadening component $\hat{q}_3$ for the HQ model.

\item Nevertheless, in Sects.\,\ref{sect:NR-LQ-nzero-nu1} and \ref{NR-HQ-nzero-nu1} we compared the parameters $\hat{q}$, i.e. $\hat{q}_2$ and $\hat{q}_3$ for the LQ and  HQ models in a magnetic field, respectively. 
\begin{itemize}
\item In Sect.\,\ref{sect:NR-LQ-nzero-nu1} we compared the $\hat{q}$ parameters $\hat{q}_2$ and $\hat{q}_3$ for the LQ model in a magnetic field, $c_B = -0.05$ \GG.  We observe a subtle but weak orientation dependence:

\begin{itemize}
 \item Fig.\,\ref{Fig:LQnu1q23new} and Fig.\,\ref{Fig:LQnu1q23new2} demonstrate that, the JQ parameter exhibits anisotropy in the presence of a magnetic field: $\log a_2 \neq \log a_3$ for $\nu = 1$. 

\item Below the first-order phase transition temperature and above the transition, $\log a_3 > \log a_2$ for $\nu = 1$.
\end{itemize}

\item  In Sect.\,\ref{NR-HQ-nzero-nu1} we compared the $\hat{q}$ parameters $\hat{q}_2$ and $\hat{q}_3$ for the HQ model in a magnetic field, $c_B=-0.05$ \GG: 
\begin{itemize}
\item For $\nu=1$, at $\mu=0.4$ GeV, and $\mu=1$ GeV  we observed $\log a_3 > \log a_2$ (see Fig.\,\ref{Fig:HQnu1q23}).
\item For $\nu=1.5$, at $\mu=1$ GeV, we observe $\log a_2 > \log a_3$. In the hadronic phase at $\mu=0.4$ and $\mu=0.6$ GeV, $\log a_2 > \log a_3$ similarly holds. However, in the QGP region, the relationship between $\log a_2$ and $\log a_3$ remains inconclusive (see Fig.\,\ref{Fig:HQ23nu15cB005}).
\item  For $\nu=4.5$, we observe $\log a_3 > \log a_2$ for $\mu=0.4$ GeV and $\mu=1$ GeV. However, at $\mu=0.6$ $\log a_2$ and $\log a_3$ nearly coincide
(see Fig.\,\ref{Fig:HQ23nu45cB005}).
\end{itemize}
\end{itemize}
\end{itemize}

Our study has analyzed the behavior of the JQ parameter under the idealized assumption that jet formation occurs within a medium consisting solely of light quarks (the LQ model) or solely of heavy quarks (the HQ model). In a physically realistic setting, however, the medium comprises both light and heavy quark species, with light quarks predominant at low energies and heavy quarks contributing significantly at high energies. 

A natural extension of this work is therefore the investigation of jets within a hybrid model that incorporates both quark types, effectively interpolating between the LQ and HQ model limits. To construct the hybrid model it is needed to chose the proper warp-factor and gauge kinetic function. The hybrid model should reproduce linear Regge trajectories that were observed experimentally and first-order phase transition lines for heavy and light quarks. This constitutes an important objective for future research.

\section*{Acknowledgments}

We thank K. Rannu and M. Usova for  discussions. 
The work of I.A. and P.S. was performed at the Steklov International Mathematical Center and supported by the Ministry of Science and Higher Education of the Russian Federation (agreement no. 075-15-2025-303). The work of A. H. was started at the Steklov International Mathematical Center and supported by the Ministry of Science and Higher Education of the Russian Federation (Agreement No. 075-15-2022-265). The work of I. A., P. S. and A.N. is also  supported by Theoretical Physics and Mathematics Advancement Foundation ``BASIS (grant No. 24-1-1-82-1, grant No. 23-1-4-43-1 and grant No. 24-2-2-4-1, respectively).
\newpage

\appendix

\section{Equations of motion  and boundary conditions} \label{app0}

Together with the variations over the scalar field and first vector
field (Maxwell field that serves a non-zero chemical potential and for
which we have chosen the electric ansatz), second vector
field (Maxwell field that serves a spatial anisotropy and for
which we have chosen the magnetic ansatz) and third vector
field (Maxwell field that serves a non-zero magnetic field and for
which we have chosen the magnetic ansatz) we get the following EOMs:
\begin{gather}
  \begin{split}
    \phi'' + \phi' \left(
      \cfrac{g'}{g} + \cfrac{3 \fb'}{2 \fb} 
      - \cfrac{\nu + 2}{\nu z} + c_B z
    \right)
    &+ \left( \cfrac{z}{L} \right)^2 \cfrac{(A_t')^2}{2 \fb g} \ 
    \cfrac{\partial f_0}{\partial \phi}
    - \left( \cfrac{L}{z} \right)^{2-\frac{4}{\nu}} 
    \cfrac{e^{-c_Bz^2} \, q_1^2}{2 \fb g} \ 
    \cfrac{\partial f_1}{\partial \phi} \ - \\
    &- \left( \cfrac{z}{L} \right)^{\frac{2}{\nu}} 
    \cfrac{q_3^2}{2 \fb g} \ \cfrac{\partial f_3}{\partial \phi}
    - \left( \cfrac{L}{z} \right)^2 \cfrac{\fb}{g} \,
    \cfrac{\partial V}{\partial \phi} = 0,
  \end{split} \label{eq:2.16} \\
  A_t'' + A_t' \left(
    \cfrac{\fb'}{2 \fb} + \cfrac{f_0'}{f_0} 
    + \cfrac{\nu - 2}{\nu z} + c_B z
  \right) = 0, \label{eq:2.17} \\
  {\bf (I)} \qquad
  g'' + g' \left(
    \cfrac{3 \fb'}{2 \fb} - \cfrac{\nu + 2}{\nu z} + c_B z
  \right)
  - \left( \cfrac{z}{L} \right)^2 \cfrac{f_0 \,  (A_t')^2}{\fb}
  - \left(\cfrac{z}{L} \right)^{\frac{2}{\nu}} \cfrac{q_3^2 f_3}{\fb}
  = 0, \label{eq:2.18} \\
  {\bf (II)} \qquad  
  \fb'' - \cfrac{3 (\fb')^2}{2 \fb} + \cfrac{2 \fb'}{z}
  - \cfrac{4 \fb}{3 \nu z^2} \left(
    1 - \cfrac{1}{\nu}
    + \left( 1 - \cfrac{3 \nu}{2} \right) c_B z^2
    - \cfrac{\nu c_B^2 z^4}{2}
  \right)
  + \cfrac{\fb \, (\phi')^2}{3} = 0, \label{eq:2.19} \\
  \begin{split}
    {\bf (III)} \quad
    2 g' \left( 1 - \cfrac{1}{\nu} \right) 
    + 3 g \left( 1 - \cfrac{1}{\nu} \right) \left(
      \cfrac{\fb'}{\fb} - \cfrac{4 \left( \nu + 1 \right)}{3 \nu z}
      + \cfrac{2 c_B z}{3}
    \right)
    + \left( \cfrac{L}{z} \right)^{1-\frac{4}{\nu}} 
    \cfrac{L \, e^{-c_Bz^2} q_1^2 \, f_1}{\fb} = 0,
  \end{split} \label{eq:2.20} \\
  \begin{split}
    {\bf  (IV)} \quad
    2 g' &\left( 1 - \cfrac{1}{\nu} + c_B z^2 \right) 
    + 3 g \left[ \Big( 1 - \cfrac{1}{\nu} + c_B z^2 \Big) 
      \left(
        \cfrac{\fb'}{\fb} - \cfrac{4}{3 \nu z} + \cfrac{2 c_B z}{3}
      \right)
      - \cfrac{4 \left( \nu - 1 \right)}{3 \nu z} \right] + \\
    + &\left( \cfrac{L}{z} \right)^{1-\frac{4}{\nu}}
    \cfrac{L \, e^{-c_Bz^2} q_1^2 \, f_1}{\fb}
    - \left( \cfrac{z}{L} \right)^{1+\frac{2}{\nu}}
    \cfrac{L \, q_3^2 \, f_3}{\fb} = 0,
  \end{split}\label{eq:2.21} 
\end{gather}

\begin{gather}
 \begin{split}
    {\bf (V)} \quad 
    \cfrac{\fb''}{\fb} &+ \cfrac{(\fb')^2}{2 \fb^2}
    + \cfrac{3 \fb'}{\fb} \left(
      \cfrac{g'}{2 g} - \cfrac{\nu + 1}{\nu z} + \cfrac{2 c_B z}{3}
    \right)
    - \cfrac{g'}{3 z g} \left( 5 + \cfrac{4}{\nu} - 3 c_B z^2 \right)
    + \\
    &+ \cfrac{8}{3 z^2}
    \left( 1 + \cfrac{3}{2 \nu} + \cfrac{1}{2\nu^2} \right)
    - \cfrac{4 c_B}{3}
    \left( 1 + \cfrac{3}{2 \nu} - \cfrac{c_B z^2}{2} \right) 
    + \cfrac{g''}{3 g} 
    + \cfrac{2}{3} \left( \cfrac{L}{z} \right)^2 \cfrac{\fb V}{g} 
    = 0.  
  \end{split} \label{eq:2.22}
\end{gather}
To solve EOMs we use the usual boundary conditions for time component of the first gauge field $A_t$ and blackening function $g$:
\bea
A_t(0)=\mu, \quad  A_t(z_h)=0,\quad g(0)=1, \quad g(z_h)=0.
\eea
The physical boundary conditions for the dilaton field are \cite{Arefeva:2024vom}:\\
\begin{itemize}
\item for the LQ model
\be
z_0=10 e^{-z_h/4}+0.1 \, ,
\label{LQ-nbc}\ee
\item for the HQ model
\be
z_0=e^{-z_h/4}+0.1 \,,
\label{QH-nbc}
\ee
\end{itemize}
where $\phi(z_0)=0$.

\section{Supplementary Figures for Light Quarks} \label{app2}

\begin{figure}[h]
  \centering
\includegraphics[scale=0.23]
{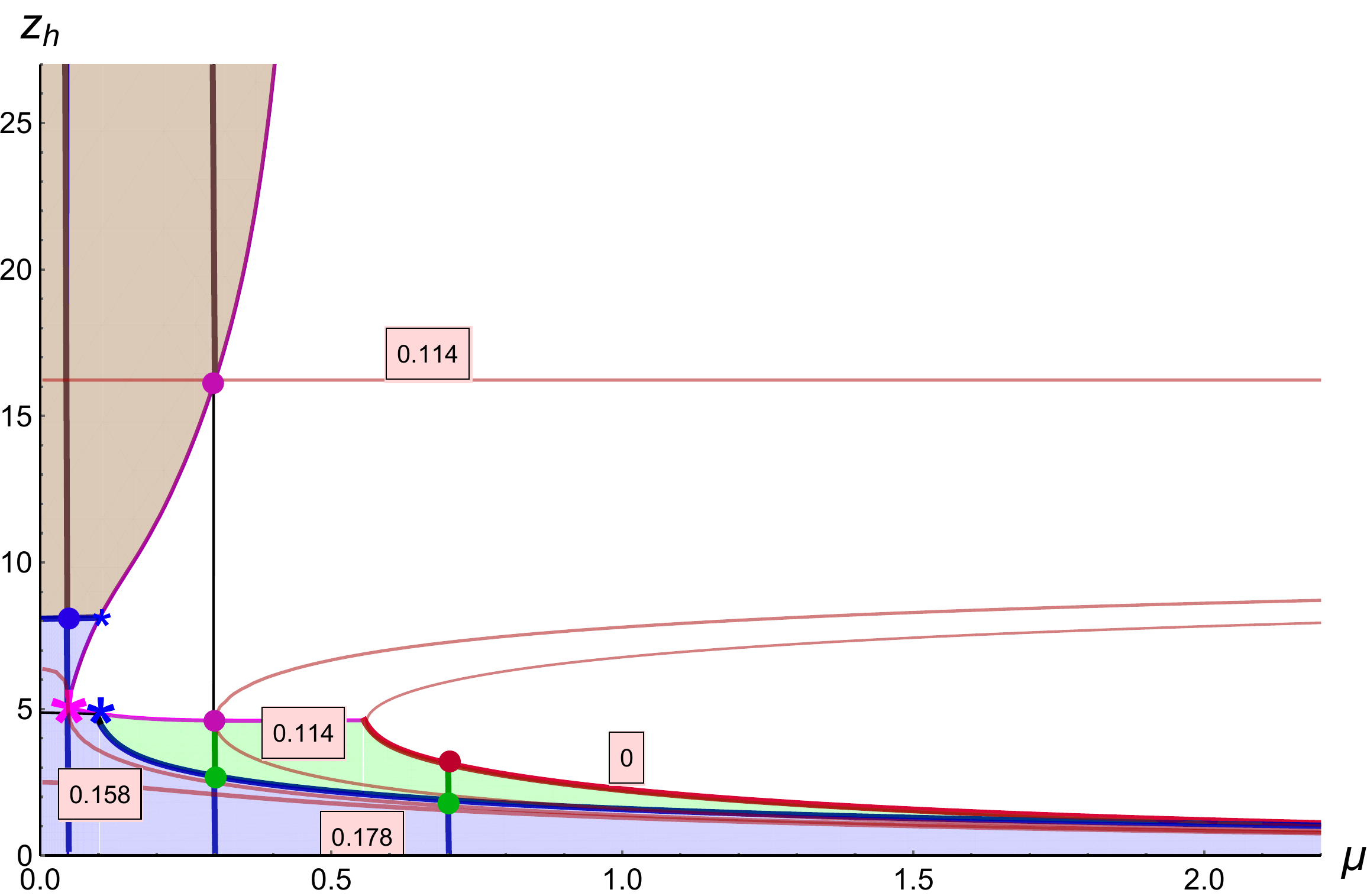}
\caption{ 2D plot in the $(\mu,z_h)$-plane
for the LQ model in isotropic case,  $c_B=0$ and $\nu=1$ with different phases, i.e. QGP, quarkyonic and hadronic corresponding to blue, green and brown regions, respectively. 
  Solid orange lines show the temperature indicated in the squares. The intersection of the confinement/deconfinement and first-order phase transition lines is denoted by the blue stars. The magenta star indicates the CEP. Vertical lines indicate the lines along which we calculate the JQ parameter presented in the next figures, i.e. Fig.\,\ref{Fig:LQQmu004}, Fig.\,\ref{Fig:LQQmu03},   and Fig.\,\ref{Fig:LQQmu07}.
 }
  \label{Fig:LQ-Iso-zh-mu}
\end{figure}

The phase structure of the LQ model  in Fig.\,\ref{Fig:LQ-Iso-zh-mu} shows different domains of phases, i.e. QGP, quarkyonic, and hadronic correspond to blue, green,
and brown regions, respectively, \cite{Arefeva:2024vom}. The solid blue lines in
Fig.\,\ref{Fig:LQ-Iso-zh-mu} correspond to the confinement/deconfinement phase
transition line  obtained via Wilson loop
calculations \cite{Li:2017tdz}.
The solid magenta lines in Fig.\,\ref{Fig:LQ-Iso-zh-mu}
correspond to the first-order phase transition line  obtained via free energy calculations \cite{Li:2017tdz}  and
the thicker dark red line corresponds to the second horizon where $T=0$, while other solid darker red curves represent constant-temperature lines \(T = T(z_h, \mu)\). Jumps for various physical quantities
appear along two thin magenta curves that bounded the unstable region. 
The first-order phase transition manifests as a discontinuity in \(z_h\) that takes place  for \(\mu\) varying from the value corresponding to the magenta star  to the zero-temperature endpoint.    The unstable region between the magenta curves is highlighted in white in Fig.\,\ref{Fig:LQ-Iso-zh-mu}. Two magenta lines  are mapping to a single transition line in the $(\mu, T)$-plane,  the magenta curve  in Fig.\,\ref{Fig:LQnu1cB0}. 
\\

\begin{figure}[h]
  \centering
 \includegraphics[scale=0.41]
 {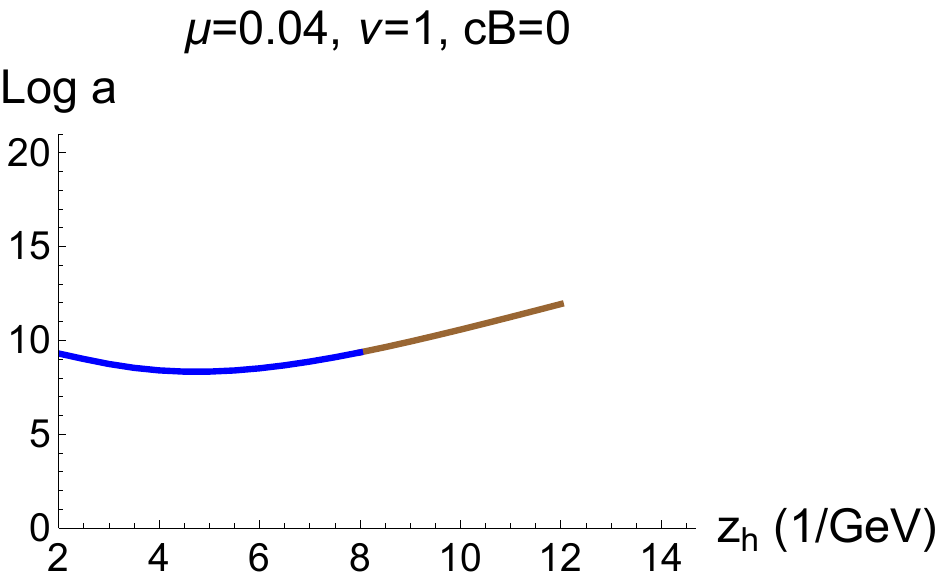}\qquad
  \includegraphics[scale=0.40]
 {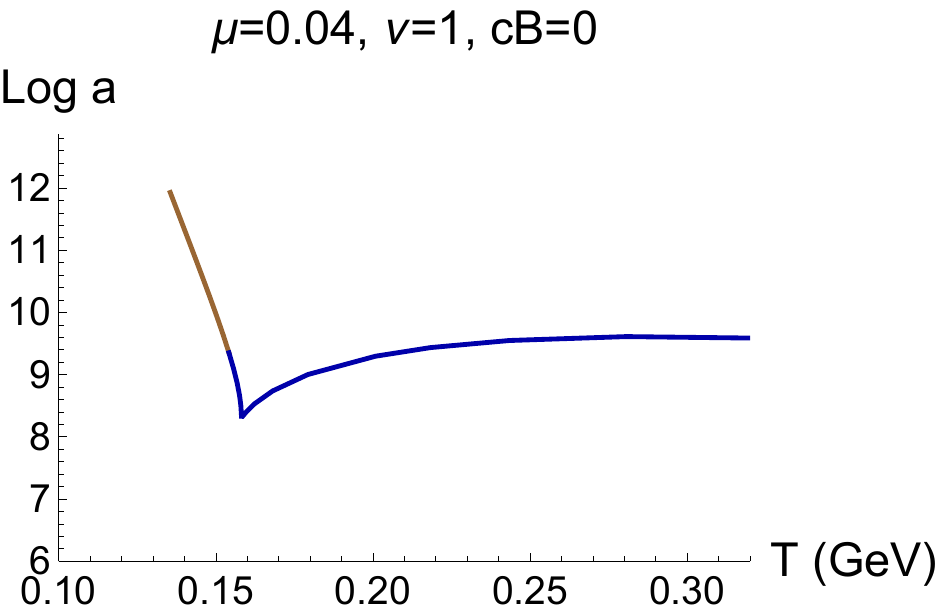}\\
 A\hspace{210pt}B
\caption{$\log a$ as a function of the size of the horizon $z_h$ (A) and the temperature $T$ (B) for the LQ model at $\mu=0.04$ GeV in crossover region for $c_B=0$ and $\nu=1$. The blue and brown lines correspond to the QGP and hadronic phases, respectively.
}
\label{Fig:LQQmu004}
\end{figure}

Fig.\,\ref{Fig:LQQmu004}A depicts the $\log a$ as a function of the size of the horizon $z_h$, and Fig.\,\ref{Fig:LQQmu004}B as a function of the temperature $T$ at $\mu=0.04$ GeV. The blue and brown lines correspond to the QGP and hadronic phases, respectively. $\log a$ decreases up to $T=0.158$ GeV  even after a continuous phase transition at $T=0.154$ GeV from the hadronic to the QGP phases, indicating an enhancement in the JQ parameter up to $T=0.158$ GeV and a decrease in the JQ parameter for $0.158$ GeV $<T<0.3$ GeV. 
For higher values of temperature, i.e. T $\gtrsim 0.3$ GeV, the JQ parameter increases; see Fig.\,\ref{Fig:LQnu1cB0}B. 


$\,$\\

\begin{figure}[h!]
  \centering
\includegraphics[scale=0.23]
  {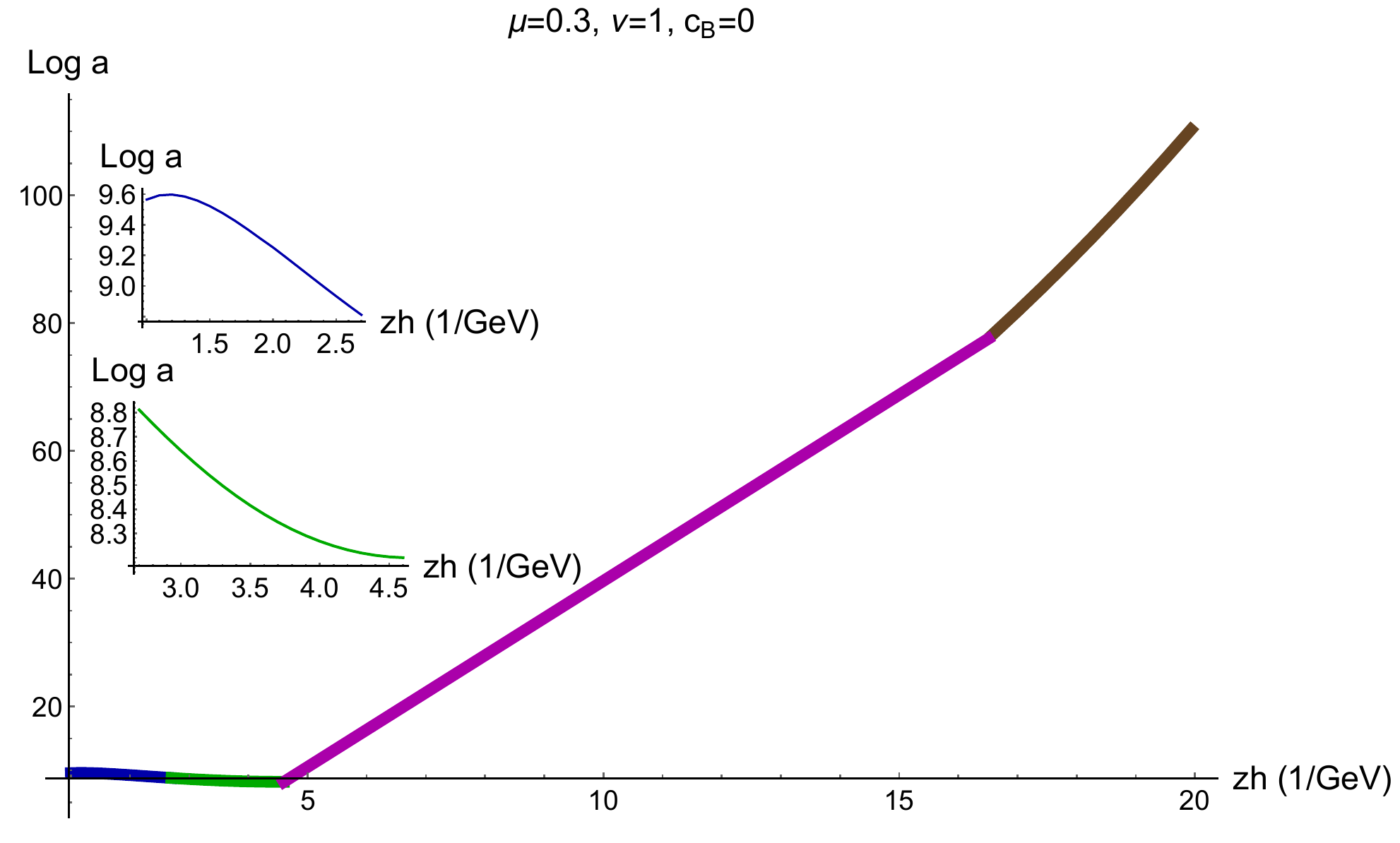}
\includegraphics[scale=0.28]
  {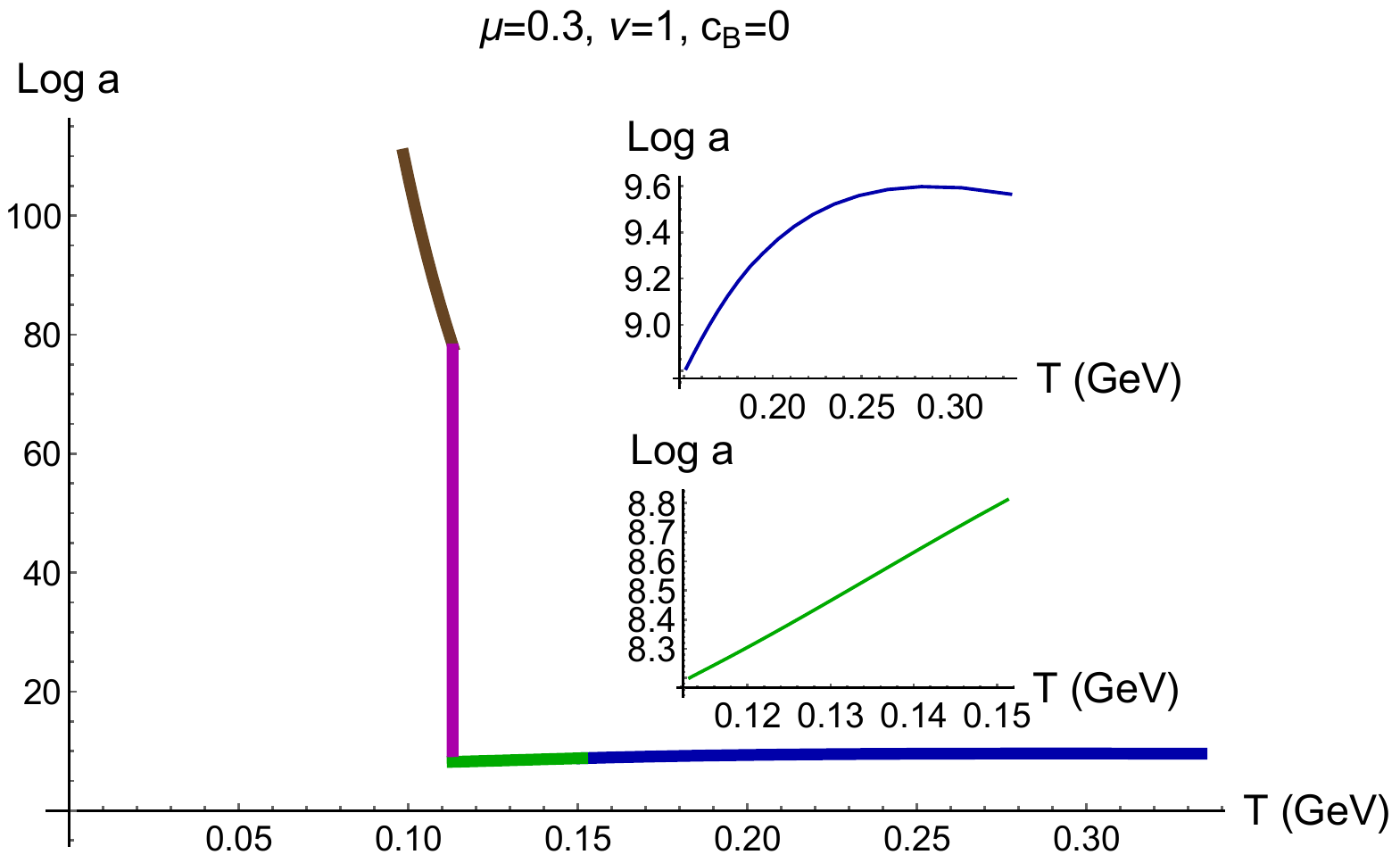}
\\A\hspace{220pt}B
\caption{(A) $\log a$ as a function of the size of the horizon $z_h$, and  (B) the temperature $T$  for the LQ model, with  $c_B=0$ and $\nu=1$,  at $\mu=0.3$ GeV. The blue, green, and brown lines correspond to the QGP, quarkyonic and hadronic phases, respectively. The magenta lines indicate the jump between quarkyonic and hadronic phases crossing the first-order phase transition line. 
}
  \label{Fig:LQQmu03} 
  \end{figure}

Fig.\,\ref{Fig:LQQmu03}A shows the $\log a$ as a function of the size of the horizon $z_h$, and Fig.\,\ref{Fig:LQQmu03}B as a function of the temperature $T$ at $\mu=0.3$ GeV for $c_B=0$ and $\nu=1$. The blue, green, and brown lines correspond to the QGP, quarkyonic and hadronic phases, respectively. The magenta lines indicate the jump between the quarkyonic and hadronic phases crossing the first-order phase transition line. In the hadronic phase, increasing temperature increases the JQ parameter, while after the first-order phase transition at $T=0.11$ GeV the JQ parameter decreases at quarkyonic and QGP phases by increasing the temperature to $T=0.28$ GeV. For higher values of $T$ in the QGP phase, that is, $T\gtrsim 0.28$ GeV the JQ parameter increases. Fig.\,\ref{Fig:LQQmu004} and Fig.\,\ref{Fig:LQQmu03} show that for zero magnetic field and spatially isotropic system, in the hadronic phase the JQ parameter increases by increasing $T$.

$\,$
\\

\begin{figure}[h!]
  \centering
 \includegraphics[scale=0.41]
  {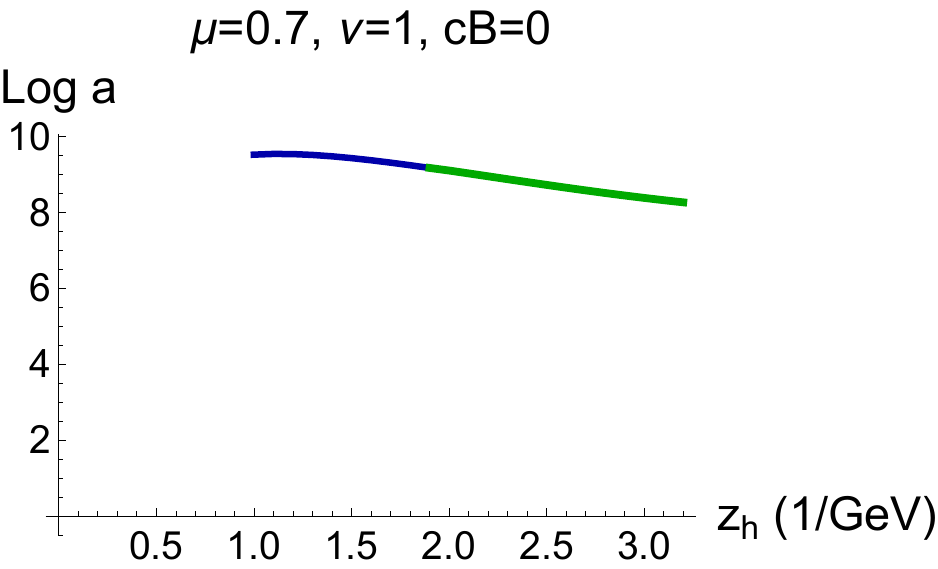} \qquad
  \includegraphics[scale=0.41]
   {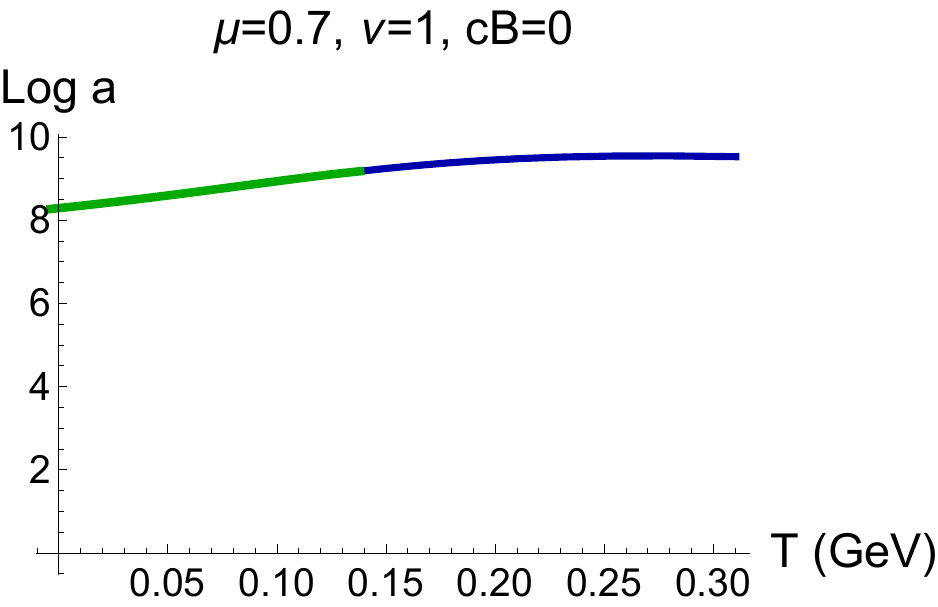}
   \\A\hspace{200pt}B
  \caption{(A) $\log a$ as a function of the size of the horizon $z_h$, and (B) the temperature $T$ for the LQ model, with  $c_B=0$ and $\nu=1$,  at $\mu=0.7$ GeV. The blue and green lines correspond to the QGP and quarkyonic phases, respectively.  
  }
\label{Fig:LQQmu07}
\end{figure}

In Fig.\,\ref{Fig:LQQmu07}A
$\log a$ as a function of the size of the horizon $z_h$, and in Fig.\,\ref{Fig:LQQmu07}B as a function of the temperature $T$ at $\mu=0.7$ GeV for $c_B=0$ and $\nu=1$ are depicted. The blue and green lines correspond to the QGP and quarkyonic phases, respectively. $\log a$ increases, with increasing temperature $T$ which indicates that the JQ parameter decreases even after the phase transition between quarkyonic and QGP phases. Fig.\,\ref{Fig:LQQmu03}B and Fig.\,\ref{Fig:LQQmu07}B show that for the zero magnetic field and spatially isotropic system, in the quarkyonic phase, the JQ parameter decreases monotonically by increasing $T$. 

$\,$

\begin{figure}[h!]
  \centering
\includegraphics[scale=0.26]
  {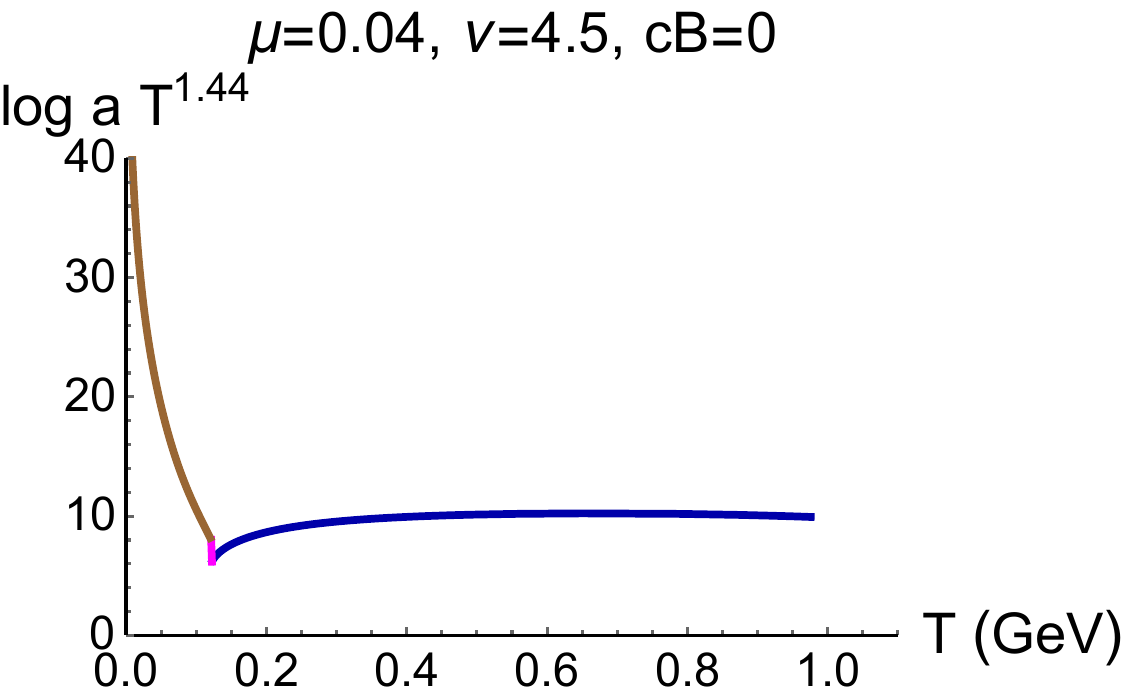}
  \includegraphics[scale=0.26]
  {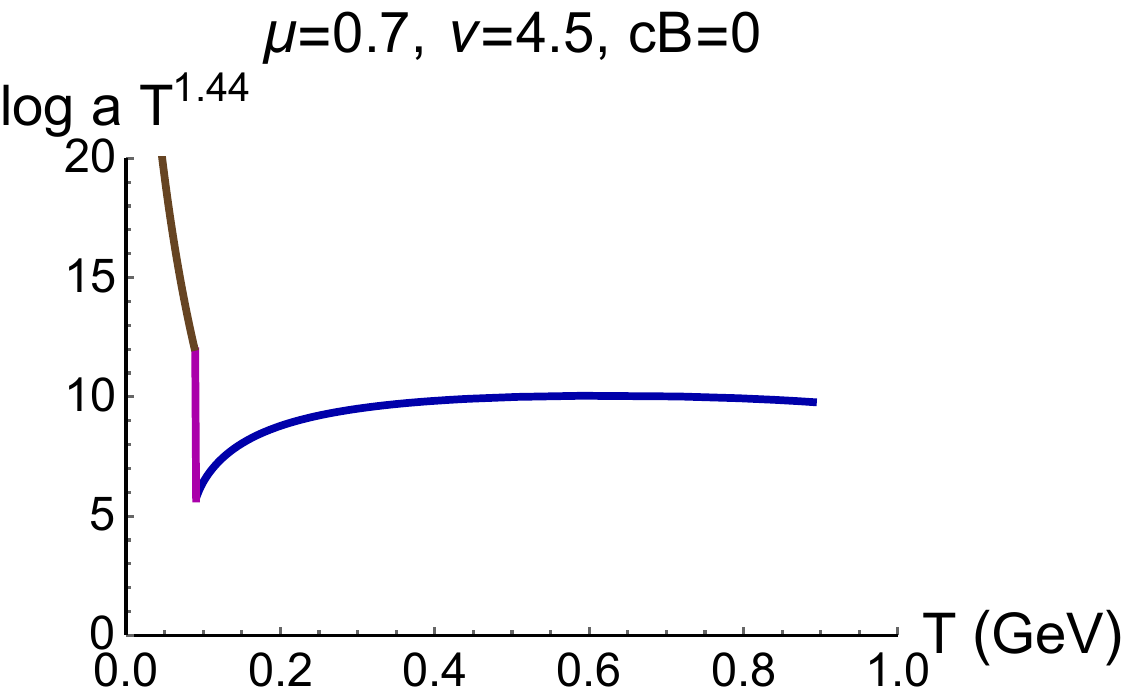}
  \includegraphics[scale=0.26]
  {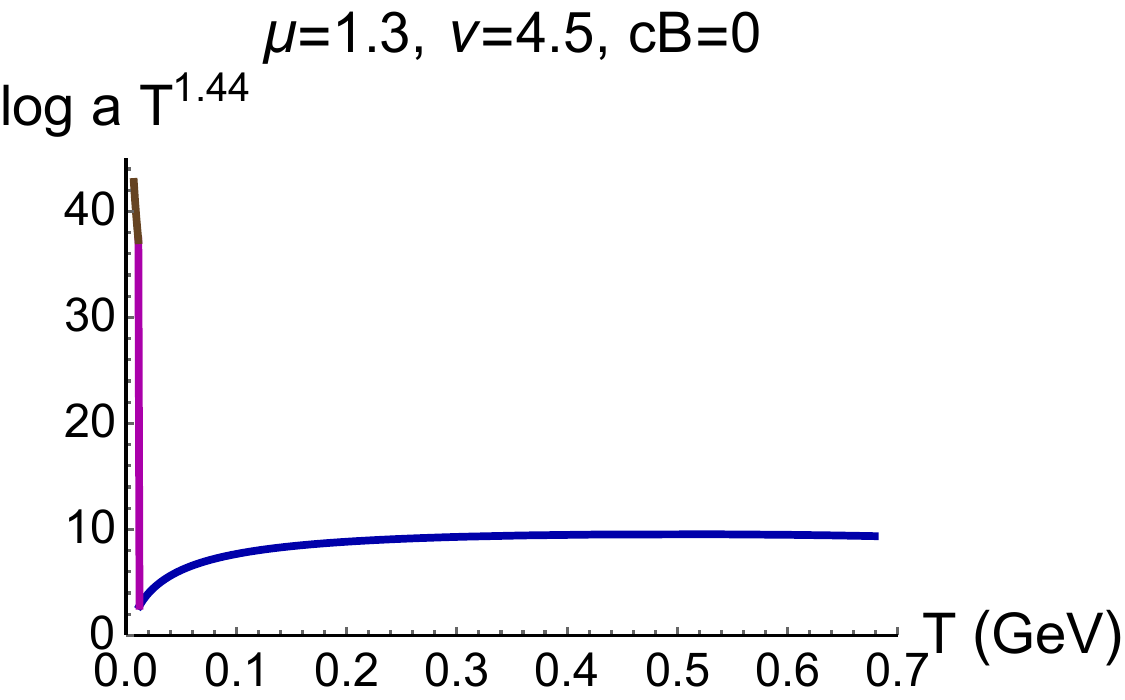}
  \\A\hspace{135pt}B \hspace{135pt}C\\$\,$\\
  \includegraphics[scale=0.29]
  {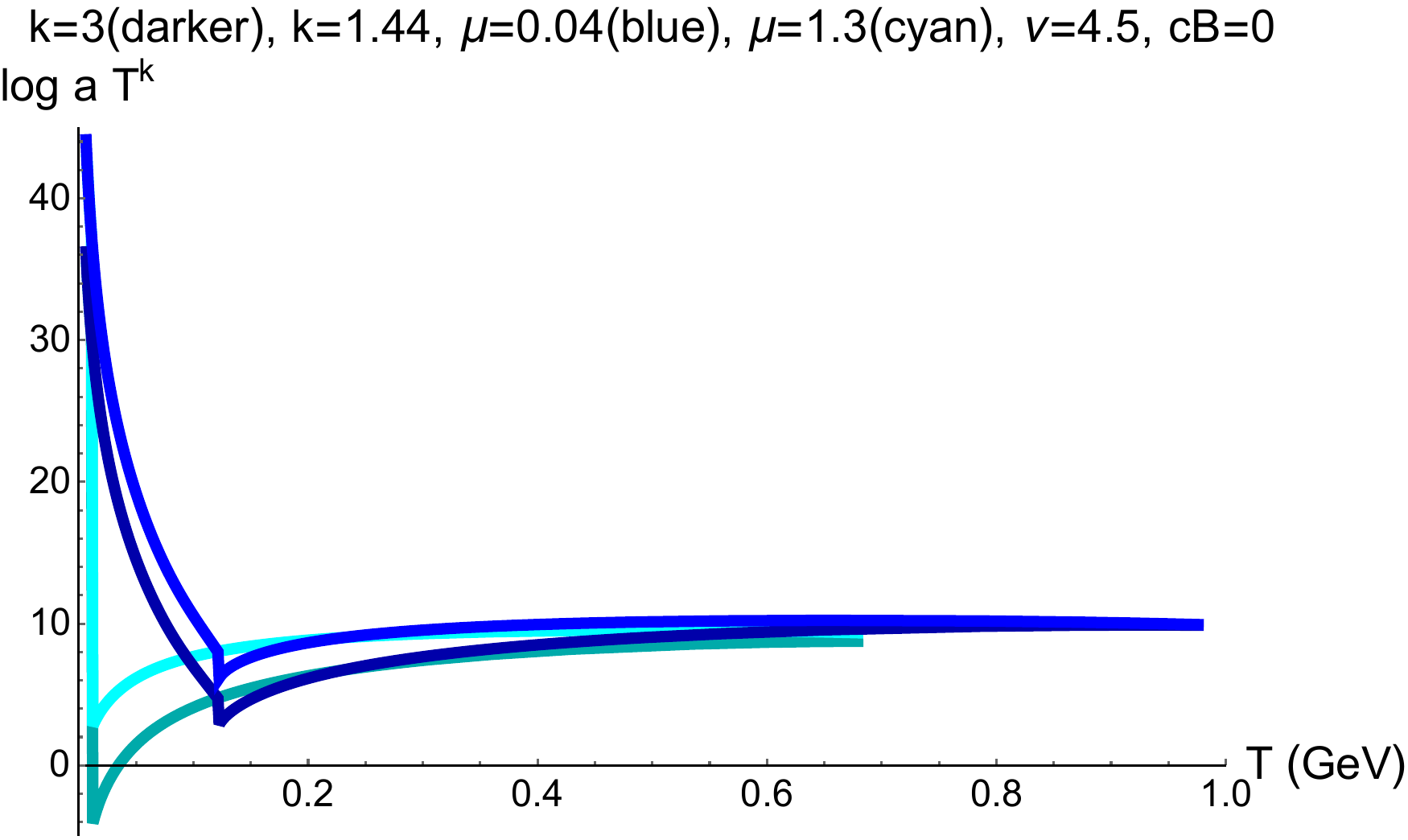}
  \\D
\caption{$\log a \,T^{1.44}$  as function of $T$ for the LQ model, with $\nu=4.5$ and $c_B=0$, at fixed chemical potentials:  (A) $\mu = 0.04$, (B) $\mu=0.7$, (C) $\mu=1.3$ (GeV). Segments of lines are colored blue, brown, and green corresponding to the QGP, hadronic, and quarkyonic phases they traverse, respectively. 
   (D) Comparison of $\log a \,T^{1.44}$ and $\log a \,T^{3}$ for the same $\nu=4.5$ and $c_B=0$,  
   $\mu = 0.04$ GeV and   $\mu = 1.3$ GeV. 
  }
\label{Fig:LQnu45cB0-c}
\end{figure}

In Fig.\,\ref{Fig:LQnu45cB0-c} $\log a \,T^{1.44}$  as function of $T$ for the LQ model, with $\nu=4.5$ and $c_B=0$, at fixed chemical potentials $\mu = 0.04$ GeV, $\mu=0.7$ GeV, and $\mu=1.3$ GeV are depicted in panels (A, B, C), respectively.  Fig.\,\ref{Fig:LQnu45cB0-c}C shows a comparison between $\log a \,T^{1.44}$ and $\log a \,T^{3}$ for the same $\nu=4.5$ and $c_B=0$,  
$\mu = 0.04$ GeV and  $\mu = 1.3$ GeV. There is the same behavior at high-temperature regime.

\begin{figure}[h!]
  \centering
\includegraphics[scale=0.34]
  {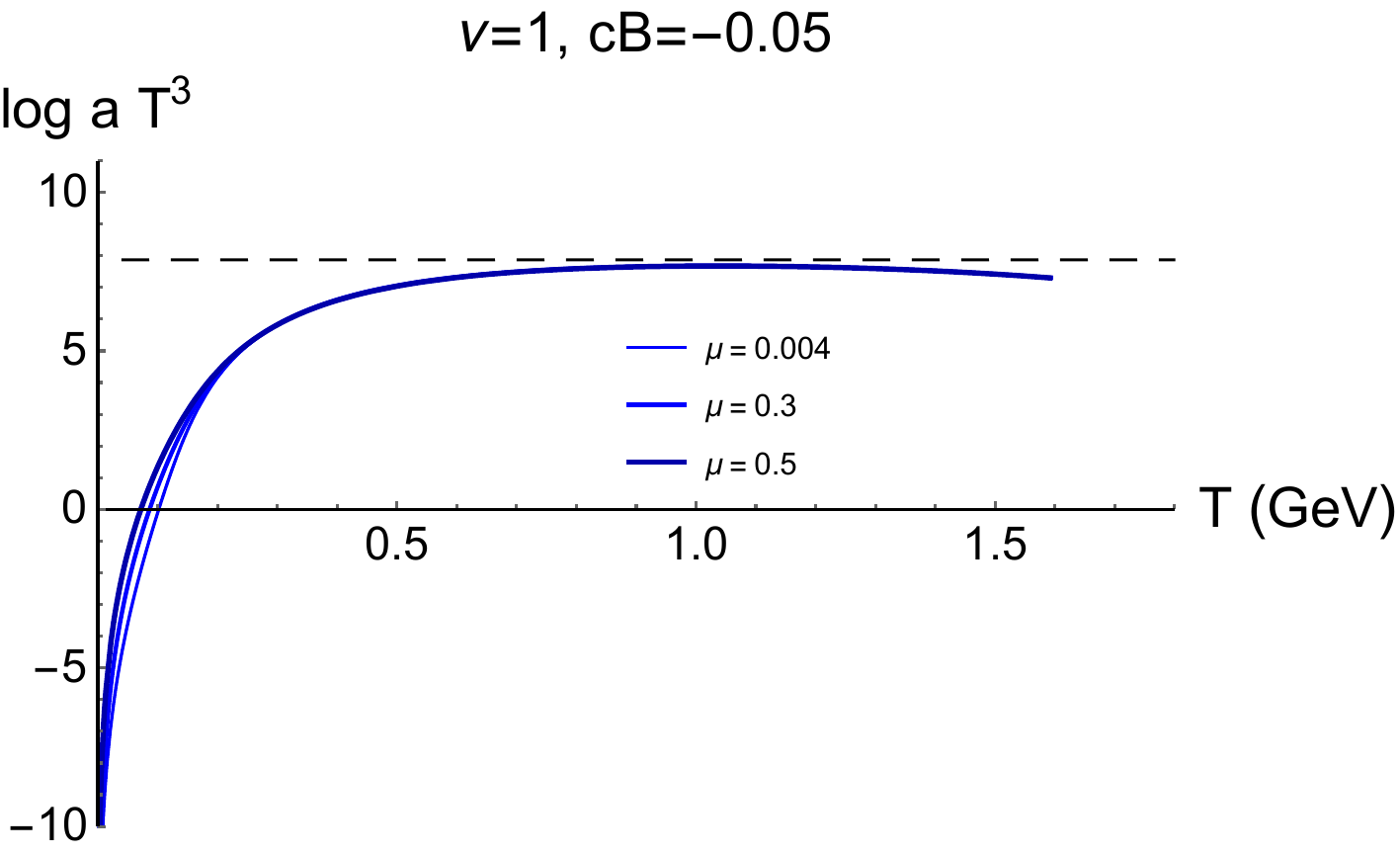}
   \caption{$\log a_2 \,T^{3}$  as a function of $T$ for the LQ model, with $\nu=1$ and $c_B=-0.05$ \GG, at fixed chemical potentials:   $\mu = 0.04$, $\mu=0.3$, and $\mu=0.5$ (GeV).
 }
  \label{Fig:LQnu1cB005m}
\end{figure}

Fig.\,\ref{Fig:LQnu1cB005m} shows that for the LQ model, with $\nu=1$ and  $c_B=-0.05$ \GG, the JQ parameter does not admit the conformal asymptotics for large $T$.

\newpage
$$\,$$

\section{Supplementary Figures for Heavy Quarks} \label{app3}

\begin{figure}[h!]
  \centering
   \includegraphics[scale=0.2]
  {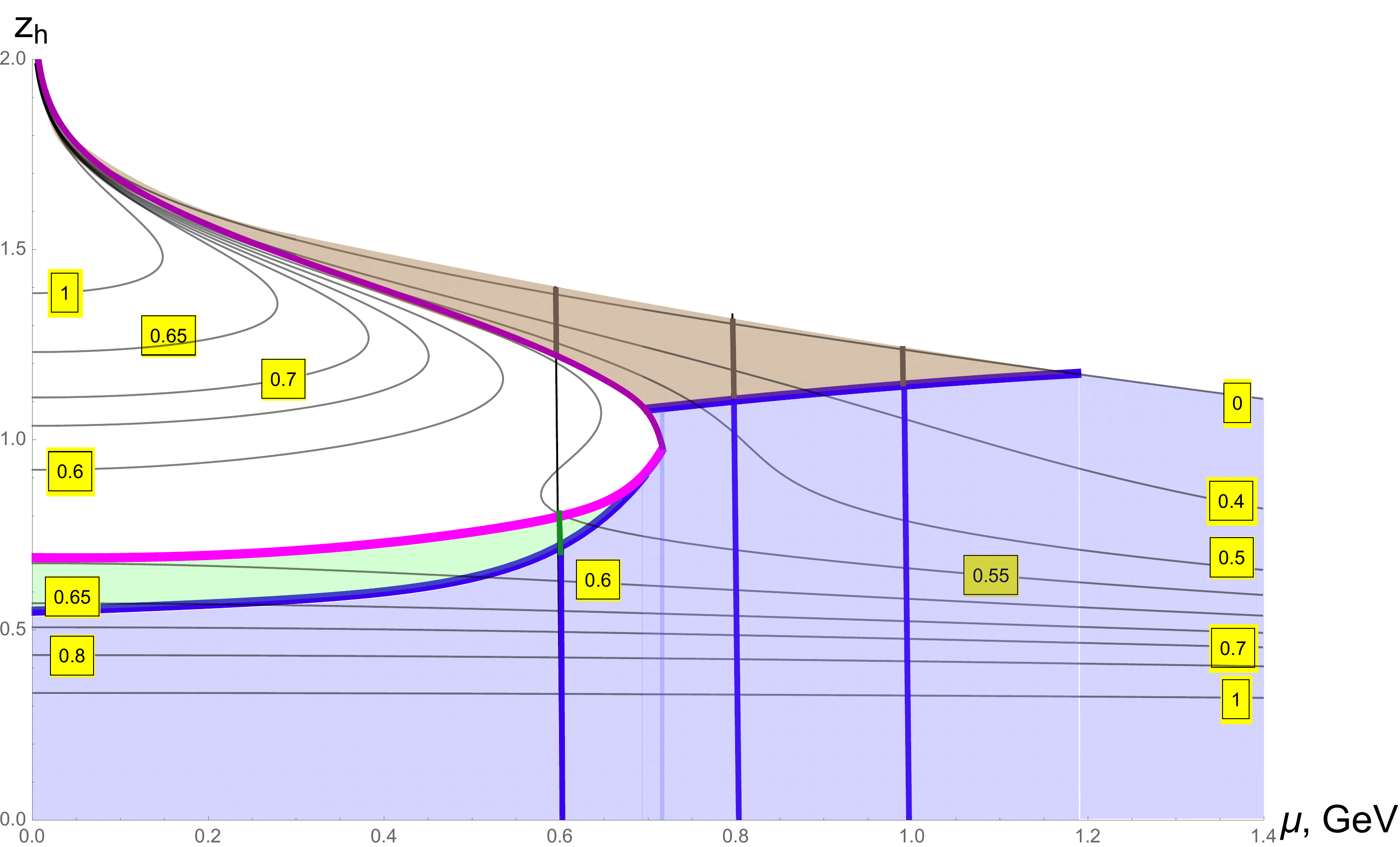}
  \caption{ The phase diagram in $(\mu,z_h)$-plane at $c_B=0$ and $\nu=1$ describing the HQ model including phase transition lines (blue for the confinement/deconfinement transition and magenta for the first-order transition line) and the thick vertical lines along which we calculate the JQ parameter; see  Fig.\,\ref{Fig:HQnu1mucB0}. The colors of regions show blue (QGP), green (quarkyonic), and brown (hadronic) phases.}
  \label{Fig:HQPT-mu-zh}
\end{figure}

The phase diagram of the HQ model in the $(\mu, z_h)$-plane describes three different states: the blue QGP region, the brown hadronic region, and quarkyonic green region; there is also  the forbidden white region 
(the region of instability), Fig.\,\ref{Fig:HQPT-mu-zh}.  Let us calculate the JQ parameter for $\nu=1$ and $c_B=0$ for different fixed $\mu$, namely for $\mu=0.6,0.8, 1$ (GeV) and for varying values of $z_h$.  We indicate these values of the chemical potential  parameter in Fig.\,\ref{Fig:HQPT-mu-zh} by thick  lines. These lines depend on the value of $\mu$ that crosses two, three, or four different colored regions.

\begin{figure}[h!]
  \centering
   \includegraphics[scale=0.22]
  {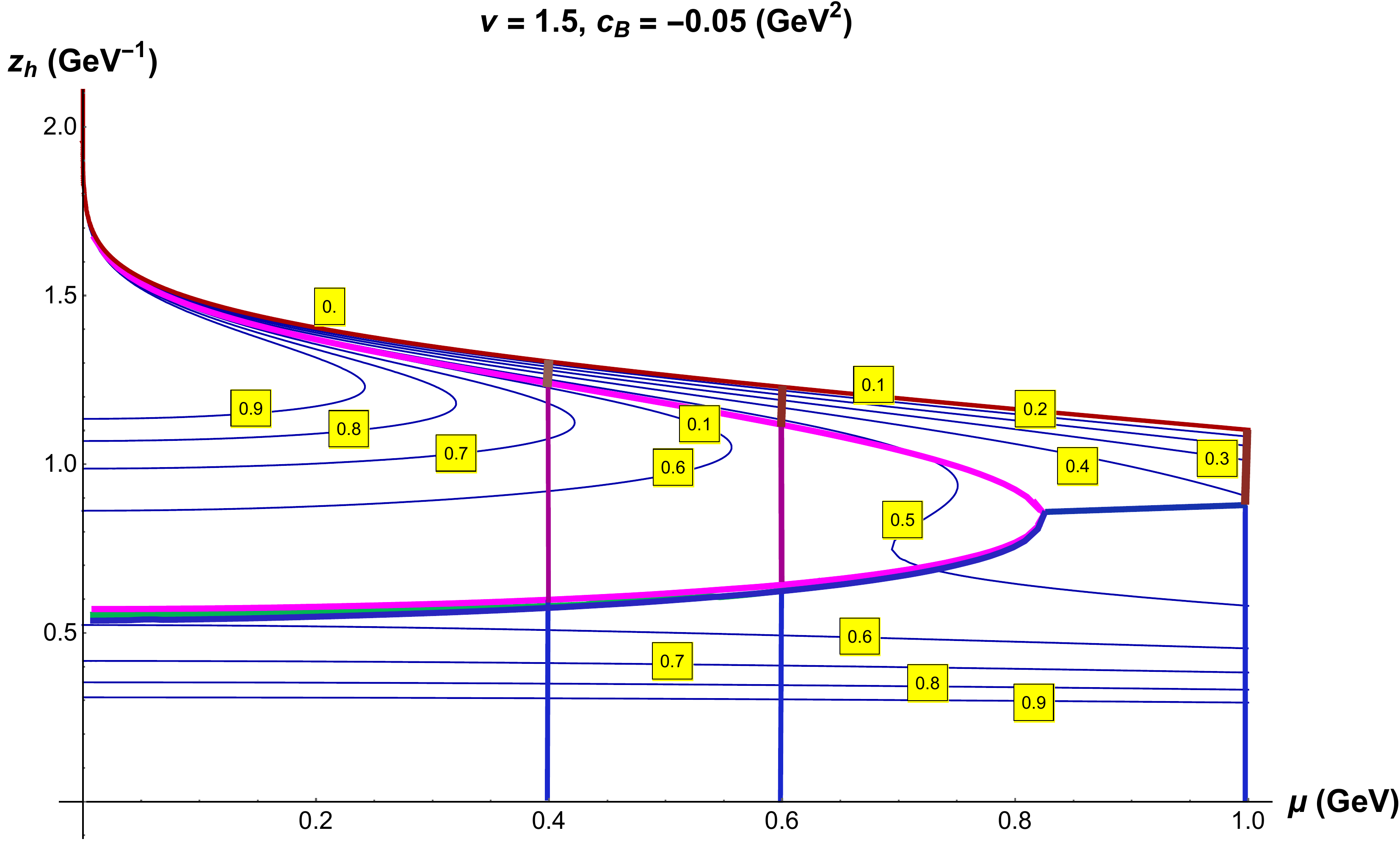} 
 \caption{
 The phase diagram for the HQ model in ($\mu, z_h$)-plane for $c_B = -0.05$ \GG \, and $\nu = 1.5$. The magenta and blue lines indicate first-order and second-order phase transitions, respectively. Additional curves, i.e. blue, purple, and brown mark paths for the JQ parameter calculations (see Fig.\,\ref{Fig:HQnu15cB005}). The green parts on lines with $\mu=0.4$ GeV and $\mu=0.6$ GeV are very small.
  }
  \label{Fig:HQnu15}
\end{figure}

\begin{figure}[h!]
  \centering
  \includegraphics[scale=0.15]
  {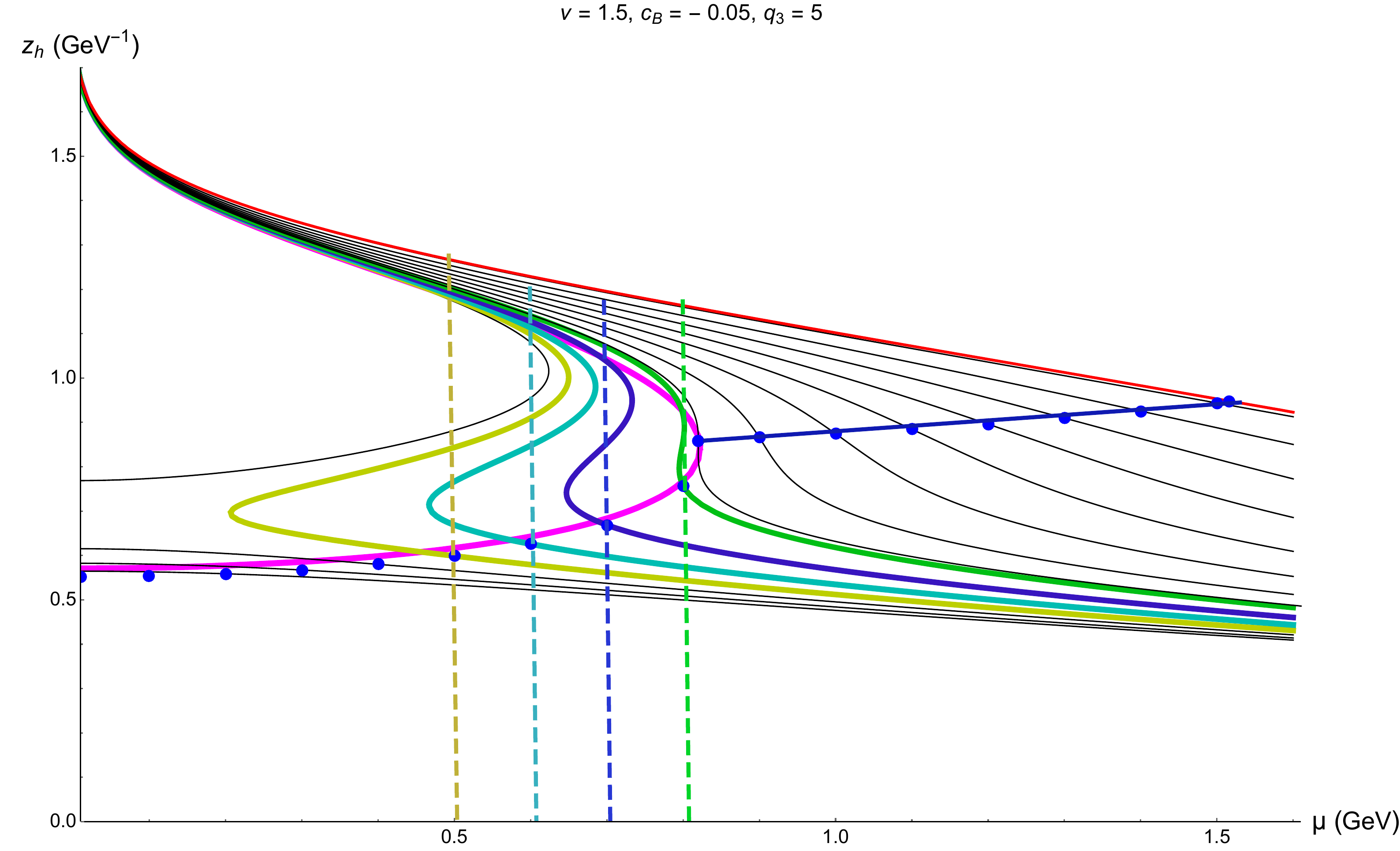}
  \includegraphics[scale=0.12]
  {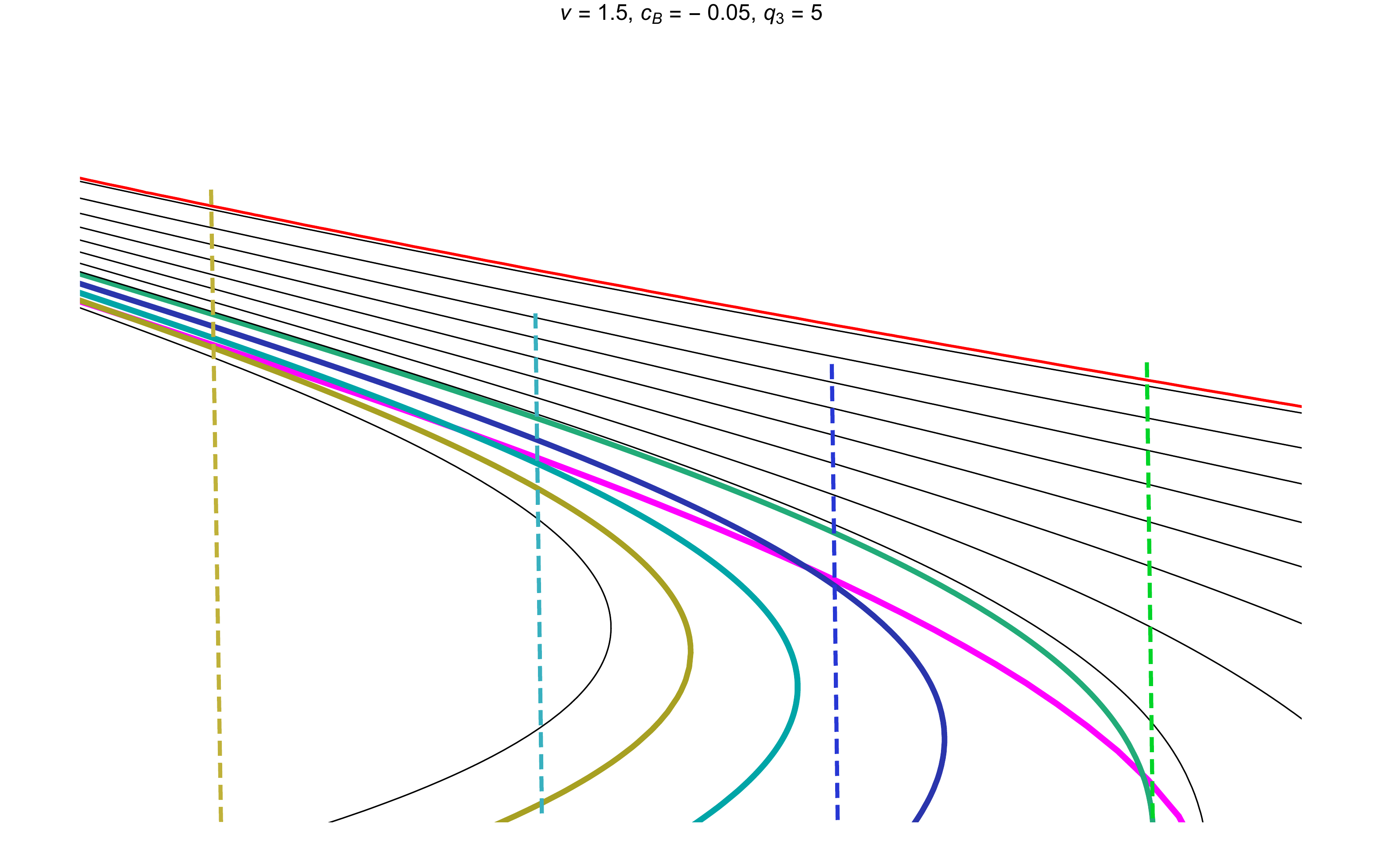}\\
\caption{The phase diagram for the HQ model in the ($\mu, z_h$)-plane for  $c_B = -0.05$ \GG \, and $\nu = 1.5$. Magenta and blue lines indicate the first-order and the second-order phase transitions, respectively. Additional curves (colored in blue, purple, brown) indicate the paths along which the JQ parameters are calculated (see Fig.\,\ref{Fig:HQnu15cB005}). 
  }
 \label{Fig:HQnu15-zoom}
\end{figure}

The phase diagram in the $(\mu,z_h)$-plane in $c_B = -0.05$ \GG \, and $\nu=1.5$ that describes the HQ model is shown in Fig.\,\ref{Fig:HQnu15}, and its zoom can be found in Fig.\,\ref{Fig:HQnu15-zoom}. Different phases of Fig.\,\ref{Fig:HQnu15} for three values of chemical potentials
$\mu = 0.4$ GeV, $\mu = 0.6$ GeV and $\mu = 1$ GeV are presented in Table\,\ref{tab:nu15cb005-table}.

\begin{table}[h]
\centering
\begin{tabular}{|l|lll|}
\hline
$\mu$ & \multicolumn{1}{c|}{QGP/QP}        & \multicolumn{1}{c|}{QP/FA}         & FA/HP         \\ \hline
0.4                & \multicolumn{1}{l|}{(0.589,0.551)} & \multicolumn{1}{l|}{(0.593,0.552)} & (1.245,0.561) \\ \hline
0.6                & \multicolumn{1}{l|}{(0.645,0.531)} & \multicolumn{1}{l|}{(0.646,0.532)} & (1.123,0.525) \\ \hline
\,1                  & \multicolumn{3}{c|}{(0.881,0.407)}                                                      \\ \hline
\end{tabular}
\caption{Points on $(\mu,z_h,T)$ corresponding to change of phases; QGP: quark gluon phase, QP: quarkyonic phase, FA: forbidden area, HP: hadron phase.}
\label{tab:nu15cb005-table}
\end{table}

\begin{figure}[h!]
  \centering
   \includegraphics[scale=0.28]
  {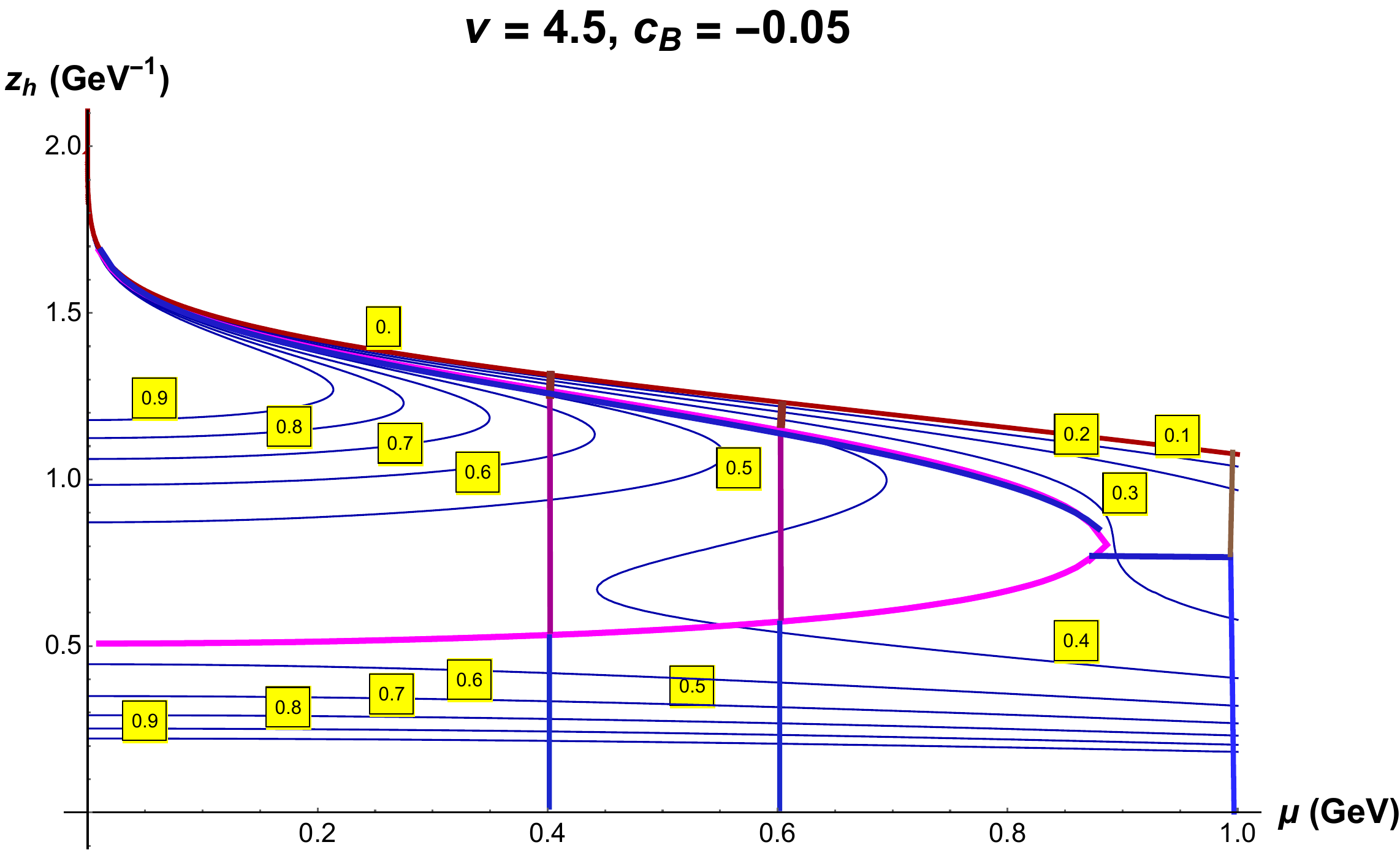} \\\,
  \caption{ The phase diagram in $(\mu,z_h)$-plane at $c_B = -0.05$ \GG \, and $\nu=4.5$ describing the HQ model. Magenta and blue lines indicate first-order and second-order phase transitions, respectively.  
  Vertical lines colored in blue, purple, and brown indicate the paths along which the JQ parameters are calculated (see Fig.\,\ref{Fig:HQPT-mu-zh}).
   } 
  \label{Fig:HQnu45}
\end{figure}

\begin{figure}[h!]
  \centering
   \includegraphics[scale=0.21]
  {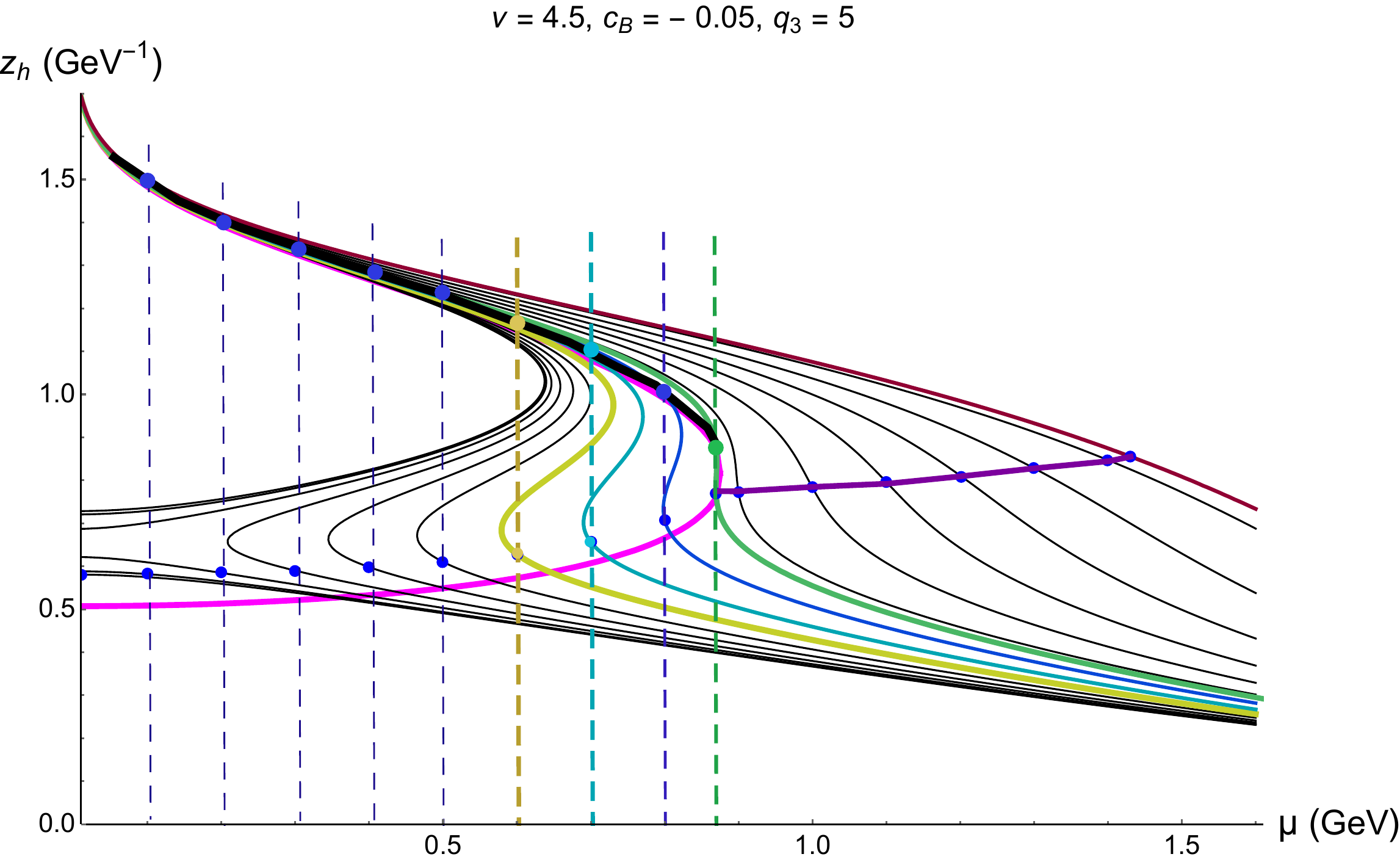} 
  \includegraphics[scale=0.15]
  {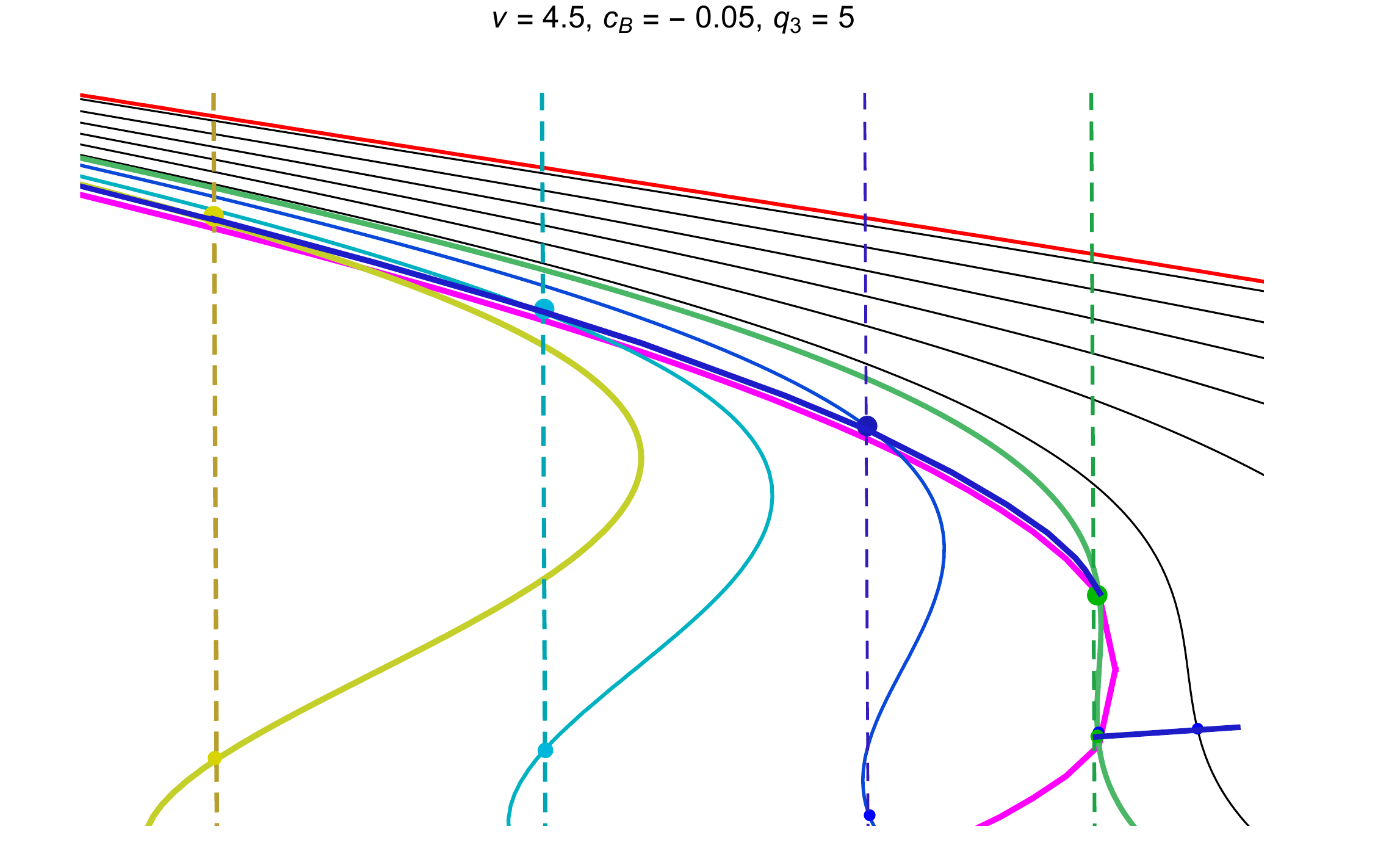}\,\\
   A\hspace{200pt}B
 \caption{ (A) The confinement/deconfinement phase transition for the HQ model in the ($\mu, z_h$)-plane, with  $c_B = -0.05$ \GG \,and $\nu = 4.5$ determined by stable points on the phase diagram. Unstable points (small dots) lie within the forbidden region. For each unstable point at given $(\mu, z_h)$-coordinates, a corresponding stable point exists at the same $(\mu$, $z_h)$ represented as a larger dots of the same color. (B) A zoomed view near the CEP.
 }
  \label{Fig:HQnu45-zoom}
\end{figure}

The phase diagram in the $(\mu,z_h)$-plane in $c_B = -0.05$ \GG \, and $\nu=4.5$ that describes the HQ model is shown in Fig.\,\ref{Fig:HQnu45}, and its zoom can be found in Fig.\,\ref{Fig:HQnu45-zoom}.

\begin{figure}[h!]
  \centering
  \includegraphics[scale=0.19]
  {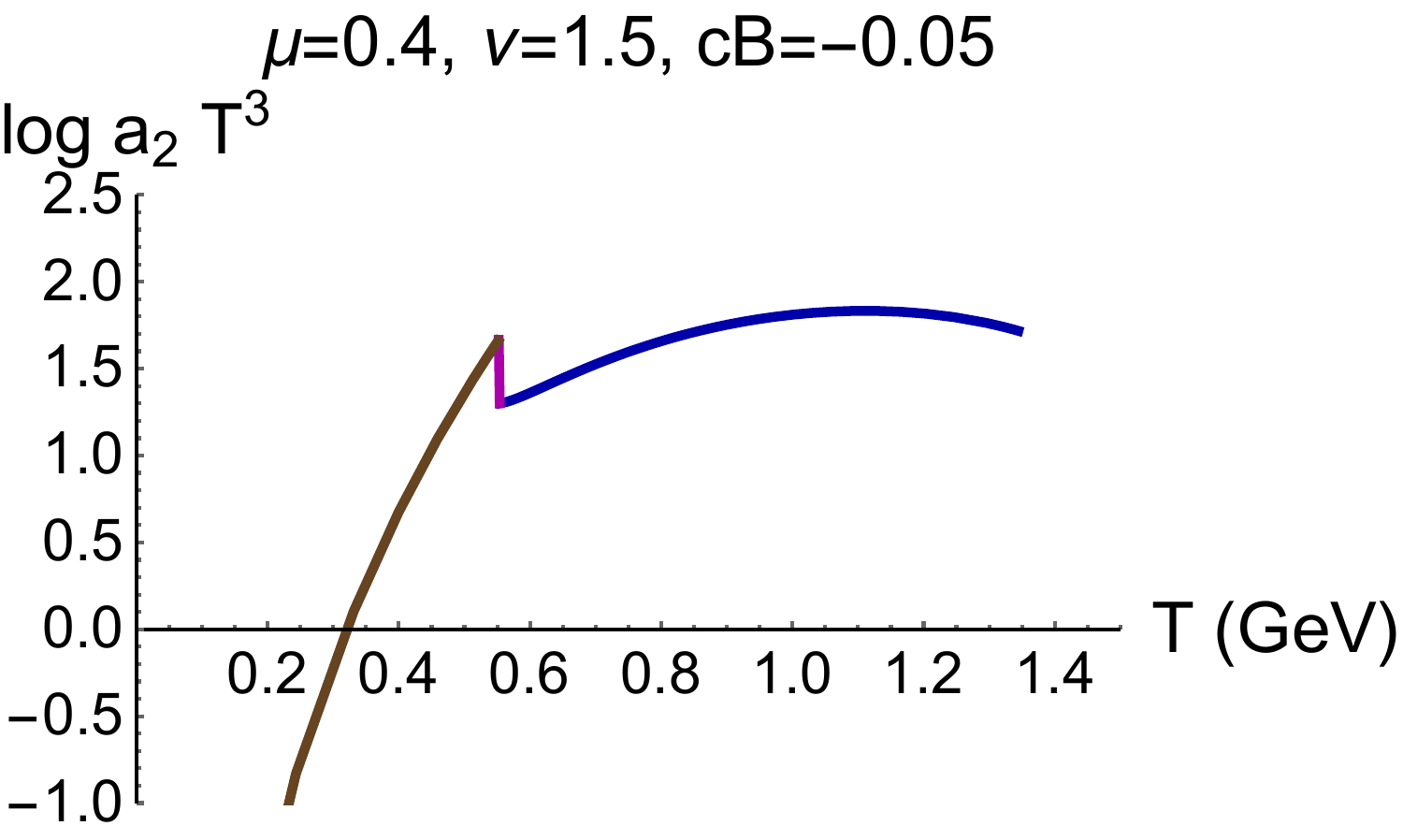}
\includegraphics[scale=0.20]
{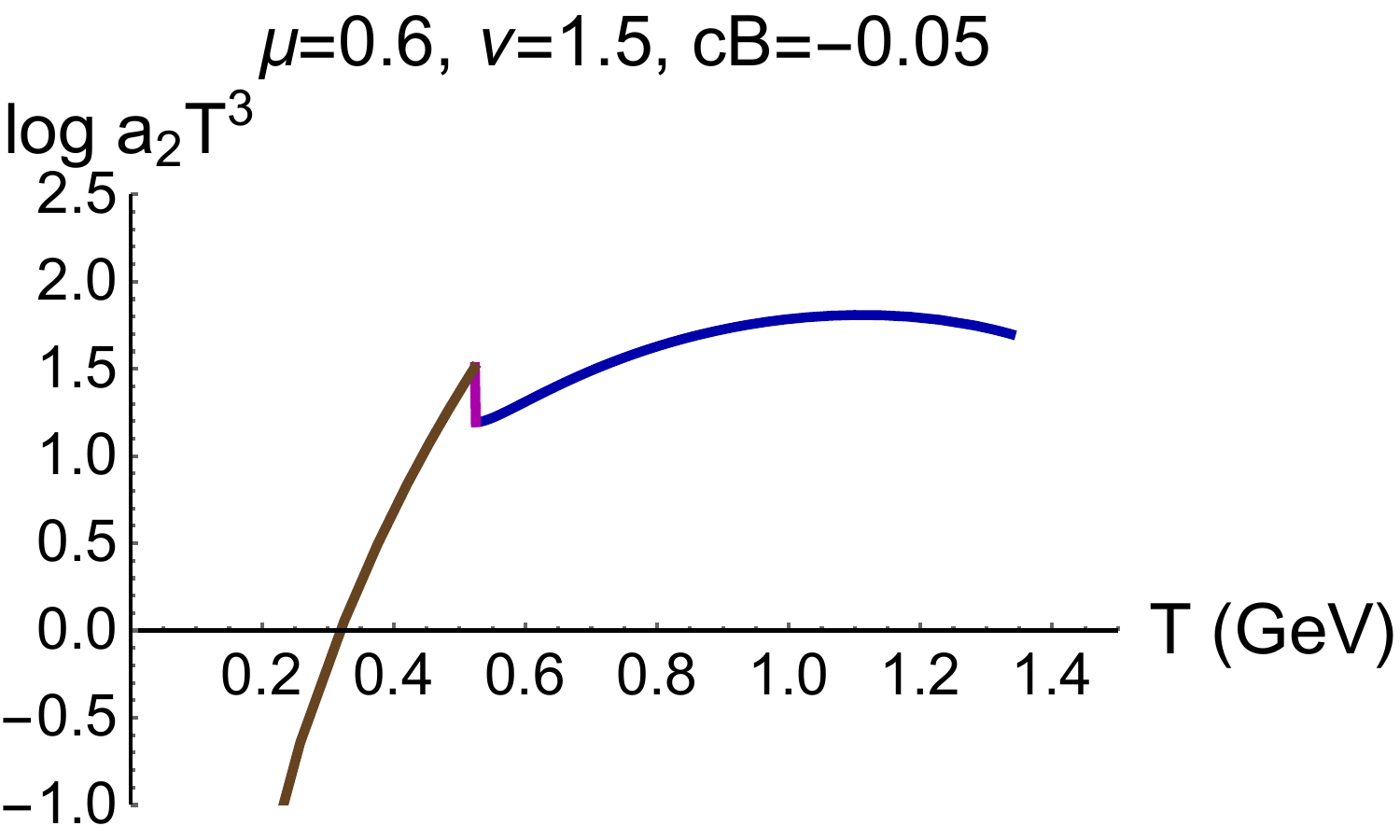} 
\includegraphics[scale=0.19]
  {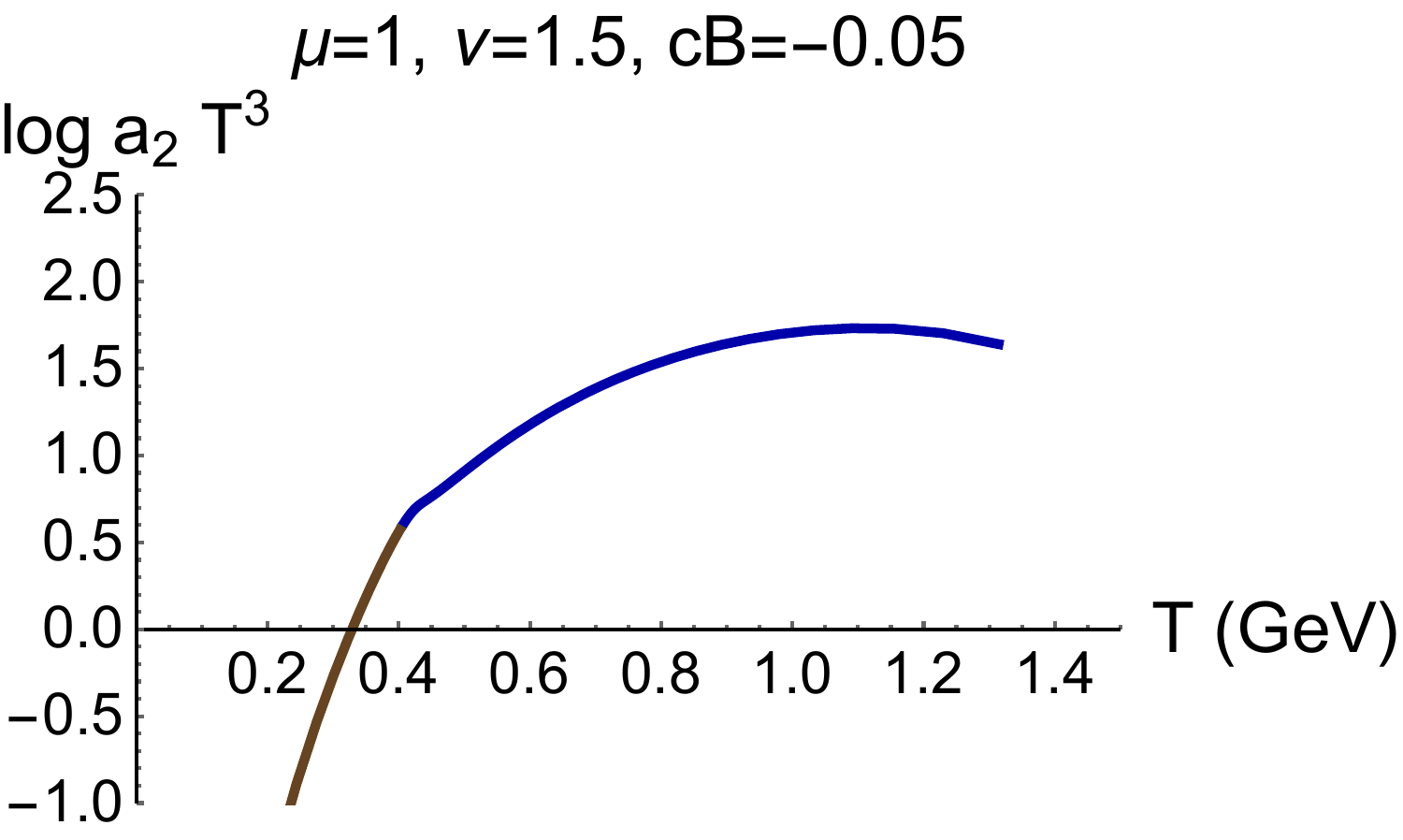}\\
  A\hspace{150pt}B\hspace{150pt}C
\caption{$\log (a_2 T^3)$ versus temperature for the HQ model, with $c_B = -0.05$ \GG \, and $\nu=1.5$, at fixed chemical potentials:  
(A) $\mu = 0.4$ GeV,  
(B) $\mu = 0.6$ GeV,  
(C) $\mu = 1$ GeV. The blue and  brown lines correspond to the QGP and  hadronic  phases, respectively. The magenta lines  correspond to unstable regions. We see  jumps in panels (A) and (B), and a smooth  change of the slopes  near the second-order phase transition at panel (C). 
  }
  \label{Fig:HQnu15cB005T3}
\end{figure}

For our model, Fig.\,\ref{Fig:HQnu15cB005T3} shows $\log a T^3$ varying slowly but non-constantly in the QGP phase contrasting with the $N=4$ SYM behavior reported in \cite{Liu:2006ug}, where it remains constant. In the hadronic phase, $\log a T^3$ increases universally with chemical potential. It is obvious that the jumps in Fig.\,\ref{Fig:HQnu15cB005T3} remain at the same temperatures as in Fig.\,\ref{Fig:HQnu15cB005}.

\begin{figure}[t!]
\includegraphics[scale=1]
 {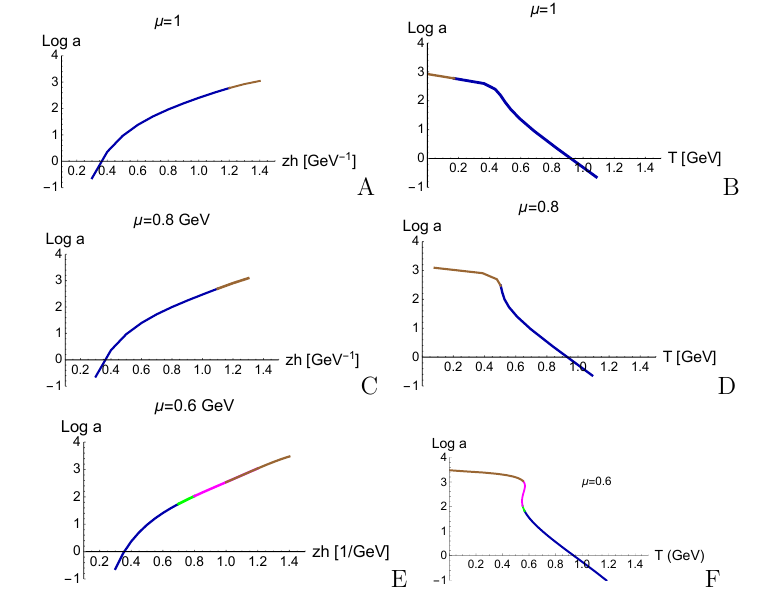}
 \caption{$\log a$ as a function of horizon size $z_h$ (panels A, C, E) and temperature $T$ (panels B, D, F) for the HQ model, with $\nu=1$ and $c_B=0$, at chemical potentials $\mu = 1, 0.8, 0.6$ GeV. The blue, brown and green lines correspond to the QGP, hadronic and quarkyonic phases, respectively. The magenta lines  correspond to unstable regions. panel (F) shows that the JQ parameter exhibits a jump at first-order phase transition. 
}
  \label{Fig:HQnu1mucB0}
\end{figure}

We compute the logarithm of the integral \eqref{a-mu} for fixed values of $\mu$ and $z_h$, and plot the results in Fig.\,\ref{Fig:HQnu1mucB0} (left panels). The curves are color-coded according to the phase regions in Fig.\,\ref{Fig:HQPT-mu-zh}:
i) brown segments  correspond to $z_h$ trajectories through brown phase regions;
ii) blue segments represent trajectories through blue regions;
iii) Magenta segments indicate forbidden white areas;
iv) green segments correspond to green phase regions.

The right panels of Fig.\,\ref{Fig:HQnu1mucB0} show the temperature dependence of $\log a$. As evident in the left panels, only one curve crosses the forbidden region specifically in Fig.\,\ref{Fig:HQnu1mucB0}E. The corresponding right panel (Fig.\,\ref{Fig:HQnu1mucB0}F) exhibits a jump at $T = T_c(\mu)$, reflecting the removal of the magenta curve segment in this graph.

\begin{figure}[h!]
  \centering
\includegraphics[scale=0.2]
{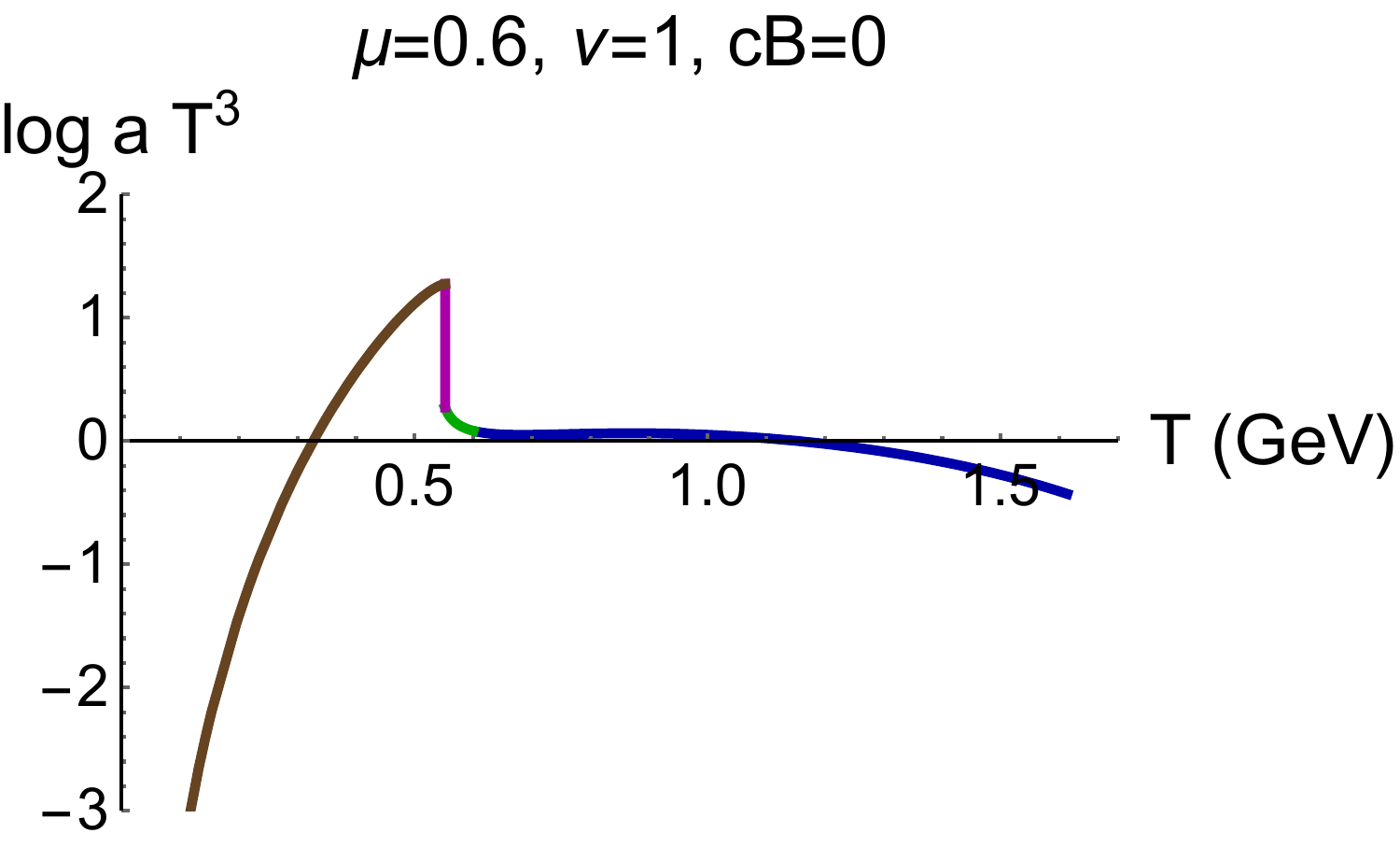}
  \includegraphics[scale=0.2]
   {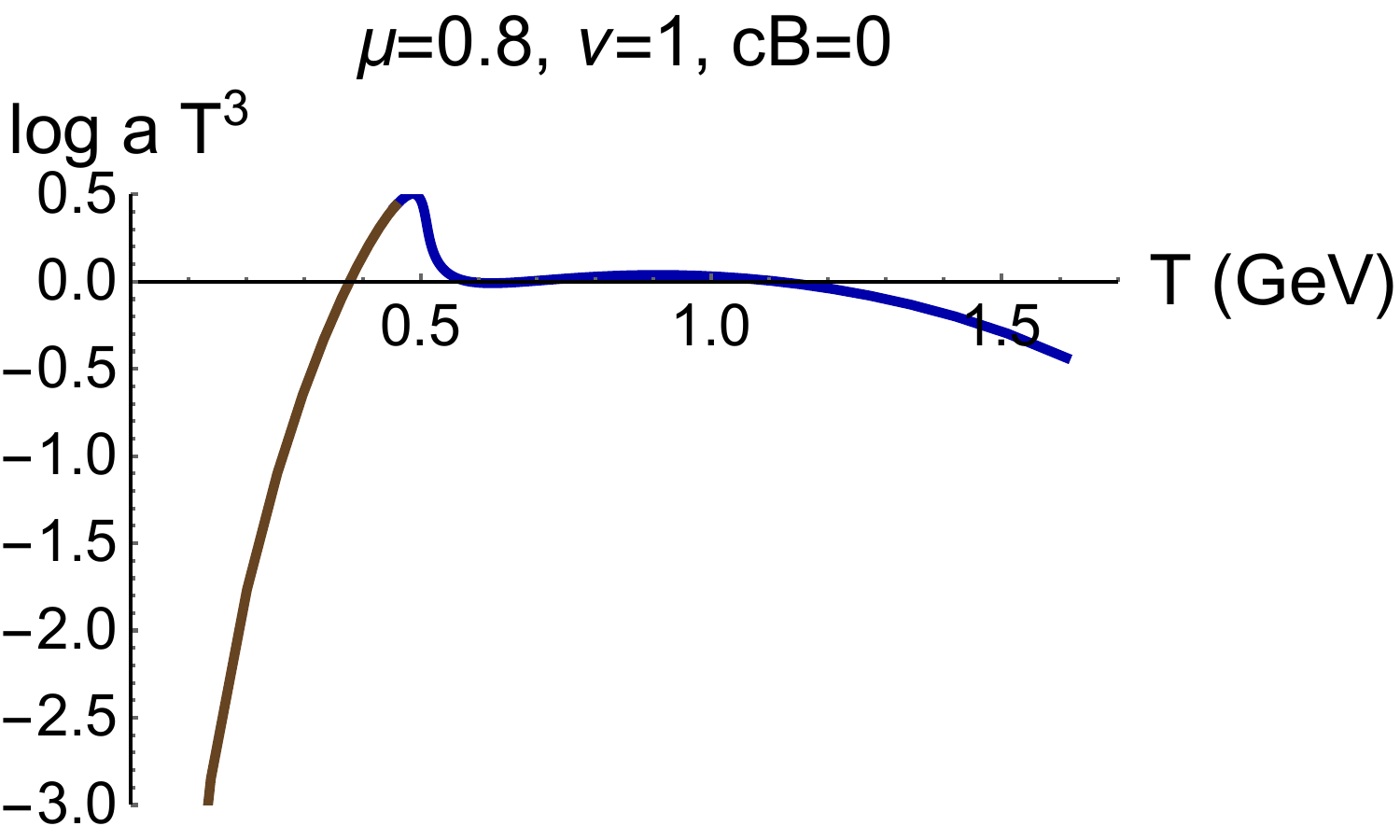}
\includegraphics[scale=0.2]
   {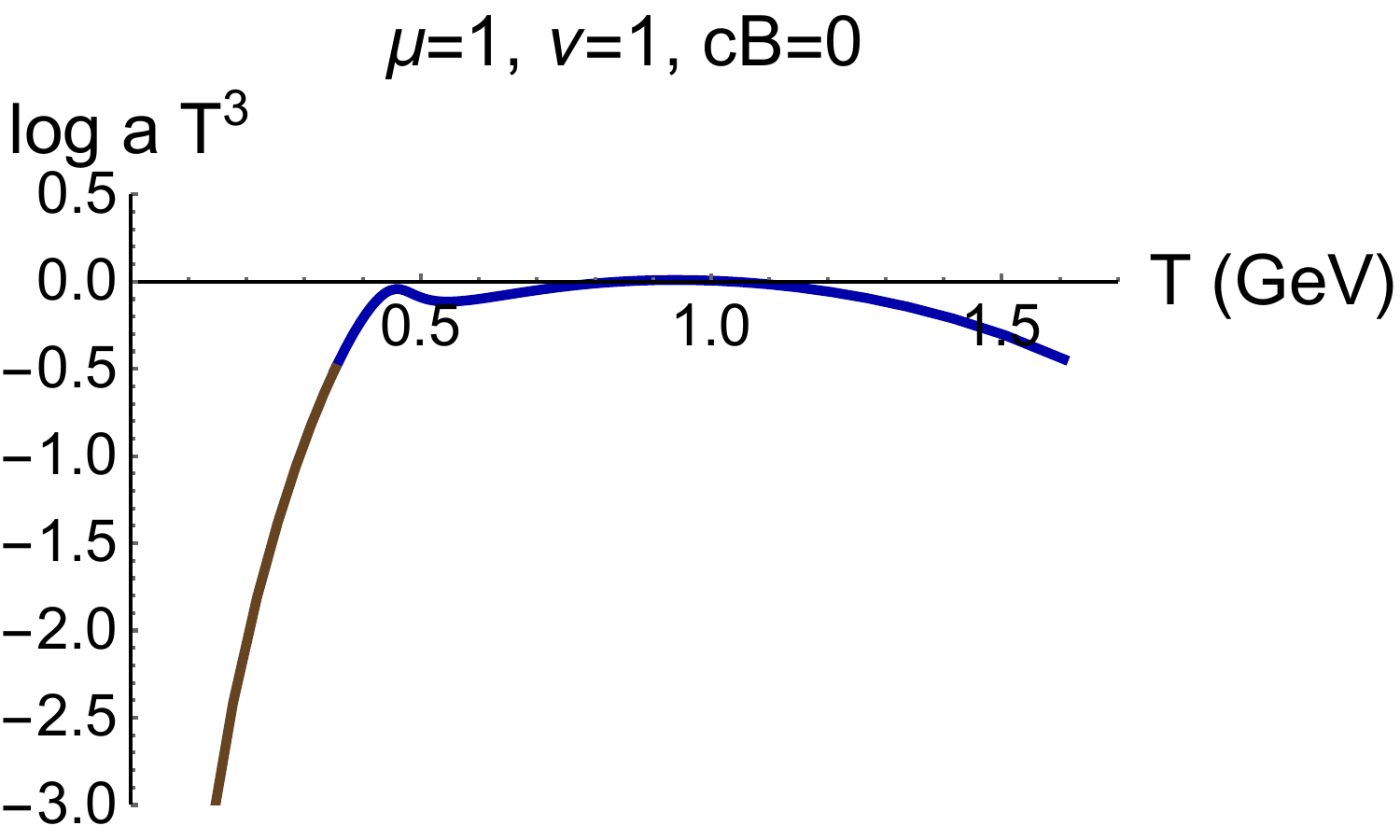}
    \\A\hspace{140pt}B\hspace{140pt}C\\$\,$\\
\includegraphics[scale=0.25]
   {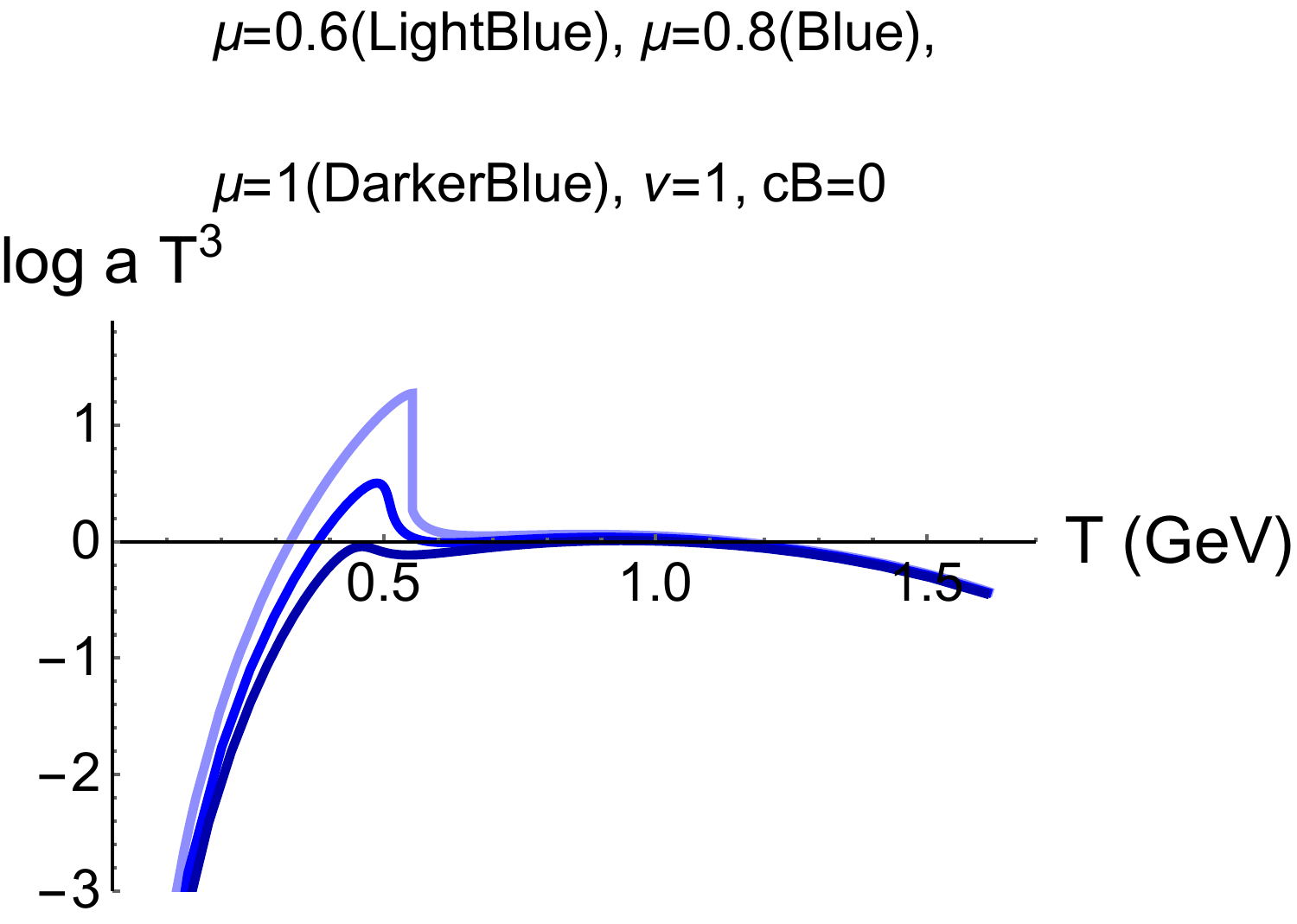}
   \\D
  \caption{The dependence of $\log a T^3$ on $T$ for the HQ model, with $\nu=1$ and $c_B=0$, at fixed chemical potentials: (A) $\mu=0.6$ GeV, (B)
  $\mu=0.8$ GeV,  (C) $\mu=1$ GeV. The segments of these lines are colored blue (QGP), brown (hadronic), and green (quarkyonic) according to the phase traversed. (D) Comparison of $\log a \,T^3$ for the same $\nu=1$ and $c_B=0$ at different $\mu$.
  }
\label{Fig:HQnu1cB0mu06081}
\end{figure}

Fig.\,\ref{Fig:HQnu1cB0mu06081} shows the $T$-dependence of $\log a T^3$ in the HQ model for fixed parameters $\nu=1$ and $c_B=0$, at chemical potentials $\mu=0.6$ GeV (panel (A)), $\mu=0.8$ GeV (panel (B)), and $\mu=1$ GeV (panel (C)). Fig.\,\ref{Fig:HQnu1cB0mu06081}D compares $\log a T^3$ for fixed parameters $\nu=1$ and $c_B=0$ at different values of $\mu$. The behavior of $\log a T^3$ in the high-temperature regime is independent of $\mu$, is not constant, and decreases as $T$ increases.

$$$$
\newpage
$$$$

\end{document}